\documentclass[fleqn,usenatbib]{mnras}
\usepackage{amsmath,amssymb,epsfig,newtxtext,newtxmath,verbatim}
\usepackage[T1]{fontenc}

\title[BEARS II: Millimetre photometry]{The Bright Extragalactic ALMA Redshift Survey (BEARS) II: Millimetre photometry of gravitational lens candidates}

\author[G. J. Bendo et al.]{
G. J. Bendo$^{1}$\thanks{E-mail: george.bendo@manchester.ac.uk}, 
S. A. Urquhart$^{2}$, 
S. Serjeant$^{2}$, 
T. Bakx$^{3,4}$, 
M. Hagimoto$^{3}$, 
P. Cox$^{5}$, 
R. Neri$^{6}$, 
\newauthor
M. D. Lehnert$^{7}$, 
H. Dannerbauer$^{8, 9}$, 
A. Amvrosiadis$^{10}$, 
P. Andreani$^{11}$, 
A. J. Baker$^{12, 13}$,
A. Beelen$^{14}$,
\newauthor
S. Berta$^{6}$,
E. Borsato$^{15}$,
V. Buat$^{14}$, 
K. M. Butler $^{16}$, 
A. Cooray$^{17}$, 
G. De Zotti$^{18}$, 
L. Dunne$^{19}$, 
S. Dye$^{20}$, 
\newauthor
S. Eales$^{19}$, 
A. Enia$^{21, 22}$, 
L. Fan$^{23, 24}$, 
R. Gavazzi$^{5, 14}$,
J. Gonz\'alez-Nuevo$^{25, 26}$,
A. I. Harris$^{27}$, 
\newauthor
C. N. Herrera$^{6}$, 
D. H. Hughes$^{28}$, 
D. Ismail$^{14}$, 
B. M. Jones$^{29}$, 
K. Kohno$^{30, 31}$, 
M. Krips$^{6}$, 
G. Lagache$^{14}$, 
\newauthor
L. Marchetti$^{32, 33}$, 
M. Massardi$^{33, 34}$, 
H. Messias$^{35, 36}$, 
M. Negrello$^{19}$,
A. Omont$^{5}$,
I. P\'erez-Fournon$^{8, 9}$, 
\newauthor
D. A. Riechers$^{29}$, 
D. Scott$^{37}$, 
M. W. L. Smith$^{19}$, 
F. Stanley$^{6}$, 
Y. Tamura$^{3}$, 
P. Temi$^{38}$,
P. van der Werf$^{16}$, 
\newauthor
A. Verma$^{39}$,
C. Vlahakis$^{40}$,
A. Wei{\ss}$^{41}$,
C. Yang$^{42}$,
and A. J. Young$^{12}$ \\
\\
$^1$ UK ALMA Regional Centre Node, Jodrell Bank Centre for Astrophysics, Department of Physics and Astronomy, The University of Manchester,\\ Oxford Road, Manchester M13 9PL, United Kingdom\\
$^{2}$ School of Physical Sciences, The Open University, Milton Keynes, MK7 6AA, United Kingdom\\
$^{3}$ Division of Particle and Astrophysical Science, Graduate School of Science, Nagoya University, Aichi 464-8602, Japan\\
$^{4}$ National Astronomical Observatory of Japan, 2-21-1, Osawa, Mitaka, Tokyo 181-8588, Japan\\
$^{5}$ Sorbonne Universit\'e, UPMC Universit\'e Paris 6 and CNRS, UMR 7095, Institut d'Astrophysique de Paris, 98bis boulevard Arago, F-75014 Paris, France \\
$^{6}$ Institut de Radioastronomie Millim\'etrique (IRAM), 300 rue de la Piscine, 38400 Saint-Martin-d'H\`eres, France\\
$^{7}$ Universit\'{e} Lyon 1, ENS de Lyon, CNRS UMR5574, Centre de Recherche Astrophysique de Lyon, F-69230 Saint-Genis-Laval, France\\
$^{8}$ Instituto de Astrof\'isica de Canarias (IAC), E-38205 La Laguna, Tenerife, Spain \\
$^{9}$ Universidad de la Laguna, Dpto. Astrof\'isica, E-38206 La Laguna, Tenerife, Spain\\
$^{10}$ Institute for Computational Cosmology, Durham University, South Road, Durham DH1 3LE, UK\\
$^{11}$ European Southern Observatory, Karl-Schwarzschild-Strasse 2, 85748 Garching, Germany \\
$^{12}$ Department of Physics and Astronomy, Rutgers, the State University of New Jersey, 136 Frelinghuysen Road, Piscataway, NJ, 08854-8019, USA \\
$^{13}$ Department of Physics and Astronomy, University of the Western Cape, Robert Sobukwe Road, Bellville 7535, South Africa\\
$^{14}$ Aix-Marseille Universit\'e, CNRS and CNES, Laboratoire d'Astrophysique de Marseille, 38, Rue Fr\'ed\'eric Joliot-Curie, 13388 Marseille, France\\
$^{15}$ Dipartimento di Fisica e Astronomia "G. Galilei", Universit\`a di Padova, vicolo dell'Osservatorio 3, Padova I-35122, Italy\\
$^{16}$ Leiden Observatory, Leiden University, PO Box 9513, 2300 RA Leiden, The Netherlands\\
$^{17}$ Department of Physics and Astronomy, University of California, Irvine CA 92697, USA\\
$^{18}$ INAF-Osservatorio Astronomico di Padova, Vicolo dell'Osservatorio 5, I-35122 Padova, Italy \\
$^{19}$ School of Physics and Astronomy, Cardiff University, Queens Building, The Parade, Cardiff, CF24 3AA, UK\\
$^{20}$ School of Physics and Astronomy, University of Nottingham, University Park, Nottingham, NG7 2RD, UK \\
$^{21}$ University of Bologna — Department of Physics and Astronomy “Augusto Righi” (DIFA), Via Gobetti 93/2, I-40129, Bologna, Italy\\
$^{22}$ INAF — Osservatorio di Astrofisica e Scienza dello Spazio, Via Gobetti 93/3, I-40129, Bologna, Italy \\
$^{23}$ CAS Key Laboratory for Research in Galaxies and Cosmology, Department of Astronomy,
University of Science and Technology of China, \\ Hefei 230026, China \\
$^{24}$ School of Astronomy and Space Science, University of Science and Technology of China, Hefei
230026, China \\
$^{25}$ Departamento de F\'isica, Universidad de Oviedo, C. Federico Garc\'ia Lorca 18, 33007, Oviedo, Spain\\
$^{26}$ Instituto Universitario de Ciencias y Tecnologias Espaciales de Asturias (ICTEA), C. Independencia 13, 33004, Oviedo, Spain \\
$^{27}$ Department of Astronomy, University of Maryland, College Park, MD 20742, USA\\
$^{28}$ Instituto Nacional de Astrof\'isica, \'Optica y Electr\'onica, Luis Enrique Erro 1, CP 72840, Tonantzintla, Puebla, M\'exico\\
$^{29}$ I. Physikalisches Institut, Universit\"at zu K\"oln, Z\"ulpicher
Strasse 77, D-50937 K\"oln, Germany\\
$^{30}$ Institute of Astronomy, Graduate School of Science, The University of Tokyo, 2-21-1 Osawa, Mitaka, Tokyo 181-0015, Japan  \\
$^{31}$ Research Center for the Early Universe, Graduate School of Science, The University of Tokyo, 7-3-1 Hongo, Bunkyo-ku, Tokyo 113-0033, Japan\\
$^{32}$ Department of Astronomy, University of Cape Town, 7701 Rondebosch, Cape Town, South Africa\\
$^{33}$ INAF - Istituto di Radioastronomia, via Gobetti 101, 40129 Bologna, Italy \\
$^{34}$ SISSA, Via Bonomea 265, 34136 Trieste, Italy\\
$^{35}$ Joint ALMA Observatory, Alonso de C\'ordova 3107, Vitacura 763-0355, Santiago de Chile, Chile\\
$^{36}$ European Southern Observatory, Alonso de C\'ordova 3107,  Vitacura, Casilla 19001, Santiago de Chile, Chile \\
$^{37}$ Department of Physics and Astronomy, University of British Columbia, 6224 Agricultural Road, Vancouver, BC V6T 1Z1, Canada \\
$^{38}$ Astrophysics Branch, NASA - Ames Research Center, MS 245-6, Moffett Field, CA 94035, USA \\
$^{39}$ Sub-department of Astrophysics, Denys Wilkinson Building, University of Oxford, Keble Road, Oxford, OX1 3RH, UK \\
$^{40}$ National Radio Astronomy Observatory, 520 Edgemont Road, Charlottesville VA 22903-2475, USA \\
$^{41}$ Max-Planck-Institut f\"ur Radioastronomie, Auf dem H\"ugel 69, D-53121 Bonn, Germany\\
$^{42}$ Department of Space, Earth and Environment, Chalmers University of Technology, Onsala Space Observatory, 439 92 Onsala, Sweden\\
}

\date{}
\pubyear{}

\begin{document}
\label{firstpage}
\pagerange{\pageref{firstpage}--\pageref{lastpage}}
\maketitle

\begin{abstract}
We present 101 and 151~GHz ALMA continuum images for 85 fields selected from {\it Herschel} observations that have 500~$\mu$m flux densities $>$80~mJy and 250-500~$\mu$m colours consistent with $z > 2$, most of which are expected to be gravitationally lensed or hyperluminous infrared galaxies.  Approximately half of the {\it Herschel} 500~$\mu$m sources were resolved into multiple ALMA sources, but 11 of the 15 brightest 500~$\mu$m {\it Herschel} sources correspond to individual ALMA sources.  For the 37 fields containing either a single source with a spectroscopic redshift or two sources with the same spectroscopic redshift, we examined the colour temperatures and dust emissivity indices.  The colour temperatures only vary weakly with redshift and are statistically consistent with no redshift-dependent temperature variations, which generally corresponds to results from other samples selected in far-infrared, submillimetre, or millimetre bands but not to results from samples selected in optical or near-infrared bands.  The dust emissivity indices, with very few exceptions, are largely consistent with a value of 2.  We also compared spectroscopic redshifts to photometric redshifts based on spectral energy distribution templates designed for infrared-bright high-redshift galaxies.  While the templates systematically underestimate the redshifts by $\sim$15\%, the inclusion of ALMA data decreases the scatter in the predicted redshifts by a factor of $\sim$2, illustrating the potential usefulness of these millimetre data for estimating photometric redshifts. 
\end{abstract}


\section{Introduction}


Gravitational lenses have several key uses in extragalactic astronomy.  The lenses magnify the light from higher redshift sources, thus allowing for the examination of the properties of galaxies at these redshifts that would otherwise be much more difficult to detect or resolve \citep[e.g.,][]{swinbank2010,dye2015,dye2022}, and the images of the lensed light can also be used to probe the dark matter content of the lensing galaxies (see \citealt{treu2010} for a review).  Additionally, statistical information about the lenses can be used to place constraints on cosmological parameters \citep{grillo2008, eales2015}.

Extragalactic surveys with the Spectral and Photometric Imaging REceiver \citep[SPIRE;][]{griffin2010} on the {\it Herschel} Space Observatory \citep{pilbratt2010}, including the {\it Herschel} Astrophysical Terahertz Large Area Survey \citep[H-ATLAS;][]{eales2010}, the {\it Herschel} Multi-tiered Extragalactic Survey \citep[HerMES;][]{oliver2012}, and the {\it Herschel} Stripe 82 Survey \citep{viero2014}, were particularly effective at finding gravitationally lensed systems and other infrared-bright high redshift galaxies.  This was not only because the dust emission was magnified but also because the dust emission from the redshifted lensed sources peaks in the {\it Herschel} 250-500~$\mu$m bands and because the negative k-correction in these bands makes it easier to detect high redshift sources.  Additionally, at 500~$\mu$m flux densities $>$100~mJy, the surface density of strongly lensed sources is expected to be higher than unlensed sources \citep[e.g.,][]{negrello2007}.

Multiple catalogues of gravitational lens candidates and other infrared-bright galaxies potentially at high redshift have been created using the data from these {\it Herschel} surveys \citep[e.g.,][]{gonzaleznuevo2012, wardlow2013, nayyeri2016, negrello2017, gonzaleznuevo2019, bakx2020viking}.  However, the angular resolution of the {\it Herschel} 250-500~$\mu$m data is 18-35~arcsec, so the sources identified in these surveys are unresolved, their exact spatial locations are poorly constrained, and it is likely that many sources are confused within the {\it Herschel} beams. 

\citet{urquhart2022}, in the Bright Extragalactic ALMA Redshift Survey (BEARS), used ALMA to observe a set of 85 gravitational lens candidates from the {\it Herschel} Bright Sources \citep[HerBS; ][]{bakx2018, bakx2020erratum} sample that lie within the South Galactic Pole field observed by H-ATLAS.  The primary goal of the observations, which were spectral scans covering most of ALMA Bands 3 and 4, was to determine the spectroscopic redshifts of the sources in these fields.  However, since the spectral line emission, when detected, is typically found within a very small fraction of the observed spectra, the rest of the data can be used for serendipitous measurements of the continuum at the observed frequencies.  We used these new ALMA continuum data to address two specific science questions in this paper.  

First, since the ALMA Band 4 data have angular resolutions of $\sim$2~arcsec, we used the images to study the multiplicities and morphologies of the sources so as to further understanding the nature of the sources within these fields.  Resolved or multiple sources could be lenses, protocluster cores or simply sources that are confused along the line of sight.  Unresolved sources could be gravitational lenses with small Einstein radii or even individual hyperluminous infrared galaxies (HLIRGs) with intrinsic luminosities of $>$$10^{13}$~L$_\odot$.  

Second, the ALMA Band 3 and 4 data are particularly useful for constraining the Rayleigh-Jeans side of the dust spectral energy distribution (SED) from these galaxies, even for galaxies at redshifts up to 5, and hence for characterizing the colour temperatures and emissivities of the dust.  This in turn can be used for comparing the properties of our galaxies to other samples and in particular to examine the relation between colour temperature and redshift reported by others \citep[e.g.,][]{magdis2012,magnelli2014, bethermin2015, bethermin2017, schreiber2018, liang2019, bouwens2020, riechers2020, chen2021, dudzeviciute2021}.  Additionally, these data can be compared to the SED templates used for calculating photometric redshifts so as to understand how well the data and templates match each other.

\section{ALMA observations and data processing}
\label{s_data}

\begin{table*}
\centering
\begin{minipage}{174mm}
\caption{ALMA image characteristics.}
\label{t_imageparam}
\begin{tabular}{@{}lccccccccccc@{}}
\hline
Array &
  Band &
  Central &
  Number of &
  Typical &
  Pixel &
  \multicolumn{2}{c}{Image size} &
  Primary &
  Typical &
  Maximum &
  Typical \\
&
  &
  frequency &
  observered &
  uv &
  scale &
  (pixels) &
  (arcsec) &
  beam &
  beam &
  recoverable &
  rms \\
&
  &
  (GHz) &
  fields &
  coverage &
  (arcsec) &
  &
  &
  diameter$^a$ &
  FWHM &
  scale &
  noise \\
&
  &
  &
  &
  (m) &
  &
  &
  &
  (arcsec) &
  (arcsec) &
  (arcsec) &
  (mJy/beam) \\
\hline
ACA &
  3 &
  101 &
  12 &
  8 - 48 &
  2.0 &
  100 $\times$ 100 &
  200 $\times$ 200 &
  151 &
  17 $\times$ 10 &
  45 &
  0.11 \\
12~m &
  3 &
  101 &
  75 &
  14 - 313 &
  0.5 &
  240 $\times$ 240 &
  120 $\times$ 120 &
  97 &
  3.6 $\times$ 2.7 &
  26 &
  0.04 \\
12~m &
  4 &
  151 &
  85 &
  13 - 311 &
  0.3 &
  240 $\times$ 240 &
  72 $\times$ 72 &
  52 &
  2.2 $\times$ 1.8 &
  18 &
  0.04 \\
\hline
\end{tabular}
$^a$ This refers to the diameter of the region where the primary beam is 0.2$\times$ its peak value, and it is also the diameter of the fields imaged by ALMA. 
\end{minipage}
\end{table*}

The details of the sample selection, observations, and the data calibration are described by \citet{urquhart2022}; we only provide brief summaries of those topics here.  However, because the continuum imaging differs from the spectral line imaging, we provide full details about the continuum imaging here.

As stated above, the BEARS sample consists of a subset of 85 fields from the HerBS sample. The HerBS sample, which was selected by \citet{bakx2018}, consists of unresolved {\it Herschel} sources with 500~$\mu$m flux densities $>$80~mJy and with photometric redshifts of $>$2 derived from fitting the {\it Herschel} data with the template from \citet{pearson2013}.  Efforts were made to remove nearby ($z \leq 0.1$) spiral galaxies and blazars that may have otherwise satisfied these selection criteria.  BEARS used a combination of the Atacama Compact Array (ACA; also called the Morita Array) and the main ALMA 12~m array to observe these fields.

Data for 12 fields were acquired with the ACA during ALMA Cycles 4 and 6 in programmes 2016.2.00133.S and 2018.1.00804.S. Each target was observed using single pointings larger than the {\it Herschel} 500~$\mu$m beam.  The observations covered a frequency range from 86.6~GHz to 115.7~GHz (with the exception of HerBS-49, which is missing data from 97.0-98.6 and 108.9-112.2~GHz) with multiple spectral windows, each of which had a bandwidth of 2~GHz and 256 channels.   

In programme 2019.1.01477, which was executed in Cycle 7, all 85 fields were observed with the 12~m array in ALMA Band 4 in the frequency range 139.0-162.2~GHz (with a small gap in coverage at 150.2-150.9~GHz).  Additionally, 75 fields that (with the exception of HerBS-37 and HerBS-39\footnote{No 101~GHz sources were detected for HerBS-37 in either the ACA or 12~m data.  In the HerBS-39 field, we only detected one source.  The flux density from the ACA data are larger than the measurement from the 12~m data by 40\%, but the signal-to-noise ratio of the ACA data are worse.  We therefore used the flux density from the 12~m data.}) had not been previously observed with the ACA were observed with the 12~m array in ALMA Band 3 within the frequency range 89.6-112.8~GHz (with a small gap in coverage at 100.8-101.5~GHz). This frequency set-up was designed to improve our efficiency in measuring robust redshifts \citep{bakx2022}.  Each field was observed using single pointing that were larger than the {\it Herschel} 500~$\mu$m beams.  Every spectral window used to cover these frequency ranges had 1920 channels and covered a bandwidth of 1.875~GHz.

The ACA visibility data were manually calibrated using the {\sc common astronomy software applications} ({\sc casa}) package version 5.6.1 \citep{mcmullin2007, casateam2022}, while the 12~m Array data were pipeline-calibrated with the same version of {\sc casa}.  This included all the standard calibration steps applied to the phases and the amplitudes of the data.  The manual calibration included visual inspections of every data set to identify and remove data with any irregular amplitude or phase values.

Continuum images were created using {\sc tclean} interactively within {\sc casa}.  The characteristics of the final images are listed in Table~\ref{t_imageparam}.  We used all data from all spectral windows excluding any channels that contained potential spectral lines.  Each field was imaged separately.  The central position of each image is the same as the phase centre of each field, which in turn is equivalent to the coordinates of the source originally identified in the {\it Herschel} images.  Different pixel scales and image sizes were used for the ACA Band 3 observations, the 12~m Array Band 3 observations, and the Band 4 observations.  In each case, the pixel scales were set to oversample the beams, and the image sizes were set to encompass the whole of the primary beams.  Natural weighting was used primarily to optimize the data for source detection.  The Hogbom algorithm \citep{hogbom1974} was used as the deconvolver because it provided the best detections for the low signal-to-noise ratio unresolved sources.  Two set of images were created with and without the primary beams corrections that adjust the signal levels for off-centre sources.  The images without the primary beam corrections were used for identifying sources and for measuring the noise levels in the ACA data; the images with the primary beam corrections were used for measuring flux densities for all of the sources.  The calibration uncertainty is expected to be 5\% \citep{privon2022}\footnote{Available from \url{https://almascience.eso.org/documents-and-tools/cycle9/alma-proposers-guide}.}.

\begin{table*}
\centering
\begin{minipage}{164mm}
\caption{ALMA photometry}
\label{t_almaphotom}
\begin{tabular}{@{}lccccccccc@{}}
\hline
Object &
  H-ATLAS &
  Number &
  Source &
  \multicolumn{2}{c}{Coordinates (J2000)$^a$} &
  \multicolumn{2}{c}{Flux density (mJy)$^b$} &
  Spectroscopic \\
&
  designation &
  of sources &
  designation &
  RA &
  Declination &
  101 GHz &
  151 GHz &
  redshift$^c$ \\
\hline
HerBS-11 &
  J012407.4-281434 & 1 & &
  01:24:07.50 &
  -28:14:34.7 &
  0.94 $\pm$ 0.02 &
  3.59 $\pm$ 0.03 &
  2.631 \\
HerBS-14 &
  J013840.5-281856 & 1 & &
  01:38:40.41 &
  -28:18:57.5 &
  1.46 $\pm$ 0.02 &
  7.43 $\pm$ 0.06 &
  3.782 \\
HerBS-18 &
  J232419.8-323927 & 1 & &
  23:24:19.82 &
  -32:39:26.5 &
  0.95 $\pm$ 0.09$^d$ &
  2.70 $\pm$ 0.03 &
  2.182 \\
  
HerBS-21 &
  J234418.1-303936 & 2 & [A+B] &
  &
  &
  0.81 $\pm$ 0.06$^d$ &
  3.94 $\pm$ 0.03 &
  3.323 \\
&
  & & A &
  23:44:18.11 &
  -30:39:38.9 &
  &
  3.01 $\pm$ 0.02 &
  \\
&
  & & B &
  23:44:18.25 &
  -30:39:34.9 &
  &
  0.93 $\pm$ 0.02 &
  \\
HerBS-22 &
  J002624.8-341738 & 2 & A &
  00:26:24.99 &
  -34:17:38.1 &
  0.66 $\pm$ 0.02 &
  3.10 $\pm$ 0.02 &
  3.050 \\
&
  & & B &
  00:26:25.56 &
  -34:17:23.3 &
  &
  0.35 $\pm$ 0.04 &
  \\
HerBS-24 &
  J004736.0-272951 & 1 & &
  00:47:36.09 &
  -27:29:52.0 &
  0.77 $\pm$ 0.03 &
  2.70 $\pm$ 0.03 &
  2.198 \\
HerBS-25 &
  J235827.7-323244 & 1 & &
  23:58:27.50 &
  -32:32:44.8 &
  0.91 $\pm$ 0.07$^d$ &
  3.46 $\pm$ 0.03 &
  2.912 \\
HerBS-27 &
  J011424.0-333614 & 1 & &
  01:14:24.01 &
  -33:36:16.5 &
  2.00 $\pm$ 0.03 &
  8.76 $\pm$ 0.04 &
  4.509 \\
HerBS-28 &
  J230815.6-343801 & 1 & &
  23:08:15.73 &
  -34:38:00.5 &
  1.61 $\pm$ 0.08$^d$ &
  5.56 $\pm$ 0.03 &
  3.925 \\
HerBS-33 &
  J224805.4-335820 & 3$^e$ & [A+B] &
  &
  &
  0.94 $\pm$ 0.05$^d$ &
  2.71 $\pm$ 0.04 &
  \\
&
  & & A &
  22:48:05.17 &
  -33:58:21.0 &
  &
  2.15 $\pm$ 0.04 &
  \\
&
  & & B &
  22:48:05.50 &
  -33:58:19.5 &
  &
  0.56 $\pm$ 0.02 &
  \\
&
  & & C &
  22:48:06.6 &
  -33:58:39 &
  0.58 $\pm$ 0.04$^d$ &
  &
  &
  \\
HerBS-36 &
  J235623.1-354119 & 1 & &
  23:56:23.08 &
  -35:41:19.5 &
  1.16 $\pm$ 0.02 &
  4.81 $\pm$ 0.03 &
  3.095 \\
HerBS-37 &
  J232623.0-342642 & 1 & &
  23:26:23.10 &
  -34:26:44.0 &
  &
  1.60 $\pm$ 0.03 &
  2.619 \\
HerBS-39 &
  J232900.6-321744 & 1 & &
  23:29:00.80 &
  -32:17:45.0 &
  0.64 $\pm$ 0.03 &
  2.98 $\pm$ 0.03 &
  3.229 \\
HerBS-40 &
  J013240.0-330907 & 1 & &
  01:32:40.28 &
  -33:09:08.0 &
  &
  0.96 $\pm$ 0.03 &
  1.971 \\
HerBS-41 &
  J000124.9-354212 &
  3$^e$ & A &
  00:01:24.79 &
  -35:42:11.0 &
  0.71 $\pm$ 0.02 &
  3.79 $\pm$ 0.03 &
  4.098 \\
&
  & & B &
  00:01:23.24 &
  -35:42:10.8 &
  &
  0.74 $\pm$ 0.05 &
  \\
&
  & & C &
  00:01:25.82 &
  -35:42:18.0 &
  &
  0.31 $\pm$ 0.02 &
  \\
HerBS-42 &
  J000007.5-334100 & 3 & [A+B+C] &
  &
  &
  0.56 $\pm$ 0.04$^d$ &
  2.95 $\pm$ 0.05$^f$ &
  3.307$^g$ \\
&
  & & A &
  00:00:07.45 &
  -33:41:03.1 &
  &
  1.84 $\pm$ 0.03 &
  \\
&
  & & B &
  00:00:07.41 &
  -33:40:55.9 &
  &
  0.54 $\pm$ 0.02 &
  \\
&
  & & C &
  00:00:07.05 &
  -33:41:03.4 &
  &
  0.39 $\pm$ 0.03 &
  \\
HerBS-45 &
  J005132.8-301848 & 2 & A &
  00:51:32.97 &
  -30:18:49.6 &
  &
  0.68 $\pm$ 0.03 &
  2.434 \\
&
  & & B &
  00:51:32.49 &
  -30:18:48.9 &
  &
  0.45 $\pm$ 0.03 &
  \\
HerBS-47 &
  J225250.7-313658 & 1 &   &
  22:52:50.76 &
  -31:36:59.9 &
  &
  1.28 $\pm$ 0.03 &
  2.433 \\
HerBS-49 &
  J230546.3-331039 & 2 & [A+B] &
  &
  &
  1.06 $\pm$ 0.06$^d$ &
  1.70 $\pm$ 0.03 \\
&
  & & A &
  23:05:46.41 &
  -33:10:38.1 &
  &
  1.21 $\pm$ 0.02 &
  2.724 \\
&
  & & B &
  23:05:46.58 &
  -33:10:43.1 &
  &
  0.49 $\pm$ 0.02 &
  2.730 \\
HerBS-55 &
  J013951.9-321446 & 1 & &
  01:39:52.08 &
  -32:14:45.5 &
  0.34 $\pm$ 0.02 &
  1.08 $\pm$ 0.03 &
  2.656 \\
HerBS-56 &
  J003207.7-303724 & 4 & A &
  00:32:07.15 &
  -30:37:13.2 &
  &
  0.59 $\pm$ 0.02 &
  \\
&
  & & B &
  00:32:08.57 &
  -30:37:31.0 &
  &
  0.51 $\pm$ 0.03 &
  \\
&
  & & C &
  00:32:07.63 &
  -30:37:35.2 &
  &
  0.32 $\pm$ 0.03 &
  2.561 \\
&
  & & D &
  00:32:07.87 &
  -30:37:32.4 &
  &
  0.33 $\pm$ 0.02 &
  \\
HerBS-57 &
  J004853.3-303110 & 1 & &
  00:48:53.38 &
  -30:31:09.9 &
  0.52 $\pm$ 0.02 &
  3.09 $\pm$ 0.03 &
  3.265 \\
HerBS-60 &
  J005724.2-273122 & 1 & &
  00:57:24.34 &
  -27:31:23.3 &
  0.56 $\pm$ 0.02 &
  2.56 $\pm$ 0.03 &
  3.261 \\
HerBS-63 &
  J005132.0-302012 & 3$^e$ & A &
  00:51:31.70 &
  -30:20:20.6 &
  0.35 $\pm$ 0.02 &
  1.07 $\pm$ 0.03 &
  2.432 \\
&
  & & B &
  00:51:31.85 &
  -30:20:04.6 &
  &
  0.30 $\pm$ 0.02 &
  \\
&
  & & C &
  00:51:32.57 &
  -30:19:48.9 &
  0.46 $\pm$ 0.03 &
  &
  \\
HerBS-67 &
  J224207.2-324159 & 1 & &
  22:42:07.20 &
  -32:42:01.9 &
  0.83 $\pm$ 0.04$^d$ &
  2.67 $\pm$ 0.04 &
  \\
HerBS-68 &
  J223753.8-305828 & 1 & $^h$ &
  22:37:53.84 &
  -30:58:27.6 &
  &
  1.62 $\pm$ 0.04 &
  2.719 \\
HerBS-69 &
  J012416.0-310500 & 2 & A &
  01:24:16.16 &
  -31:04:59.5 &
  &
  0.69 $\pm$ 0.02 &
  2.075 \\
&
  & & B &
  01:24:15.87 &
  -31:05:05.1 &
  &
  0.53 $\pm$ 0.02 &
  2.073 \\
HerBS-73 &
  J012853.0-332719 & 1 & &
  01:28:53.07 &
  -33:27:19.1 &
  0.44 $\pm$ 0.02 &
  2.08 $\pm$ 0.03 &
  3.026 \\
HerBS-75 &
  J011823.8-274404 & 3 & A &
  01:18:23.61 &
  -27:44:11.5 &
  &
  0.40 $\pm$ 0.03 &
  \\
&
  & & B &
  01:18:24.25 &
  -27:44:02.7 &
  &
  0.30 $\pm$ 0.02 &
  \\
&
  & & C &
  01:18:23.84 &
  -27:44:15.0 &
  &
  0.24 $\pm$ 0.02 &
  \\
HerBS-77 &
  J005629.6-311206 & 2 & A &
  00:56:29.25 &
  -31:12:07.5 &
  &
  1.08 $\pm$ 0.03 &
  2.228 \\
&
  & & B &
  00:56:30.52 &
  -31:12:15.7 &
  &
  0.51 $\pm$ 0.05 &
  \\
HerBS-80 &
  J230002.6-315005 & 3 & A &
  23:00:02.54 &
  -31:50:08.9 &
  &
  0.28 $\pm$ 0.02 &
  2.231 \\
&
  & & B &
  23:00:02.86 &
  -31:50:08.0 &
  &
  0.29 $\pm$ 0.02 &
  1.968 \\
&
  & & C &
  23:00:02.91 &
  -31:50:02.0 &
  &
  0.20 $\pm$ 0.02 &
  \\
HerBS-81 &
  J002054.6-312752 & 2 & A &
  00:20:54.20 &
  -31:27:57.4 &
  0.20 $\pm$ 0.02 &
  0.76 $\pm$ 0.03 &
  3.160 \\
&
  & & B &
  00:20:54.74 &
  -31:27:50.8 &
  &
  0.68 $\pm$ 0.02 &
  2.588 \\
HerBS-84 &
  J224400.8-340031 & 1 & &
  22:44:01.10 &
  -34:00:32.5 &
  0.25 $\pm$ 0.02 &
  0.90 $\pm$ 0.03 &
  \\
HerBS-86 &
  J235324.7-331111 & 1 & &
  23:53:24.56 &
  -33:11:11.8 &
  0.26 $\pm$ 0.02 &
  1.53 $\pm$ 0.02 &
  2.564 \\
HerBS-87 &
  J002533.6-333826 & 1 & &
  00:25:33.67 &
  -33:38:26.3 &
  &
  1.24 $\pm$ 0.02 &
  \\
\hline
\end{tabular}
\end{minipage}
\end{table*}

\addtocounter{table}{-1}
\begin{table*}
\centering
\begin{minipage}{164mm}
\caption{ALMA photometry (continued)}
\begin{tabular}{@{}lccccccccc@{}}
\hline
Object &
  H-ATLAS &
  Number &
  Source &
  \multicolumn{2}{c}{Coordinates (J2000)$^a$} &
  \multicolumn{2}{c}{Flux density (mJy)$^b$} &
  Spectroscopic \\
&
  designation &
  of sources &
  designation &
  RA &
  Declination &
  101 GHz &
  151 GHz &
  redshift$^c$ \\
\hline
HerBS-90 &
  J005659.4-295039 & 2 & A &
  00:56:59.28 &
  -29:50:39.3 &
  0.69 $\pm$ 0.03 &
  3.23 $\pm$ 0.03 &
  3.992 \\
&
  & & B &
  00:57:00.31 &
  -29:50:40.7 &
  &
  0.36 $\pm$ 0.03 &
  \\
HerBS-93 &
  J234750.5-352931 & 1 & &
  23:47:50.44 &
  -35:29:30.2 &
  0.16 $\pm$ 0.01 &
  1.37 $\pm$ 0.02 &
  2.400 \\
HerBS-94 &
  J000950.5-353829 & 2 & A &
  00:09:50.23 &
  -35:38:26.4 &
  0.38 $\pm$ 0.02 &
  1.43 $\pm$ 0.03 &
  \\
&
  & & B &
  00:09:51.15 &
  -35:38:35.0 &
  &
  0.23 $\pm$ 0.01 &
  \\
HerBS-97 &
  J224027.8-343135 & 2 & A &
  22:40:28.54 &
  -34:31:33.0 &
  &
  0.54 $\pm$ 0.03 &
  \\
&
  & & B &
  22:40:27.71 &
  -34:31:38.1 &
  &
  0.39 $\pm$ 0.03 &
  \\
HerBS-98 &
  J001030.1-330622 & 2$^e$ & A &
  00:10:30.59 &
  -33:06:04.8 &
  &
  0.76 $\pm$ 0.06 &
  \\
&
  & & B &
  00:10:30.03 &
  -33:06:11.1 &
  &
  0.50 $\pm$ 0.03 &
  \\
HerBS-101 &
  J011246.5-330611 & 2 & A &
  01:12:46.52 &
  -33:06:10.5 &
  0.56 $\pm$ 0.02 &
  1.55 $\pm$ 0.03 &
  \\
&
  & & B &
  01:12:46.10 &
  -33:06:12.4 &
  &
  0.50 $\pm$ 0.03 &
  \\
HerBS-102 &
  J233024.1-325032 & 2 & A &
  23:30:24.43 &
  -32:50:32.3 &
  0.34 $\pm$ 0.02 &
  1.38 $\pm$ 0.04 &
  3.287 \\
&
  & & B &
  23:30:23.52 &
  -32:50:43.4 &
  &
  1.23 $\pm$ 0.05 &
  \\
HerBS-103 &
  J225324.2-323504 & 1 & &
  22:53:24.24 &
  -32:35:04.2 &
  0.44 $\pm$ 0.03 &
  1.42 $\pm$ 0.03 &
  2.942 \\
HerBS-104 &
  J001838.7-354133 & 2 & A &
  00:18:39.47 &
  -35:41:48.0 &
  &
  0.64 $\pm$ 0.03 &
  \\
&
  & & B &
  00:18:38.84 &
  -35:41:33.1 &
  &
  0.52 $\pm$ 0.02 &
  \\
HerBS-106 &
  J001802.2-313505 & 2 & A &
  00:18:02.46 &
  -31:35:05.1 &
  0.29 $\pm$ 0.02 &
  1.45 $\pm$ 0.03 &
  2.369 \\
&
  & & B &
  00:18:01.14 &
  -31:35:08.0 &
  &
  0.87 $\pm$ 0.05 &
  \\
HerBS-107 &
  J014520.0-313835 & 1 & &
  01:45:20.07 &
  -31:38:32.5 &
  &
  1.01 $\pm$ 0.04 &
  2.553 \\
HerBS-111 &
  J223942.4-333304 & 1 & &
  22:39:42.34 &
  -33:33:04.1 &
  0.22 $\pm$ 0.02 &
  1.32 $\pm$ 0.04 &
  2.371 \\
HerBS-114 &
  J012209.5-273824 & 1 & &
  01:22:09.38 &
  -27:38:25.5 &
  &
  1.08 $\pm$ 0.03 &
  \\
HerBS-117 &
  J000806.8-351205 & 2 & A &
  00:08:07.20 &
  -35:12:05.0 &
  0.79 $\pm$ 0.02 &
  3.46 $\pm$ 0.02 &
  4.526 \\
&
  & & B &
  00:08:06.86 &
  -35:12:10.0 &
  &
  0.48 $\pm$ 0.02 &
  \\
HerBS-118 &
  J232200.1-355622 & 2$^e$ & A &
  23:21:59.43 &
  -35:56:21.0 &
  0.18 $\pm$ 0.02 &
  1.13 $\pm$ 0.02 &
  \\
&
  & & B &
  23:22:01.66 &
  -35:56:05.0 &
  0.39 $\pm$ 0.03 &
  0.82 $\pm$ 0.04 &
  \\
HerBS-120 &
  J012222.3-274456 & 2 & A &
  01:22:22.44 &
  -27:44:53.7 &
  &
  1.21 $\pm$ 0.03 &
  3.125 \\
&
  & & B &
  01:22:22.13 &
  -27:44:59.0 &
  0.39 $\pm$ 0.02 &
  1.28 $\pm$ 0.03 &
  3.124 \\
HerBS-121 &
  J223615.2-343301 & 2 & A &
  22:36:15.31 &
  -34:33:02.3 &
  0.33 $\pm$ 0.02 &
  1.97 $\pm$ 0.06 &
  3.741 \\
&
  & & B &
  22:36:15.01 &
  -34:32:56.6 &
  &
  0.35 $\pm$ 0.03 &
  \\
HerBS-122 &
  J003717.0-323307 & 2 & A &
  00:37:16.71 &
  -32:32:57.4 &
  &
  0.37 $\pm$ 0.03 &
  2.883 \\
&
  & & B &
  00:37:16.87 &
  -32:33:09.3 &
  &
  0.15 $\pm$ 0.01 &
  \\
HerBS-123 &
  J233037.3-331218 & 1 & &
  23:30:37.45 &
  -33:12:16.8 &
  &
  1.24 $\pm$ 0.05 &
  2.170 \\
HerBS-131 &
  J225339.1-325550 & 2 & A$^h$ &
  22:53:38.45 &
  -32:55:48.2 &
  0.38 $\pm$ 0.05 &
  1.08 $\pm$ 0.04 &
  \\
&
  & & B &
  22:53:39.50 &
  -32:55:52.3 &
  &
  0.79 $\pm$ 0.04 &
  2.197 \\
HerBS-132 &
  J231205.2-295027 & 1 & &
  23:12:05.31 &
  -29:50:26.5 &
  0.15 $\pm$ 0.01 &
  0.87 $\pm$ 0.02 &
  2.473 \\
HerBS-135 &
  J225611.7-325653 & 2 & A &
  22:56:11.79 &
  -32:56:52.0 &
  0.27 $\pm$ 0.02 &
  0.82 $\pm$ 0.03 &
  2.401 \\
&
  & & B &
  22:56:11.42 &
  -32:56:52.1 &
  &
  0.21 $\pm$ 0.01 &
  \\
HerBS-138 &
  J011730.3-320719 & 2 & [A+B]$^h$ &
  01:17:30.56 &
  -32:07:20.9 &
  &
  0.90 $\pm$ 0.03 &
  1.407$^i$\\
HerBS-141 &
  J224759.7-310135 & 1 & $^h$&
  22:47:59.75 &
  -31:01:36.0 &
  &
  0.73 $\pm$ 0.05 &
  2.085\\
HerBS-144 &
  J222629.4-321112 & 2 & A &
  22:26:28.62 &
  -32:11:08.1 &
  &
  1.01 $\pm$ 0.04 &
  \\
&
  & & B &
  22:26:30.28 &
  -32:11:10.5 &
  &
  0.56 $\pm$ 0.04 &
  \\
HerBS-145 &
  J012335.1-314619 & 2 & A &
  01:23:34.65 &
  -31:46:23.6 &
  &
  0.72 $\pm$ 0.03 &
  2.730 \\
&
  & & B &
  01:23:35.75 &
  -31:46:25.4 &
  &
  0.28 $\pm$ 0.02 &
  \\
HerBS-146 &
  J232210.9-333749 & 2 & A &
  23:22:10.94 &
  -33:37:48.9 &
  &
  0.50 $\pm$ 0.03 &
  \\
&
  & & B &
  23:22:10.62 &
  -33:37:58.4 &
  &
  0.40 $\pm$ 0.02 &
  2.003 \\
HerBS-148 &
  J224026.5-315155 & 1 & &
  22:40:26.55 &
  -31:51:54.1 &
  0.26 $\pm$ 0.02 &
  1.22 $\pm$ 0.04 &
  \\
HerBS-151 &
  J012530.5-302509 & 2 & A &
  01:25:30.78 &
  -30:25:11.7 &
  &
  0.48 $\pm$ 0.04 &
  \\
&
  & & B &
  01:25:29.83 &
  -30:24:55.5 &
  &
  0.30 $\pm$ 0.03 &
  \\
HerBS-155 &
  J000330.7-321136 & 2 & A &
  00:03:30.65 &
  -32:11:35.1 &
  0.29 $\pm$ 0.01 &
  2.16 $\pm$ 0.04 &
  3.077 \\
&
  & & B &
  00:03:30.06 &
  -32:11:39.3 &
  &
  0.54 $\pm$ 0.03 &
  \\
HerBS-156 &
  J002144.8-295218 & 2 & A &
  00:21:44.48 &
  -29:52:17.7 &
  0.20 $\pm$ 0.01 &
  0.79 $\pm$ 0.02 &
  \\
&
  & & B &
  00:21:45.63 &
  -29:52:17.2 &
  &
  0.41 $\pm$ 0.03 &
  \\
HerBS-159 &
  J235122.0-332902 & 2 & A &
  23:51:21.74 &
  -33:29:00.4 &
  &
  0.83 $\pm$ 0.03 &
  2.236 \\
&
  & & B &
  23:51:22.36 &
  -33:29:08.0 &
  &
  0.23 $\pm$ 0.03 &
  2.235 \\
HerBS-160 &
  J011014.5-314814 & 1 & &
  01:10:14.46 &
  -31:48:15.9 &
  0.91 $\pm$ 0.02 &
  4.20 $\pm$ 0.04 &
  3.955 \\
HerBS-163 &
  J000745.8-342014 & 3 & A &
  00:07:46.23 &
  -34:20:03.0 &
  &
  0.43 $\pm$ 0.02 &
  3.140 \\
&
  & & B &
  00:07:45.93 &
  -34:20:16.2 &
  &
  0.47 $\pm$ 0.02 &
  \\
&
  & & C &
  00:07:45.45 &
  -34:20:17.4 &
  &
  0.35 $\pm$ 0.02 &
  \\
HerBS-166 &
  J222503.8-304848 & 2 & A &
  22:25:03.53 &
  -30:48:48.4 &
  &
  0.40 $\pm$ 0.03 &
  \\
&
  & & B &
  22:25:04.32 &
  -30:48:33.2 &
  &
  0.27 $\pm$ 0.03 &
  \\
HerBS-168 &
  J225045.5-304719 & 2 & A &
  22:50:45.48 &
  -30:47:20.3 &
  0.67 $\pm$ 0.04 &
  2.66 $\pm$ 0.03 &
  2.583 \\
&
  & & B &
  22:50:45.78 &
  -30:47:13.4 &
  &
  0.19 $\pm$ 0.01 &
  \\
\hline
\end{tabular}
\end{minipage}
\end{table*}

\addtocounter{table}{-1}
\begin{table*}
\centering
\begin{minipage}{164mm}
\caption{ALMA photometry (continued)}
\begin{tabular}{@{}lccccccccc@{}}
\hline
Object &
  H-ATLAS &
  Number &
  Source &
  \multicolumn{2}{c}{Coordinates (J2000)$^a$} &
  \multicolumn{2}{c}{Flux density (mJy)$^b$} &
  Spectroscopic \\
&
  designation &
  of sources &
  designation &
  RA &
  Declination &
  101 GHz &
  151 GHz &
  redshift$^c$ \\
\hline
HerBS-170 &
  J000455.4-330812 & 1 & $^h$ &
  00:04:55.44 &
  -33:08:12.8 &
  0.81 $\pm$ 0.02 &
  3.50 $\pm$ 0.03 &
  \\
HerBS-174 &
  J003728.7-284125 & 2 & A &
  00:37:29.03 &
  -28:41:28.6 &
  &
  0.31 $\pm$ 0.02 &
  \\
&
  & & B &
  00:37:28.37 &
  -28:41:25.4 &
  &
  0.26 $\pm$ 0.02 &
  \\
HerBS-178 &
  J011850.1-283642 & 4 & A &
  01:18:50.27 &
  -28:36:44.0 &
  &
  0.77 $\pm$ 0.02 &
  2.658 \\
&
  & & B &
  01:18:50.10 &
  -28:36:40.5 &
  &
  0.55 $\pm$ 0.02 &
  2.655 \\
&
  & & C &
  01:18:49.98 &
  -28:36:43.2 &
  &
  0.35 $\pm$ 0.02 &
  2.656 \\
&
  & & D &
  01:18:50.18 &
  -28:36:42.9 &
  0.52 $\pm$ 0.03 &
  & 
  \\
HerBS-181 &
  J005850.0-290122 & 2 & A &
  00:58:49.78 &
  -29:01:18.0 &
  &
  0.21 $\pm$ 0.02 &
  \\
&
  & & B &
  00:58:50.65 &
  -29:01:13.8 &
  &
  0.19 $\pm$ 0.02 &
  \\
HerBS-182 &
  J230538.5-312204 & 1 & &
  23:05:38.80 &
  -31:22:05.6 &
  0.20 $\pm$ 0.01 &
  1.09 $\pm$ 0.03 &
  2.227 \\
HerBS-184 &
  J234955.7-330833 & 1 & &
  23:49:55.66 &
  -33:08:34.4 &
  0.46 $\pm$ 0.02 &
  1.47 $\pm$ 0.02 &
  2.507 \\
HerBS-186 &
  J013217.0-320953 & 2$^e$ & A &
  01:32:17.23 &
  -32:09:55.4 &
  0.30 $\pm$ 0.02 &
  1.92 $\pm$ 0.03 &
  \\
&
  & & B &
  01:32:15.55 &
  -32:09:39.0 &
  &
  0.42 $\pm$ 0.03 &
  \\
HerBS-189 &
  J225600.7-313232 & 1 & &
  22:56:00.74 &
  -31:32:33.0 &
  &
  1.89 $\pm$ 0.05 &
  3.300 \\
HerBS-192 &
  J222628.8-304421 & 1 & &
  22:26:28.94 &
  -30:44:23.3 &
  0.13 $\pm$ 0.02 &
  0.73 $\pm$ 0.03 &
  \\  
HerBS-198 &
  J222235.8-324528 & 1 & &
  22:22:35.89 &
  -32:45:23.8 &
  0.26 $\pm$ 0.02 &
  1.00 $\pm$ 0.03 &
  \\
HerBS-200 &
 J014313.2-332633 & 1 & &
  01:43:13.30 &
  -33:26:33.1 &
  0.24 $\pm$ 0.02 &
  0.79 $\pm$ 0.02 &
  2.151 \\
HerBS-207 &
  J005506.5-300027 & 1 & &
  00:55:06.51 &
  -30:00:28.3 &
  0.24 $\pm$ 0.02 &
  0.88 $\pm$ 0.02 &
  1.569 \\
HerBS-208 &
  J225744.6-324231 & 2 & [A+B] &
  &
  &
  0.32 $\pm$ 0.03$^j$ &
  1.43 $\pm$ 0.04 \\
&
  & & A &
  22:57:44.58 &
  -32:42:33.0 &
  &
  0.86 $\pm$ 0.03 &
  2.478 \\
&
  & & B &
  22:57:44.84 &
  -32:42:32.9 &
  &
  0.57 $\pm$ 0.03 &
  2.483 \\
HerBS-209 &
  J224920.6-332940 & 2 & A &
  22:49:21.04 &
  -33:29:41.5 &
  &
  0.65 $\pm$ 0.04 &
  2.272 \\
&
  & & B &
  22:49:20.53 &
  -33:29:40.9 &
  &
  0.28 $\pm$ 0.02 &
  \\
\hline
\end{tabular}
$^a$ Based on the information from \citet{privon2022}, the coordinates for most sources are expected to be accurate to within 0.10~arcsec.  The positions of Band 3 (101 GHz) sources with no Band 4 (151 GHz) counterparts should be accurate to within 0.16~arcsec except for HerBS-33C, which was identified in ACA data and should have a position accurate to within 0.65~arcsec. \\
$^b$ The uncertainties in the flux densities do not include the calibration uncertainties, which are 5\%.  Because of unit conversions applied when measuring the flux densities, the typical uncertainties in mJy are slightly lower than those reported in mJy beam$^{-1}$ in Table~\ref{t_imageparam}.\\
$^c$ These spectroscopic redshifts (for the millimetre sources) come from \citet{urquhart2022}.\\
$^d$ This measurement is from ACA data.\\
$^e$ One of these sources falls outside the 35~arcsec beam of the {\it Herschel} 500~$\mu$m data.\\
$^f$ The central 12~arcsec of the 151~GHz image for the HerBS-42 field contains three closely-spaced sources detected at the $\geq$5$\sigma$ level along with a fourth source in between these three sources which has a peak measured at the $\geq$4$\sigma$ level.  Consequently, the 151~GHz photometry measurement listed for A+B+C is higher than for the individual sources.\\
$^g$ Spectral line emission in ALMA Band 4 was only detected for the A and B components of HerBS-42, but since the line emission was unresolved in the Band 3 data, one redshift is reported for all sources in the field.\\
$^h$ These sources consist of two peaks separated by less than 3~arcsec (or $\sim$3.5~arcsec in the case of HerBS-170), although the two peaks may only be apparent in the higher resolution 151~GHz data.  For all of these sources except the ones in HerBS-138, it is unclear whether the two peaks are part of one elongated object, whether they are two objects that are physically associated with each other, or whether they are two unassociated sources that just happen to lie close to each other along the line of sight.  For HerBS-138, the spectra indicate that the two peaks correspond to two objects at different redshifts.  The coordinates for each source correspond to the brighter peak in the emission.\\
$^i$ This redshift is for HerBS-138B.  The spectrum measured by \citet{urquhart2022} for HerBS-138A indicate that it is at a different redshift, but since only one line was detected, its redshift could not be determined.\\
$^j$ The 151~GHz emission from the two sources in the HerBS-208 field is sufficiently resolved and detected at a sufficiently high signal-to-noise level that it is possible to measure the 151~GHz  flux densities from the two sources independently.  However, the sources are detected at a lower signal-to-noise level in the 101~GHz image and tend to blur together, so we reported a single flux density for the two sources in that band.\\
\end{minipage}
\end{table*}

\section{ALMA photometry and morphology}

\subsection{Photometric measurements}

We identified sources as locations in the images without primary beam corrections where the surface brightnesses peaked at $\geq$5$\sigma$ ($\geq$0.20 mJy beam$^{-1}$ in the 12~m data and $\geq$0.55 mJy beam$^{-1}$ in the ACA data).  We also measured 151~GHz flux densities for any source with associated line emission detected at the $\geq$5$\sigma$ level by \citet{urquhart2022}.  Note that some sources observed with the ACA at 101~GHz were separated into multiple sources when observed with the 12~m array at 151~GHz.

Flux densities were measured using aperture photometry within the images with the primary beam corrections.  For most sources, we used elliptical apertures with axis ratios and position angles equivalent to the beam shape for each image.  However, for sources that appeared significantly extended relative to the beam in the 151~GHz data, we used apertures with axis ratios and position angles equivalent to the shape of the source convolved with the beam, and for sources that appeared double-lobed (generally with two point sources separated by $\lesssim$3~arcsec), we used circular measurement apertures (and these objects are discussed more in Section~\ref{s_extended}).  The sizes of the apertures were manually adjusted for each source to maximize the measured flux density while excluding excess background noise and other nearby sources; the width of the apertures are typically 2-4$\times$ the beam sizes.  When two or more sources were located more than 3~arcsec away from each other but close enough that the measurement apertures would overlap, we measured the flux densities for each source from pixels that both fell within its aperture and that fell on its side of dividing lines we used to separate the emission from the sources.

In the 12~m data, nine detected sources are located $>$15~arcsec from the centres of the primary beams where the sensitivity levels drop notably relative to the central regions.  To measure the local background noise levels around each source within the primary beam corrected images, we used relatively small circular annuli centered on each source.  The diameters of these annuli were set to 20-25~arcsec for the Band 3 data and 15-20~arcsec for the Band 4 data.  

In the ACA data, all detected sources lie relatively close to the centres of the fields where the responsivity of the telescope is still $\gtrsim$85\%, but because of the size of the beam, it is not possible to measure the background noise in regions with the same responsivity.  We therefore measured the background noise levels in ACA images without the primary beam corrections using circular annuli with diameters of 90-120~arcsec positioned at the centre of each image, which should yield representative noise levels for sources centred in these fields.

While we used aperture photometry to measure the flux densities, we fitted the sources with Gaussian functions to measure the positions of the sources and to determine whether the sources are significantly extended relative to the beam size.  These extended sources are discussed in Section~\ref{s_extended}.

Table~\ref{t_almaphotom} lists each field and the positions and flux densities measured for each detected source within each field.  The individual Band 3 and Band 4 images are shown in the supplemental online material.  In fields with multiple sources, we have labelled them alphabetically in descending order of flux density\footnote{This alphabetical labelling was set before the publication of \cite{urquhart2022}, but adjustments were made to the continuum flux densities afterwards.  Relabelling the sources with published redshifts would have caused confusion, so we avoided doing this.  Consequently, the sources in the HerBS-56, HerBS-80, HerBS-120, and HerBS-163 fields are not labelled in terms of decreasing 151~GHz flux density (as reported in Table~\ref{t_almaphotom}).}.  Some fields have two or more sources that lie $<$2$\times$ the full width at half-maximum (FWHM) apart from each other or that are connected by extended structures detected at the $>$3$\sigma$ level.  In these cases, Table~\ref{t_almaphotom} reports integrated flux densities for these sources.

\subsection{101~GHz sources without 151~GHz counterparts}

\begin{figure}
\begin{center}
\includegraphics[width=8cm]{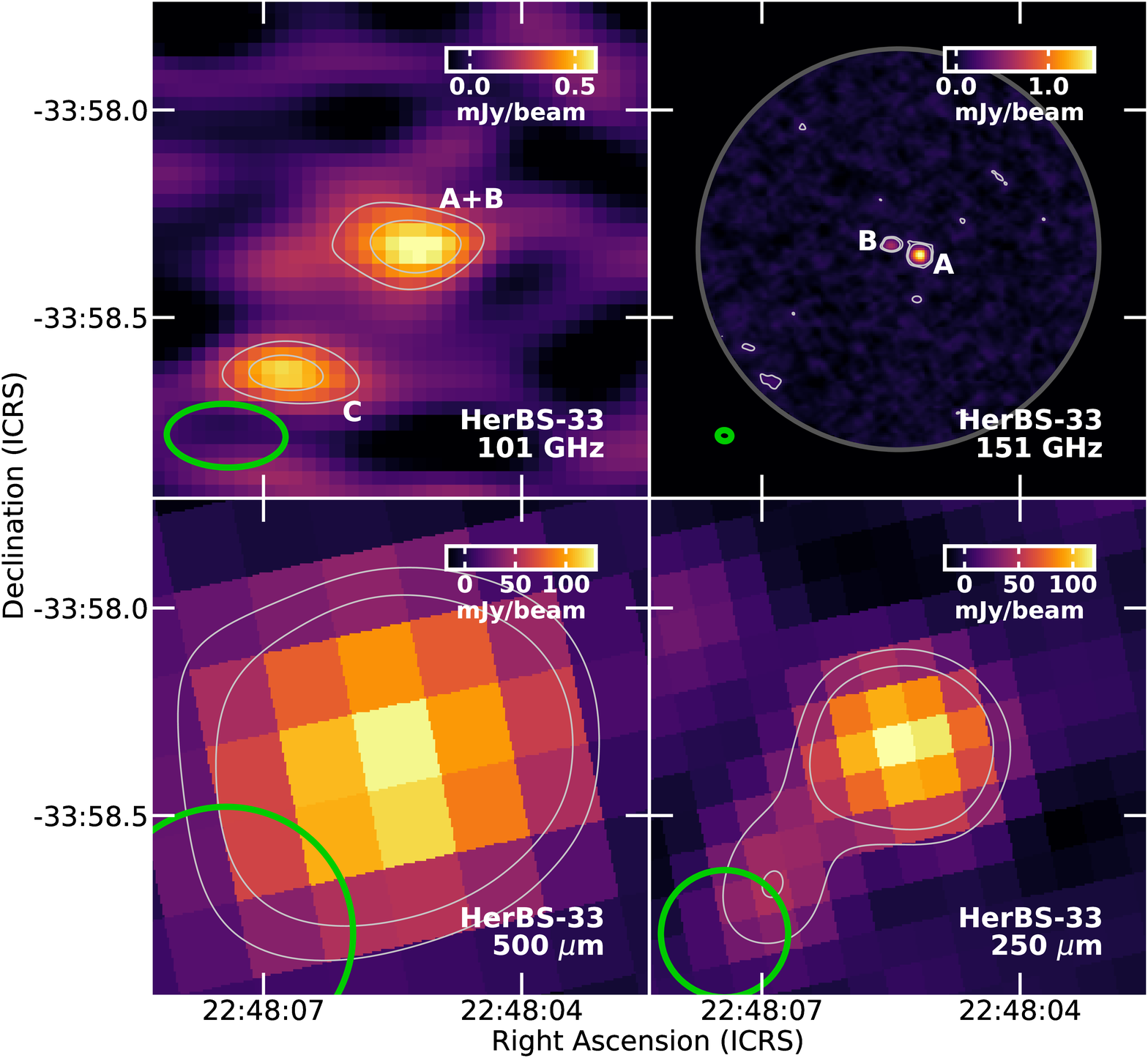}
\includegraphics[width=8cm]{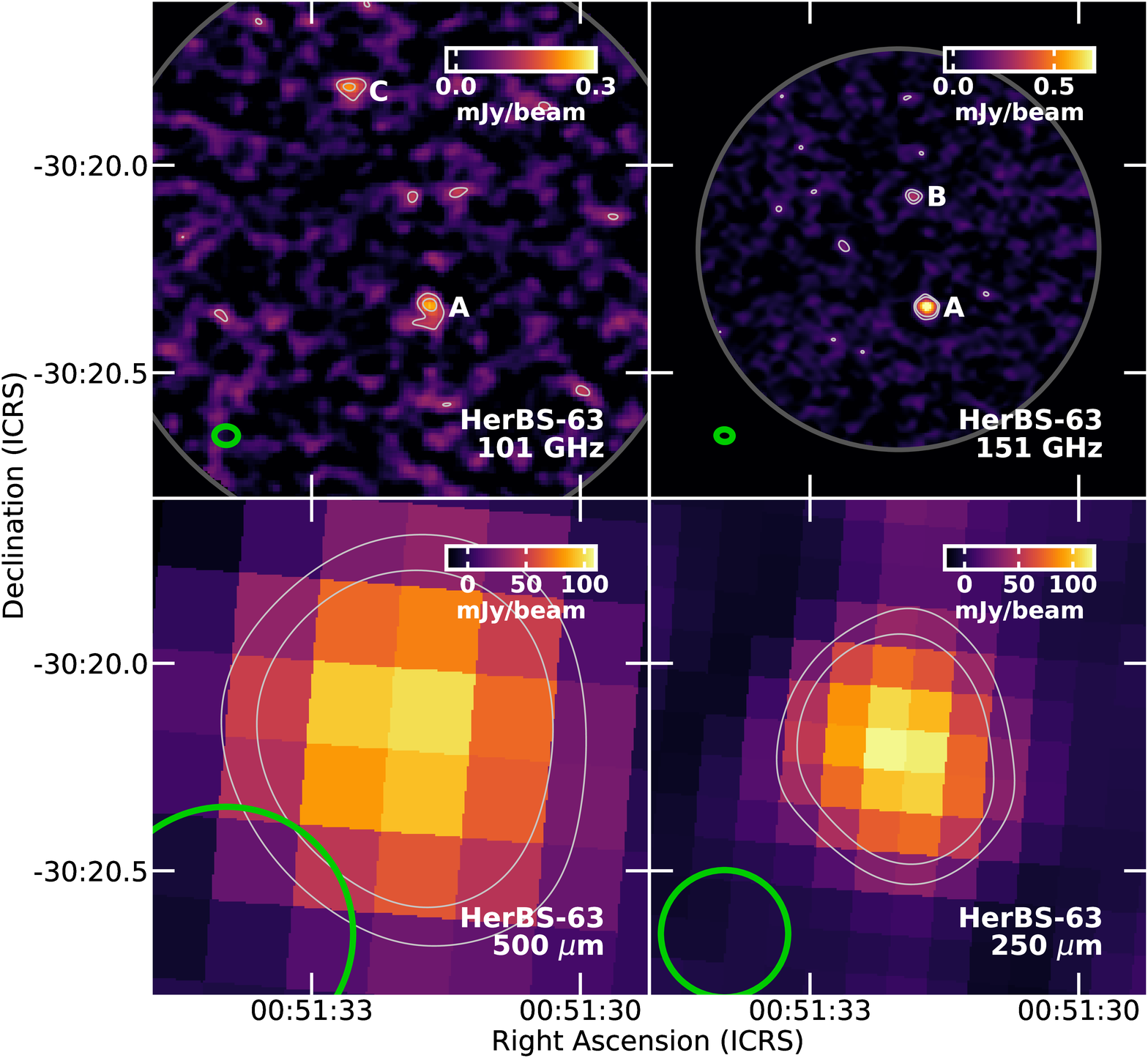}
\end{center}
\caption{Image of the HerBS-33 and HerBS-63 fields, which are the two fields that contain 101~GHz sources with no corresponding 151~GHz sources.  The colours are scaled linearly.  The contours show the emission detected at the $\geq$3 and $\geq$5$\sigma$ levels.  The green ellipses in the bottom left corner of each panel show the different beam sizes of the data.  The grey circles in the 101 and 151 GHz images show the spatial extent of the imaged regions.  The letters in the 101 and 151~GHz images correspond to the labels for the sources given in Table~\ref{t_almaphotom}.}
\label{f_herbs33-63}
\end{figure}

\begin{figure}
\begin{center}
\includegraphics[width=7cm]{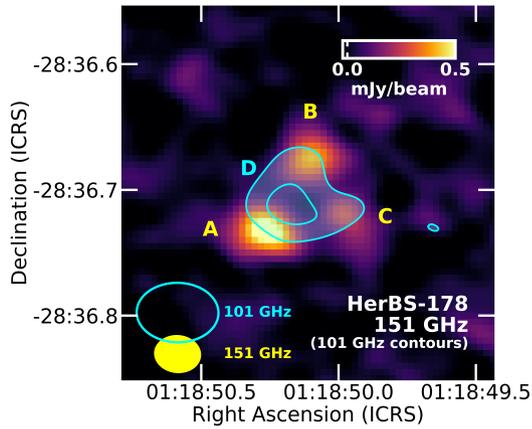}
\end{center}
\caption{Image of the sources in the HerBS-178 field.  The colour scale map shows the 151~GHz emission, while the contours show the 101~GHz emission detected at the $\geq$3 and $\geq$5$\sigma$ levels.  The two ellipses in the bottom left corner show the different beam sizes of the data.  The letters correspond to the labels for the sources given in Table~\ref{t_almaphotom}.}
\label{f_herbs178}
\end{figure}

Most continuum sources are detected in the 151~GHz images.  It is common for sources to appear fainter or to not be detected in the 101~GHz images, even though the same sensitivity levels are achieved in both the 101 and 151~GHz images from the 12~m array.  This is mainly because we are observing the rest-frame thermal dust emission, which should scale as approximately $\nu^4$ in these bands.  However, the HerBS-33 and HerBS-63 fields contain sources located significantly off-centre that are detected only at 101~GHz.  Images of these sources are visible in Figure~\ref{f_herbs33-63}.  In the case of HerBS-33, it is apparent that the off-centre source (labelled C) is visible as a separate source in the {\it Herschel} 250~$\mu$m image but that its emission is blended together with the central two sources (A and B) in the {\it Herschel} 500~$\mu$m image.  However, the off-centre source in the HerBS-63 field (labelled C) does not have a clear 250~$\mu$m counterpart.  Since both 101~GHz sources lie at the edge of the imaged 151~GHz field (where the interferometer drops to 20\% of the sensitivity at the centre of the field), it is possible that HerBS-33C and HerBS-63C were not detected at 151~GHz because of sensitivity issues.  Nevertheless, we cannot immediately rule out the possibility that the 101~GHz emission is simply brighter.  For example, it is possible that HerBS-33C and HerBS-63C are $z<2$ galaxies with AGN that produce synchrotron emission with a spectral index of $\lesssim$-0.3 that would be detectable in the 101~GHz band but not in the 151~GHz band.  The Very Large Array Sky Survey \citep[VLASS;][]{gordon2021} reported a 3~GHz source with a flux density of 1.3~mJy at the position of HerBS-63C, and the spectral slope from 3 to 101~GHz is consistent with a spectral index of $\sim$-0.3.  However, no source from that survey corresponded to the position of HerBS-33C, although the 3~GHz emission would be expected to be $\gtrsim$1.7~mJy and should have been detectable by the VLASS.

Also worth discussing is the exotic situation in the HerBS-178 field, which is shown in detail in Figure~\ref{f_herbs178}.  The 151~GHz image contains three point-like sources (labelled A, B, and C) located within 4.5~arcsec of each other.  All three of these sources have similar redshifts of $z \cong 2.656$ \citep{urquhart2022}.  However, the single detected source in the 101~GHz image lies between these three other sources and has no measured redshift or detectable line emission.  This does not appear to be an astrometry problem, as the spectral line emission in Band 3 corresponds to the location of the 151~GHz continuum emission.  To test whether the offset between the 101 and 151~GHz sources is a consequence of the difference in beam sizes between the two images, we convolved the 151~GHz image with a Gaussian function to match the beam to the 101~GHz data, but we were still able to resolve the three separate sources in the convolved image, and we did not reproduce emission that peaked in the location of the 101~GHz continuum source.  Hence, we conclude that the 101~GHz emission originates from a different location than the 151~GHz emission.

It is unclear how the 101~GHz source in the HerBS-178 field is related to the three 151~GHz sources in that field.  It could be that the D source is a foreground lensing object and that A, B, and C are all parts of an Einstein ring.  Much less likely but still possible is that A, B, and C are part of a cluster that are in front of and lensing the emission from D.  Alternately, it could be possible that the 101 and 151~GHz emission originate from different objects at the same redshift that are physically associated with each other.  The fourth data release of the Kilo-Degree Survey \citep{kuijken2019} and the fourth data release of the VISTA Kilo-degree Infrared Galaxy Survey \citep{edge2013} have reported optical and near-infrared sources within 1~arcsec of the A and B sources and $\sim$1~arcsec south of the C source, but no optical or near-infrared counterparts have been reported closer than 2~arcsec (i.e., closer than the A source) to D.  Additionally, no VLASS detection is reported near any of the sources.  More observations would be needed to understand the nature of how these objects are associated with each other.

\subsection{Extended emission}
\label{s_extended}

Most of the detected sources were unresolved, which would be expected for high-redshift objects in data with our angular resolutions.  Any arcs or ring-like features from gravitational lensing may be too small to resolve in these data.  However, a few objects do appear significantly more extended than the beam.  Since we do not see any gravitational lensing structures and since the profiles for the extended extended emission still appears approximately Gaussian, we checked the spatial extent of all of the sources detected in the 151~GHz images by fitting Gaussian functions to them using the {\sc casa} tool {\sc imfit}.  A total of 15 objects were identified as single-peaked sources with diffuse, extended structures on the basis that the FWHM of the major axes of the observed sources were both 3$\sigma$ greater and 0.6~arcsec (or two pixels) greater than the FWHM of the major axes of the beams.  The second condition was selected to avoid issues with pixelization effects that could make the beam broader, and it effectively limits us to reporting sources with deconvolved major axis FWHMs of $\gtrsim$1.7~arcsec.   A couple of examples of these types of sources are shown in Figure~\ref{f_extended} (with images of all of the sources presented in the supplemental online material), and the deconvolved dimensions for the sources are listed in Table~\ref{t_diffuse}.  Higher angular resolution observations would be needed to understand the nature of the extended emission of these sources.

We also examined whether we could calculate any approximate lensing parameters for unresolved background galaxies.  Galaxy-galaxy lenses can be characterized by the singular isothermal sphere model, for which the critical Einstein radius is given by
\begin{equation}
\frac{\theta_{\rm E}}{\rm arcsec} \simeq  1.4\times \left ( \frac{\sigma_{\rm V}}{220\,{\rm km\,s}^{-1}}  \right )^2 \frac{D_{\rm LS}}{D_{\rm S}}
\end{equation}
where $\sigma_{\rm V}$ is the velocity dispersion, $D_{\rm LS}$ is the lens-source angular diameter distance, and $D_{\rm S}$ is the angular diameter distance from Earth to the source \citep[e.g., ][]{serjeant2010}. Unless the lens is close to the source, $D_{\rm LS}\simeq D_{\rm S}$ \citep[e.g., ][]{serjeant2012}, so galaxy-galaxy gravitational lensing will tend to yield Einstein radii of $\sim$0.5-2.0~arcsec regardless of other factors.  For foreground galaxy clusters acting as gravitational lenses, the critical radii will be much larger, and highly magnified objects will appear as arcs that would be resolvable in our ALMA data. It is not yet possible to determine whether our resolved sources are extended unlensed systems, massive galaxy-galaxy lensing systems, or gravitationally lensed arcs, but one interpretation of the inferred magnification distribution in \cite{urquhart2022} is that some fields contain lensing galaxy clusters.

\begin{figure}
\begin{center}
\includegraphics[width=7cm]{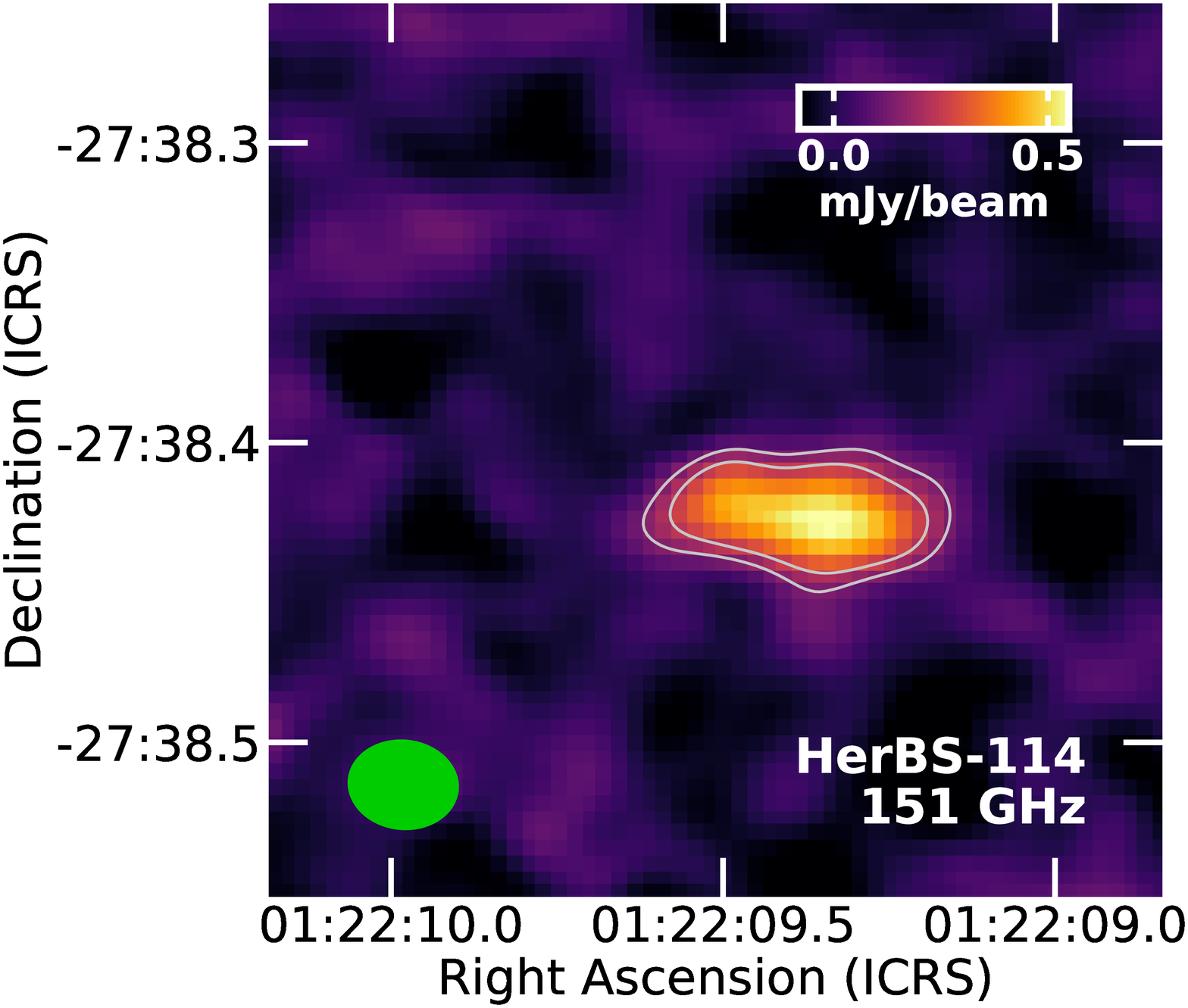}
\\
\ \\
\includegraphics[width=7cm]{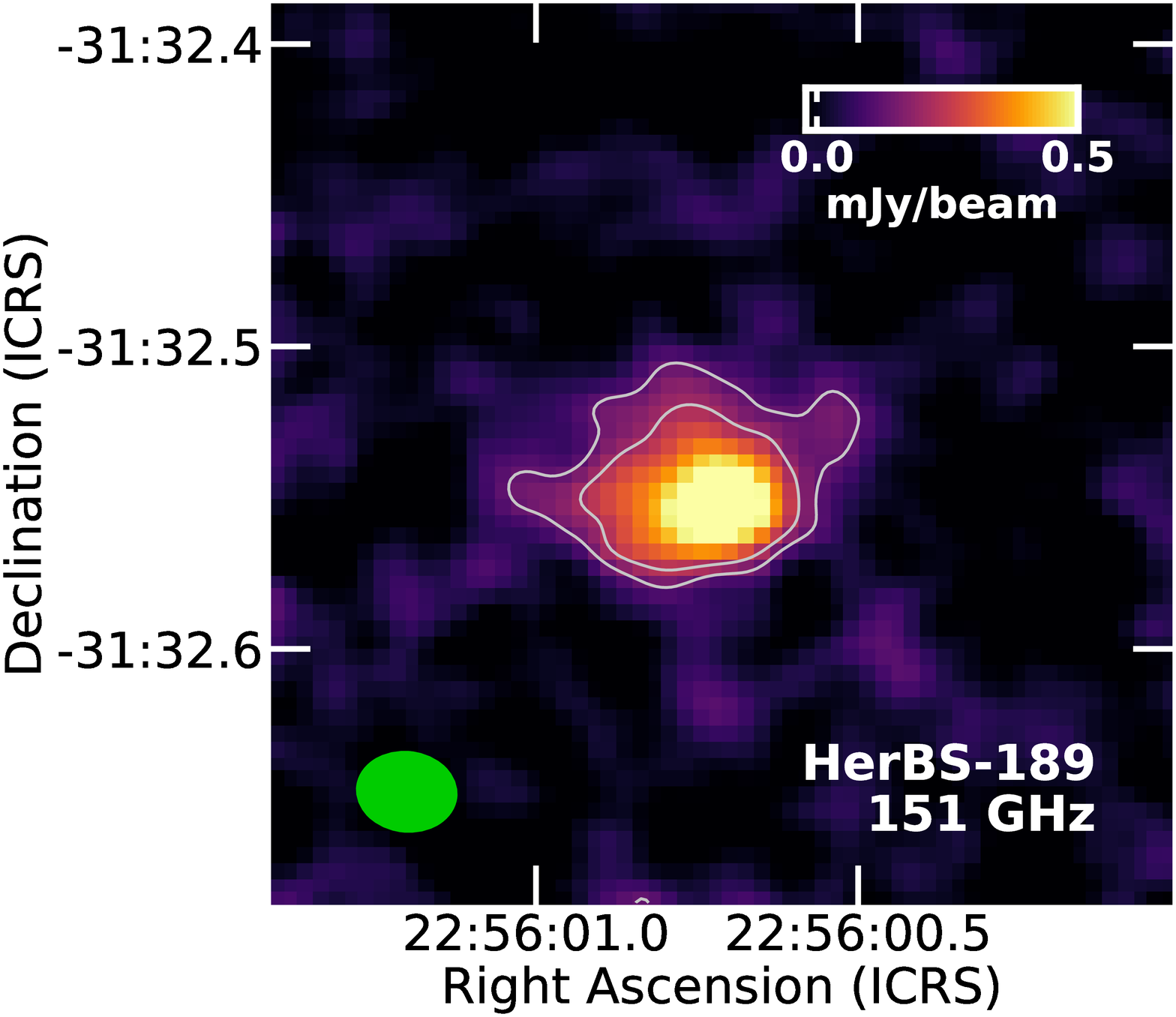}
\end{center}
\caption{Two 151~GHz images of example sources (HerBS-114 and HerBS-189) with emission that appear significantly more extended than the $\sim$2~arcsec beam size.  The colours are scaled linearly.  The contours show the emission that is detected at the $\geq$3 and $\geq$5$\sigma$ levels.  The green ellipses in the bottom left corners of the images show the beam sizes.}
\label{f_extended}
\end{figure}

\begin{table}
\caption{Dimensions of single-peaked sources with diffuse, extended emission}
\label{t_diffuse}
\begin{center}
\begin{tabular}{@{}lcc@{}}
\hline
Source &
  Gaussian &
  Physical \\
&
  FWHM &
  dimensions \\
&
  (arcsec)$^a$ &
  (kpc)$^b$ \\
\hline
HerBS-21A &
  2.0 $\times$ 0.5 &
  15 $\times$ 4 \\
HerBS-37 &
  2.1 $\times$ 1.0 &
  17 $\times$ 8 \\
HerBS-39 &
  2.2 $\times$ 1.4 &
  17 $\times$ 11 \\
HerBS-41C &
  2.1 $\times$ 0.4 &
  \\
HerBS-56D &
  2.0 $\times$ 1.0 &
  \\
HerBS-80B &
  1.8 $\times$ 0.7 &
  16 $\times$ 6 \\
HerBS-97A &
  2.6 $\times$ 1.7 &
  \\
HerBS-102B &
  1.9 $\times$ 0.2 &
  \\
HerBS-107 &
  2.2 $\times$ 0.6 &
  18 $\times$ 5 \\
HerBS-114 &
  2.8 $\times$ 0.5 &
  \\
HerBS-123 &
  2.4 $\times$ 1.8 &
  21 $\times$ 16 \\
HerBS-159A &
  2.5 $\times$ 1.6 &
  21 $\times$ 13 \\
HerBS-163B &
  2.7 $\times$ 0.7 &
  \\
HerBS-189 &
  2.0 $\times$ 1.4 &
  16 $\times$ 11 \\
HerBS-209A &
  2.5 $\times$ 1.0 &
  21 $\times$ 8 \\
\hline
\end{tabular}
\end{center}
$^a$ This is the FWHM of the deconvolved sources.  The uncertainties are $\lesssim$0.3~arcsec except for HerBS-209A, where the uncertainties are 0.4~arcsec for the major axis and 0.7~arcsec for the minor axis.\\
$^b$ These dimensions are calculated using the Gaussian FWHMs and a spatially flat $\Lambda$CDM comology with $H_0$=67.4 km s$^{-1}$ Mpc$^{-1}$ and $\Omega_M$=0.315 \citep{planck2020}.
\end{table}

Five sources (HerBS-68, HerBS-131A, HerBS-138, HerBS-141, and HerBS-170) have two peaks that are separated by $\sim$3~arcsec  (or $\sim$3.5 arcsec in the case of HerBS-170), which is $\sim$1.5$\times$ the FWHM of the beam.  Figure~\ref{f_doublelobe} shows two examples of these sources (with images of the other sources presented in the supplemental online material).  In these situations, it is generally unclear whether the two point sources are part of one larger structure, if they are two physically associated but distinctly separate objects that lie at the same redshift, or if they are two sources at different redshifts that happen to lie along the same line of sight.  The one double-peaked source where we have a clear understanding of the relation between the two peaks is HerBS-138.  \citet{urquhart2022} detected different lines at different frequencies for the two peaks, which indicates that they correspond to two different, unassociated sources at different redshifts (although the line detections only allowed for the determination of an accurate redshift for source B).

\begin{figure}
\begin{center}
\includegraphics[width=7cm]{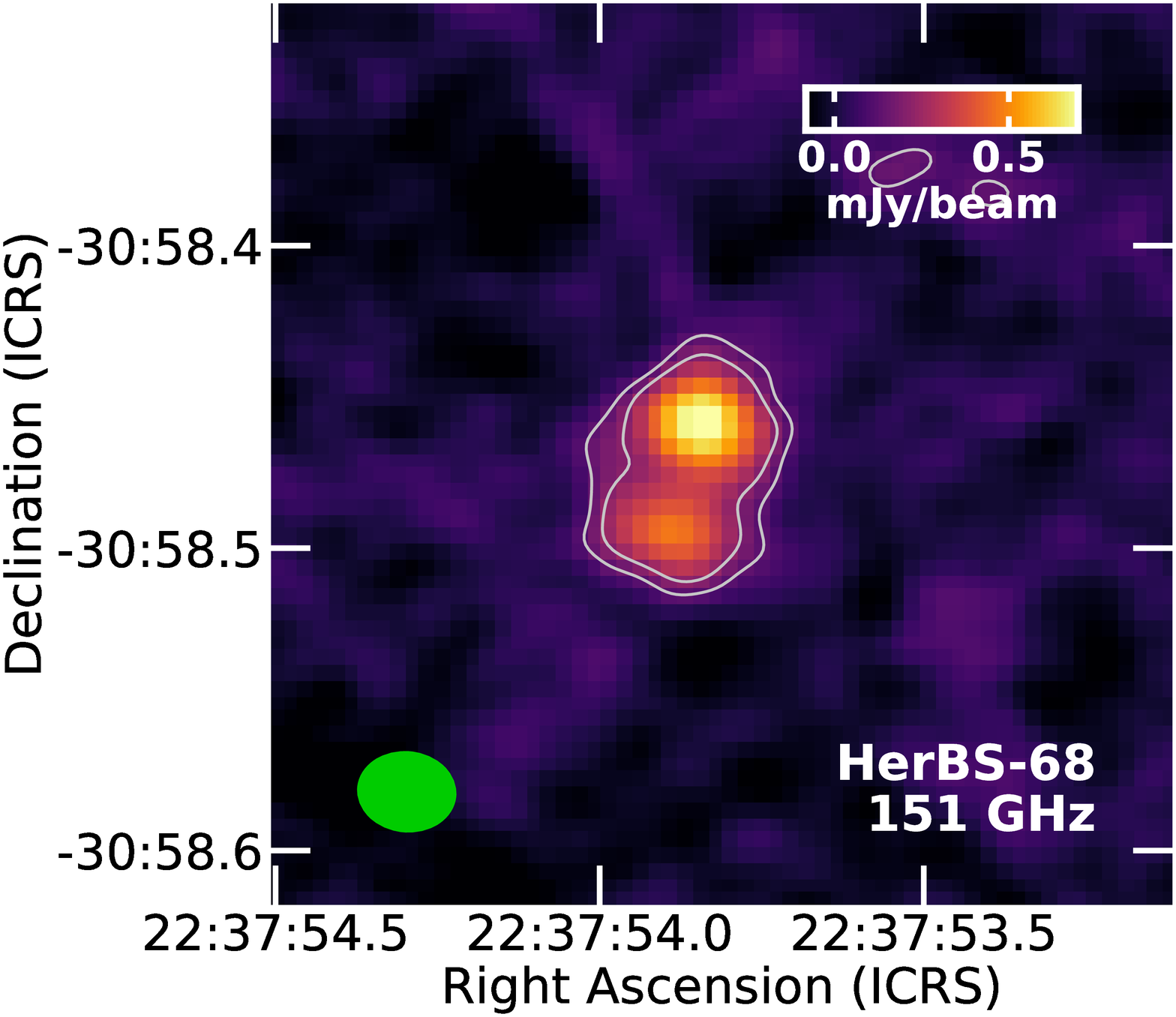}
\\
\ \\
\includegraphics[width=7cm]{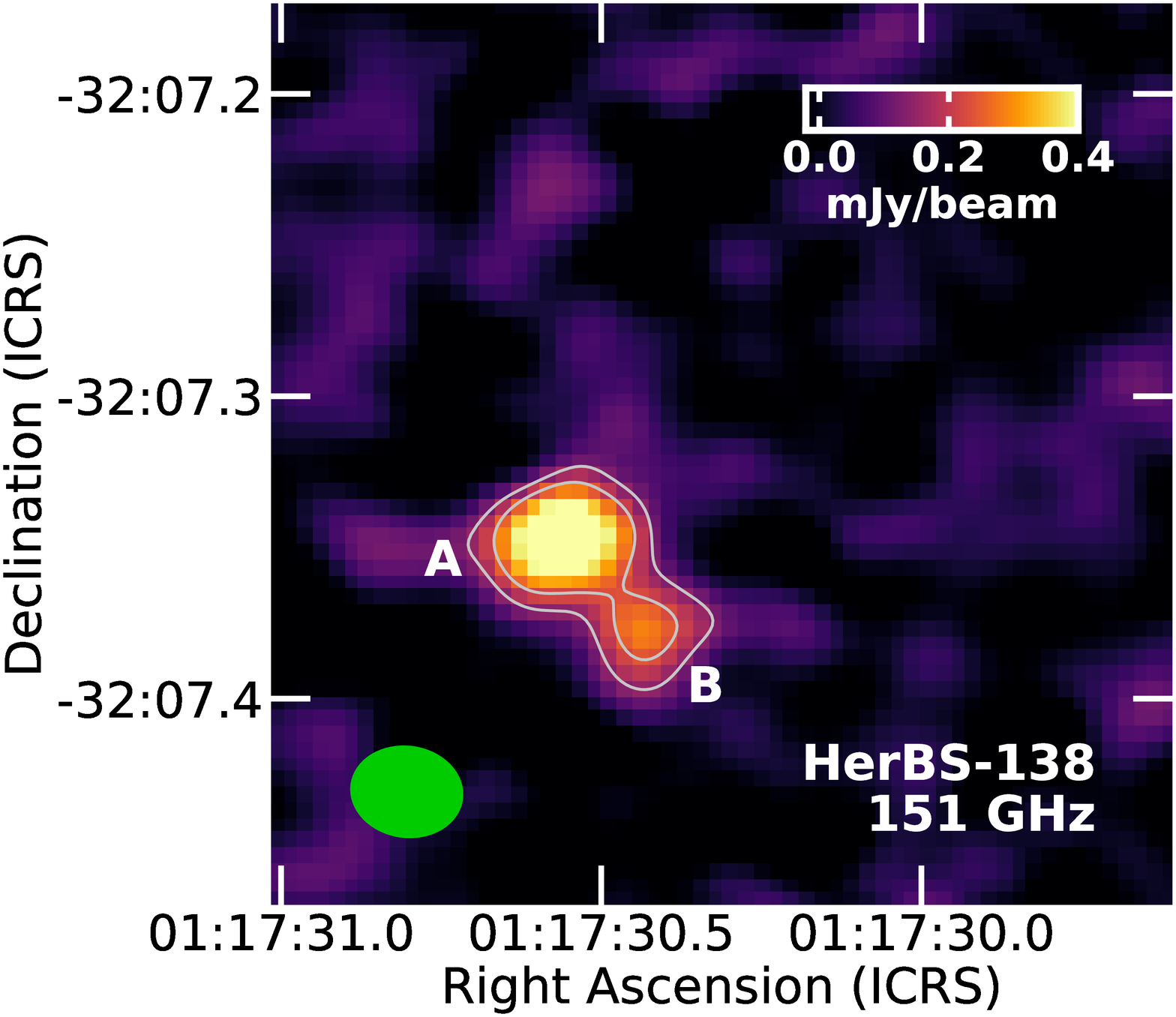}
\end{center}
\caption{Two example 151~GHz images of fields with double-lobed sources.  In most cases, it is ambiguous whether the lobes are two sources at different redshifts, are two separate but physically associated objects, or are two parts of one larger structure.  However, in the case of HerBS-138, the ALMA spectra demonstrated that the two lobes are at different redshifts (and hence they are labelled as A and B to indicate that they are distinct sources).  The colours are scaled linearly.  The contours show the emission that is detected at the $\geq$3 and $\geq$5$\sigma$ levels.  The green ellipses in the bottom left corners of the images show the beam sizes.}
\label{f_doublelobe}
\end{figure}

\begin{figure}
\begin{center}
\includegraphics[width=7cm]{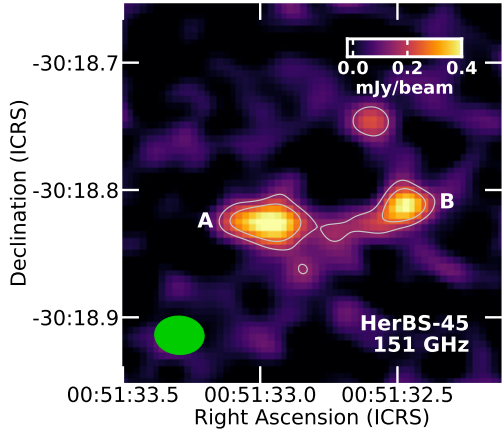}
\end{center}
\caption{The 151~GHz image of the HerBS-45 field, where the two brightest sources (labelled A and B) are connected by a thin structure detected at the $>$3$\sigma$ level.  The colours are scaled linearly.  The contours show the emission that is detected at the $\geq$3 and $\geq$5$\sigma$ levels.  The green ellipses in the bottom left corner of the image shows the beam sizes.}
\label{f_herbs45}
\end{figure}

It is also worth noting that, as shown in Figure~\ref{f_herbs45} the centre of HerBS-45 contains two sources (labelled A and B) separated by $\sim$6.5~arcsec that appear to be connected by a thin, filamentary structure detected at the $>$3$\sigma$ level in the 151~GHz image.  It is not clear if the two objects are actually physically connected.  Note that the only object in this field with a measured redshift is A; line emission was not detected from the B source or the filamentary structure.

\section{Multiplicities}
\label{s_mult}

Submillimetre and millimetre interferometers, including ALMA, have been essential for locating or resolving individual infrared and millimetre sources that had been detected with single-dish telescopes, including {\it Herschel}, the Atacama Pathfinder Experiment, the Atacama Submillimeter Telescope Experiment, the James Clerk Maxwell Telescope, the Large Millimeter Telescope, and the South Pole Telescope.  When the infrared, submillimetre, and millimetre sources initially detected in deep fields with $\gtrsim 15$ arcsec resolutions are then re-observed using interferometers with angular resolutions of $\lesssim$3~arcsec, the new data often show that a significant fraction (potentially up to 70\%) of the $\gtrsim$15~arcsec submillimetre sources have multiple counterparts \citep[e.g.,][]{hodge2013, karim2013, bussmann2015, cowie2018, hill2018, stach2018}, although this depends on the sensitivity and angular resolutions of the interferometers used in the follow-up observations.  Additional observational and theoretical studies have demonstrated how confusion could affect at least 50\% of submillimetre galaxies identified in data with $\gtrsim$15~arcsec beams \citep[e.g.,][]{hayward2013,hayward2018,scudder2016,scudder2018}, and this has implications for analyses looking at the overall properties of this class of sources, such as their luminosity functions \citep{karim2013}.

Importantly, the HerBS sources observed in our study were not blindly selected by their infrared or submillimetre flux densities like the sources in most other deep surveys.  Instead, the HerBS sources were selected using both a flux density threshold and a photometric redshift threshold, with additional steps applied to remove blazars and foreground galaxies \citep{bakx2018}.  These steps are intended to optimize for the selection of the brightest high-redshift infrared sources, including gravitational lenses (which will mainly be unresolved in our data), individual HLIRGs, and infrared-bright protoclusters \citep{negrello2017, bakx2018}.  Models of the HerBS fields specifically indicate that between 57 and 82\% of the fields with high {\it Herschel} 500~$\mu$m flux densities are expected to correspond to gravitational lenses \citet{bakx2018, bakx2020viking}.  Consequently, the multiplicity results for the BEARS fields are expected to differ from other surveys.

\subsection{Statistics from the BEARS fields}

Table~\ref{t_multstat} lists the number of sources we detected within the {\it Herschel} 500~$\mu$m beam in each field.  These are generally sources with peak brightnesses detected at above the 5$\sigma$ level (or above $\sim$0.2 mJy beam$^{-1}$) in either the 101 or 151~GHz images, although it also includes two sources (HerBS-80B and HerBS-122B) with peaks only detected at the 3-5$\sigma$ threshold but that have spectral line counterparts listed by \citet{urquhart2022}.  Double-lobed sources are treated as single sources except for HerBS-138, where the two lobes are known to correspond to objects at different redshifts.  Fields with one source are subdivided as to whether they have a spectroscopic redshift.  Fields with two sources are subdivided into four subgroups based on whether both sources have measured spectroscopic redshifts and whether the source are at the same or different redshifts\footnote{\citet{urquhart2022} only list redshifts for one of the two sources in the HerBS-122, HerBS-135, HerBS-138, and HerBS-146 fields and list no redshifts for either source in the HerBS-144 field.  Spectral lines were detected for both sources in each of these fields, but the detected lines were insufficient for unambiguously identifying the redshifts of at least one of the sources.  However, because the lines for the sources in each of these fields are at very different frequencies, it is clear that the sources are at different redshifts, so we can still list them in Table~\ref{t_multstat} as such.}.  The fields with two sources are also subdivided by the ratio of the brighter source to the fainter source, which is important for interpreting the contributions of the fainter source to the SED integrated across the field.  Fields with three or more sources are subdivided into similar groups, although we do not have any fields with three or more sources that all have measured redshifts.

\begin{table}
\caption{Multiplicity information}
\label{t_multstat}
\begin{center}
\begin{tabular}{@{}lc@{}}
\hline
Number of detected sources within ALMA &
  Number\\
~ ~ fields &
  of fields\\
\hline
1 & 39 \\
~ ~ (source with $z_{\text{spec}}$) & 31 \\
~ ~ (source without $z_{\text{spec}}$) & 8 \\
\noalign{\vskip 0.75em}
2 & 34 \\
~ ~ (associated $z_{\text{spec}}$) & 6 \\
~ ~ (different $z_{\text{spec}}$) & 6  \\
~ ~ ($z_{\text{spec}}$ for only one source) & 13 \\
~ ~ (no $z_{\text{spec}}$) & 9 \\
\noalign{\vskip 0.75em}
~ ~ ($\geq$80\% of total 151~GHz emission from brighter source) & 7 \\
~ ~ (67-80\% of total 151~GHz emission from brighter source) & 9 \\
~ ~ ($\leq$67\% of total 151~GHz emission from brighter source)$^a$ & 18 \\
\noalign{\vskip 0.75em}
$\geq$ 3 & 6 \\
~ ~ (associated $z_{\text{spec}}$ for $\geq$ 2 sources) & 2 \\
~ ~ (different $z_{\text{spec}}$ among the sources) & 1 \\
~ ~ ($z_{\text{spec}}$ for only one source) & 3 \\
~ ~ (no $z_{\text{spec}}$) & 1 \\
\noalign{\vskip 0.75em}
$[$At least one detected source outside the & 6 \\
~ ~ central 35~arcsec diameter region$]$ & \\
\hline
\end{tabular}
\end{center}
$^a$ The relative fraction of the total emission from the brightest (A) source in the HerBS-138 field was estimated based on the ratio of the peak brightnesses of the two sources.
\end{table}

Approximately half of the fields contain just one detected source in the ALMA data.  These fields are largely consistent with what would be expected if the sources are unresolved gravitationally lensed galaxies or are HLIRGs, although additional observations at higher angular resolutions would be needed to confirm the nature of these sources.  Eight fields contain two or more objects that have similar spectroscopic redshifts.  These could be physically associated infrared-bright galaxies or single sources lensed by a foreground object, potentially a cluster.  Seven fields contain sources at different redshifts which are more likely to be chance alignments.  As for the fields with multiple sources with incomplete redshift information, it is unclear exactly how to interpret the nature of these sources.

\begin{figure}
\includegraphics[width=8.5cm]{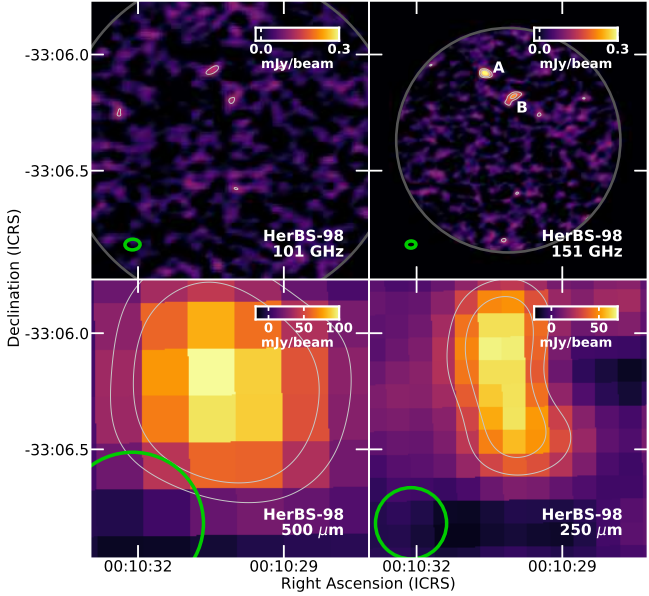}
\caption{Images of the HerBS-98 field, which contains two detected 151~GHz sources located at the northern edge of the field observed by ALMA.  The centre of the field corresponds to the coordinates of the 250~$\mu$m source from the H-ATLAS catalogue, even though the 500~$\mu$m is clearly offset from this position. The colours are scaled linearly.  The contours show the emission that is detected at the $\geq$3 and $\geq$5$\sigma$ levels.  The green ellipses in the bottom left corner of each panel show the different beam sizes of the data.  The grey circles in the 101 and 151 GHz images show the spatial extent of the imaged regions.  The letters in the 151~GHz image correspond to the labels for the sources given in Table~\ref{t_almaphotom}.}
\label{f_herbs98}
\end{figure}

The six fields where we identified sources within the ALMA images but outside the central 35~arcsec diameter region (corresponding to the FWHM of the {\it Herschel} 500~$\mu$m beam) are listed in a separate category in Table~\ref{t_multstat}.  This is because it is not always clear whether all of the sources detected in the ALMA images were confused in the {\it Herschel} 500~$\mu$m beam or how to interpret the ALMA results for the multiplicity analysis.  In the HerBS-41, HerBS-63, HerBS-118, and HerBS-186 fields, the {\it Herschel} 250~$\mu$m images appear to contain single sources, but in the HerBS-33 and HerBS-98 fields, it is possible to see emission in the {\it Herschel} 250~$\mu$m images corresponding to the ALMA sources outside the central 35~arcsec, and it is also apparent that the emission from the multiple sources was blended together in the {\it Herschel} 500~$\mu$m beam.  HerBS-33 is already shown in Figure~\ref{f_herbs33-63}; HerBS-98 in shown in Figure~\ref{f_herbs98}.  The coordinates for HerBS-98 from the H-ATLAS catalogues are based on the locations of the sources detected at 250~$\mu$m \citep{valiante2016}, so while ALMA observed the central coordinates of HerBS-98 (J001030.1-330622) listed by the H-ATLAS catalogue, the 500~$\mu$m source and the detected 151~GHz sources are offset from this position.  Note that HerBS-98 is the only field in our sample with this specific coordinate issue.

\begin{figure}
\includegraphics[width=8cm]{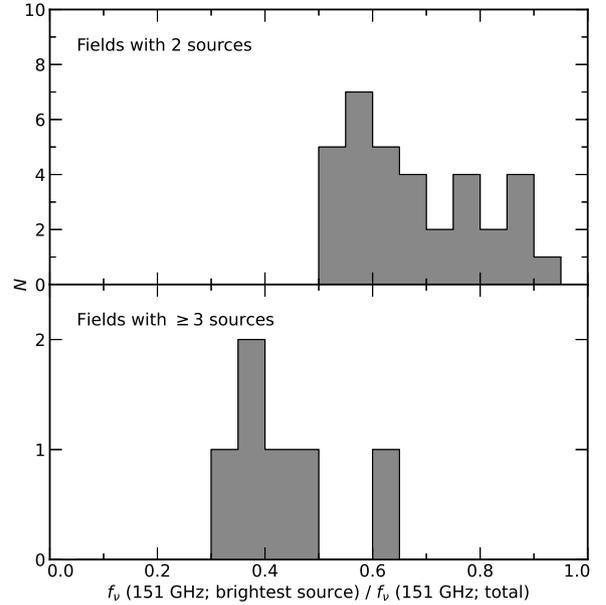}
\caption{Histograms of the fraction of the integrated 151~GHz emission that comes from the brightest source in fields with two detected 151~GHz sources (top) and fields with three or more detected 151~GHz sources (bottom).  The six ALMA fields with sources falling outside the central 35~arcsec were excluded from these histograms.}
\label{f_multratiohist}
\end{figure}

When identifying fields as containing multiples, we have not placed any restrictions on the 151~GHz flux density ratios of the second and first brightest sources or the ratio of the brightest source to the total flux density.  This means that the multiplicity results could depend on the sensitivities achieved in our data, with more fields appearing to contain multiples when the sensitivities improve.  Hence, it would be useful to assess the significance of the emission from the other sources detected in any field as compared to the brightest source.  This is why we separated the fields with two detected 151~GHz sources in Table~\ref{t_multstat} into three groups based on the ratio of the 151~GHz flux density of the brightest source to the total 151~GHz emission.  We also have provided histograms of this ratio in Figure~\ref{f_multratiohist}.  In the fields with two sources, the median ratio of the brighter source to the total emission is 0.65 (or, alternately, the average ratio of the brightnesses of the brighter source to the fainter source in these fields is $<$2/1 for half of these fields).  In contrast, we only found seven fields with two sources where the ratio of the brighter source emission to total emission is $>$0.80 (or where the ratio of the emission from the brighter source to the fainter source is $>$4/1).  In fields with three or more sources, however, the A source never produces more than 62\% of the total 151~GHz emission; the emission in these fields is truly fragmented.  To summarize, in the vast majority of fields we have identified as containing multiple sources, the A source only produces $<$80\% of the total 151~GHz emission, with other sources producing a significant amount of emission in at least the ALMA bands.

\begin{figure}
\begin{center}
\includegraphics[width=8cm]{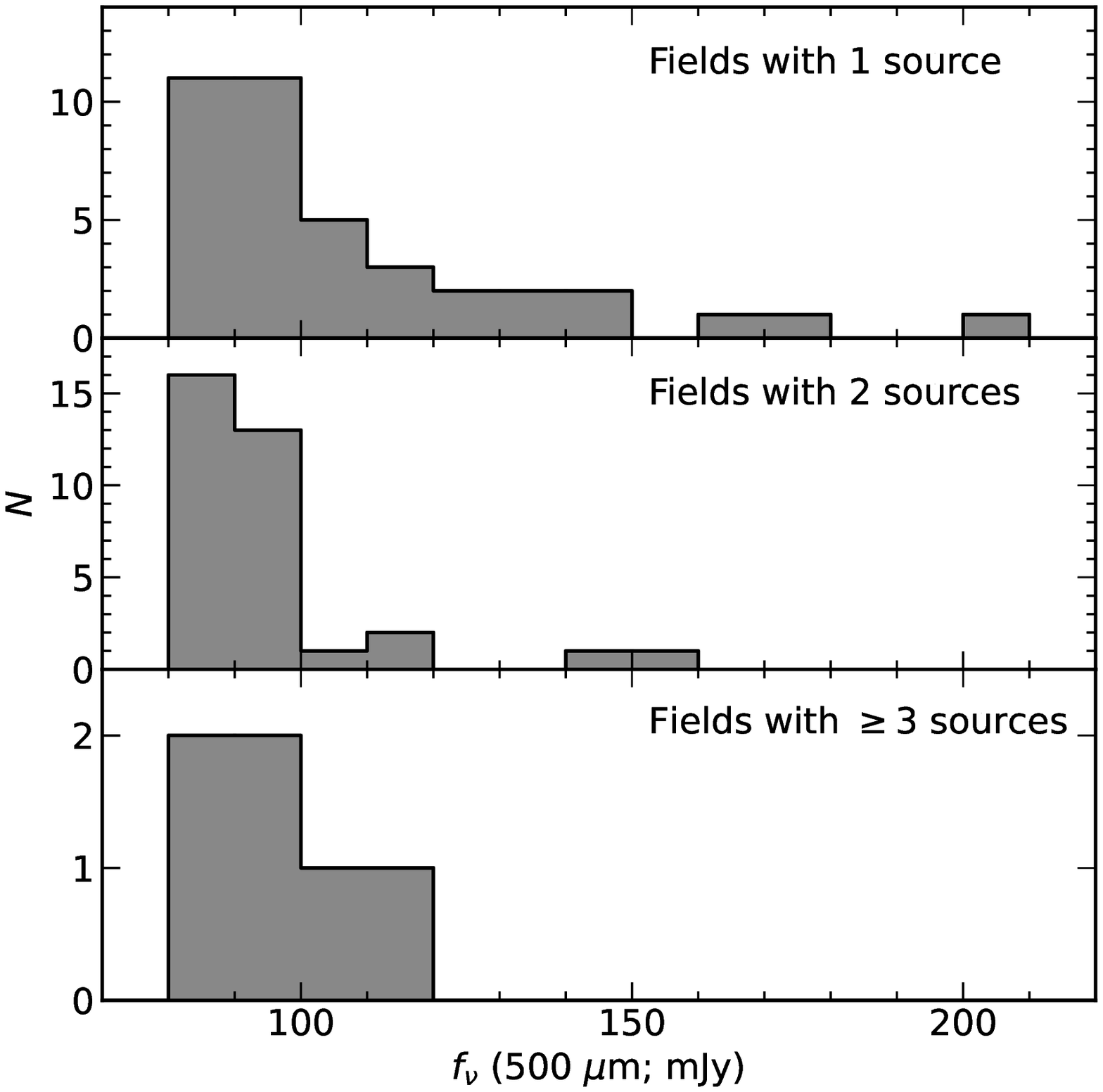}
\end{center}
\begin{center}
\includegraphics[width=8cm]{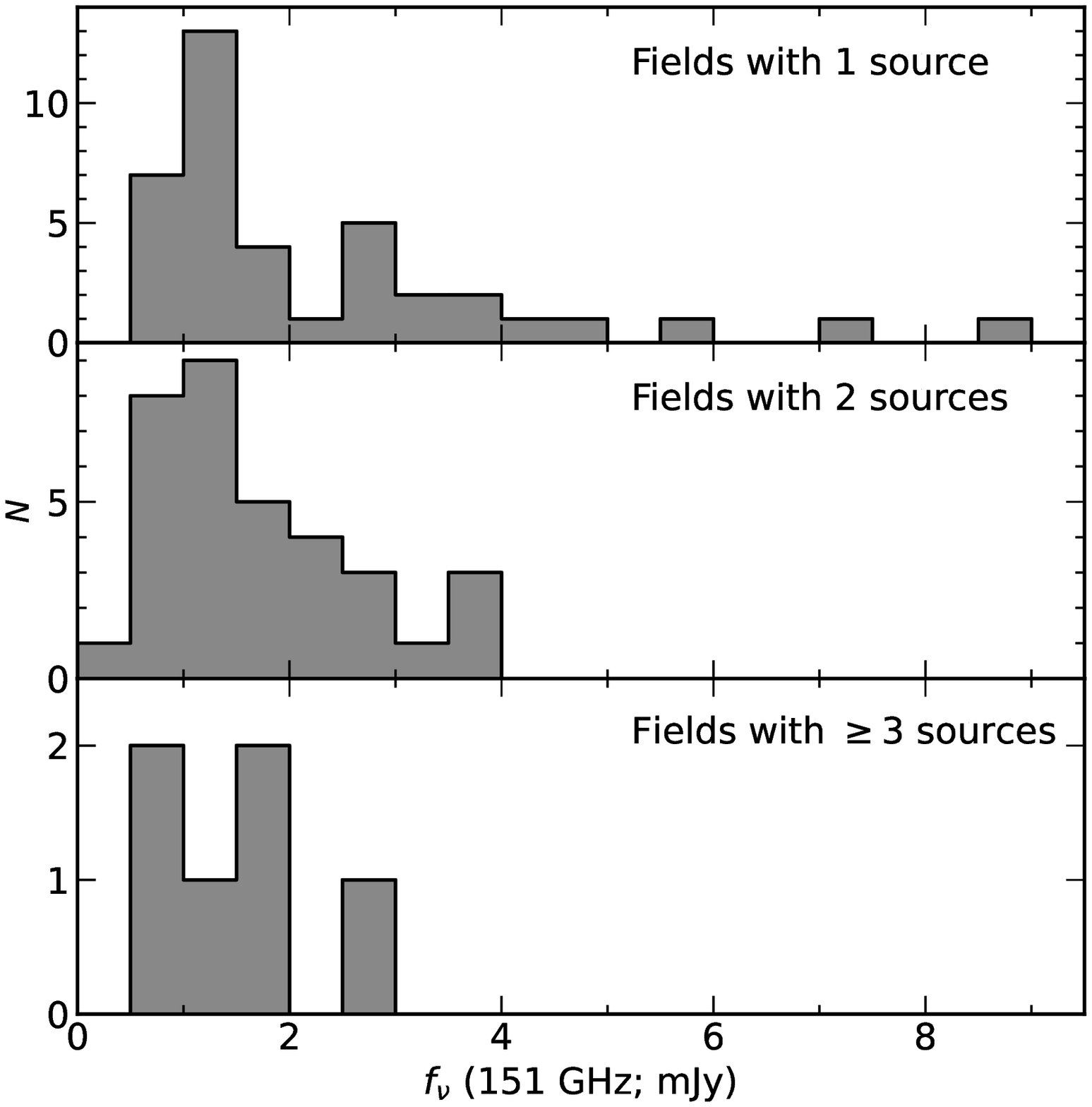}
\end{center}
\caption{Histograms of the total {\it Herschel} 500~$\mu$m flux densities (top) and total 151~GHz flux densities (bottom) measured within the central 35~arcsec diameter regions for different subsets of fields based on the number of sources detected in those fields within the ALMA data.  The six ALMA fields with sources falling outside the central 35~arcsec were excluded from these histograms.  The minimum limits on the histogram values are 80~mJy at 500~$\mu$m (which is set by the sample selection criteria) and 0.20 mJy at 151~GHz (which is the 5$\sigma$ detection limit).}
\label{f_multhist}
\end{figure}

To understand source confusion as a function of the integrated flux densities within each field, we created separate histograms in Figure~\ref{f_multhist} of the 500~$\mu$m and 151~GHz flux densities (as measured for all sources within the central 35~arcsec) for fields with one, two, or three or more sources (excluding the fields with one detected source outside the central 35~arcsec diameter region).  The distribution of integrated flux densities for the three separate sets of fields appear somewhat similar in this plot, although the fields with higher flux densities tend to contain single sources.  Applying Kolomorov-Smirnov tests to the distributions, we calculated a 7\% probability that the fields with 1 source and fields with multiple sources have the same distributions of 500~$\mu$m flux densities and a 63\% probability that the data for fields with 1 source and the data for fields with multiple sources are drawn from the same distributions of 151~GHz flux densities.  These results indicate that the majority, but not all, of the fields with the highest flux densities contain single ALMA sources.

\subsection{Comparisons to other multiplicity studies}

The multiplicity results from our sample do not quite match the results from most other surveys for various reasons.  First, the use of photometric redshifts and the removal of foreground galaxies and blazars when constructing the HerBS sample should help for identifying gravitational lenses (which should be mostly unresolved in our data) or HLIRGs \citet{negrello2017, bakx2018}.  The {\it Herschel} sources that meet both our flux density and colour criteria are relatively rare and would be unlikely to appear in other surveys.  For reference, our sample of 85 sources comes from a field that is 290 deg$^2$ in size.  Secondly, differences in the beam sizes of the single-dish data used to identify the locations of bright submillimetre sources would lead to variations in the multiplicities observed by interferometers.  Thirdly, differences in the beams used in the interferometric follow-up observations could affect whether multiple sources are resolved or unresolved in those data.  Fourth, choosing to either deblend multilobed sources or treating them as single sources could affect the multiplicity results.  Since, with the exception of HerBS-138, we report multi-lobed sources as single sources, the fraction of fields that we identify with multiple sources may be lower than what is identified in studies that deblend such sources.  Finally, differences in observing depth achieved between our ALMA observations and other surveys would affect how many sources could be detected in any field.  We have reported detections of $\geq$0.20 mJy at 151~GHz, which is equivalent to $\sim$3-6~mJy at 345~GHz (870~$\mu$m) for a modified blackbody with an emissivity index $\beta$ between 1 and 2.  Other multiplicity studies are based on objects selected at $\sim$345~GHz with detection limits ranging from 0.7 to 6~mJy.

Our finding that $\sim$50\% of our fields contain multiple sources is consistent with the results for 870~$\mu$m selected sources from APEX observations studied by \citet{hodge2013}, even though they would have been more sensitive to fainter sources.  Additionally, our results are consistent with the 850~$\mu$m selected sources with flux densities $\geq$9~mJy from the JCMT studied by \citet{stach2018}, although the multiplicity fraction was lower for JCMT sources with fainter flux densities, probably because of source detection issues.  In contrast, \citet{cowie2018} and \citet{hill2018} found that only $\sim$15\% of their fields had multiple sources.  Both worked with fields selected from JCMT 850~$\mu$m observations that have a 14~arcsec beam, and the smaller beam size would contribute to the lower multiplicities.  The integrated 850~$\mu$m flux densities in many of the fields observed by \citet{cowie2018} were relatively close to the detection threshold in their ALMA data, and they indicated that, if that emission was divided into multiple ALMA sources, they may have had difficulty identifying all of those sources, which could also explain their lower multiplicities fraction.  \citet{hill2018} achieved sensitivity levels effectively equivalent to ours when adjusting for frequency, but they applied a requirement that fields would only be identified as multiples if the ratio of the brightest to second brightest source was $>$2, which would also explain their lower multiplicities fraction.  

Meanwhile, \citet{bussmann2015}, \citet{scudder2016}, and \citet{scudder2018} found a significantly higher fraction of their fields contained multiple sources.  The beams of the data from the follow-up observations in all of these studies are smaller than the beams in our data, which is one reason why these other studies measured higher multiplicity fractions.  The beams in the \citet{bussmann2015} data were 0.45~arcsec, and they also deblended multi-lobed structures in their ALMA data while we generally counted such structures as single objects.  Both of these aspects of their source identification led them to identifying a higher percentage ($\sim$70\%) of their fields as containing multiples.  If the \citet{bussmann2015} fields were observed using a 2~arcsec beam and sources within 3~arcsec of each other were not deblended, then the fraction of fields that would be identified as containing multiple sources would only be 34\%, which would actually be below what we measured.  Also note that \citet{bussmann2015} selected fields based on {\it Herschel} 500~$\mu$m data with the same beam size as ours, but while the catalogue that is the basis for our sample was created using the 250~$\mu$m source positions to effectively deblend the 500~$\mu$m emission, the sample used by \citet{bussmann2015} did not deblend the 500~$\mu$m emission before selecting their ALMA targets, which would also contribute to the higher percentage of fields with multiple sources that they identified.  \citet{scudder2016} and \citet{scudder2018} observed fields identified in {\it Herschel} 250~$\mu$m data, which has a smaller beam and should be resolved into fewer multiples.  However, their multiplicity results are based on identifying counterparts in 3.6 and 24~$\mu$m {\it Spitzer} Space Telescope data.  While the beam sizes of these mid-infrared data are not significantly better than ours, their 3.6~$\mu$m data in particular generally contained more detections per unit area, which could be why they obtain the relatively high number of 95\% of fields containing multiple sources.

Other aspects of the nature of the multiplicity in our sample are also different from what has been found in samples selected purely by flux density.  Most notably, many of the brightest sources in our sample, including 11 of the 15 brightest at 500~$\mu$m, are single sources.  If a second source below our 5$\sigma$ detection threshold is present in any of these 11 fields, it would contribute $\lesssim$20\% of the total 151~GHz emission.  In contrast, \citet{hodge2013} and \citet{karim2013} in their 870 $\mu$m selected sources from APEX (which has a 19~arcsec beam) and \citet{stach2018}, in their 850~$\mu$m selected sources from the JCMT (which has a 15~arcsec beam) found a strong tendency for the brightest sources in their sample to be multiple systems.  Additionally, the models by \citet{hayward2013} indicate that, in a 15~arcsec beam, all sources with 850~$\mu$m flux densities $>$8~mJy (which would be equivalent to $\sim$0.4-0.8~mJy at 151~GHz) should be multiple systems.  It is particularly notable that our sources were selected using a 35~arcsec beam and that we placed no restrictions on the ratio of the brightest to second brightest sources when counting multiples. This should bias our results towards identifying fields with more multiples in cases where the brightest detected sources have very high flux densities relative to the noise levels.  However, in fields where the brightest sources are detected at just above the 5$\sigma$ level, we may not be able to identify additional sources that may only be slightly fainter (for example, have flux densities that are between 50 and 100\% of the flux densities of the brightest sources) but that fall below our detection threshold.

\citet{hodge2013} and \citet{stach2018} also reported non-detections in ALMA observations of a significant fraction (15\%-20\%) of their fields selected at 850~$\mu$m with the JCMT or 870~$\mu$m with APEX, and they explain how this could occur if the total submillimetre flux is divided into several sources that fall below their detection threshold.  However, we always detected at least one 151~GHz source in every field that we observed.  Since our sample was selected in {\it Herschel} 500~$\mu$m data with a larger beam, we should have been more strongly biased towards finding fields that contain such blends of multiple fainter sources, and our detection threshold (when scaled from 151 to 345 GHz) is effectively higher than the ones used by \citet{hodge2013} and \citet{stach2018}, which should have made our observations more prone to non-detections in general.  On the other hand, our field selection is based on 500~$\mu$m flux densities that (when scaled to 850 or 870~$\mu$m) are higher than those used by \citet{hodge2013} and \citet{stach2018}, which would improve our chances of source detection in any field.  It is possible that, if our sources were observed with a smaller beam like the $\sim$0.6~arcsec beam used by \citet{stach2018}, they would be resolved into multiple sources that could fall below our detection threshold, but this explanation does not apply to the \citet{hodge2013} results, which observed their fields using a 1.5~arcsec beam size that is not that different from our 151~GHz beam.

Overall, the multiple differences between our study and other surveys are most likely related to our sample selection criteria, which were originally designed to identify infrared-bright gravitational lenses \citet{negrello2017} and would generally be expected to be unresolved single sources in our data.   As stated above, between 57 and 82\% of the BEARS fields were expected to correspond to gravitational lenses that would be unresolved in our data \citet{bakx2018, bakx2020viking}.  Also note that the types of sources that meet both our flux density and colour criteria are relatively rare and would have been unlikely to appear in other multiplicity studies.

\section{Spectral energy distributions}
\label{s_sed}

For a subset of the objects where the spectroscopic redshifts were measured by \citet{urquhart2022}, we constructed SEDs based on the {\it Herschel} 250-500~$\mu$m flux densities (from H-ATLAS Data Release 2) and the ALMA 151 and 101~GHz flux densities (as integrated over the entire fields) and then performed an analysis of the SEDs.  The 101 and 151~GHz data in particular can be used to constrain the emissivity index $\beta$ of the dust emission, but the SED data can also be used to search for temperature variations versus redshift and to compare the results of photometric and spectroscopic redshifts.

The SEDs for fields with single sources will represent single objects and are therefore straightforward to work with.  The flux densities for these fields from {\it Herschel} correspond to the same sources seen in the ALMA data (assuming that any lower redshift galaxies in the foreground contribute negligible amounts of emission at these wavelengths), so  the {\it Herschel} and ALMA data can be straightforwardly combined to create individual SEDs for individual sources.  However, we can also work with fields containing two sources that are at the same redshift.  The {\it Herschel} data can be combined with the sum of the ALMA flux densities for the sources in those fields to create SEDs that are still useful for determining the average dust temperatures and emissivities of the detected objects, although note that, if the sources in these fields have significantly different average dust temperatures, the SEDs may appear unusually broad.

With fields containing two or more sources with different or unidentified redshifts, combining the integrated ALMA flux densities with {\it Herschel} data will not produce SEDs that have any meaningful physical interpretation.  It is also very difficult to accurately disentangle the contributions of the different sources in these fields to the total integrated flux densities measured in the {\it Herschel} data.  While it might be possible to perform a spatial decomposition analysis of the {\it Herschel} data using the positions from ALMA, that is beyond the scope of this paper.  Given this, we will focus on characterizing the SEDs of the fields with single sources with spectroscopic redshifts (31 fields) and with two sources at the same redshift (6 fields).

\subsection{SED fits}
\label{s_sedfit}

\begin{table}
\caption{Colour temperatures and emissivities from SED fits to observed-frame 250-2970~$\mu$m data and 151/101~GHz ratios}
\label{t_sedfit}
\begin{center}
\begin{tabular}{@{}lccccc@{}}
\hline
Field &
  \multicolumn{3}{c}{250-2970~$\mu$m fit results}  &
  $f_{151\text{GHz}}$ \\
&
  $\beta$=2 &
  \multicolumn{2}{c}{$\beta$=variable} &
    /$f_{101\text{GHz}}~^a$ \\
&
  $T$ (K) &
  $T$ (K) &
  $\beta$ &
  \\
\hline
\multicolumn{5}{l}{Fields with single sources} \\
HerBS-11 &
  32.0 $\pm$ 0.8 &
  35.7 $\pm$ 2.0 &
  1.78 $\pm$ 0.10 &
  3.8 $\pm$ 0.3 \\
HerBS-14 &
  32.4 $\pm$ 1.1 &
  36.6 $\pm$ 4.1 &
  1.71 $\pm$ 0.24 &
  5.1 $\pm$ 0.4 \\
HerBS-18 &
  28.7 $\pm$ 0.9 &
  30.7 $\pm$ 3.6 &
  1.86 $\pm$ 0.23 &
  2.8 $\pm$ 0.3 \\
HerBS-24 &
  27.4 $\pm$ 0.8 &
  31.2 $\pm$ 2.4 &
  1.74 $\pm$ 0.14 &
  3.5 $\pm$ 0.3 \\
HerBS-25 &
  29.7 $\pm$ 0.7 &
  32.9 $\pm$ 2.0 &
  1.76 $\pm$ 0.13 &
  3.8 $\pm$ 0.4 \\
HerBS-27 &
  33.0 $\pm$ 1.4 &
  41.2 $\pm$ 4.5 &
  1.45 $\pm$ 0.23 &
  4.4 $\pm$ 0.3 \\
HerBS-28 &
  32.0 $\pm$ 1.2 &
  39.6 $\pm$ 1.4 &
  1.49 $\pm$ 0.07 &
  3.5 $\pm$ 0.3 \\
HerBS-36 &
  29.5 $\pm$ 1.2 &
  37.5 $\pm$ 2.6 &
  1.48 $\pm$ 0.12 &
  4.2 $\pm$ 0.3 \\
HerBS-37 &
  34.8 $\pm$ 0.9 &
  31.4 $\pm$ 0.5 &
  2.23 $\pm$ 0.04 &
  $>$2.9  \\
HerBS-39 &
  33.6 $\pm$ 0.8 &
  37.3 $\pm$ 2.5 &
  1.78 $\pm$ 0.13 &
  4.7 $\pm$ 0.4 \\
HerBS-40 &
  30.2 $\pm$ 2.9 &
  21.4 $\pm$ 1.0 &
  2.82 $\pm$ 0.13 &
  $>$4.8  \\
HerBS-47 &
  33.1 $\pm$ 1.3 &
  28.9 $\pm$ 2.6 &
  2.31 $\pm$ 0.21 &
  $>$2.3  \\
HerBS-55 &
  34.5 $\pm$ 1.8 &
  33.2 $\pm$ 6.7 &
  2.08 $\pm$ 0.39 &
  3.2 $\pm$ 0.3 \\
HerBS-57 &
  34.4 $\pm$ 1.2 &
  36.3 $\pm$ 5.6 &
  1.89 $\pm$ 0.31 &
  5.9 $\pm$ 0.5 \\
HerBS-60 &
  31.9 $\pm$ 0.5 &
  32.9 $\pm$ 2.0 &
  1.93 $\pm$ 0.13 &
  4.6 $\pm$ 0.4 \\
HerBS-68 &
  34.8 $\pm$ 0.4 &
  36.1 $\pm$ 1.4 &
  1.92 $\pm$ 0.07 &
  $>$3.0  \\
HerBS-73 &
  34.3 $\pm$ 0.6 &
  36.8 $\pm$ 2.3 &
  1.86 $\pm$ 0.12 &
  4.7 $\pm$ 0.4 \\
HerBS-86 &
  30.2 $\pm$ 1.2 &
  26.5 $\pm$ 2.9 &
  2.30 $\pm$ 0.26 &
  5.9 $\pm$ 0.6 \\
HerBS-93 &
  30.6 $\pm$ 2.7 &
  22.7 $\pm$ 4.2 &
  2.70 $\pm$ 0.48 &
  8.6 $\pm$ 0.8 \\
HerBS-103 &
  36.2 $\pm$ 1.2 &
  41.3 $\pm$ 5.7 &
  1.75 $\pm$ 0.23 &
  3.2 $\pm$ 0.3 \\
HerBS-107 &
  34.0 $\pm$ 1.7 &
  29.3 $\pm$ 4.6 &
  2.33 $\pm$ 0.37 &
  $>$5.0  \\
HerBS-111 &
  31.6 $\pm$ 1.0 &
  28.1 $\pm$ 2.5 &
  2.25 $\pm$ 0.19 &
  6.0 $\pm$ 0.7 \\
HerBS-123 &
  29.5 $\pm$ 0.7 &
  28.7 $\pm$ 3.4 &
  2.07 $\pm$ 0.26 &
  $>$6.2  \\
HerBS-132 &
  34.9 $\pm$ 2.2 &
  26.2 $\pm$ 1.7 &
  2.60 $\pm$ 0.15 &
  5.8 $\pm$ 0.6 \\
HerBS-141 &
  33.5 $\pm$ 2.2 &
  26.5 $\pm$ 1.8 &
  2.52 $\pm$ 0.17 &
  $>$3.6  \\
HerBS-160 &
  31.4 $\pm$ 1.0 &
  35.3 $\pm$ 4.0 &
  1.72 $\pm$ 0.24 &
  4.6 $\pm$ 0.3 \\
HerBS-182 &
  30.8 $\pm$ 1.0 &
  26.0 $\pm$ 0.9 &
  2.34 $\pm$ 0.08 &
  5.5 $\pm$ 0.5 \\
HerBS-184 &
  29.8 $\pm$ 1.1 &
  35.5 $\pm$ 4.8 &
  1.67 $\pm$ 0.22 &
  3.3 $\pm$ 0.3 \\
HerBS-189 &
  37.7 $\pm$ 1.1 &
  44.5 $\pm$ 2.1 &
  1.65 $\pm$ 0.09 &
  $>$5.0  \\
HerBS-200 &
  32.3 $\pm$ 1.5 &
  31.4 $\pm$ 6.1 &
  2.05 $\pm$ 0.34 &
  3.3 $\pm$ 0.4 \\
HerBS-207 &
  25.7 $\pm$ 1.0 &
  23.2 $\pm$ 3.3 &
  2.19 $\pm$ 0.28 &
  3.7 $\pm$ 0.4 \\

\noalign{\vskip 0.5em}
\multicolumn{5}{l}{Fields with multiple sources at the same redshift} \\
HerBS-21 &
  33.4 $\pm$ 0.4 &
  34.9 $\pm$ 1.4 &
  1.89 $\pm$ 0.09 &
  4.9 $\pm$ 0.5 \\
HerBS-49 &
  28.5 $\pm$ 3.1 &
  39.0 $\pm$ 21.5 &
  1.40 $\pm$ 0.84 &
  1.6 $\pm$ 0.1 \\
HerBS-69 &
  30.9 $\pm$ 0.8 &
  27.6 $\pm$ 0.2 &
  2.24 $\pm$ 0.01 &
  $>$3.0  \\
HerBS-120 &
  30.3 $\pm$ 0.4 &
  32.4 $\pm$ 0.1 &
  1.84 $\pm$ 0.00 &
  $>$4.2  \\
HerBS-159 &
  30.4 $\pm$ 1.0 &
  27.4 $\pm$ 3.2 &
  2.22 $\pm$ 0.27 &
  $>$2.6  \\
HerBS-208 &
  29.0 $\pm$ 0.3 &
  28.3 $\pm$ 1.1 &
  2.05 $\pm$ 0.08 &
  4.5 $\pm$ 0.5 \\
\hline
\end{tabular}
\end{center}
$^a$ Lower limits in the 151/101~GHz ratios are given for fields where at least one 151~GHz source is not detected at 101~GHz.  The limits for fields with single source are calculated using 101 GHz flux densities equivalent to 5 times the rms noise levels listed in Table~\ref{t_imageparam}, which are effectively the 5$\sigma$ detection limits for point sources and could overestimate the limit for extended sources.  Since HerBS-189 is notably extended, this does not work, so the 101~GHz 5$\sigma$ upper limit is multiplied by 1.9, which is based on the expected convolved source size (using the data from Table~\ref{t_diffuse}) to the beam size.  The limits HerBS-69 and HerBS-159 are calculated using 101 GHz flux densities equivalent to 5 times the rms noise levels multiplied by 2 (for the number of undetected sources).  The A source in the HerBS-120 field was detected but the B source was not, so for calculating the lower limit in the 151/101~GHz ratio for this field, the 101~GHz flux density of the A source was added to 5 times the rms noise level (the assumed upper limit for the B source).  
\end{table}

To understand how the dust colours vary within the sample, we can fit the (observed wavelength) 250-2970~$\mu$m data with single optically thin modified blackbodies with a fixed $\beta$ value.  We set $\beta$ to 2, which is similar to what is used in the models from \citet{draine2003} that are based on Milky Way observations and which is similar to the value of 1.8 found by \citet{planck2011} for the Milky Way.  While the single modified blackbodies do not necessarily accurately characterize the physical dust temperatures or the details of the dust emission processes, they are still useful for characterizing the overall colours.  However, for examining the dust emissivities, we also fit the data with single modified blackbodies with variable $\beta$ values.  Allowing $\beta$ to vary leads to degeneracies between $\beta$ and temperature \citep[e.g.,][]{casey2014}, which adds confusion to any comparison of dust temperatures fit with such SEDs, which is why a second set of fits with a fixed $\beta$ value are needed for comparing the colours of the sources in our sample.  

\begin{figure*}
\includegraphics[width=17cm]{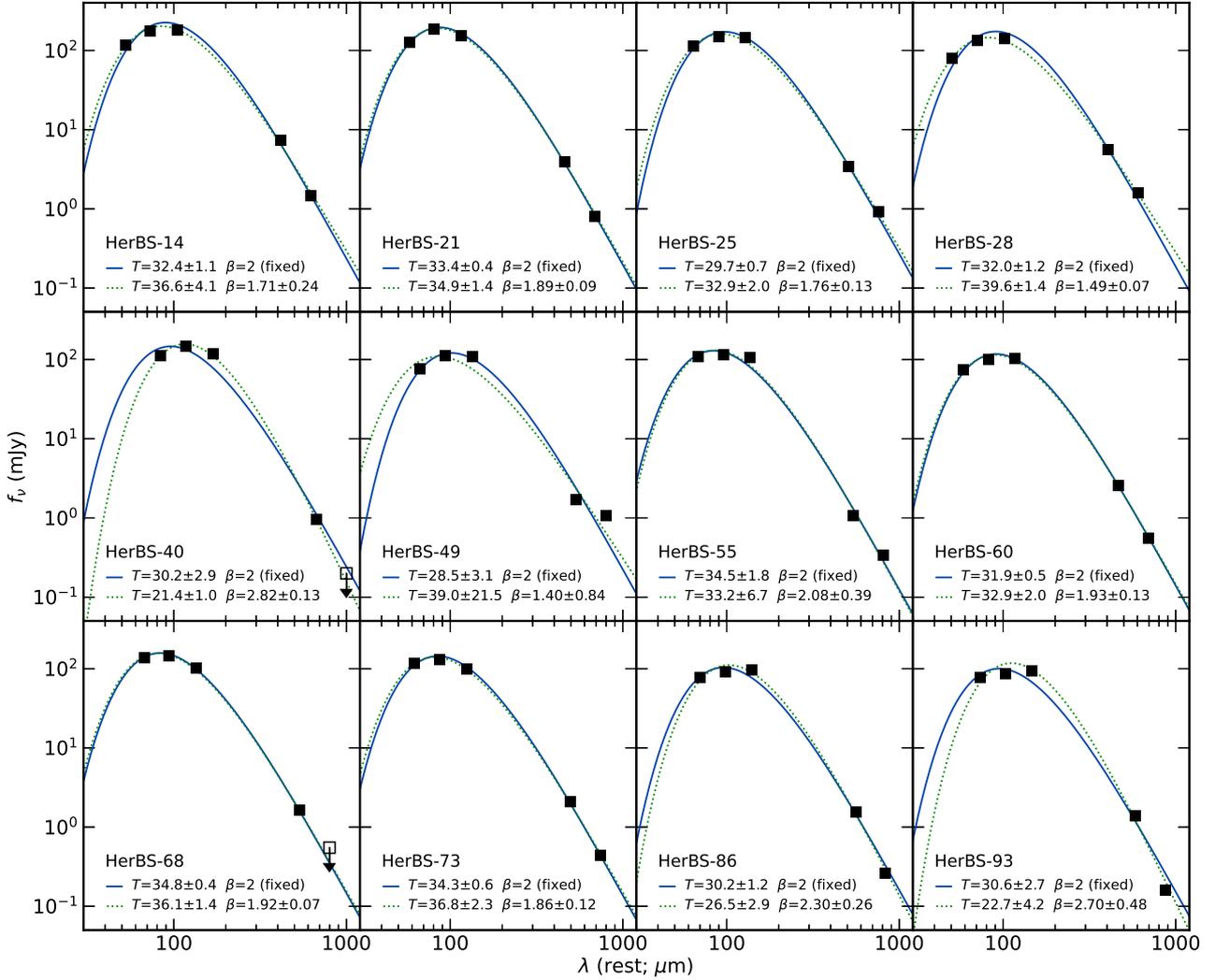}
\caption{SEDs for 16 of the sample fields along with the best-fitting modified blackbody functions where $\beta$ is fixed to 2 and where $\beta$ is allowed to vary.  The parameters from the fits are listed at the bottom of each panel. All data are plotted as a function of their rest wavelengths calculated using the spectroscopic redshifts from \citet{urquhart2022}. HerBS-21, HerBS-25, HerBS-60, HerBS-68, and HerBS-73 are shown as examples of where, when $\beta$ is allowed to vary, the best-fitting $\beta$ value is close to 2.  HerBS-14 and HerBS-28 are examples of fields with low fitted $\beta$ values.  HerBS-40 and HerBS-86 are examples of fields with high fitted $\beta$ values.  HerBS-49, HerBS-55, and HerBS-93 are examples of SEDs where the single modified blackbodies do not accurately fit the data.  The uncertainties are equivalent to or smaller than the data points in these plots.  Open symbols represent 5$\sigma$ upper limits.}
\label{f_sed}
\end{figure*}

The fits were performed using a standard Levenberg Marquardt algorithm, and both the measurement and calibration uncertainties were used to calculate the input flux density uncertainties.  The calibration uncertainties for the {\it Herschel} data are set to 4\% \citep{bendo2013}.  Colour corrections of 1.03, 1.005, and 0.985 were applied to the {\it Herschel} 250, 350, and 500~$\mu$m data, respectively, based on the tables from The Spectral and Photometric Imaging Receiver (SPIRE) Handbook \citep{valtchanov2018}\footnote{The SPIRE Handbook is available at \url{https://www.cosmos.esa.int/web/herschel/legacy-documentation-spire}.  The 250-500~$\mu$m data for the subsample discussed in Section~\ref{s_sed} have colours consistent with a modified blackbody at $z=0$ (without a CMB correction) with $\beta=1.5$ and temperature of 9.6$\pm$0.9~K or with $\beta=2$ and temperature of 8.5$\pm$0.7~K , so we used colour corrections for point sources consistent with those modified blackbodies.}  The redshifts of our sample extend up to $\sim$4.5, where the temperature of the cosmic microwave background (CMB) is$\sim$15~K.  This potentially affects the temperatures and $\beta$ from or SED fits, so we applied a correction for the CMB when fitting the data (see \citealt{dacunha2013} for an overview).  When one or more objects in these fields were not detected at 101~GHz, we only performed fits to the {\it Herschel} and 151~GHz data, although we still display 5$\sigma$ upper limits for the 101~GHz data (based on the rms noise values in Table~\ref{t_imageparam}) in the plots.  Table~\ref{t_sedfit} lists the colour temperatures and $\beta$ derived from these fits as well as integrated infrared luminosities (with no correction for magnification).  Additionally, a subset of the SEDs are shown in Figure~\ref{f_sed}.

For most of the fields, one or both of the SED fits work reasonably well in describing the shape of the SED.  Quite a few of the fields have SEDs that can be described using $\beta$$\approx$2, as illustrated by how the two modified blackbody curves (for $\beta$ fixed to 2 and for variable $\beta$ values) overlap in the SED plots for HerBS-21, HerBS-25, HerBS-60, HerBS-68, and HerBS-73 in Figure~\ref{f_sed}.  This is also seen in the $\beta$ values derived when $\beta$ was allowed to vary as a free parameter.  About half of these values are either within 10\% of 2 or are statistically equivalent to 2, and the mean fitted value is 2.0 with a standard deviation of 0.3.

The fits with the variable $\beta$ have shown that some SEDs are consistent with either a notably shallow or notably steep SED where $\beta$ deviates significantly from 2.  HerBS-27, HerBS-28, and HerBS-36 are the best examples of sources with low fitted $\beta$ values, while HerBS-40, HerBS-132, and HerBS-141 are the best examples of sources with high fitted $\beta$ values.  Notably, the three example sources with low fitted $\beta$ values are at $z>3.0$, while the three example sources with high $\beta$ values are at $z<2.5$;  we discuss this more in Section~\ref{s_151-101ratios}.

In three fields (HerBS-49, HerBS-55, and HerBS-93), the single modified blackbodies simply do not fit the data well (i.e., one or more of the data points deviate by $>$3$\sigma$ from the best-fitting modified blackbodies).  The slopes between the observed 500 and 1970~$\mu$m data points in the HerBS-49 and HerBS-55 fields are too steep to be consistent with a single modified blackbody with $\beta$=2, but the slope of the ALMA data points is much shallower in comparison to the slope between the 500 and 1970~$\mu$m data.  These could be cases where either the ALMA or the {\it Herschel} data are affected by noise or other measurement issues, but the most likely physical explanation for these SED shapes would be that the fields contain sources with both dust with high $\beta$ values and with submillimetre emission from sources other than $\gtrsim$10~K dust.  This is discussed more in Section~\ref{s_151-101ratios}.  HerBS-93 is a case where the ALMA data are significantly steeper than would be expected given the shape of the curve defined by the {\it Herschel} data, but assuming again that no technical issues affected the data, it would be possible to describe the SED using a sum of modified blackbodies with high $\beta$ values.  Additional continuum measurements would be needed to define the SEDs more precisely before we could proceed with trying to interpret the physics of the dust emission from these sources.

\subsection{Analysis of the 151/101 GHz flux density ratios}
\label{s_151-101ratios}

SEDs produced by dust at multiple temperatures can be fit by a single modified blackbody with a $\beta$ lower than the actual $\beta$ of the dust \citep[e.g.,][]{dunne2000, klaas2001, bendo2003}.  The ratios of the 151 to 101~GHz emission, which are also listed in Table~\ref{t_sedfit}, measure the slope of the Rayleigh-Jeans side of the SEDs for these objects, which will primarily be affected by the physical $\beta$ of the dust and which will be relatively insensitive to dust temperature.  It would therefore be useful to compare the results from these two different metrics of dust emissivity.

\begin{figure}
\begin{center}
\includegraphics[width=7cm]{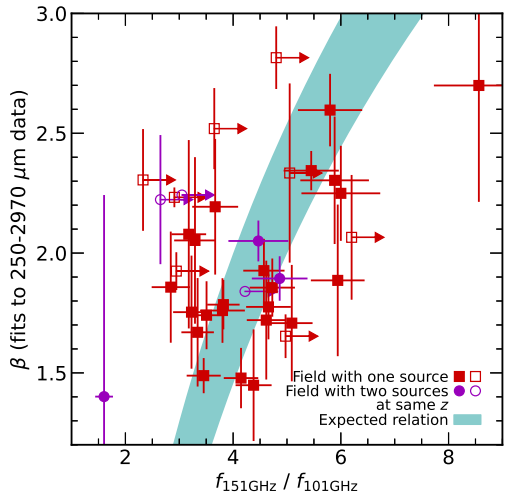}
\end{center}
\caption{Plots of the $\beta$ values derived from fits to the 250-2970~$\mu$m data (where $\beta$ was treated as a free parameter) to the 151/101~GHz flux density ratios for the subset of sources from Table~\ref{t_sedfit} with 101~GHz flux density detections.  The shaded area shows the expected relation for dust with temperatures between 20 and 50~K and for redshifts between 1 and 5.  The open symbols represent data points based on 101~GHz 5$\sigma$ upper limits.}
\label{f_ratvsbeta}
\end{figure}

Figure~\ref{f_ratvsbeta} shows the $\beta$ values derived from the 250-2970~$\mu$m SED fits to the 151/101~GHz ratios.  Although the data roughly follow the relation expected for modified blackbodies with temperatures and redshifts consistent with those values in our sample (as shown by the shaded region), the relation shows some scatter.  Notably, data with large uncertainties in the fitted $\beta$ values tend to lie further away from the range of expected values.  Issues related to blended emission from dust at different temperatures will drive many data points downwards in Figure~\ref{f_ratvsbeta}, while the degeneracy between temperature and $\beta$ in the SED fits will cause additional scatter.

\begin{figure}
\begin{center}
\includegraphics[width=7cm]{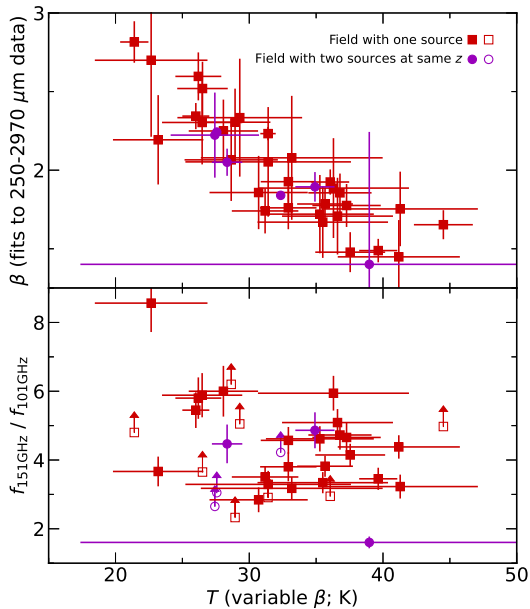}
\end{center}
\caption{Plots of $\beta$ (from SED fits where the $\beta$ was treated as a free parameter) and the 151/101~GHz ratios as a function of temperature.  The open symbols represent data points based on 101~GHz 5$\sigma$ upper limits.  The data point with the extra large error bars is HerBS-49, which was fit poorly by the single modified blackbody.}
\label{f_ratiobetavstemp}
\end{figure}

Figure~\ref{f_ratiobetavstemp} shows the $\beta$ values from the SED fits and the 151/101~GHz ratios plotted as a function of temperature.  If the $\beta$ from the SED fits purely reflected emissivity variations, then both $\beta$ and the 151/101~GHz ratios should vary similarly as a function of the best-fitting temperature.  We found that $\beta$ varies strongly as a function of temperature as a result of the well-known degeneracy between temperature and $\beta$ \citep[e.g.,][]{casey2014}; the Pearson correlation coefficient for the relation is -0.88.  Meanwhile, the 151/101~GHz ratios do not exhibit such a strong dependence on temperature; the Pearson correlation coefficient for that relation is -0.50.  The difference between these correlation coefficients reveals biases in the $\beta$ from the SED fits.

\begin{figure}
\begin{center}
\includegraphics[width=7cm]{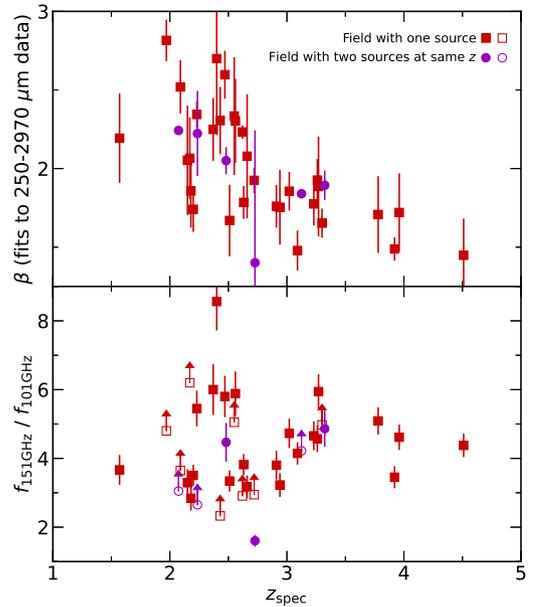}
\end{center}
\caption{Plots of $\beta$ (from SED fits where the $\beta$ was treated as a free parameter) and the 151/101~GHz ratios as a function of spectroscopic redshift.  The open symbols represent data points based on 101~GHz 5$\sigma$ upper limits.}
\label{f_ratiobetavsz}
\end{figure}

Figure~\ref{f_ratiobetavsz} plots the $\beta$ values from the SED fits and the 151/101~GHz ratios as a function of spectroscopic redshift.  A relation is seen for the $\beta$ values for the SED fits (with a Pearson correlation coefficient of -0.65).  However, no relation is seen at all with the 151/101~GHz ratios (with a Pearson correlation coefficient of 0.04).  This indicates that the slopes of the Rayleigh-Jeans sides of the SEDs are not varying with redshift and that the redshift variations in the $\beta$ values from the SED fits are an artefact of the fitting process.

The $\beta$ derived from the fits are influenced strongly by the rest wavelengths sampled by the data.  As indicated in Section~\ref{s_sedfit}, the fits with the variable $\beta$ may yield shallower emissivity functions because of the blending of emission from multiple dust components with different temperatures.  When progressing from $z=2$ to $z=4.5$, the 250~$\mu$m and 350~$\mu$m bands will increasingly include more emission from hotter dust, including very small grains not at thermal equilibrium \citep[e.g.,][]{li2001, popescu2011}.  Given this, modified blackbodies with variable $\beta$ fit to SEDs at the same observed wavelengths would be expected to yield both higher temperatures and lower $\beta$ for objects at higher redshifts.  Hence, the $\beta$ values from these fits should not be relied upon for characterizing the physical emissivities of the dust emission.

\begin{figure*}
\begin{center}
\includegraphics[width=15cm]{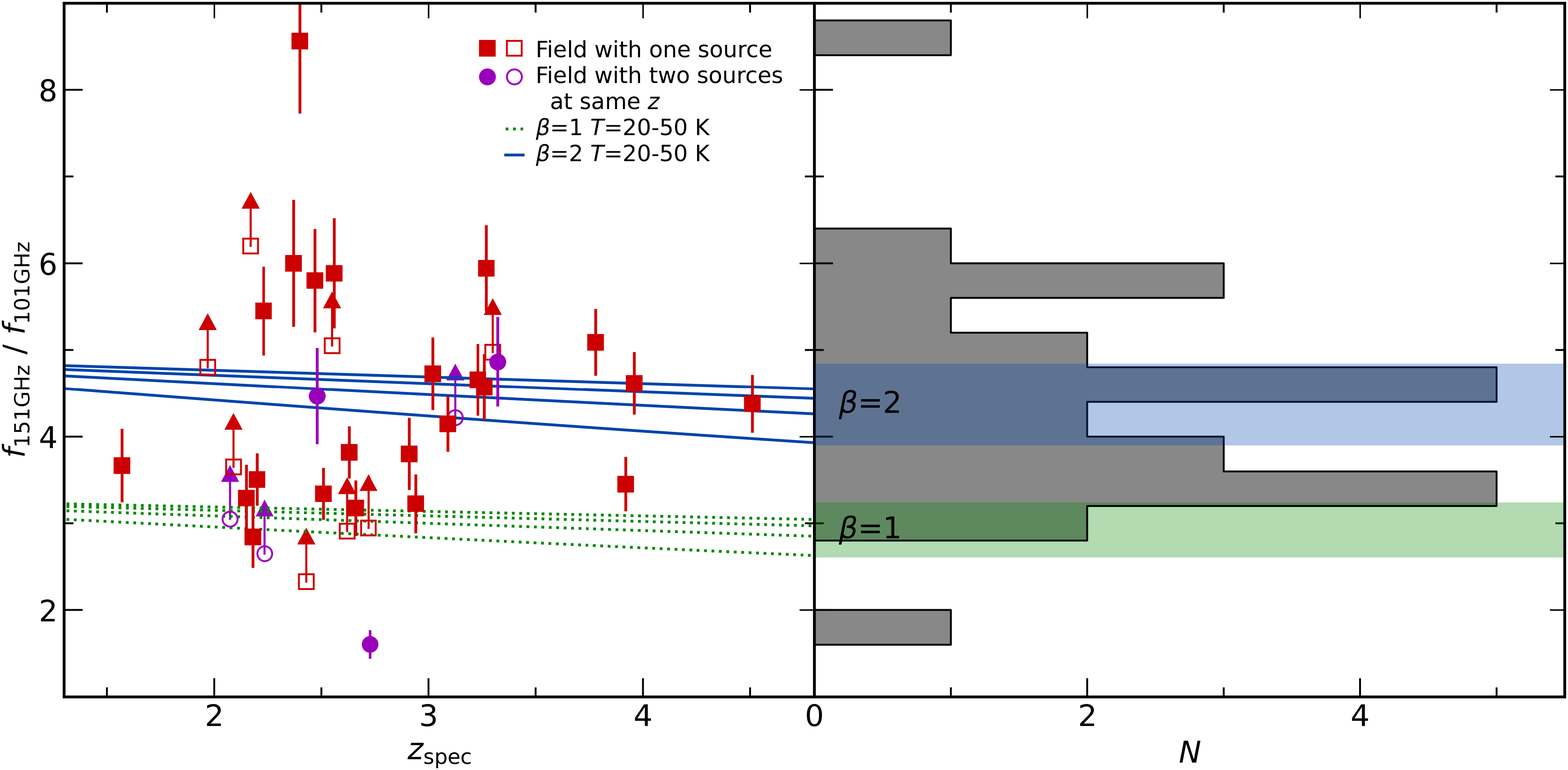}
\end{center}
\caption{The left-hand panel shows a plot of the (observed frame) 151/101 GHz flux density ratios from Table~\ref{t_sedfit} as a function of redshift along with the ratios expected for modified blackbodies with temperatures between 20 and 50~K and $\beta$ of either 1 or 2.  The open symbols represent data points based on 101~GHz 5$\sigma$ upper limits.  The right-hand panel shows a histogram of the distribution of these ratios along with the ranges of these ratios consistent with these modified blackbodies over the redshift range of 1 to 5.  HerBS-49 is the data point with the lowest 151/101 GHz ratio, while HerBS-93 is the data point with the highest ratio.}
\label{f_ratio}
\end{figure*}

Having compared the 151/101~GHz ratios to the $\beta$ values derived from the SED fits and having concluded that the 151/101~GHz ratios may be more indicative of the physical emissivity of the dust grains themselves, we can now focus on analysing and interpreting the 151/101~GHz ratios.  Figure~\ref{f_ratio} shows the ratios as a function of redshift and a histogram of these ratios.  We also calculated the range of 151/101~GHz ratios expected for modified blackbodies (corrected for the effects of the CMB) with temperatures ranging from 20 to 50~K, $\beta$ of either 1 or 2, and redshifts from 1.5 to 4.6.  This redshift range encompasses the range of the sources listed in Table~\ref{t_sedfit}.  Shaded regions in Figure~\ref{f_ratio} show the range of ratios consistent with the two $\beta$ values over these temperature and redshift ranges.

The mean ratio that we measure is 4.4, and the standard deviation is 1.3.  For comparison, the mid-point of the range of ratios for $\beta$=2 is also 4.4, while the mid-point of the range of ratios for $\beta$=1 is 2.9.  While some ratios are measured relatively precisely, others have relatively large uncertainties.  Still, these numbers as well as Figure~\ref{f_ratio} demonstrate that the 151/101~GHz ratios for the subsample as a whole are largely consistent with $\beta$ values of close to 2.  Few studies have performed measurements of $\beta$ for high redshift galaxies, and these studies have primarily been limited to shorter rest wavelengths.  Our $\beta$ are consistent with those for the $z\sim5.5$ galaxies studied by \citet{faisst2020}, although they only have measurements at rest wavelengths of $<$200~$\mu$m, and they assume that $\tau=1$ near the peak of their SEDs.  \citet{bakx2021} measure a $\beta$ of 1.6 for a $z=7.13$ galaxy, but they too only have data at rest wavelengths of $<$200~$\mu$m.

However, we see some outliers.  HerBS-93 has a 151/101~GHz ratio of 8.7$\pm$0.8, which would indicate that $\beta$ is $\sim$3.5 for this source.  This is the only field with a ratio higher than 6.5.  We cannot identify any issues with the data processing or the flux density measurements for this source, although the peak 101~GHz emission is very close to the 5$\sigma$ surface brightness detection limit we used when reporting flux density measurements.  If the redshift was poorly defined, then it would be possible that our correction for CMB effects was inaccurate, which could affect the 151/101~GHz ratio.  However the redshift has been very strongly constrained by two spectral lines (corresponding to CO(3-2) and the [C{\small I}]($^3P_1$-$^3P_0$) emission), placing the object at $z=2.400$; no other combination of lines could reproduce the spectra observed by \citet{urquhart2022}.  Aside from either heretofore unidentified issues with the 101~GHz flux density measurement for this specific field or the possibility that the dust emissivity is inherently steep for this object, we have no explanation for why the 151/101~GHz ratio is so high.  Also note that some but not all of the data points based on 101~GHz 5$\sigma$ upper limits may also be consistent with $\beta$ values significantly higher than 2, although either deeper observations at $\sim$101~GHz or additional measurements at higher frequencies would be needed to verify that the objects in these field have such steep spectral slopes.

Fields with 151/101~GHz ratios of $\lesssim$3.5 are potentially the more interesting because the low ratios could point to the presence of emission sources other than thermal dust emission.  The field with the lowest ratio is HerBS-49, while other fields of potential interest include HerBS-28, HerBS-55, HerBS-184, and HerBS-200.  In Section~\ref{s_sedfit}, HerBS-49, HerBS-55 and HerBS-200 were identified as being fit relatively poorly by the single modified blackbodies, and part of the reason was that the steep slope between the (observed frame) 500~$\mu$m and 151~GHz data points was inconsistent with the shallower slope between the 151 and 101~GHz data.  HerBS-28 was a case where the whole of the SED was consistent with a single modified blackbody with a relatively low $\beta$ of 1.49.  The SED of HerBS-184 is actually fit reasonably well by the modified blackbody where $\beta$ is fixed to 2, but interestingly, that curve falls below the 101~GHz data point.

As we have stated, it is likely that at least the 101~GHz band but also possibly the 151~GHz band as well contains emission produced by physical processes other than thermal dust emission, but it is not clear from these data alone what the alternate emission mechanisms are.  The most obvious possibility is synchrotron emission, which would most likely be associated with previously unidentified AGN.  However, none of the sources with low 151/101~GHz ratios are associated with radio sources detected in the VLASS \citep{gordon2021}.  Free-free emission is a possibility but unlikely given that, in nearby starburst galaxies, it is not seen as a dominant source of emission at (rest frame) $<$1~mm \citep[e.g.,][]{condon1992, peel2011, bendo2015, bendo2016}.  Very cold ($<$5~K) dust has been suggested as a possibility for such submillimetre excesses in some nearby galaxies \citep[e.g.,][]{galliano2005}, but it seems extremely unlikely for any sources in our sample since the dust would be colder than the CMB at these redshifts.  Anomalous microwave emission from spinning dust and other exotic phenomena involving dust grains with unusual properties are possible, but additional data would be needed to identify the emission mechanisms.

\subsection{Variations in colour temperatures with redshift}
\label{s_tvsz}

For the subset of galaxies discussed in this section, the dust colour temperatures range from 26 to 38~K when $\beta$ is fixed to 2.  This is warmer than the range of 15 to 30~K seen in typical nearby spiral galaxies when using fits with $\beta$ set to 2 \citep[e.g.,][]{boselli2012, kirkpatrick2014}.  Several recent studies have indicated that dust colour temperatures increase with redshift \citep{magdis2012, magnelli2014, bethermin2015, schreiber2018, liang2019, bouwens2020, chen2021, dudzeviciute2021, sommovigo2022}.  However, a few other studies that mainly worked with galaxies selected at far-infrared or submillimeter wavelengths found either no trend in colour temperature with wavelength or notable outliers from this relation \citep[e.g.,][]{jin2019, dudzeviciute2020, reuter2020, magdis2021, drew2022}. 

\begin{figure}
\begin{center}
\includegraphics[width=8.5cm]{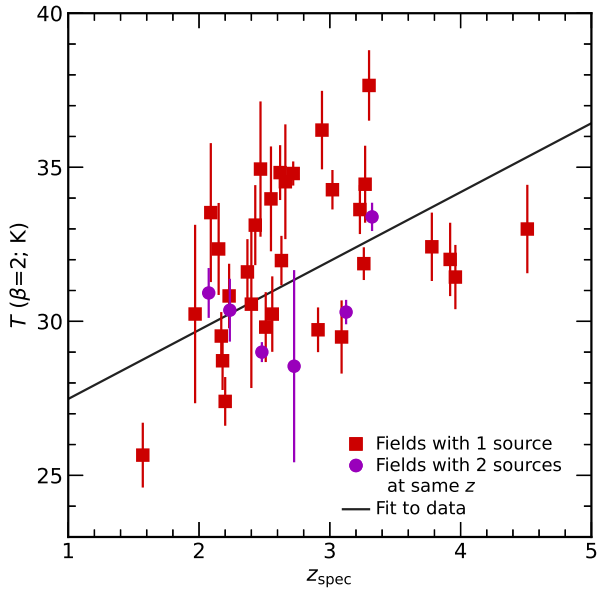}
\end{center}
\caption{Plot of colour temperature versus redshift for for fields with single sources with measured redshifts and for fields with multiple sources all measured to be at similar redshifts.}
\label{f_tvsz}
\end{figure}

Unfortunately, it is not straightforward to directly compare our dust colour temperatures to those obtained from other references, including those from \citet{schreiber2018}, \citet{bouwens2020}, \citet{reuter2020}, and \citet{dudzeviciute2021}.  First of all, these different studies used different values of $\beta$ ranging from 1.5 to 2.0, which affects the overall scale of the colour temperatures.  Secondly, if the dust is treated as becoming optically thick at far-infrared wavelengths, as was done by \citet{reuter2020}, then the resulting temperatures will be scaled to higher values (see also \citealt{cortzen2020} for a discussion of this topic).  Additional complications related to the handling of dust emission at $\leq$50~$\mu$m could also affect the resulting colour temperatures.  However, we can still examine whether our colour temperatures still show any change relative to redshift to examine whether such a relation actually exists, at least in our sample.  Therefore, in Figure~\ref{f_tvsz}, we plotted the colour temperatures for $\beta$ fixed to 2 from Table~\ref{t_sedfit} as a function of redshift and also performed some analyses on these data. 

We do find a trend in Figure~\ref{f_tvsz}, but the trend is notably weak.  Our best-fitting relation can be described by
\begin{equation}
T(K) = (29.7 \pm 0.7) + (2.2 \pm 0.8)(z-2),
\end{equation}
although the slope of this relation is only inconsistent with no evolution at the 2.75$\sigma$ level, and the data points between redshifts of 2 and 3.5 exhibit a lot of scatter around this relation.  The Pearson correlation coefficient for the two values is 0.39, which indicates that 15\% of the variance in dust temperature can be described by our relation.  This would indicate that infrared-bright high-redshift sources like the ones in our sample do not necessarily exhibit any strong relation between temperature and redshift.  However, note that the relation in Figure~\ref{f_tvsz} is affected by the colour criterion that was applied when selecting the data, which would potentially exclude objects that are significantly warmer than our best-fitting line \footnote{Unfortunately, it is not possible to calculate a strict temperature limit above which we would not detect sources.  As described in Section~\ref{s_data}, the data needed to be consistent with a photometric redshift of $z\geq2$ based on the photometric template from \citet{pearson2013}, but as discussed in Section~\ref{s_template}, the objects in our sample have warmer temperatures than what is predicted by the template.  It is still clear that this criterion biased our sample towards objects with colder colour temperatures; it just is not straightforward to characterize this bias.}, although including such objects would potentially only increase the scatter in the relation.  Additionally, note that, when objects are selected at far-infrared or submillimetre wavelengths, the selection would tend to be biased towards warmer objects at higher redshifts.

Among other studies that find a relation between dust temperature and redshift, the statistical significance of the slopes of their resulting relations are often much stronger.  For these comparisons, we mainly focused on relations found using samples with redshifts that overlapped those of our sample.  The slope of the relations found by \citet{schreiber2018} and \citet{bouwens2020} are measured at greater than the 10$\sigma$ level.  Unfortunately, they do not provide any data on the strength of the correlations in their data.  \citet{dudzeviciute2020} presented a relation between temperature and redshift with no slope, and the Pearson correlation coefficient that we calculated using their data was 0.06, indicating virtually no dependence of the temperatures on redshift.  Of the two relations presented by \citet{dudzeviciute2021}, the one derived for 450~$\mu$m selected galaxies has a slope measured at the $\sim$6$\sigma$ level, while the other (a subset of the galaxies from \citealt{dudzeviciute2020}) is measured at $<$3$\sigma$. Using the data for their 450~$\mu$m selected sample, we calculated a Pearson correlation coefficient of 0.34, which is similar to what we obtained for our sample. Notably, \citet{reuter2020} found a slope in their data at the $\sim$3$\sigma$ level but also reported a high level of scatter in their relation, and various statistical tests indicated that a line with no slope was favoured for their data.  

Comparing the distribution of our colour temperatures and our relation between colour temperature and redshift to the relations from other studies is difficult because different studies used different $\beta$ values when fitting their data, and the selection of a specific $\beta$ will affect the derived temperature.  Additionally, the change in best-fitting temperature from $\beta=2$ to the best-fitting temperature from another value of $\beta$ also depends on the rest wavelengths at which the SED is measured (and hence on the redshift of the source) and the relative uncertainties in the SED measurements.  This means that we cannot straightforwardly rescale relations from other studies that use different $\beta$ to correspond to $\beta=2$.  The best that we can do is simply look at how the distribution of our colour temperatures when we fit the data using other $\beta$ values (although for succinctness, we will not list those alternate temperatures in this paper).

If we fit our data using $\beta$ fixed to 1.6, which matches the values used by \citet{schreiber2018} and \citet{bouwens2020}, then the temperature distribution of our data shifts to a range spanning from 33 to 45~K with a mean of 39~K.  The relations between temperature and redshift derived by \citet{schreiber2018} and \citet{bouwens2020} would pass through the lower end of the distribution of our colour temperatures, with some data points in the $2\leq z \leq 3$ range offset above either of their relations by $\sim$9~K and some data points at $z\geq 3.5$ falling below either of their relations by $\sim$6~K.  If we fit our data using $\beta$ fixed to 1.8, which is what is used by \citet{dudzeviciute2020} and \citet{dudzeviciute2021}, then our derived temperatures shift to the range 29 to 41~K with a mean of 35~K.  The relations from these papers pass through this distribution of our data points; our data with lower temperatures are more consistent with the Dudzevi{\v{c}}i{\={u}}t{\.{e}} et al. relations for their 850~$\mu$m selected samples, while our data with the higher temperatures are consistent with the Dudzevi{\v{c}}i{\={u}}t{\.{e}} et al. relation for their 450~$\mu$m selected sample.

In comparing all of these results, one of the main issues seems to be the waveband used to select the data.  Many of the observational papers working with samples selected at optical or near-infrared wavelengths (which tend to be samples composed of main sequence galaxies) report the presence of relatively strong or well-defined relations between dust colour temperature and redshift \citep[e.g.,][]{schreiber2018, bouwens2020} or otherwise present results in agreement with such a relation being present \citep[e.g.,][]{magdis2012, bethermin2015}.  Meanwhile, observational papers based on galaxies selected at infrared, submillimetre, or millimetre wavelengths tend to have measured relations with more scatter or slopes with lower statistical significance \citep[e.g.,][and our results]{reuter2020, dudzeviciute2020, dudzeviciute2021} or otherwise found results indicating that colour temperature does not necessarily increase with redshift \citep[e.g.,][]{jin2019, drew2022}.  Nevertheless, it is possible to find exceptions.  For example, while \citet{bethermin2015} reported a relation between radiation field intensity (which would be directly related to colour temperature) and redshift for main sequence galaxies, they found no such relation for strong starbursts, even though both subsets of galaxies were selected from near-infrared data.  Additionally, \citet{magdis2021} reported no relation between dust temperature and redshift even though they selected their sample based on a near-infrared magnitude limit, although their sample is designed to contain specifically quiescent galaxies.  

None the less, it seems like sample selection strongly influences the trend that is measured in dust colour temperature with redshift; different types of galaxies are generally selected in different bands.  Since many of the objects selected in optical or near-infrared data, such as those from \citet{schreiber2018}, tend to be main sequence galaxies, such galaxies at any given redshift may be expected to have relatively uniform properties because they are selected to lie upon a specific relation.  However, galaxies selected at far-infrared or submillimetre wavelengths, such as those in our sample or those from \citet{dudzeviciute2020} and \citet{dudzeviciute2021}, are more extreme, dusty objects that may naturally be expected to deviate from the main sequence in general terms.  Consequently, the data from these samples exhibited a much higher level of dispersion in plots of temperature versus redshift.  This explanation would be consistent with the finding specifically by \citet{bethermin2015} in which the colour temperatures varied with redshift for main sequence galaxies but not for starbursts.

\subsection{Comparisons to existing SED templates}
\label{s_template}

Since multiple SED templates are still used for determining the photometric redshifts of deep field sources, it would be a useful test to compare photometric redshifts derived from these templates to our spectroscopic redshifts.  \citet{urquhart2022} already presented a comparison of spectroscopic redshifts for this sample to photometric redshifts derived by \citet{ivison2016} and \citet{bakx2018}, but those photometric redshifts did not incorporate the ALMA continuum measurements, which, as we have discussed above, are very effective at constraining the Rayleigh-Jeans side of the dust emission.  

We used five different SED templates to derive photometric redshifts for comparison to our spectroscopic redshifts.  Two of the templates (\citealt{pearson2013} and \citealt{bakx2018}) are based on the sums of two modified blackbodies and were derived from H-ATLAS sources with known redshifts.  The \citet{pearson2013} template was derived using just {\it Herschel} data, while the \citet{bakx2018} model was derived using {\it Herschel} and JCMT 850~$\mu$m data.  The third template is based on functions fitted to the empirical SED of the well-studied gravitational lens SMM J2135-0102 \citep{ivison2010, swinbank2010}; this object is also called the Cosmic Eyelash, and we refer to its SED as the Eyelash template.  This was one of three SED models that was found to work very effectively when applied to the 250-850~$\mu$m data for a sample of 69 gravitational lens candidates studied by \citet{ivison2016}.  The other two templates that were also recommended by \citet{ivison2016} for SED fitting are based on composites of multiple submillimetre galaxy SEDs.  One of these is the \citet{pope2008} template, which was based on a sample of $z\sim2$ submillimetre galaxies selected at mid-infrared wavelengths.  These templates were built using (observed wavelength) 16, 24, 70, and 850~$\mu$m photometry as well as mid-infrared spectroscopy covering polycyclic aromatic hydrocarbon spectral features at rest wavelengths.  The other template was created by \citet{swinbank2014} using 24-870~$\mu$m and 1.4~GHz observations for 99 submillimetre galaxies from the ALMA LABOCA ECDFS Submillimeter Survey (ALESS), which we refer to as the ALESS template.

\begin{figure*}
\begin{center}
\includegraphics[width=16cm]{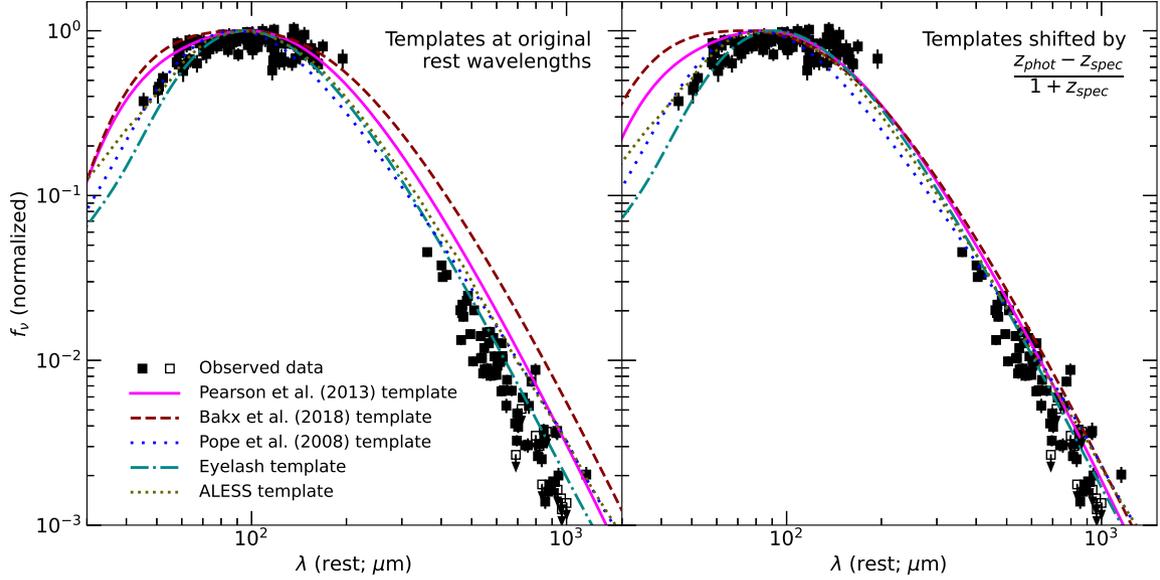}
\end{center}
\caption{Plot of the SED data for fields with single sources with measured redshifts and for fields with multiple sources all measured to be at similar redshifts alongside the five SED templates examined in the analysis in Section~\ref{s_template}.  For visualization purposes, all of the observed data have been shifted to the rest wavelength frame based on their spectroscopic redshifts and have been normalized based on the peak of the best-fitting modified blackbody functions with $\beta$ fixed to 2, and all templates have been normalized so that their peak values are equal to 1.  Open symbols represent 5$\sigma$ upper limits.  The left-hand panel shows the templates at their original rest wavelengths, while the right-hand panel shows the templates shifted by the mean $(z_{\text{phot}}-z_{\text{spec}})/(1+z_{\text{spec}})$ values listed in Table~\ref{t_templvsspec} so that they match up with the spectroscopic data better.  The templates in this plot have also been adjusted to account for CMB effects at $z=2.6$, which is the median spectroscopic redshift for the sources in this subsample.}
\label{f_template}
\end{figure*}

\begin{figure*}
\begin{center}
\includegraphics[width=17.5cm]{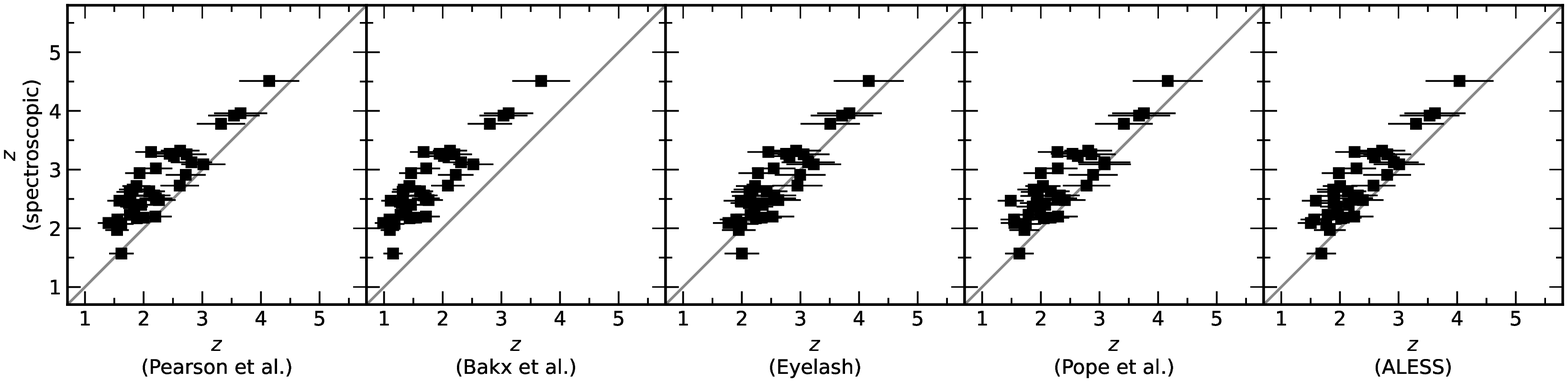}
\end{center}
\caption{Comparisons of the photometric redshifts determined using five SED templates to the spectroscopic redshifts determined for fields with single sources with measured redshifts and for fields with multiple sources all measured to be at similar redshifts.  The grey lines show where the photometric and spectroscopic redshifts are equal.}
\label{f_zcomp}
\end{figure*}

These templates were fit to the (observed wavelength) 250-2970~$\mu$m data in logarithmic space while applying corrections for CMB effects \citep{dacunha2013}.  Table~\ref{t_template} lists the photometric redshifts from these templates as well as spectroscopic redshifts.  Again, we only performed this analysis for fields with single sources that have spectroscopic redshifts and for fields with multiple detected sources that all have the same spectroscopic redshift.  Table~\ref{t_templvsspec} lists the means and standard deviations of the ratios of the photometric redshifts to the spectroscopic redshifts as well as the metric $(z_{\text{phot}}-z_{\text{spec}})/(1+z_{\text{spec}})$.  Additionally, Figure~\ref{f_template} shows, for all of the objects with spectroscopic redshifts, normalized flux densities plotted at rest wavelengths along with two versions of the templates: one set of templates plotted at their original rest wavelengths and another set of templates shifted by $(z_{\text{phot}}-z_{\text{spec}})/(1+z_{\text{spec}})$. Figure~\ref{f_zcomp} shows comparisons of the photometric and spectroscopic redshifts.

\begin{table*}
\centering
\begin{minipage}{105mm}
\caption{Photometric redshifts (based on template fits to 250-2970~$\mu$m data) and spectroscopic redshifts for fields with spectroscopic redshifts$^a$}
\label{t_template}
\begin{tabular}{@{}lcccccc@{}}
\hline
Field &
  \multicolumn{6}{c}{$z$} \\
&
  Pearson et &
  Bakx et al. &
  Eyelash &
  Pope et al. &
  ALESS &
  Spectro- \\
&
  al. template &
  template &
  template &
  template &
  template &
  scopic \\  
\hline
\multicolumn{6}{l}{Fields with single sources} \\
HerBS-11 &
  2.1 $\pm$ 0.3 &
  1.6 $\pm$ 0.2 &
  2.4 $\pm$ 0.3 &
  2.2 $\pm$ 0.3 &
  2.1 $\pm$ 0.3 &
  2.630 \\
HerBS-14 &
  3.3 $\pm$ 0.4 &
  2.8 $\pm$ 0.4 &
  3.5 $\pm$ 0.5 &
  3.4 $\pm$ 0.5 &
  3.3 $\pm$ 0.5 &
  3.780 \\
HerBS-18 &
  2.0 $\pm$ 0.2 &
  1.5 $\pm$ 0.2 &
  2.3 $\pm$ 0.3 &
  2.2 $\pm$ 0.3 &
  2.1 $\pm$ 0.3 &
  2.180 \\
HerBS-24 &
  2.2 $\pm$ 0.3 &
  1.7 $\pm$ 0.2 &
  2.5 $\pm$ 0.4 &
  2.3 $\pm$ 0.3 &
  2.2 $\pm$ 0.3 &
  2.200 \\
HerBS-25 &
  2.7 $\pm$ 0.3 &
  2.2 $\pm$ 0.3 &
  3.0 $\pm$ 0.4 &
  2.9 $\pm$ 0.4 &
  2.8 $\pm$ 0.4 &
  2.910 \\
HerBS-27 &
  4.1 $\pm$ 0.5 &
  3.7 $\pm$ 0.5 &
  4.2 $\pm$ 0.6 &
  4.2 $\pm$ 0.6 &
  4.0 $\pm$ 0.6 &
  4.510 \\
HerBS-28 &
  3.5 $\pm$ 0.4 &
  3.0 $\pm$ 0.4 &
  3.7 $\pm$ 0.5 &
  3.7 $\pm$ 0.5 &
  3.5 $\pm$ 0.5 &
  3.920 \\
HerBS-36 &
  3.0 $\pm$ 0.4 &
  2.5 $\pm$ 0.3 &
  3.2 $\pm$ 0.5 &
  3.1 $\pm$ 0.4 &
  3.0 $\pm$ 0.4 &
  3.090 \\
HerBS-37 &
  1.8 $\pm$ 0.2 &
  1.3 $\pm$ 0.2 &
  2.1 $\pm$ 0.3 &
  1.9 $\pm$ 0.3 &
  1.9 $\pm$ 0.3 &
  2.620 \\
HerBS-39 &
  2.5 $\pm$ 0.3 &
  2.0 $\pm$ 0.3 &
  2.8 $\pm$ 0.4 &
  2.6 $\pm$ 0.4 &
  2.6 $\pm$ 0.4 &
  3.230 \\
HerBS-40 &
  1.5 $\pm$ 0.2 &
  1.1 $\pm$ 0.1 &
  1.9 $\pm$ 0.3 &
  1.7 $\pm$ 0.2 &
  1.8 $\pm$ 0.3 &
  1.970 \\
HerBS-47 &
  1.7 $\pm$ 0.2 &
  1.3 $\pm$ 0.2 &
  2.1 $\pm$ 0.3 &
  1.9 $\pm$ 0.3 &
  1.9 $\pm$ 0.3 &
  2.430 \\
HerBS-55 &
  1.8 $\pm$ 0.2 &
  1.3 $\pm$ 0.2 &
  2.2 $\pm$ 0.3 &
  1.9 $\pm$ 0.3 &
  1.9 $\pm$ 0.3 &
  2.660 \\
HerBS-57 &
  2.4 $\pm$ 0.3 &
  1.9 $\pm$ 0.3 &
  2.8 $\pm$ 0.4 &
  2.5 $\pm$ 0.4 &
  2.6 $\pm$ 0.4 &
  3.270 \\
HerBS-60 &
  2.7 $\pm$ 0.3 &
  2.2 $\pm$ 0.3 &
  3.1 $\pm$ 0.4 &
  2.9 $\pm$ 0.4 &
  2.8 $\pm$ 0.4 &
  3.260 \\
HerBS-68 &
  1.9 $\pm$ 0.2 &
  1.4 $\pm$ 0.2 &
  2.2 $\pm$ 0.3 &
  2.0 $\pm$ 0.3 &
  2.0 $\pm$ 0.3 &
  2.720 \\
HerBS-73 &
  2.2 $\pm$ 0.3 &
  1.7 $\pm$ 0.2 &
  2.5 $\pm$ 0.4 &
  2.3 $\pm$ 0.3 &
  2.3 $\pm$ 0.3 &
  3.020 \\
HerBS-86 &
  2.2 $\pm$ 0.3 &
  1.7 $\pm$ 0.2 &
  2.6 $\pm$ 0.4 &
  2.3 $\pm$ 0.3 &
  2.3 $\pm$ 0.3 &
  2.560 \\
HerBS-93 &
  2.0 $\pm$ 0.2 &
  1.5 $\pm$ 0.2 &
  2.4 $\pm$ 0.3 &
  2.1 $\pm$ 0.3 &
  2.2 $\pm$ 0.3 &
  2.400 \\
HerBS-103 &
  1.9 $\pm$ 0.2 &
  1.5 $\pm$ 0.2 &
  2.3 $\pm$ 0.3 &
  2.0 $\pm$ 0.3 &
  2.0 $\pm$ 0.3 &
  2.940 \\
HerBS-107 &
  1.8 $\pm$ 0.2 &
  1.3 $\pm$ 0.2 &
  2.1 $\pm$ 0.3 &
  1.9 $\pm$ 0.3 &
  1.9 $\pm$ 0.3 &
  2.550 \\
HerBS-111 &
  1.8 $\pm$ 0.2 &
  1.4 $\pm$ 0.2 &
  2.2 $\pm$ 0.3 &
  1.9 $\pm$ 0.3 &
  2.0 $\pm$ 0.3 &
  2.370 \\
HerBS-123 &
  1.9 $\pm$ 0.2 &
  1.4 $\pm$ 0.2 &
  2.2 $\pm$ 0.3 &
  2.1 $\pm$ 0.3 &
  2.0 $\pm$ 0.3 &
  2.170 \\
HerBS-132 &
  1.6 $\pm$ 0.2 &
  1.1 $\pm$ 0.1 &
  2.0 $\pm$ 0.3 &
  1.5 $\pm$ 0.2 &
  1.6 $\pm$ 0.2 &
  2.470 \\
HerBS-141 &
  1.4 $\pm$ 0.2 &
  1.0 $\pm$ 0.1 &
  1.8 $\pm$ 0.2 &
  1.5 $\pm$ 0.2 &
  1.5 $\pm$ 0.2 &
  2.090 \\
HerBS-160 &
  3.7 $\pm$ 0.4 &
  3.1 $\pm$ 0.4 &
  3.8 $\pm$ 0.5 &
  3.8 $\pm$ 0.5 &
  3.6 $\pm$ 0.5 &
  3.960 \\
HerBS-182 &
  1.8 $\pm$ 0.2 &
  1.3 $\pm$ 0.2 &
  2.1 $\pm$ 0.3 &
  1.8 $\pm$ 0.3 &
  1.8 $\pm$ 0.3 &
  2.230 \\
HerBS-184 &
  2.2 $\pm$ 0.3 &
  1.7 $\pm$ 0.2 &
  2.6 $\pm$ 0.4 &
  2.3 $\pm$ 0.3 &
  2.2 $\pm$ 0.3 &
  2.510 \\
HerBS-189 &
  2.1 $\pm$ 0.3 &
  1.7 $\pm$ 0.2 &
  2.5 $\pm$ 0.3 &
  2.3 $\pm$ 0.3 &
  2.3 $\pm$ 0.3 &
  3.300 \\
HerBS-200 &
  1.6 $\pm$ 0.2 &
  1.1 $\pm$ 0.1 &
  1.9 $\pm$ 0.3 &
  1.5 $\pm$ 0.2 &
  1.6 $\pm$ 0.2 &
  2.150 \\
HerBS-207 &
  1.6 $\pm$ 0.2 &
  1.2 $\pm$ 0.1 &
  2.0 $\pm$ 0.3 &
  1.6 $\pm$ 0.2 &
  1.7 $\pm$ 0.2 &
  1.570 \\

\noalign{\vskip 0.5em}
\multicolumn{6}{l}{Fields with multiple sources at the same redshift} \\
HerBS-21 &
  2.6 $\pm$ 0.3 &
  2.1 $\pm$ 0.3 &
  2.9 $\pm$ 0.4 &
  2.8 $\pm$ 0.4 &
  2.7 $\pm$ 0.4 &
  3.323 \\
HerBS-49 &
  2.6 $\pm$ 0.3 &
  2.1 $\pm$ 0.3 &
  2.9 $\pm$ 0.4 &
  2.8 $\pm$ 0.4 &
  2.6 $\pm$ 0.4 &
  2.727 \\
HerBS-69 &
  1.6 $\pm$ 0.2 &
  1.2 $\pm$ 0.2 &
  2.0 $\pm$ 0.3 &
  1.7 $\pm$ 0.2 &
  1.8 $\pm$ 0.2 &
  2.074 \\
HerBS-120 &
  2.8 $\pm$ 0.3 &
  2.3 $\pm$ 0.3 &
  3.1 $\pm$ 0.4 &
  3.1 $\pm$ 0.4 &
  2.9 $\pm$ 0.4 &
  3.124 \\
HerBS-159 &
  1.8 $\pm$ 0.2 &
  1.4 $\pm$ 0.2 &
  2.2 $\pm$ 0.3 &
  2.0 $\pm$ 0.3 &
  2.0 $\pm$ 0.3 &
  2.236 \\
HerBS-208 &
  2.3 $\pm$ 0.3 &
  1.8 $\pm$ 0.2 &
  2.6 $\pm$ 0.4 &
  2.4 $\pm$ 0.3 &
  2.4 $\pm$ 0.3 &
  2.480 \\
\hline
\end{tabular}
$^a$ The photometric redshifts in this table may differ from those published for the same sources in prior papers that did not incorporate ALMA continuum measurements into their SEDs \citep[e.g.,][]{bakx2020erratum, urquhart2022}.  The uncertainties for the photometric redshift incorporate both the uncertainties from fitting the templates to the data and the accuracies of these templates as published by \citet{pearson2013}, \citet{ivison2016}, and \citet{bakx2018}.  The spectroscopic redshifts have uncertainties of $<$0.001.
\end{minipage}
\end{table*}

\begin{table}
\caption{Statistics of the comparisons of photometric to spectroscopic redshifts}
\label{t_templvsspec}
\begin{center}
\begin{tabular}{@{}lcccc@{}}
\hline
SED model&
  \multicolumn{2}{c}{$\frac{z_{\text{phot}}}{z_{\text{spec}}}$} &
  \multicolumn{2}{c}{$\frac{z_{\text{phot}}-z_{\text{spec}}}{1+z_{\text{spec}}}$} \\
&
  Mean &
  $\sigma$ &
  Mean &
  $\sigma$ \\
\hline
Pearson et al. &
  0.81 &
  0.11 &
  -0.14 &
  0.08 \\
Bakx et al. &
  0.63 &
  0.11 &
  -0.27 &
  0.07 \\
Eyelash &
  0.94 &
  0.11 &
  -0.04 &
  0.08 \\
Pope et al. &
  0.86 &
  0.11 &
  -0.10 &
  0.08 \\
ALESS &
  0.84 &
  0.10 &
  -0.11 &
  0.07 \\
\hline
\end{tabular}
\end{center}
\end{table}

Overall, these data show that these photometric templates systematically underestimate the actual redshifts of these sources by typically $\sim$15\%, although the Eyelash template performs slightly better than the other templates and the Bakx template performs notably worse.  The panel on the left in Figure~\ref{f_template} illustrates that this is because the SED templates are all slightly colder than the dust that we observe from our sources.  Four of the templates are based on modified blackbodies with temperatures between 20 and 30~K, while the \citet{pope2008} template has colours consistent with a modified blackbody with a temperature of 32~K and a $\beta$ of 2.  In contrast, the sources in our sample have colour temperatures ranging from 26 to 38~K when $\beta$ is fixed to 2.  This temperature difference is not apparent when looking at just the {\it Herschel} data, which sample the peak of the SED, but it is easy to see that all of the templates lie redwards of the ALMA data.  Additionally, the $\beta$ of 1.83 used by \citet{bakx2018} makes the Rayleigh-Jeans side of the dust emission in that specific template look broader than observed in our sources, and this creates an additional offset between the measurements and the template that has a notable effect on the photometric redshifts.

It would be tempting to interpret the results from Figure~\ref{f_template} as providing evidence that the dust emissivity is steeper than what is assumed in these templates, but except for the \citet{bakx2018} template, this is not the case.  The results of the SED fits in Section~\ref{s_sedfit} showed that they are largely consistent with modified blackbodies with $\beta$ values of 2, and the analysis of the 151/101~GHz ratios in Section~\ref{s_151-101ratios} demonstrated that they too are generally consistent with $\beta$ of 2.  This is also the $\beta$ used in the ALESS, Eyelash, \citet{pearson2013}, and \citet{pope2008} templates, and in fact, our ALMA data lie parallel to but offset from these templates.  Additionally, when the SED templates are shifted to match the spectroscopic redshifts better, as seen in the panel on the right in Figure~\ref{f_template}, the templates replicate the slope from the {\it Herschel} data to the ALMA data much better (although, with this correction, the \citet{pearson2013} and \citet{bakx2018} templates predict significantly higher emission at $\leq$60~$\mu$m then what is indicated by our data).  Hence, it is more likely that the offset between our data and these templates is related to dust temperatures.

The differences between the SEDs of our sample and the SED templates potentially relate to how the various samples were selected for creating the SED templates and how they differ from our sample.  \citet{pearson2013} used a large number of sources at $z<1$ to build their template, which is a significantly lower redshift than what we measured for sources in our sample.  In using these closer objects, it may be possible that \citet{pearson2013} were biased towards selecting objects with colder dust.  The \citet{pope2008} and the ALESS templates used sources selected solely by their submillimetre flux densities, while our sample was also selected by the photometric redshifts inferred from their 250-500~$\mu$m colours, and our colour selection criteria yielded a sample that was slightly more distant and that may also have warmer dust than what is found in typical submillimetre galaxies.  Notably, the Eyelash template, which is based on a gravitational lens at $z=2.3$ and which is therefore comparable to many of the objects in our sample, actually performs reasonably well compared to the other templates, although it still systematically predicts low photometric redshifts.  The only template that should not have been significantly affected by sample selection criteria is the \citet{bakx2018} template, which produced the parent sample that was the basis for ours.  However, they did not constrain the Rayleigh-Jeans side of the SED well when constructing their template, and both the lower temperatures and  $\beta$ used by \citet{bakx2018} caused the discrepancies between the photometric redshifts based on their template and the spectroscopic redshifts.

Although the templates all systematically undermeasure the redshifts to our sample galaxies, the low standard deviation of 0.07-0.10 in the $(z_{\text{phot}}-z_{\text{spec}})/(1+z_{\text{spec}})$ values indicates that the results are notably precise.  In fits to SED measurements at observed wavelengths $\leq$850~$\mu$m using the same templates that we have used, the standard deviation in this metric is typically 0.12-0.14 \citep{pearson2013, ivison2016, bakx2018}.  This indicates that including data at $\sim$2-3~mm ($\sim$100-150~GHz) to constrain the Rayleigh-Jeans side of the SED will also lead to more precise photometric redshift measurements even if they still have accuracy issues.

Given the issues with photometric redshifts derived using older SED templates, it is clear that any specific SED template for high-redshift far-infrared or submillimetre sources may not be universally applicable or even reliable.  To measure an accurate photometric redshift to a specific galaxy or class of galaxies, the best results will potentially be obtained when using SED templates derived from the same class of galaxies.  However, spectroscopic redshifts are imperative to confirm those measurements.

\section{Conclusions}

We have presented 101 and 151~GHz (ALMA Band 3 and 4) photometry for 85 fields originally selected from the H-ATLAS observations of the South Galactic Pole as potentially containing gravitational lenses based on their 500~$\mu$m flux densities and their relatively red {\it Herschel} colours.  We detected 151~GHz continuum emission within every targeted field, and we detected 101~GHz continuum emission within 55 fields in the survey.

21 of these fields contained either double-lobed or extended sources, some with relatively complex morphologies.  About half of the fields contained multiple sources within the region described by the {\it Herschel} 500~$\mu$m beam, which is consistent with some prior surveys of some bright sources selected using infrared or submillimetre single-dish telescopes \citep{hodge2013, stach2018} but either higher or lower than results from other surveys \citep{bussmann2015, scudder2016, cowie2018, montana2021}.   Notably, we also find that many of the fields in our sample with the brightest submillimetre flux densities contain single bright objects, which is expected for our sample (which was optimized for selecting gravitationally lensed sources that would be unresolved in our ALMA data) but which contrasts with the results from the fields selected from APEX data by \citet{hodge2013} and \citet{karim2013} and the fields selected from JCMT data by \citet{stach2018}.  These variations between our results and others as well as among the published results in the literature appear to be a consequence of difference in the sample selection criteria; both the waveband used to identify high-redshift sources and the application of colour selection criteria can potentially affect the multiplicity results.

For the subset of fields that either contain a single detected source with a spectroscopic redshift or that contain two detected sources with the same spectroscopic redshift (as based on the data from \citet{urquhart2022}), we performed some analyses on the SEDs of the {\it Herschel} and ALMA data.  The SEDs for this subset are largely consistent with dust described by single modified blackbodies with colour temperatures ranging from 26 to 38~K and dust emissivity indices $\beta$ of 2.  We also demonstrated that the ALMA (observed frame) 151/101~GHz ratios provided a more reliable measurement of $\beta$ than the single modified blackbodies that are fitted to the {\it Herschel} and ALMA data.  With rare exceptions, we found no evidence of other sources of emission in the ALMA bands.

We found that the relation between colour temperature and redshift for our sample was relatively weak.  We measured a slope of $2.2 \pm 0.8$ K $z^{-1}$ that is only inconsistent with no evolution at the $\sim$2.75$\sigma$ level, and the correlation coefficient for the relation was 0.39.  Our relatively weak relation is largely consistent with studies based on samples selected at far-infrared, submillimetre, and millimetre wavelengths, which generally found either weak relations or no relations \citep[e.g.,][]{reuter2020, dudzeviciute2021}, but it is generally inconsistent with the stronger relations found using samples based on samples selected at optical and near-infrared wavelengths \citep[e.g.,][]{schreiber2018, bouwens2020}.  The differences among these results seem largely driven by selection effects where samples selected from optical and near-infrared bands seem to be missing galaxies with relatively cold but large dust masses.

We also tested the performance of photometric redshifts derived from five SED templates versus the spectroscopic redshifts from \citet{urquhart2022}, and we generally found that all of these photometric redshifts were systematically lower than the spectroscopic redshifts, although the template based on the SED of the Cosmic Eyelash \citep{ivison2010, swinbank2010} performed best.  The colour temperatures of the dust in these SED templates are generally colder than what we found in our sample galaxies, which again may point to differences between our sample of galaxies and the galaxies used to create these templates.  These results demonstrate that SED templates are not universally applicable to all galaxies and also imply that the best photometric redshifts to specific galaxies may be obtained when using SED templates derived from the same class of galaxies.  However, also note that the relative scatter that we measured between the photometric and spectroscopic redshifts, as measured using $(z_{\text{phot}}-z_{\text{spec}})/(1+z_{\text{spec}})$, is generally $\sim$2$\times$ lower than what had previously been measured in other comparisons of photometric and spectroscopic redshifts \citep[e.g.,][]{pearson2013,ivison2016,bakx2018}.  This indicates that SED fits that include data at $\lesssim$150~GHz (which covers ALMA Bands 3 and 4) are very useful for constraining the Rayleigh-Jeans side of the SEDs of high-redshift sources and help to improve the precision of the photometric redshifts from SED templates.

These observations represent a significant step in understanding the phenomenology of the bright, red sources found in H-ATLAS, but the results are also clearly applicable to similar sources from other surveys such as HerMES.  Both these photometric results and the spectroscopic data from \citet{urquhart2022} have allowed us, to some degree, to separate individual infrared-bright objects at high redshift, associated galaxies at high redshift, and confused sources.  The individual infrared-bright high-redshift objects are potentially gravitationally lensed galaxies or HLIRGs and should be studied further in follow-up ALMA observations to confirm the phenomenology of these objects, while some of the associated high-redshift sources could actually be parts of protoclusters and should be examined more carefully at multiple wavelengths to understand more about how these structures are forming.

\section*{Acknowledgements}

The authors thank the anonomous reviewer as well as Robert Ivison and Shuowen Jin for their comments on this paper.  GJB acknowledges support from STFC Grant ST/T001488/1.  SS was partly supported by the ESCAPE project; ESCAPE - The European Science Cluster of Astronomy \& Particle Physics ESFRI Research Infrastructures has received funding from the European Union's Horizon 2020 research and innovation programme under Grant Agreement no. 824064. SS also thanks the Science and Technology Facilities Council for financial support under grant ST/P000584/1. SU would like to thank the Open University School of Physical Sciences for supporting this work. TB acknowledges funding from NAOJ ALMA Scientific Research Grant Numbers 2018-09B and JSPS KAKENHI No. 17H06130.  HD acknowledges financial support from the Agencia Estatal de Investigación del Ministerio de Ciencia e Innovación (AEI-MCINN) under grant (La evolución de los cíumulos de galaxias desde el amanecer hasta el mediodía cósmico) with reference (PID2019-105776GB-I00/DOI:10.13039/501100011033) and acknowledge support from the ACIISI, Consejería de Economía, Conocimiento y Empleo del Gobierno de Canarias and the European Regional Development Fund (ERDF) under grant with reference PROID2020010107.  {\it Herschel} is an ESA space observatory with science instruments provided by European-led Principal Investigator consortia and with important participation from NASA.  This paper makes use of the following ALMA data: ADS/JAO.ALMA\#2016.2.00133.S, 2018.1.00804.S, and 2019.1.01477.S. ALMA is a partnership of ESO (representing its member states), NSF (USA) and NINS (Japan), together with NRC (Canada), MOST and ASIAA (Taiwan), and KASI (Republic of Korea), in cooperation with the Republic of Chile. The Joint ALMA Observatory is operated by ESO, AUI/NRAO and NAOJ.

\section*{Data Availability}

The {\it Herschel} SPIRE data can be downloaded from \url{https://www.h-atlas.org}\,, while the reduced, calibrated and science-ready ALMA data is available from the ALMA Science Archive at \url{https://almascience.eso.org/alma-data/archive}\,.

\bibliographystyle{mnras}
\bibliography{bendogj}

\appendix

\section{ALMA 101 and 151~GHz images}
\label{a_images}

	\begin{figure*}
		\begin{center}
			\includegraphics[width=7cm]{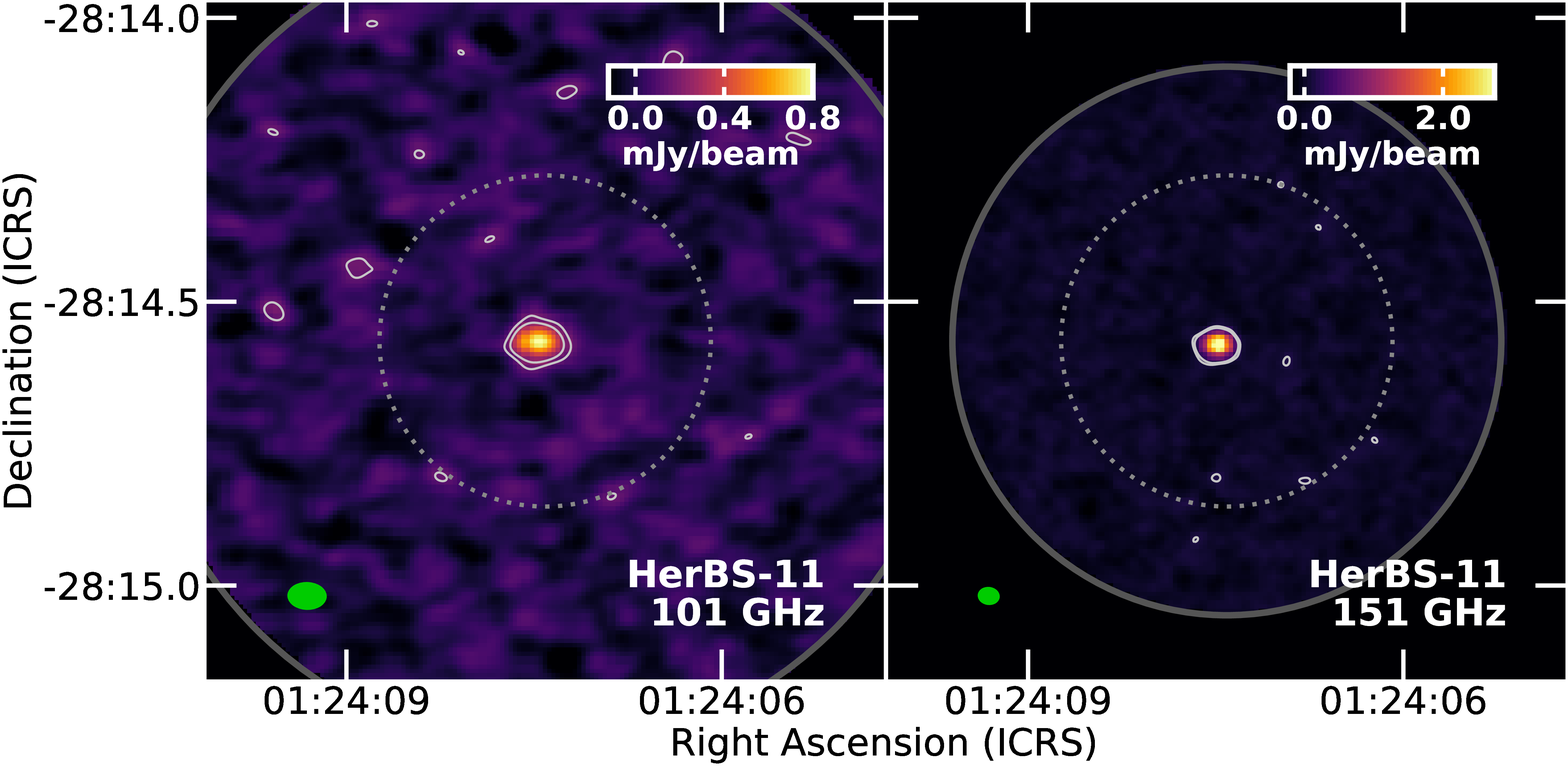}
			\ \ \ \ \ \
			\includegraphics[width=7cm]{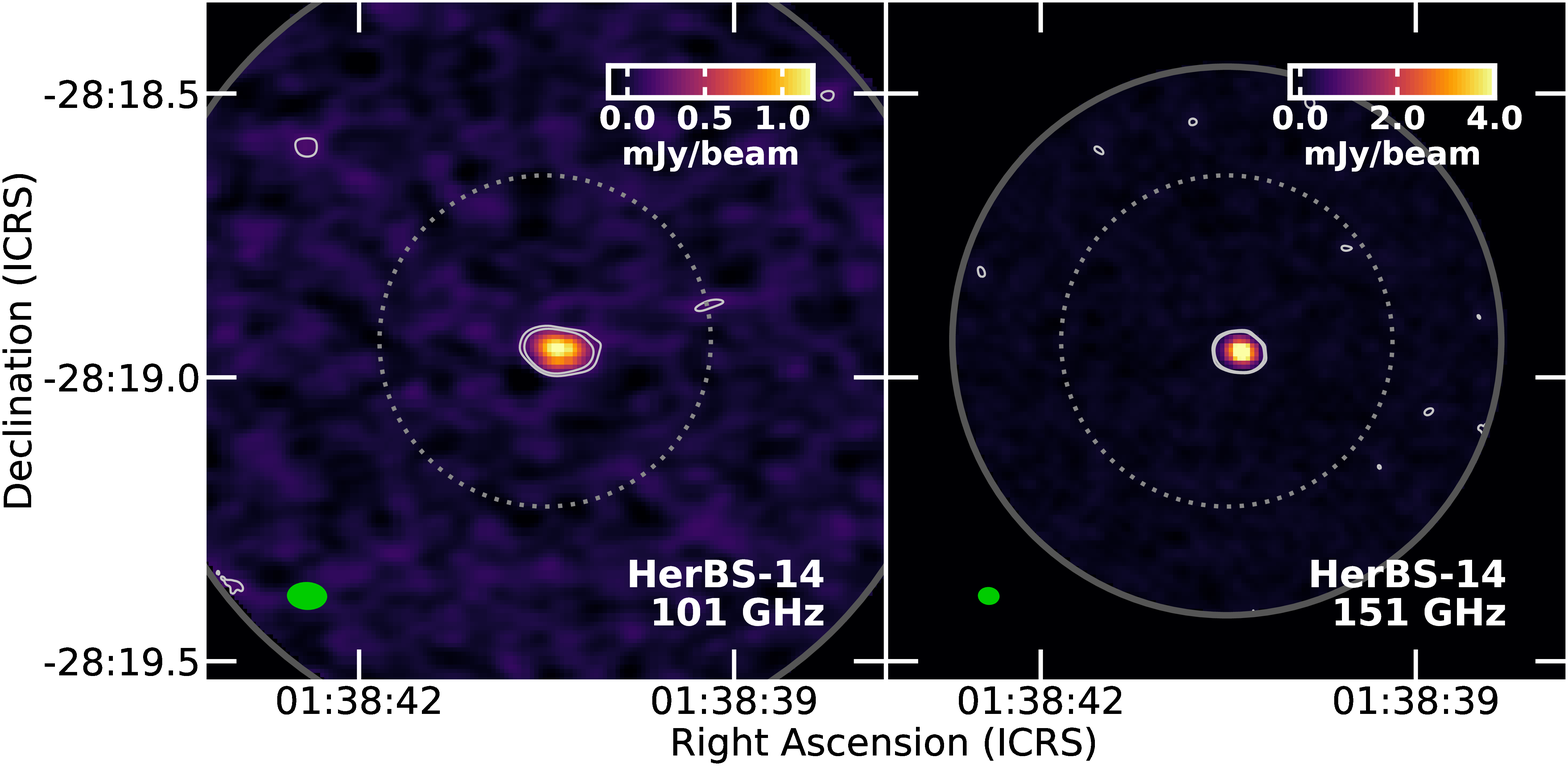}
			\vspace{0.5em}\\
			\includegraphics[width=7cm]{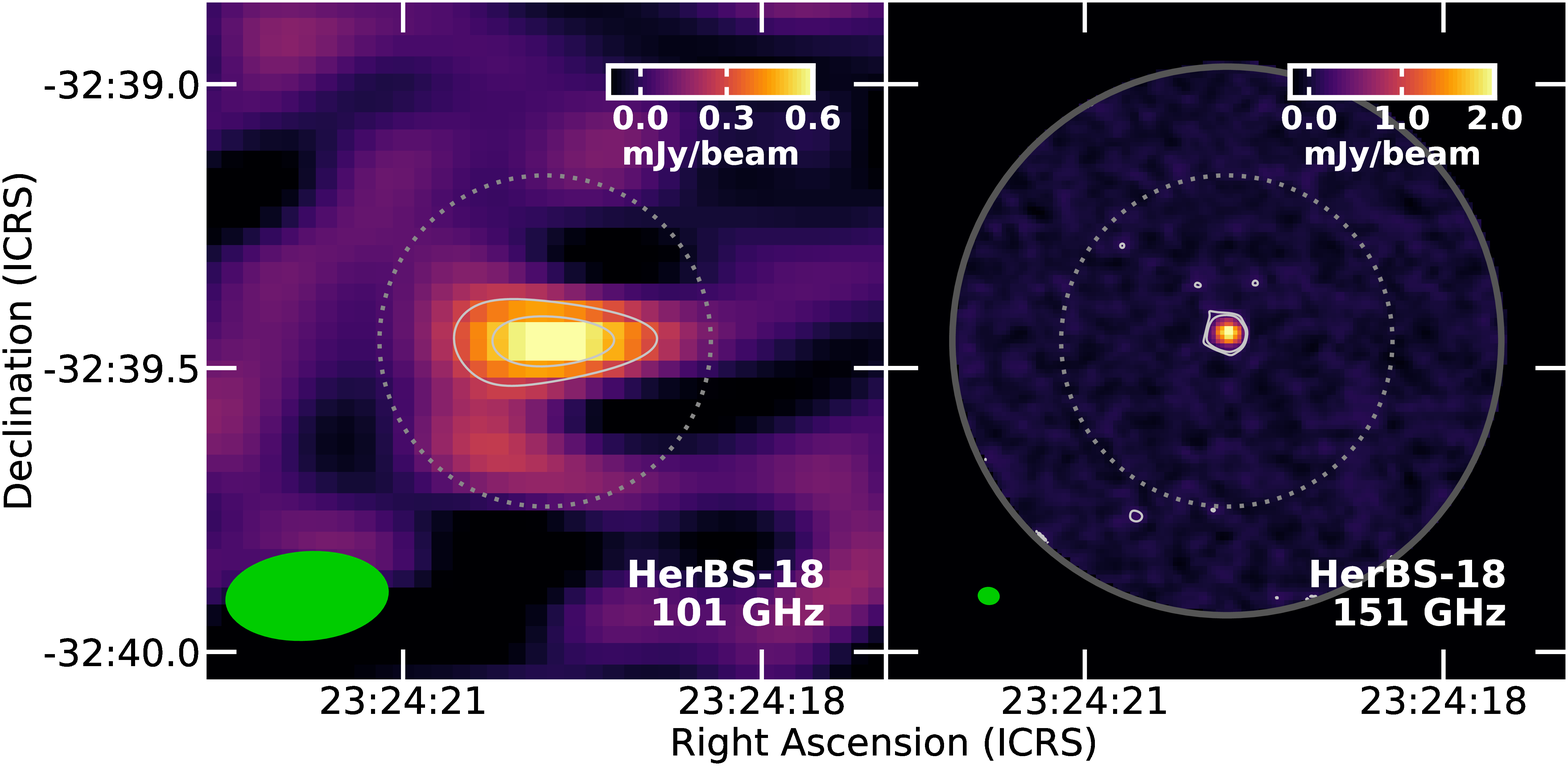}
			\ \ \ \ \ \
			\includegraphics[width=7cm]{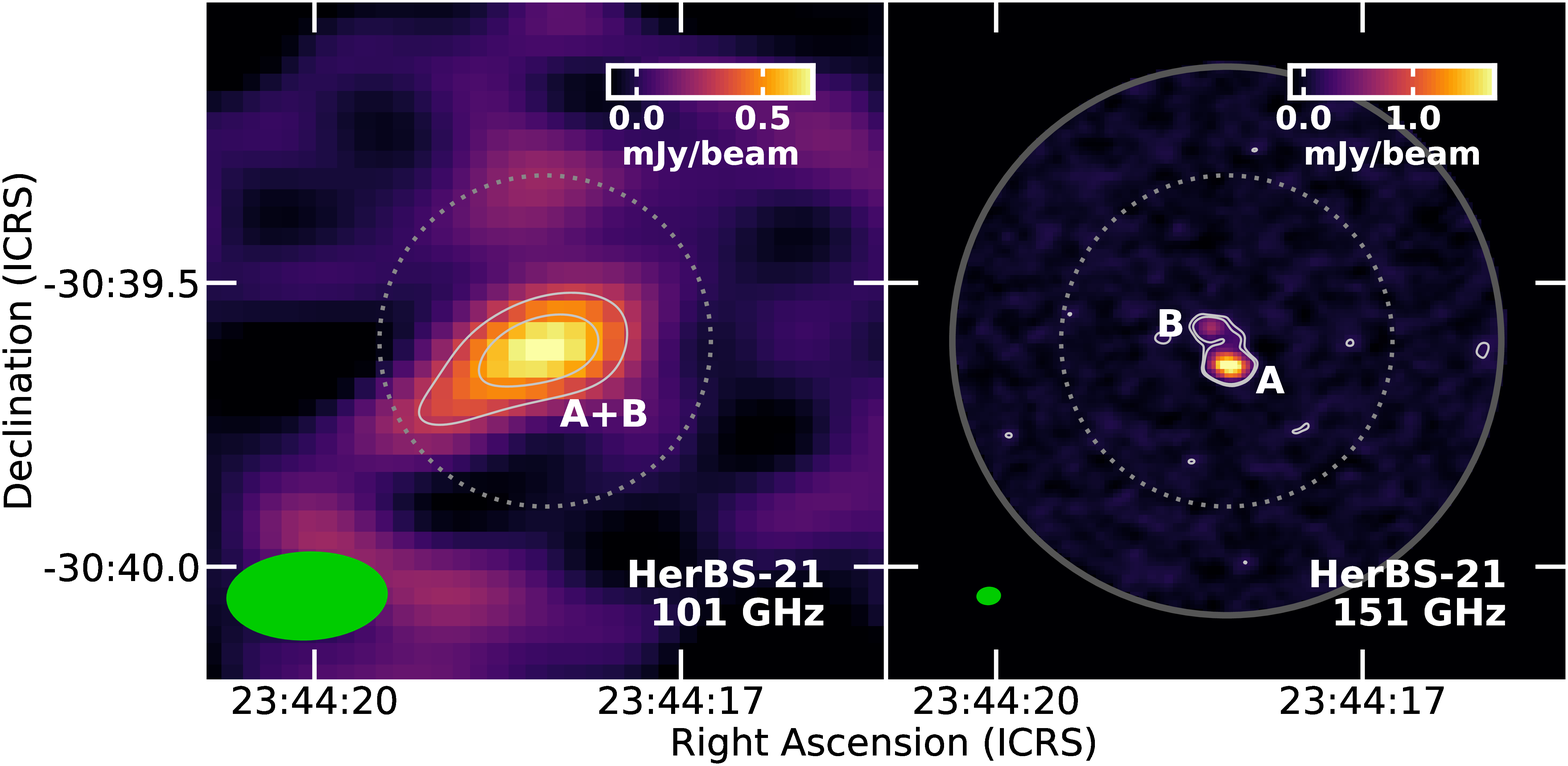}
			\vspace{0.5em}\\
			\includegraphics[width=7cm]{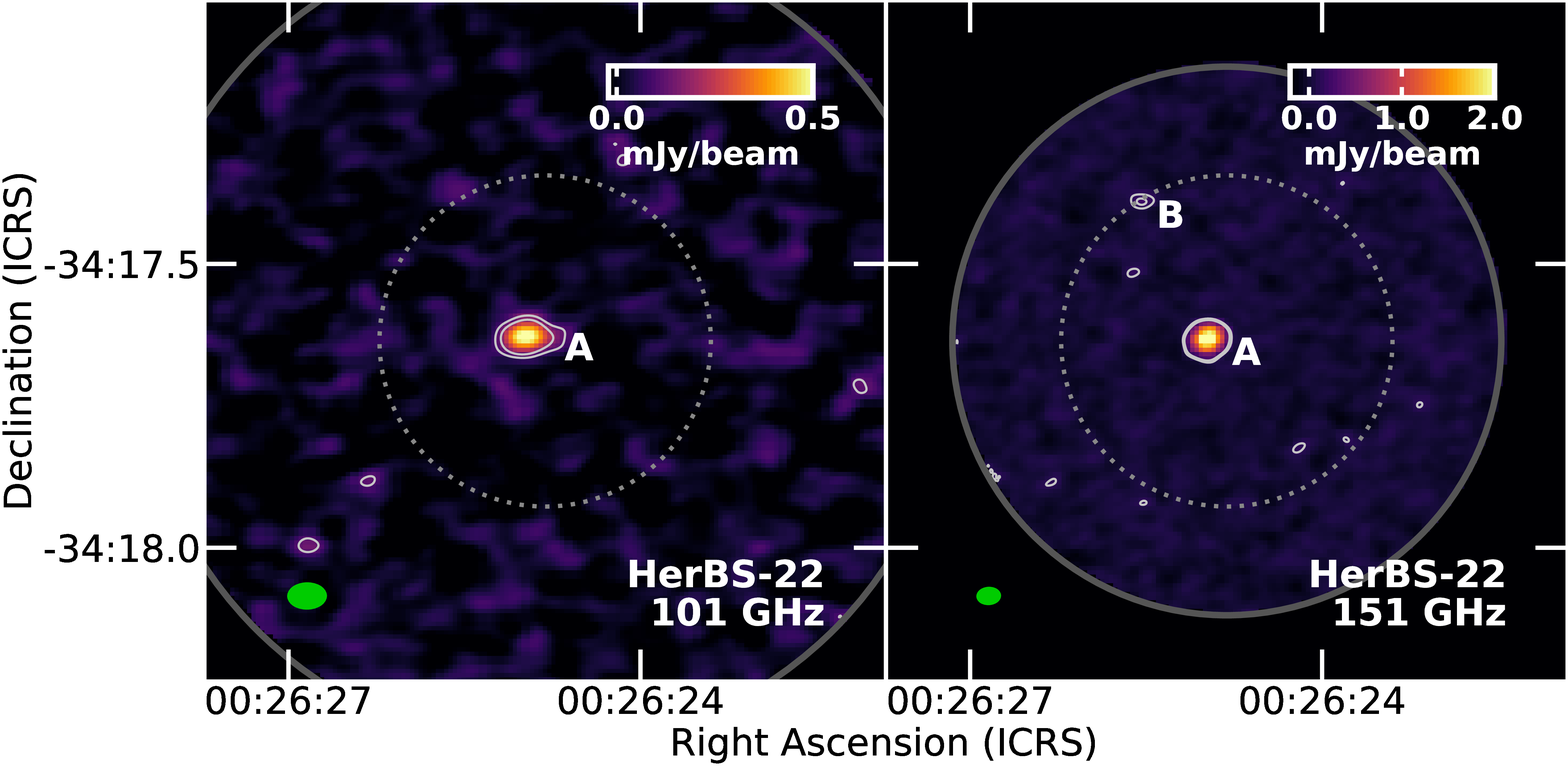}
			\ \ \ \ \ \
			\includegraphics[width=7cm]{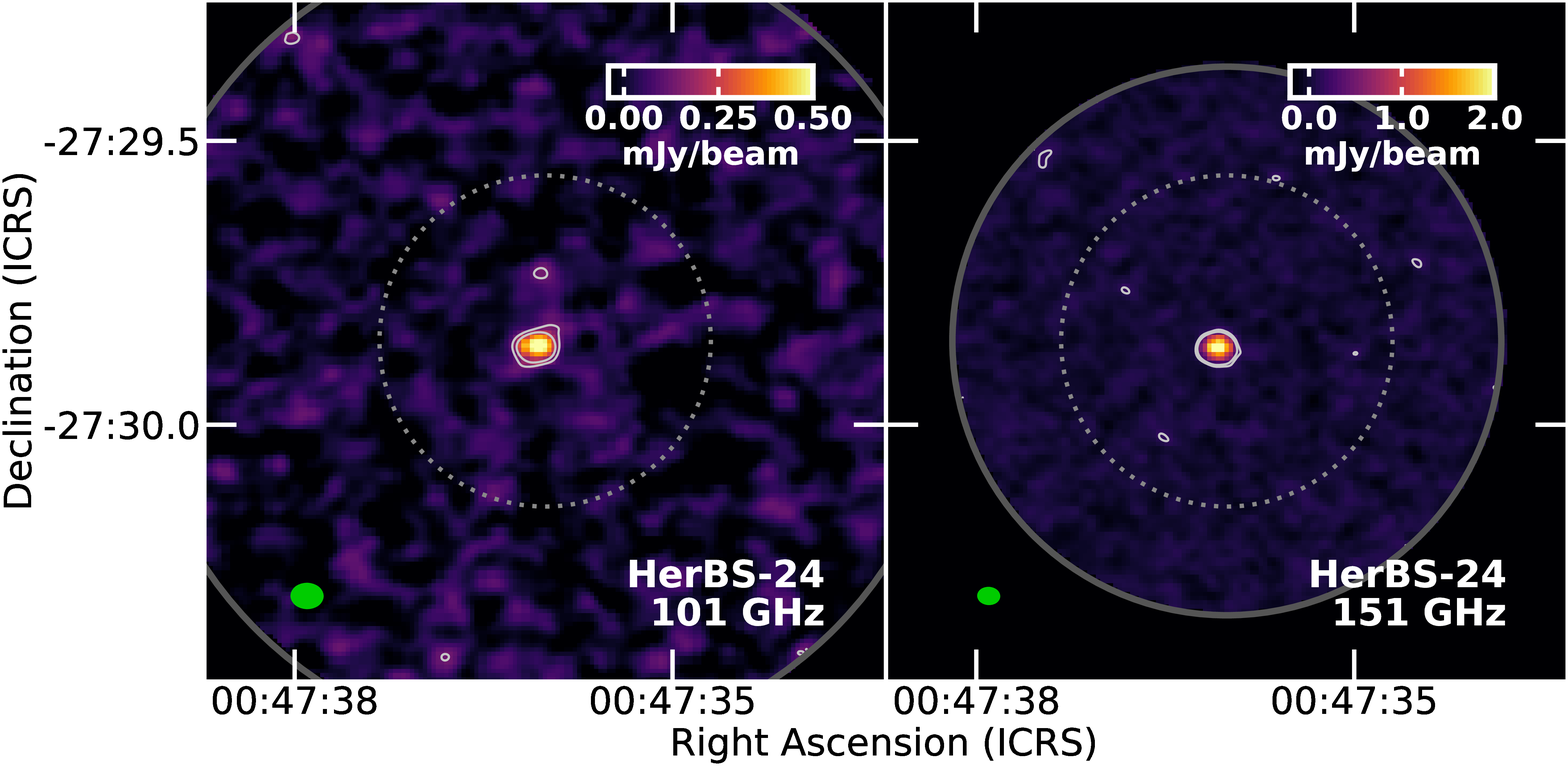}
			\vspace{0.5em}\\
			\includegraphics[width=7cm]{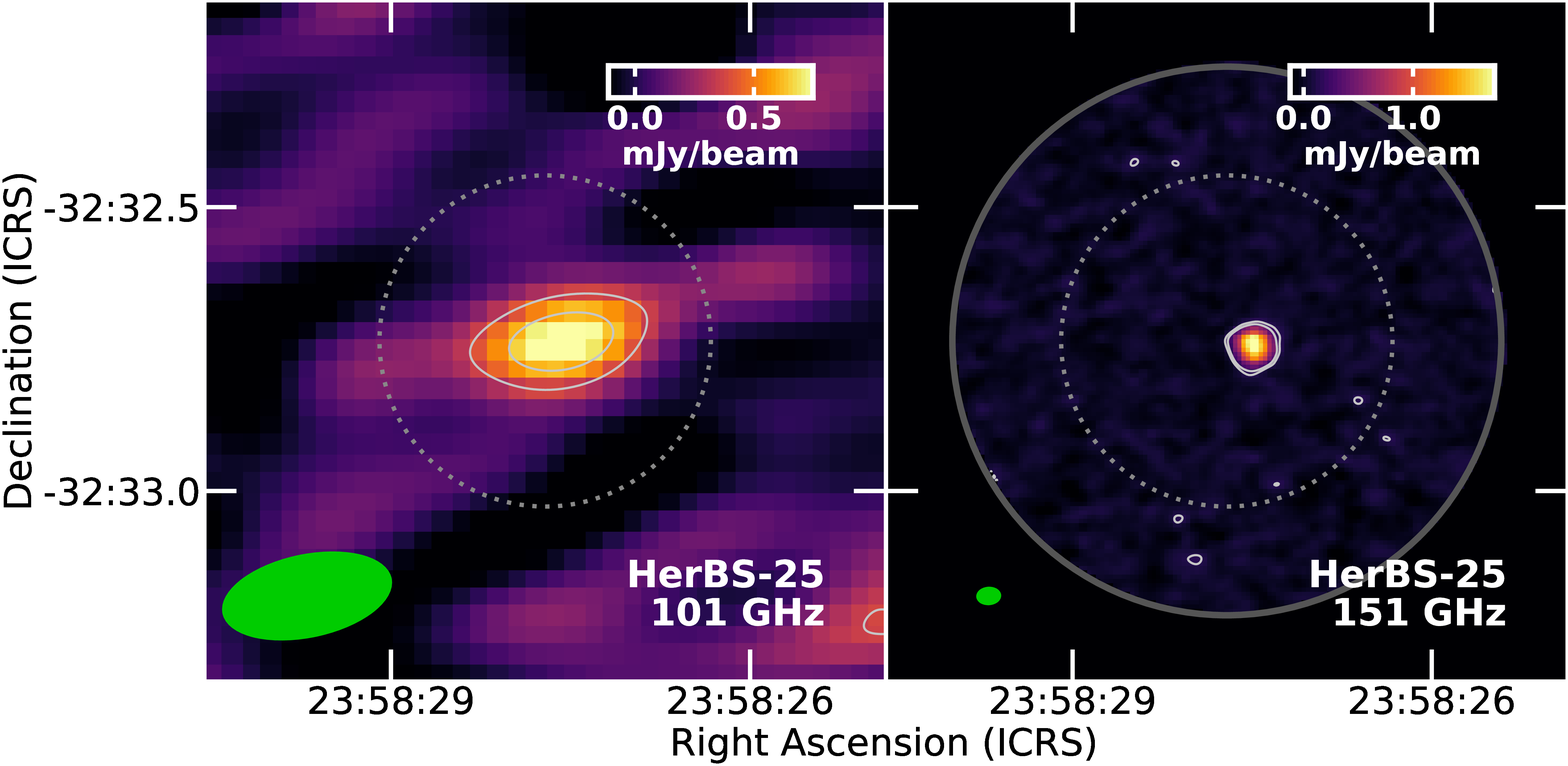}
			\ \ \ \ \ \
			\includegraphics[width=7cm]{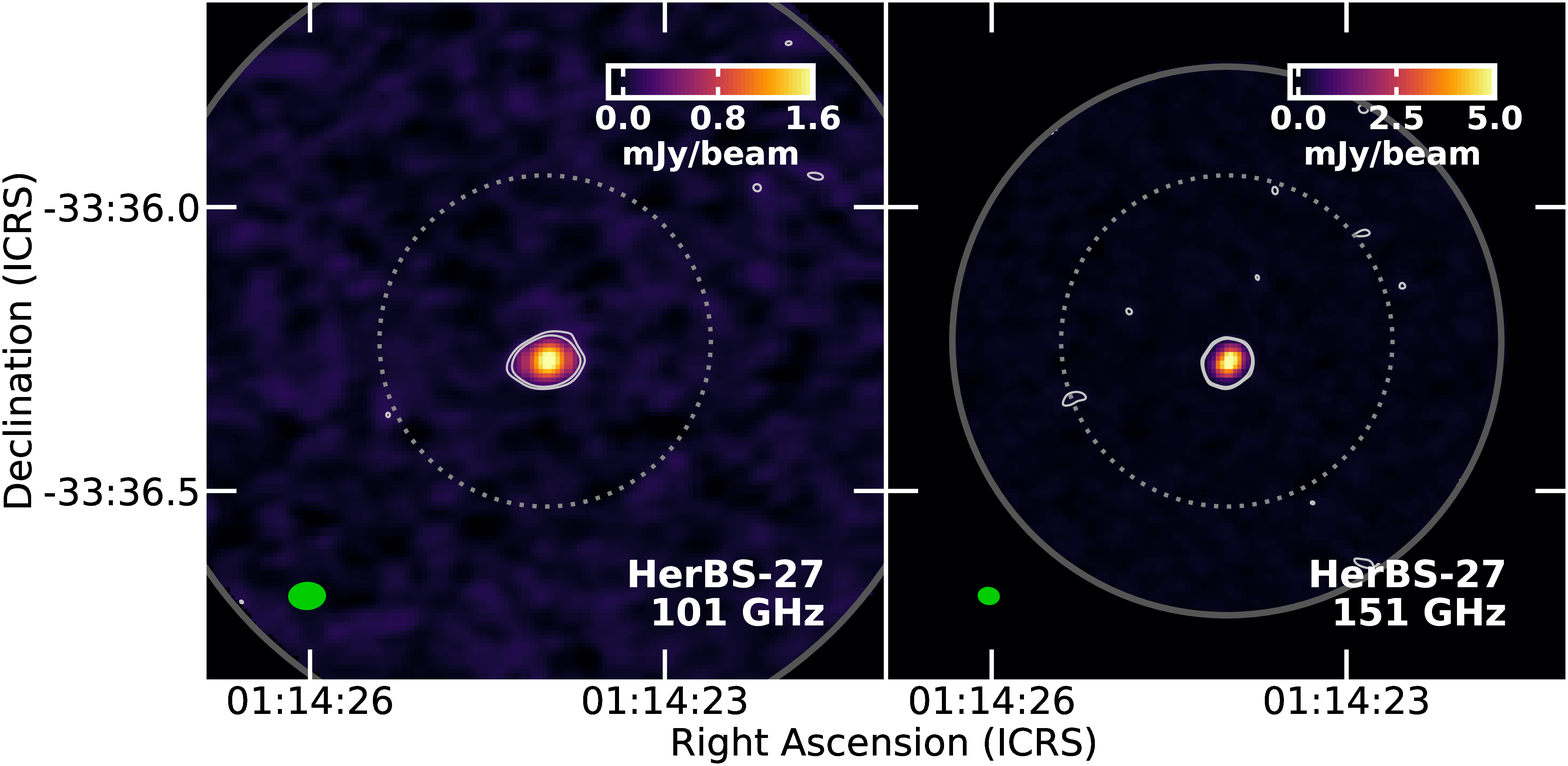}
			\vspace{0.5em}\\
			\includegraphics[width=7cm]{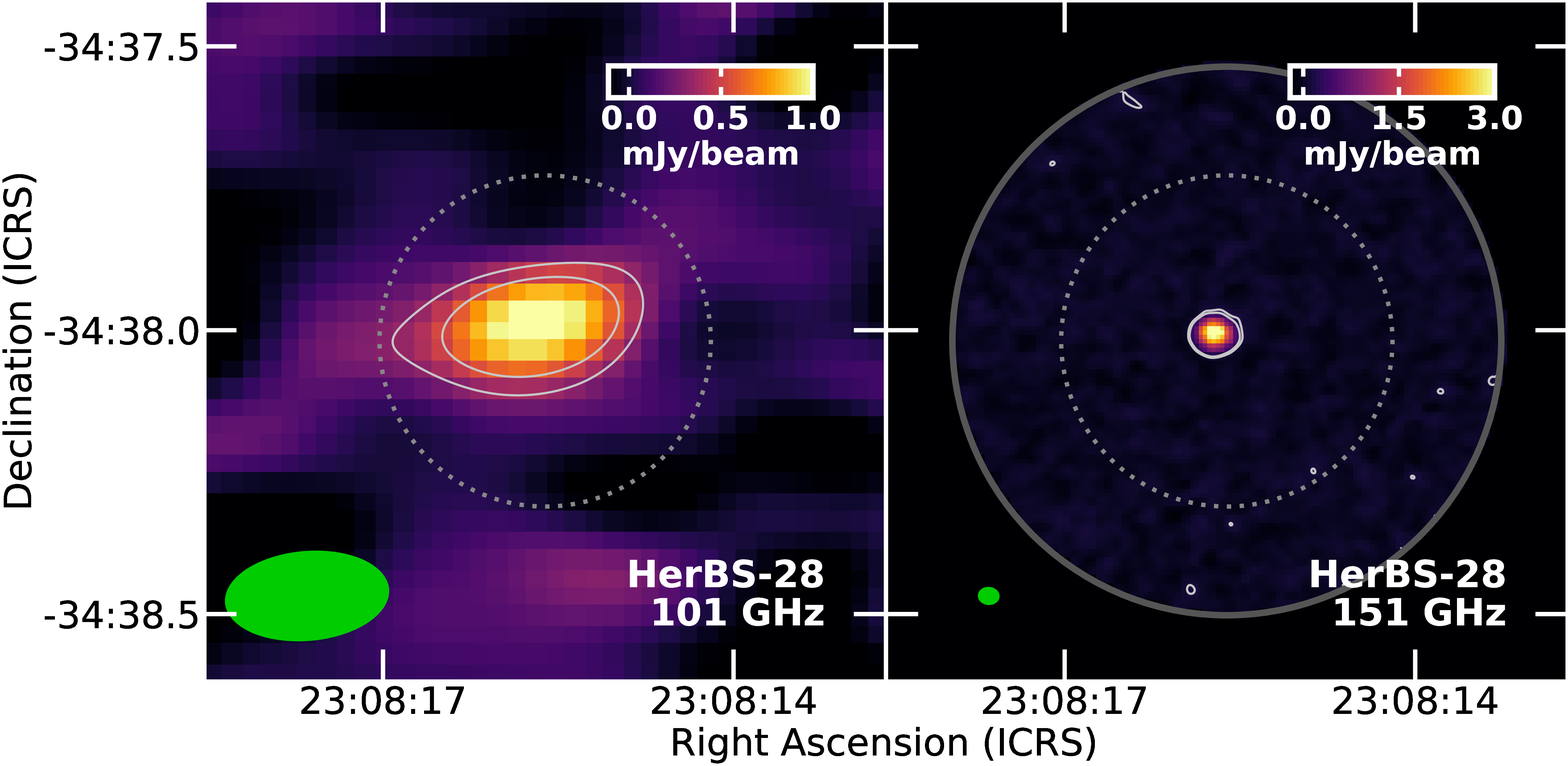}
			\ \ \ \ \ \
			\includegraphics[width=7cm]{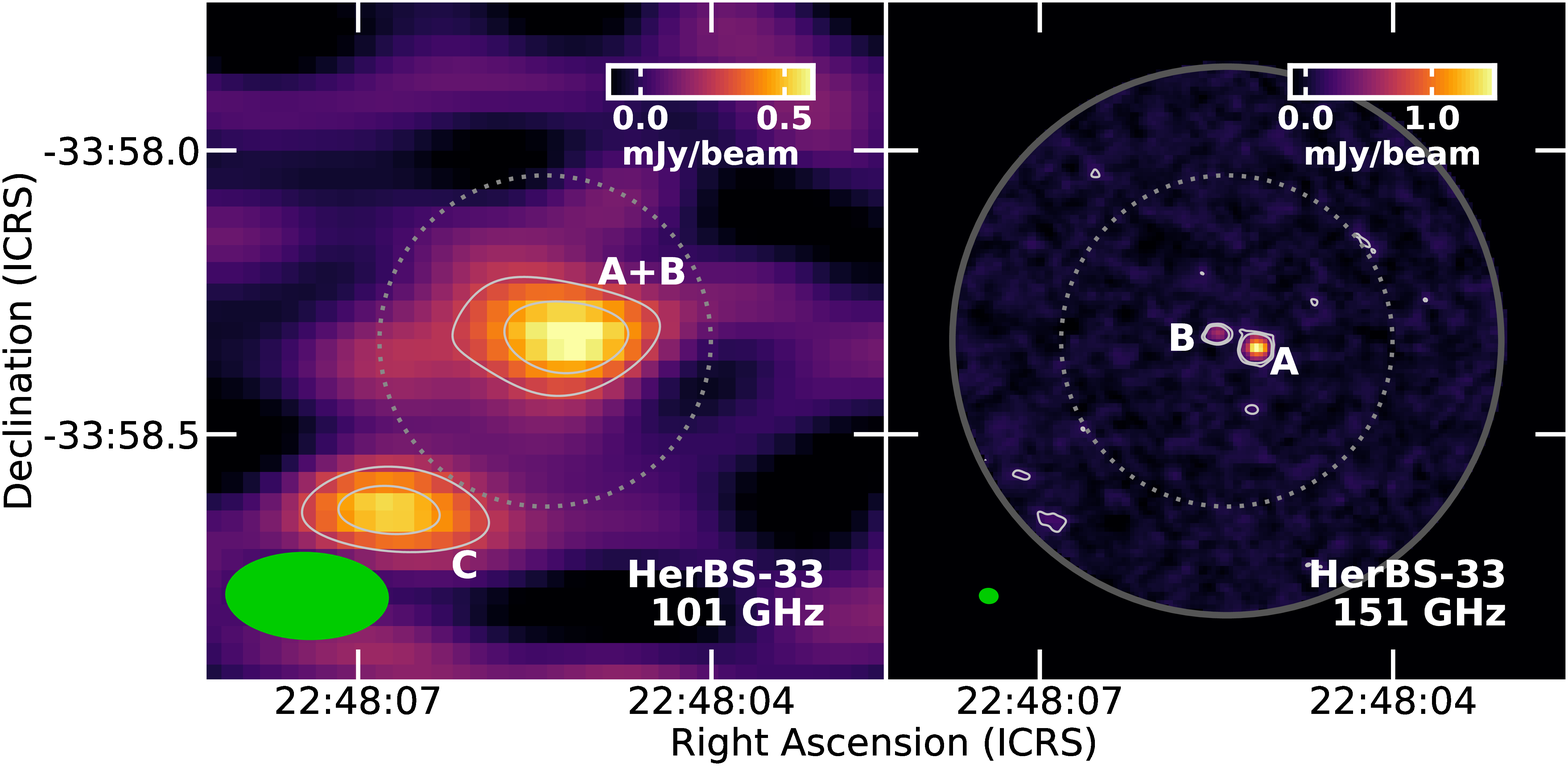}
			\vspace{0.5em}\\
			\includegraphics[width=7cm]{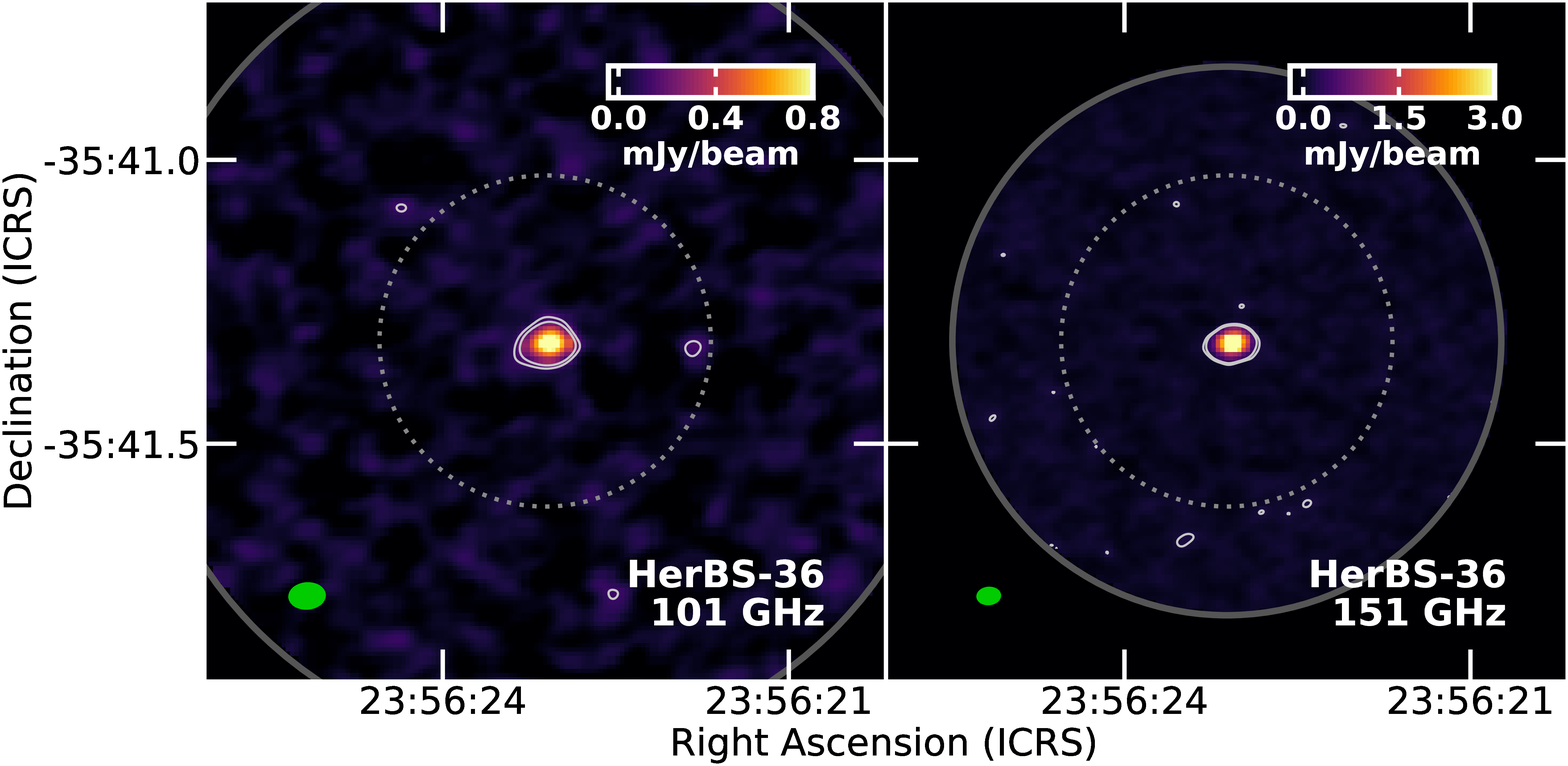}
			\ \ \ \ \ \
			\includegraphics[width=7cm]{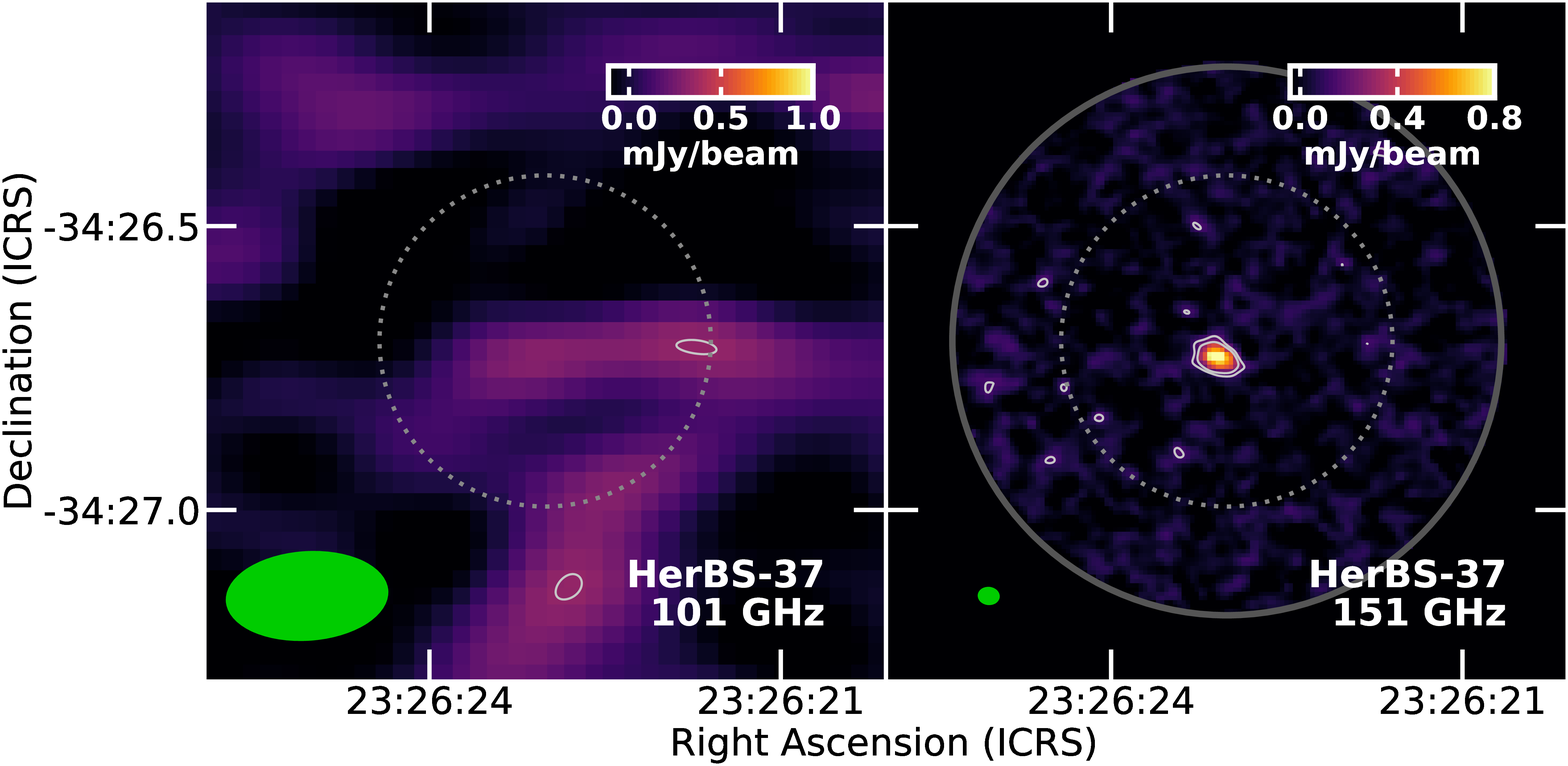}
		\end{center}
		\caption{101 and 151 GHz images of all of the fields imaged in the Bright Extragalactic ALMA Redshift Survey (BEARS) survey.  The colours are scaled linearly.  The contours show the $\geq$3 and $\geq$5$\sigma$ detection levels in the images.  The green ellipses at the bottom left of each image show the size of the beam.  The grey solid circle shows the imaged region in each field (corresponding to where the primary beam is $\geq$20\% of the peak sensitivity), and the grey dotted circle shows the 35~arcsec FWHM of the {\it Herschel} 500~$\mu$m beam.}
		\label{f_maps}
	\end{figure*}
	
	\addtocounter{figure}{-1}
	
	\begin{figure*}
		\begin{center}
			\includegraphics[width=7cm]{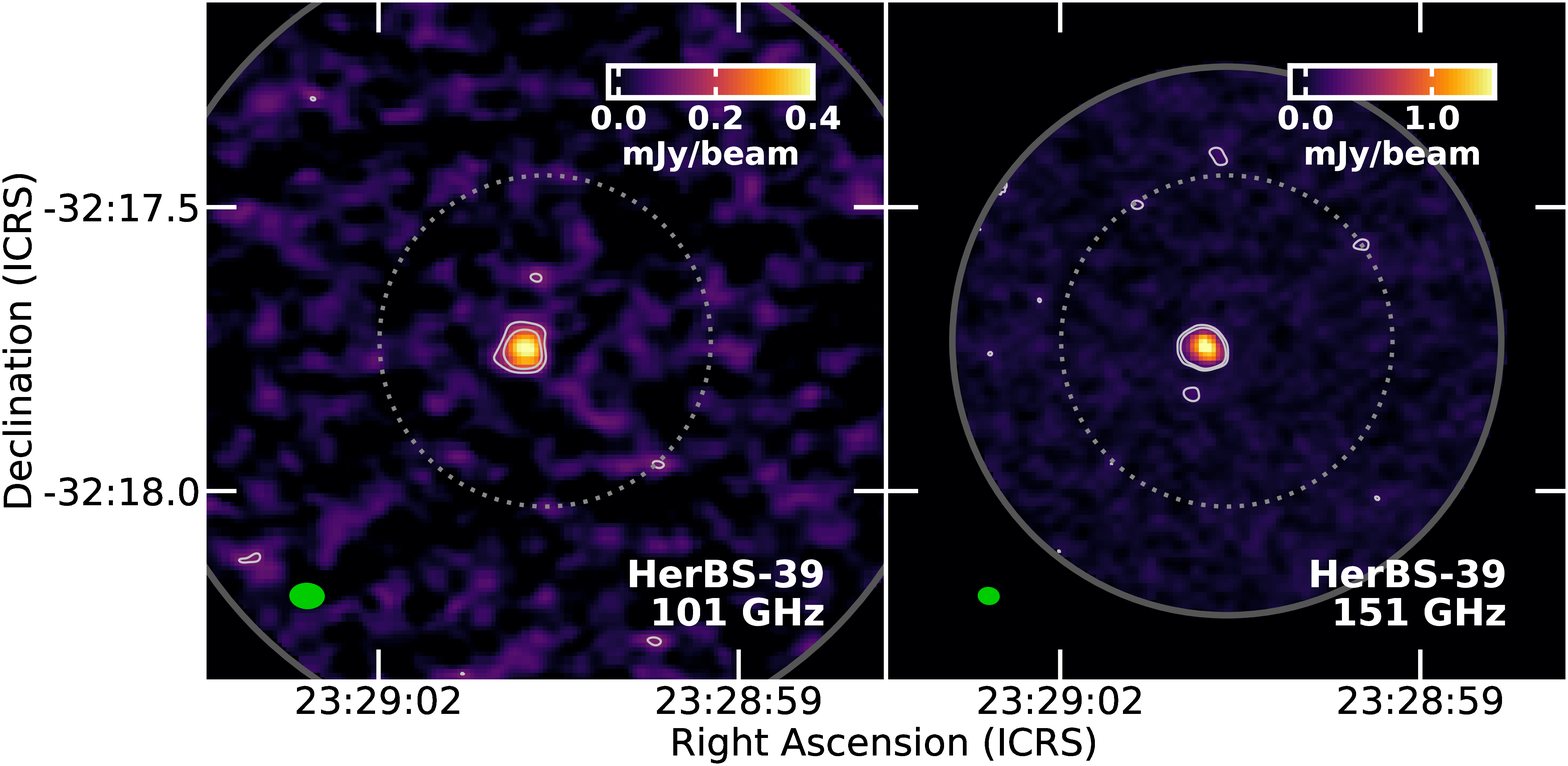}
			\ \ \ \ \ \
			\includegraphics[width=7cm]{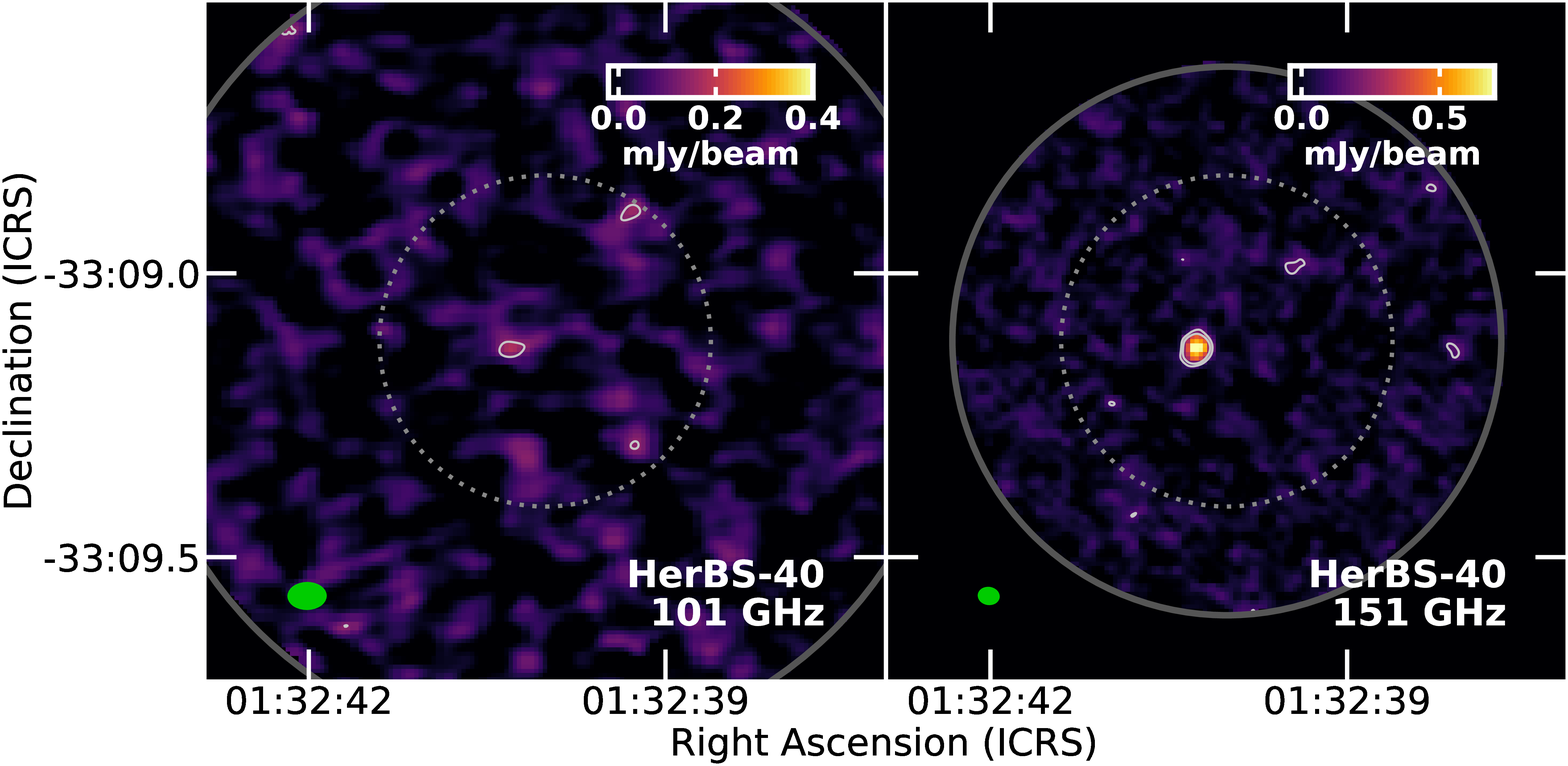}
			\vspace{0.5em}\\
			\includegraphics[width=7cm]{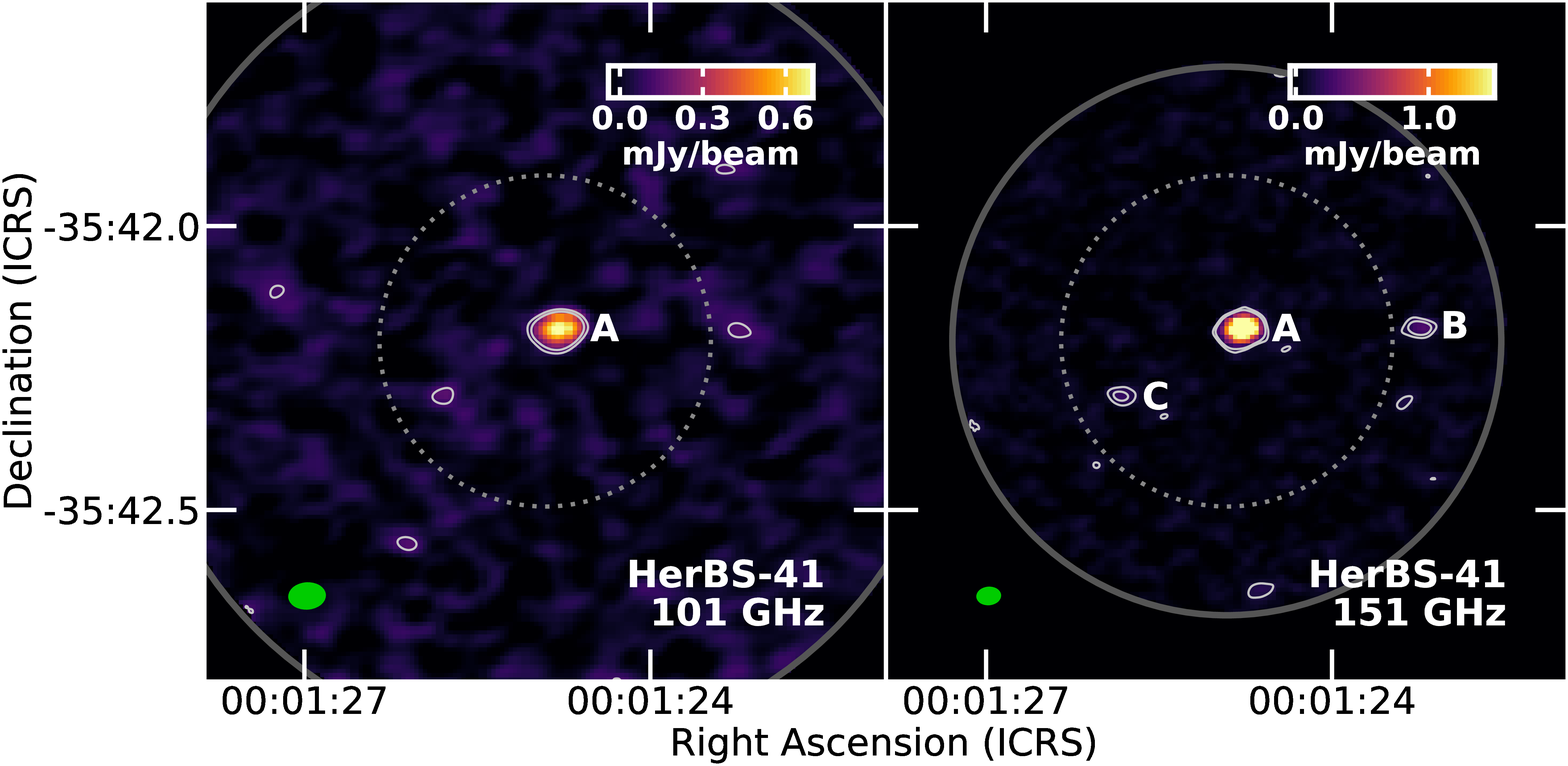}
			\ \ \ \ \ \
			\includegraphics[width=7cm]{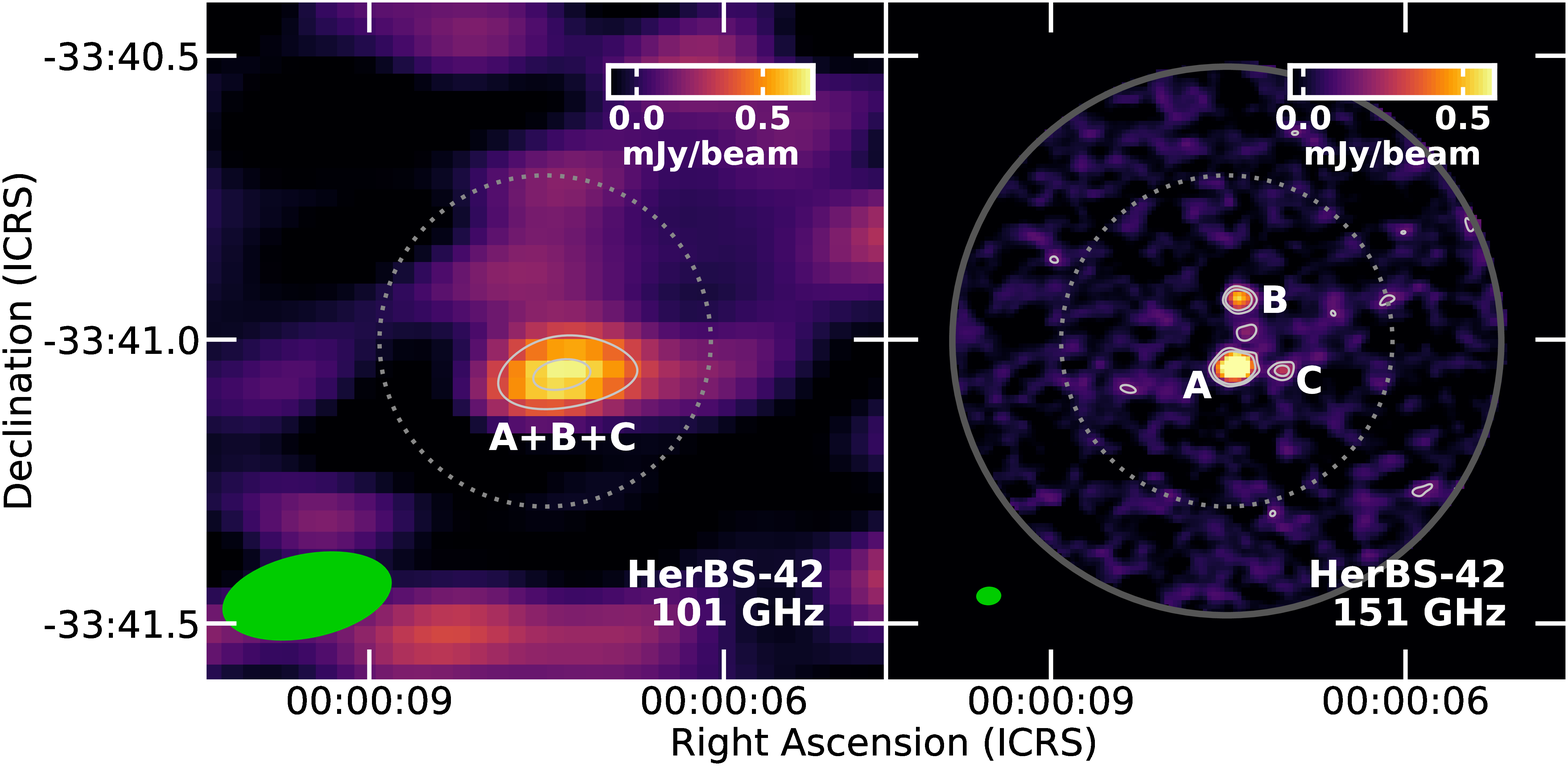}
			\vspace{0.5em}\\
			\includegraphics[width=7cm]{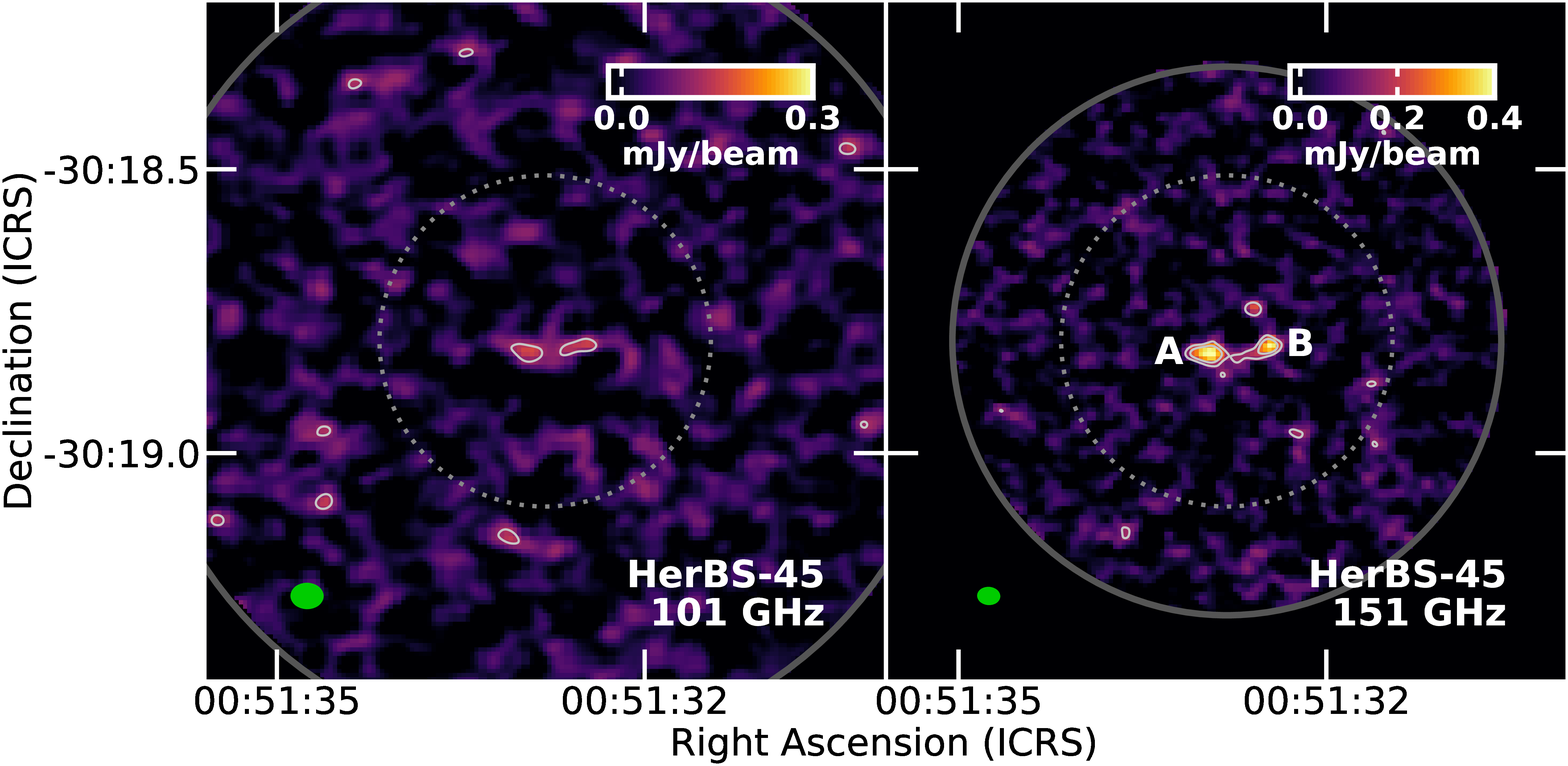}
			\ \ \ \ \ \
			\includegraphics[width=7cm]{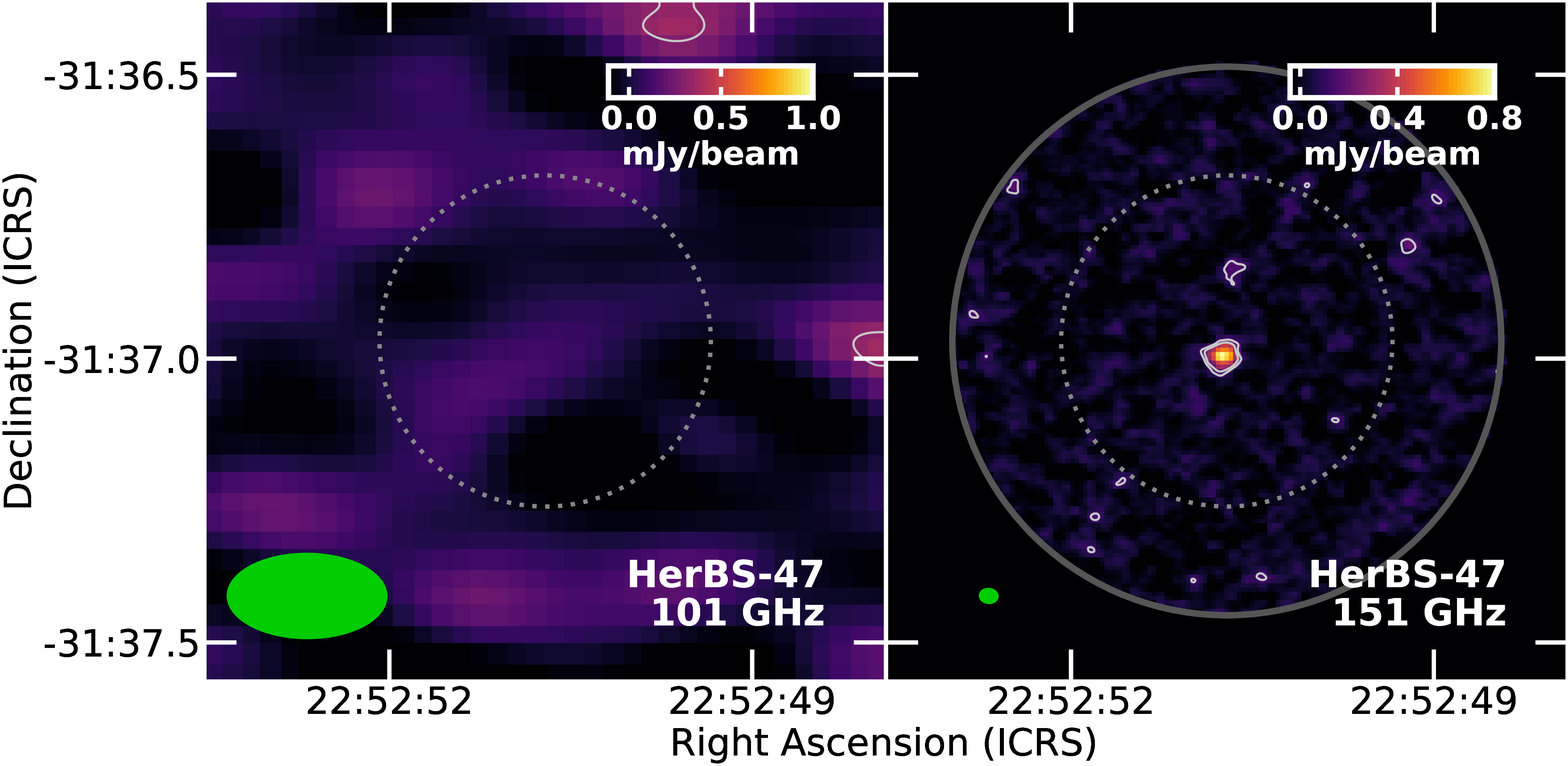}
			\vspace{0.5em}\\
			\includegraphics[width=7cm]{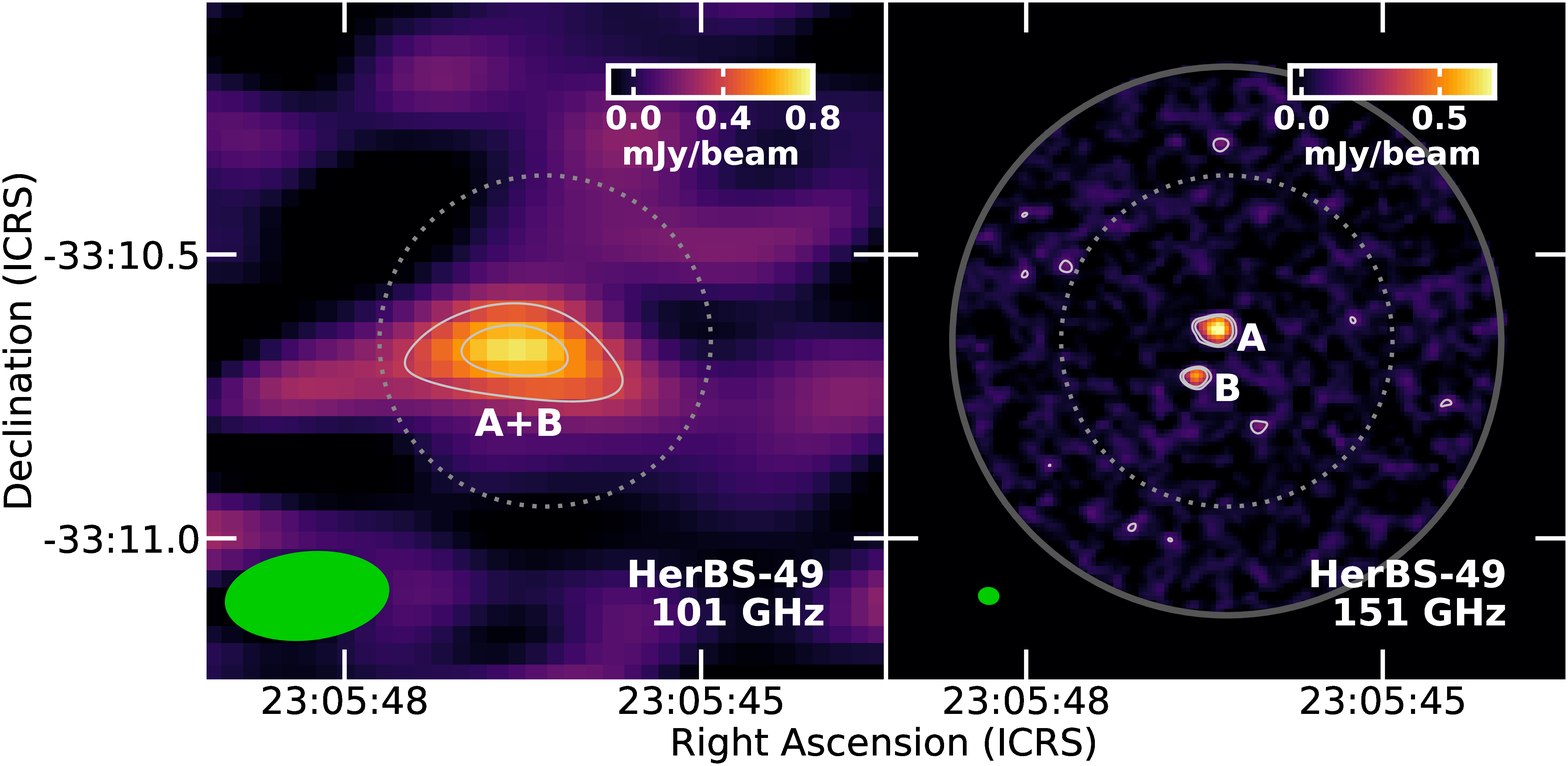}
			\ \ \ \ \ \
			\includegraphics[width=7cm]{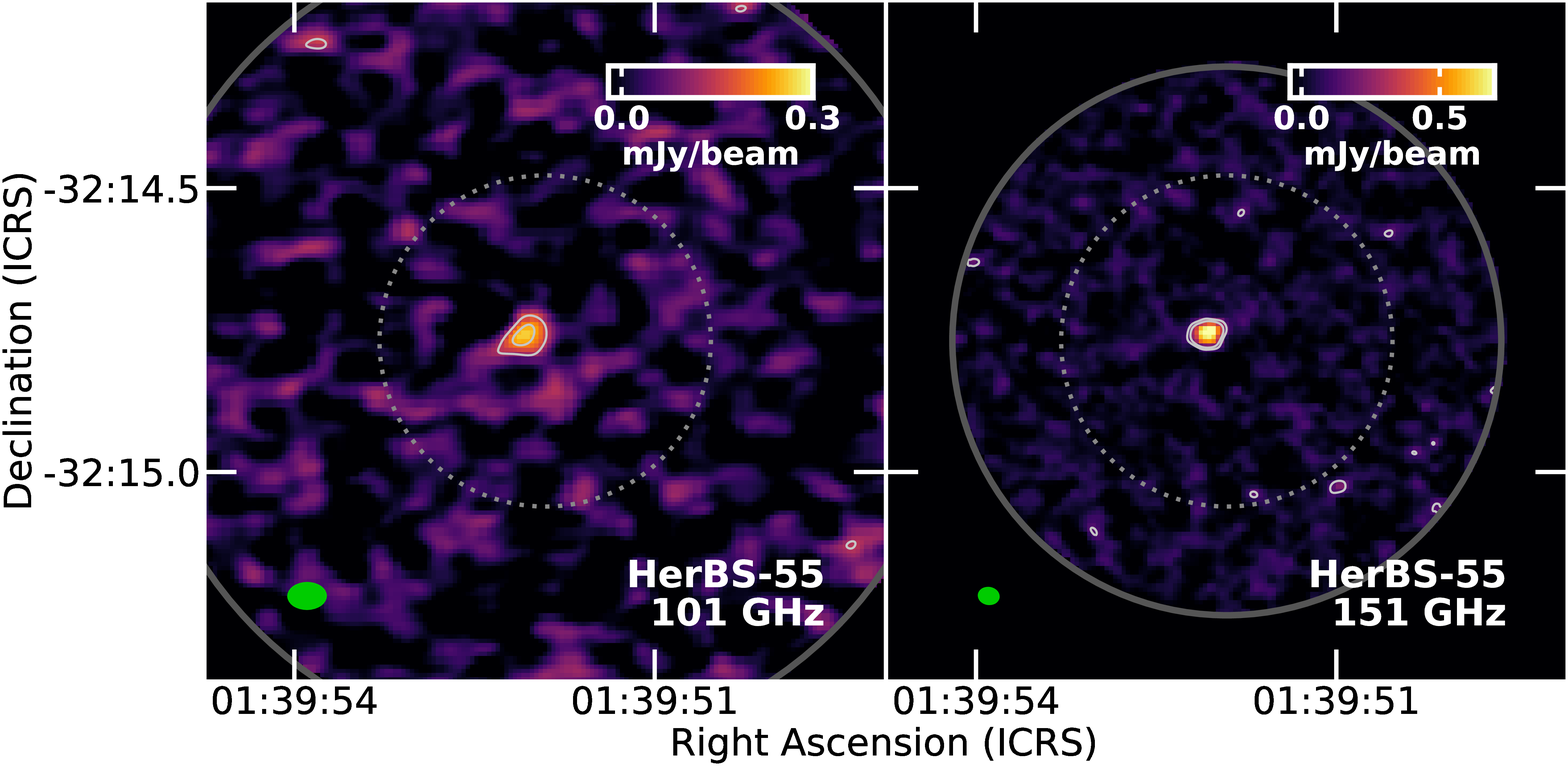}
			\vspace{0.5em}\\
			\includegraphics[width=7cm]{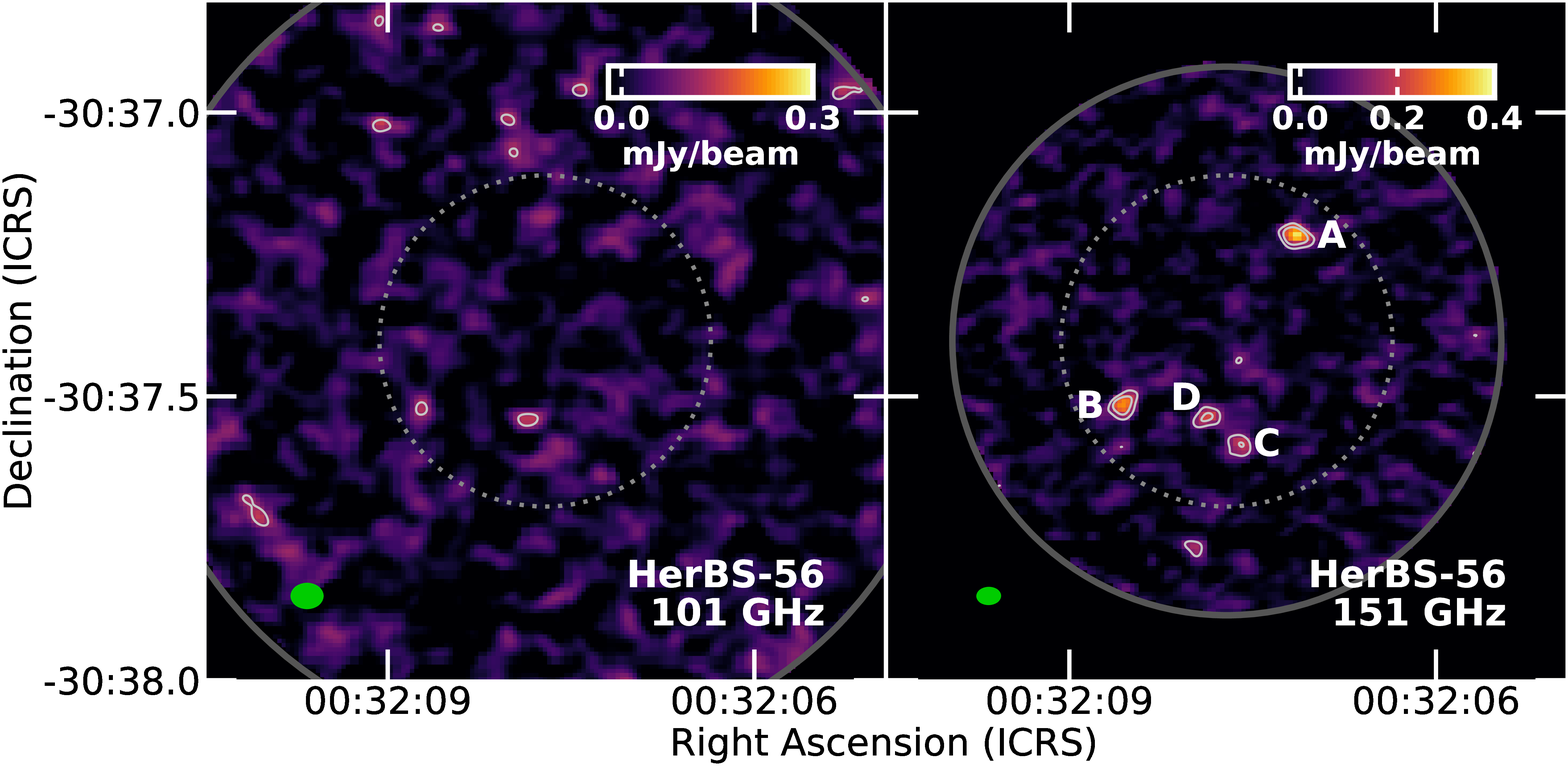}
			\ \ \ \ \ \
			\includegraphics[width=7cm]{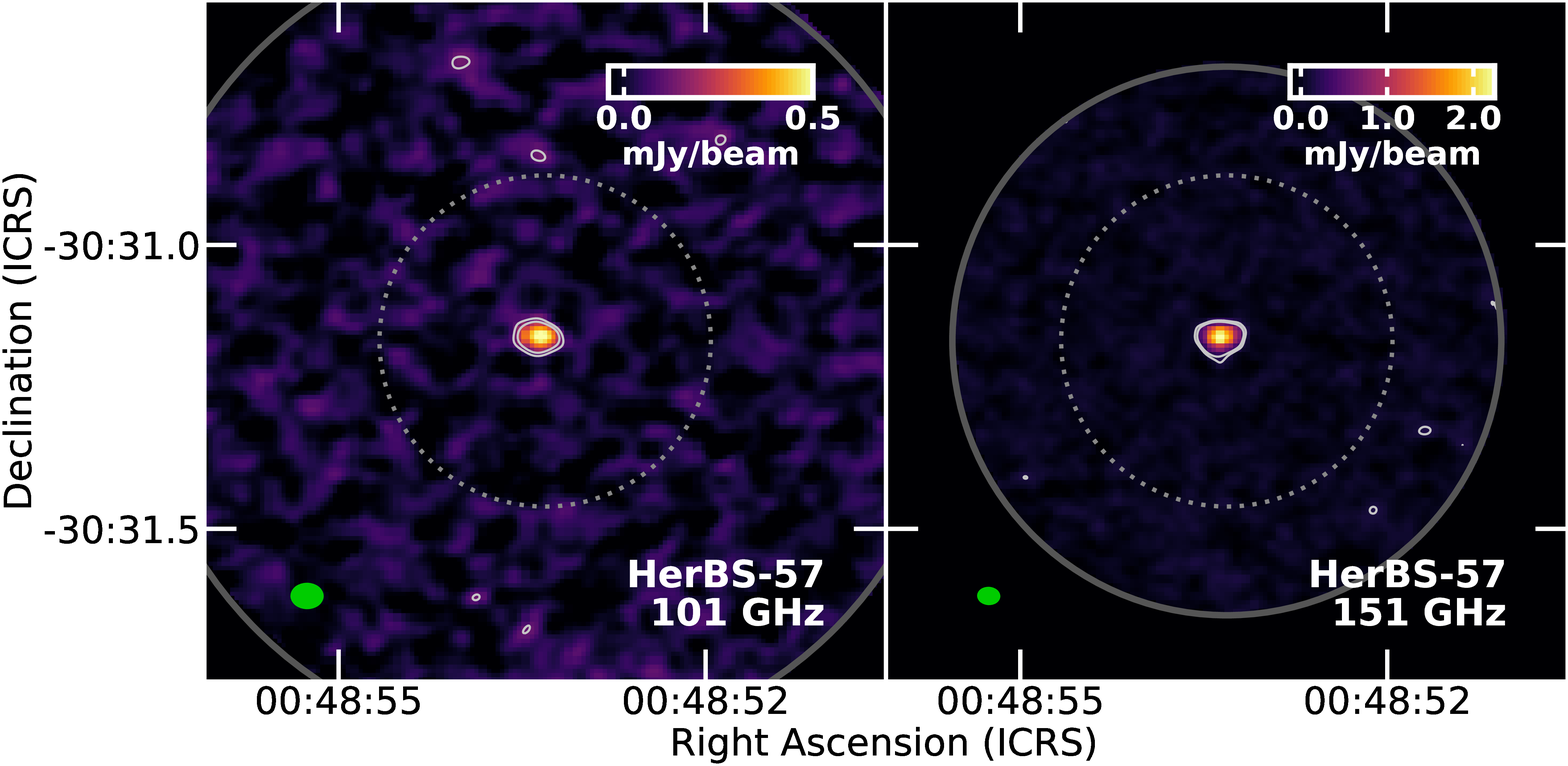}
			\vspace{0.5em}\\
			\includegraphics[width=7cm]{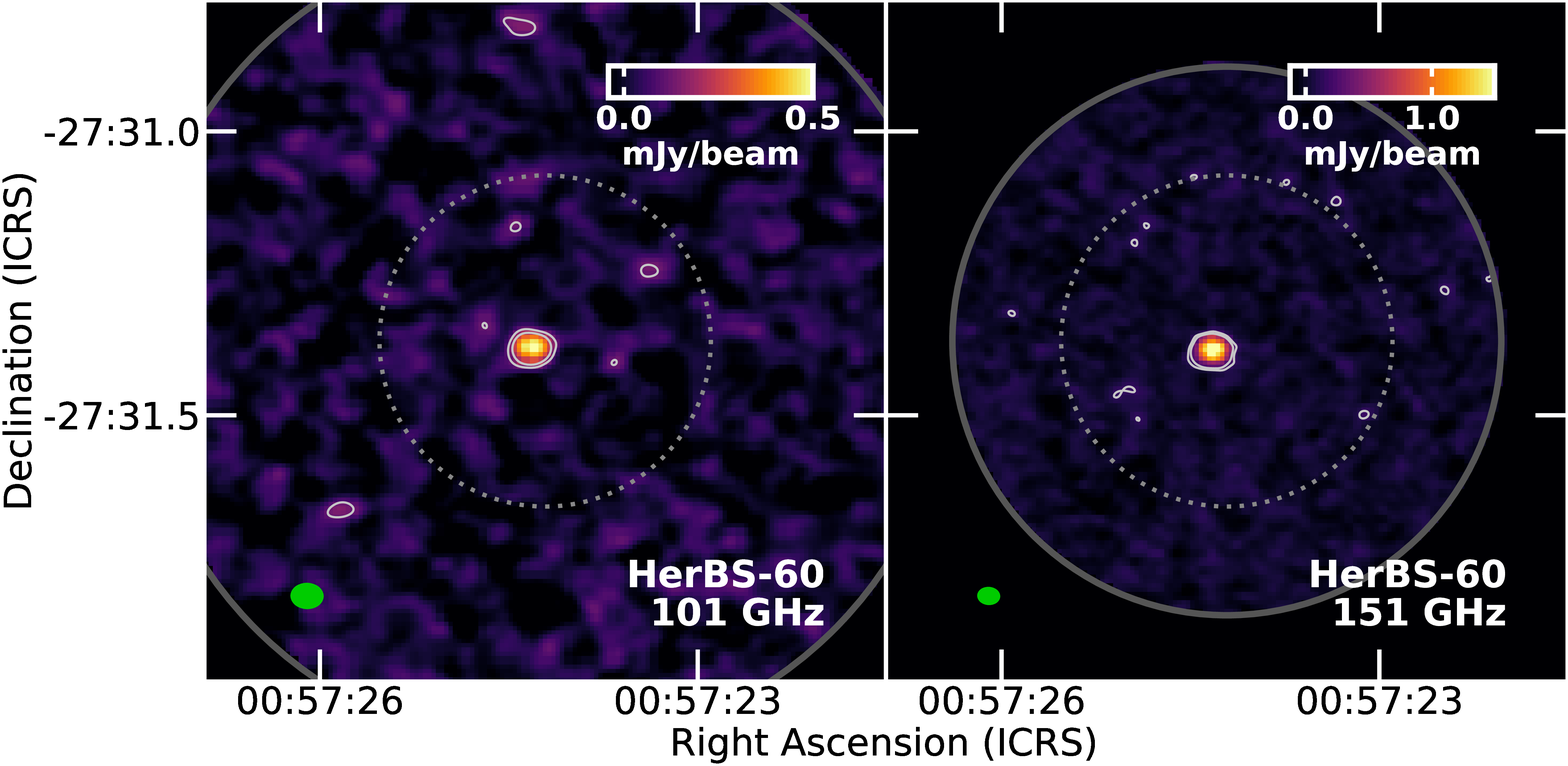}
			\ \ \ \ \ \
			\includegraphics[width=7cm]{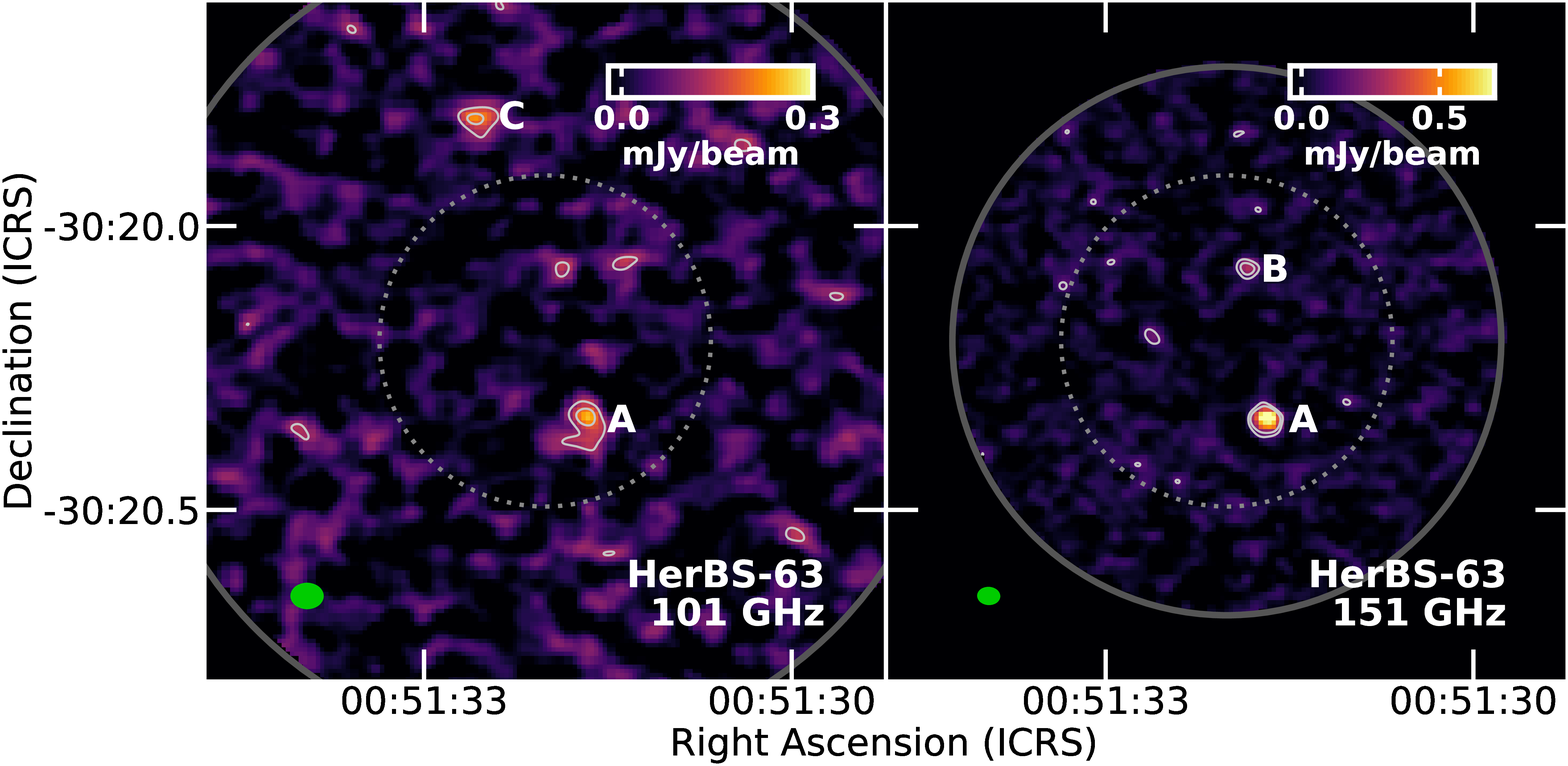}
			\vspace{0.5em}\\
		\end{center}
		\caption{Continued.}
	\end{figure*}
	
	\addtocounter{figure}{-1}
		
	\begin{figure*}
		\begin{center}
			\includegraphics[width=7cm]{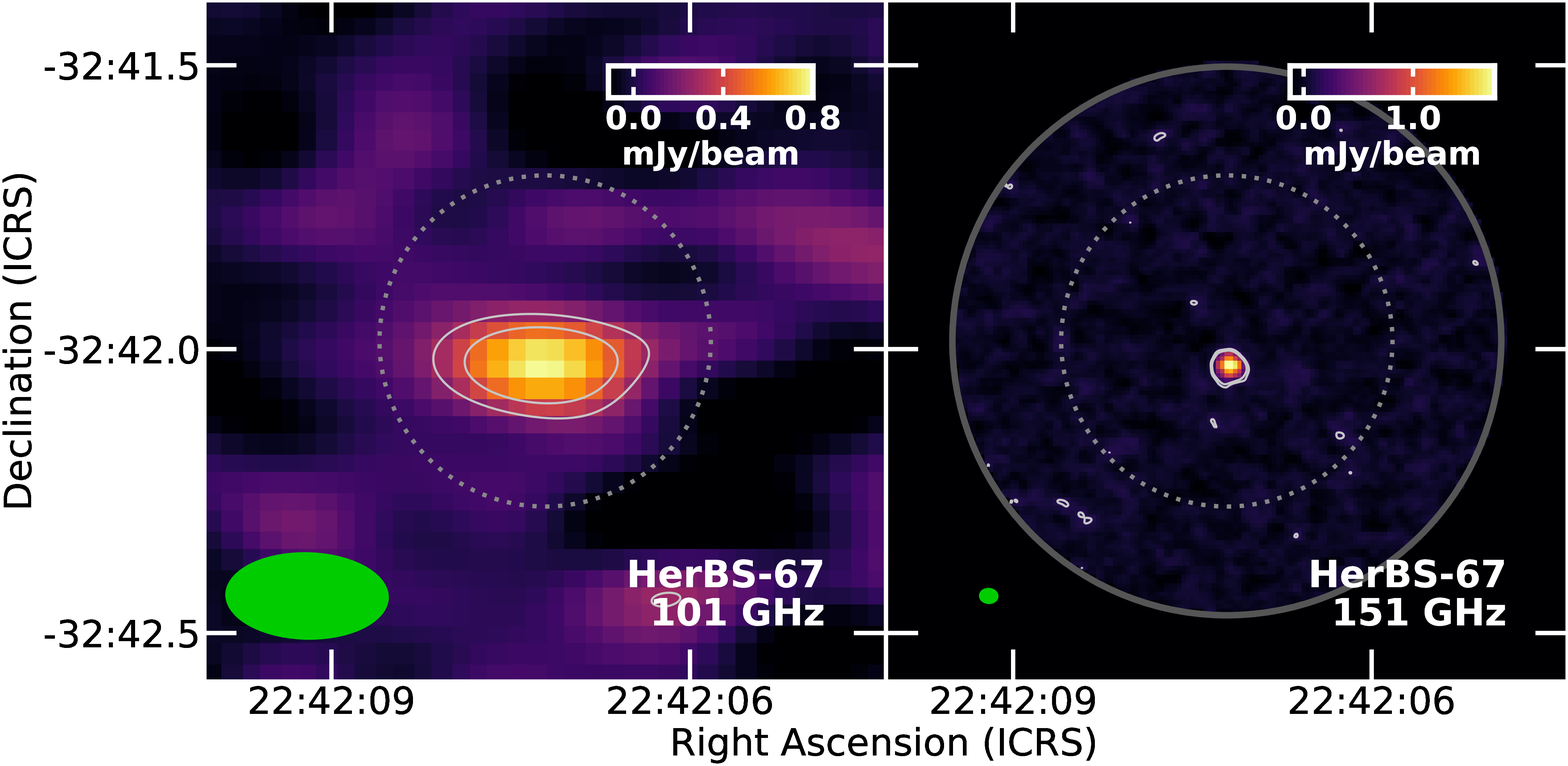}
			\ \ \ \ \ \
			\includegraphics[width=7cm]{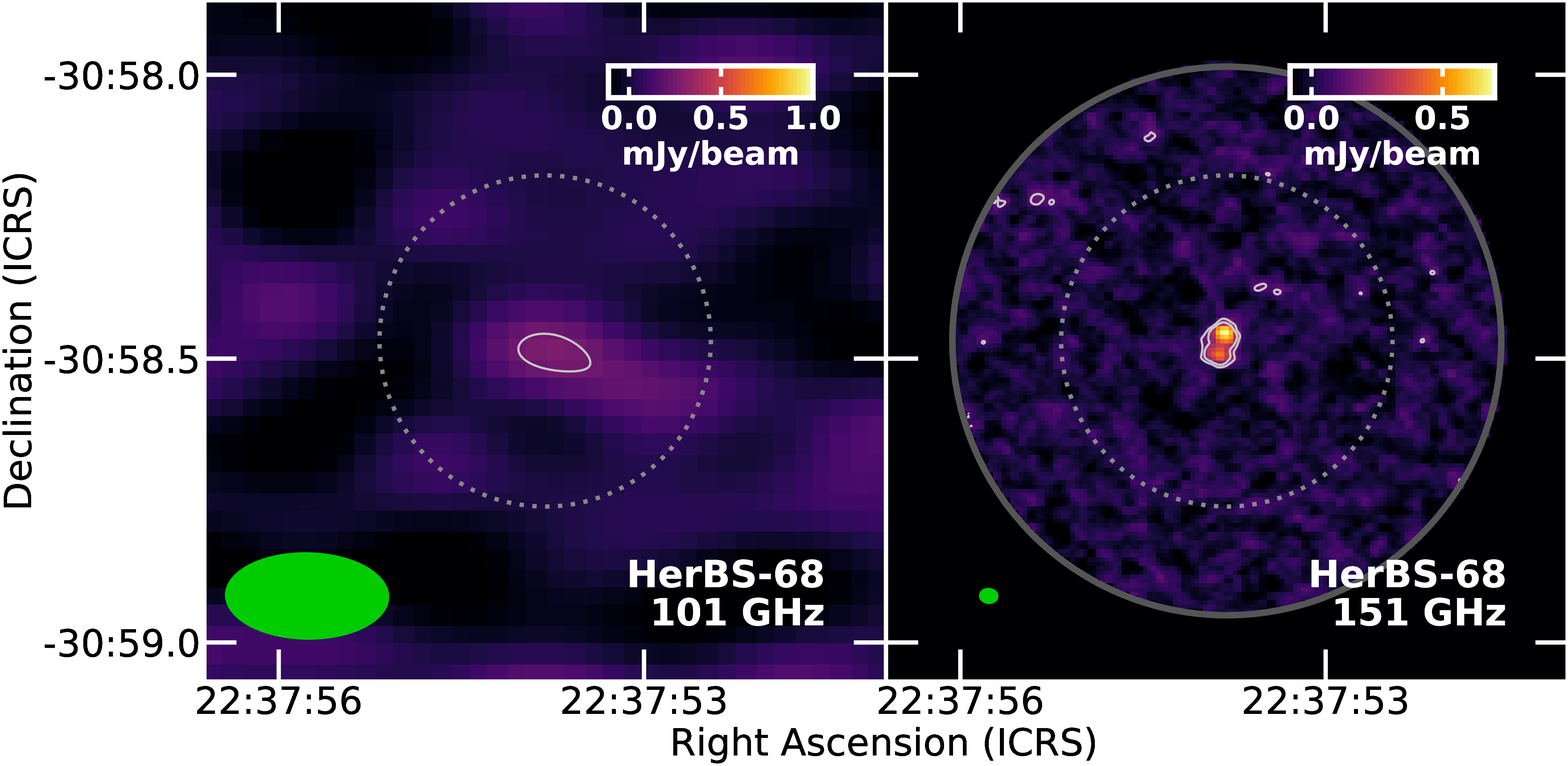}
			\vspace{0.5em}\\
			\includegraphics[width=7cm]{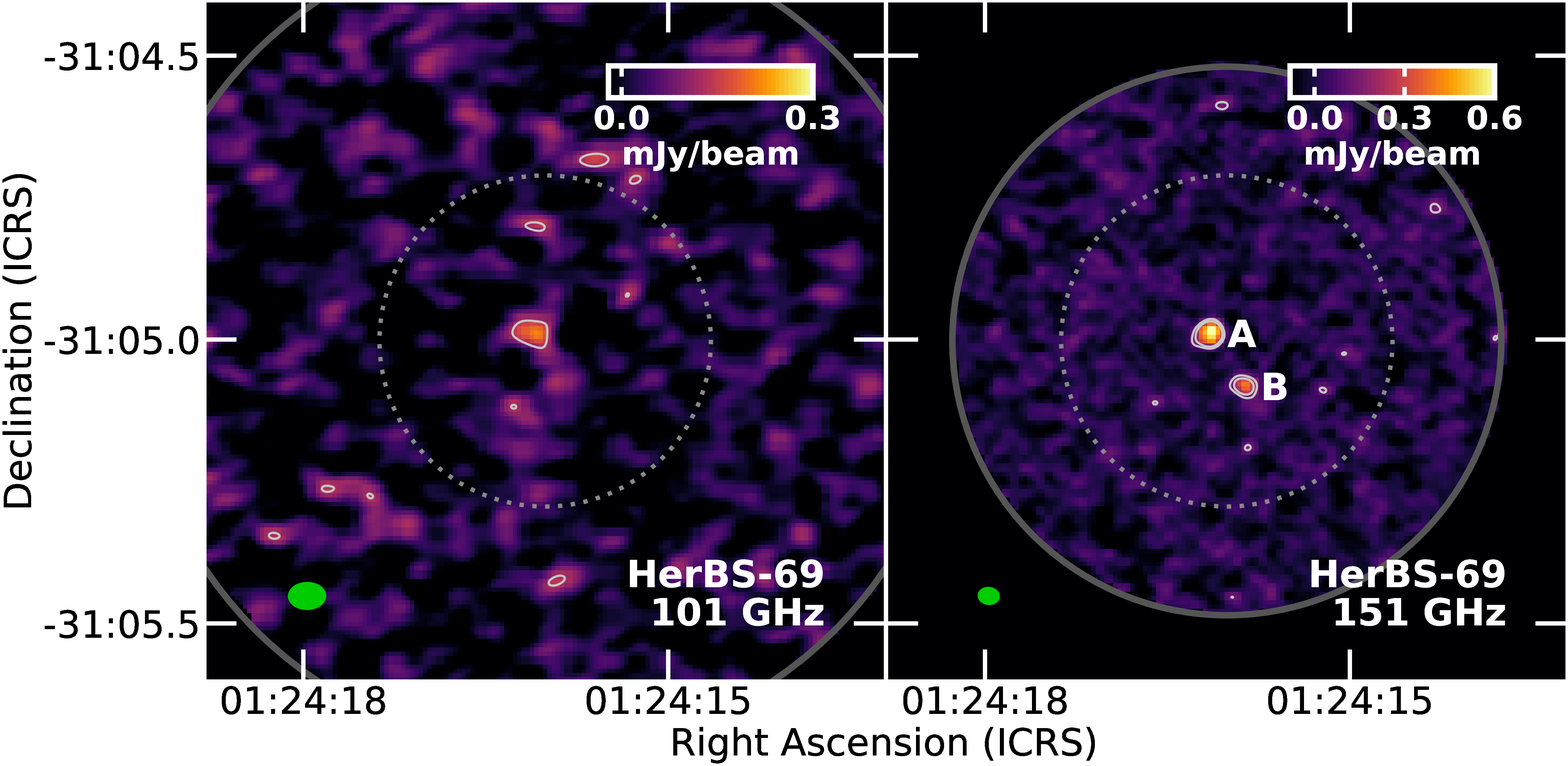}
			\ \ \ \ \ \
			\includegraphics[width=7cm]{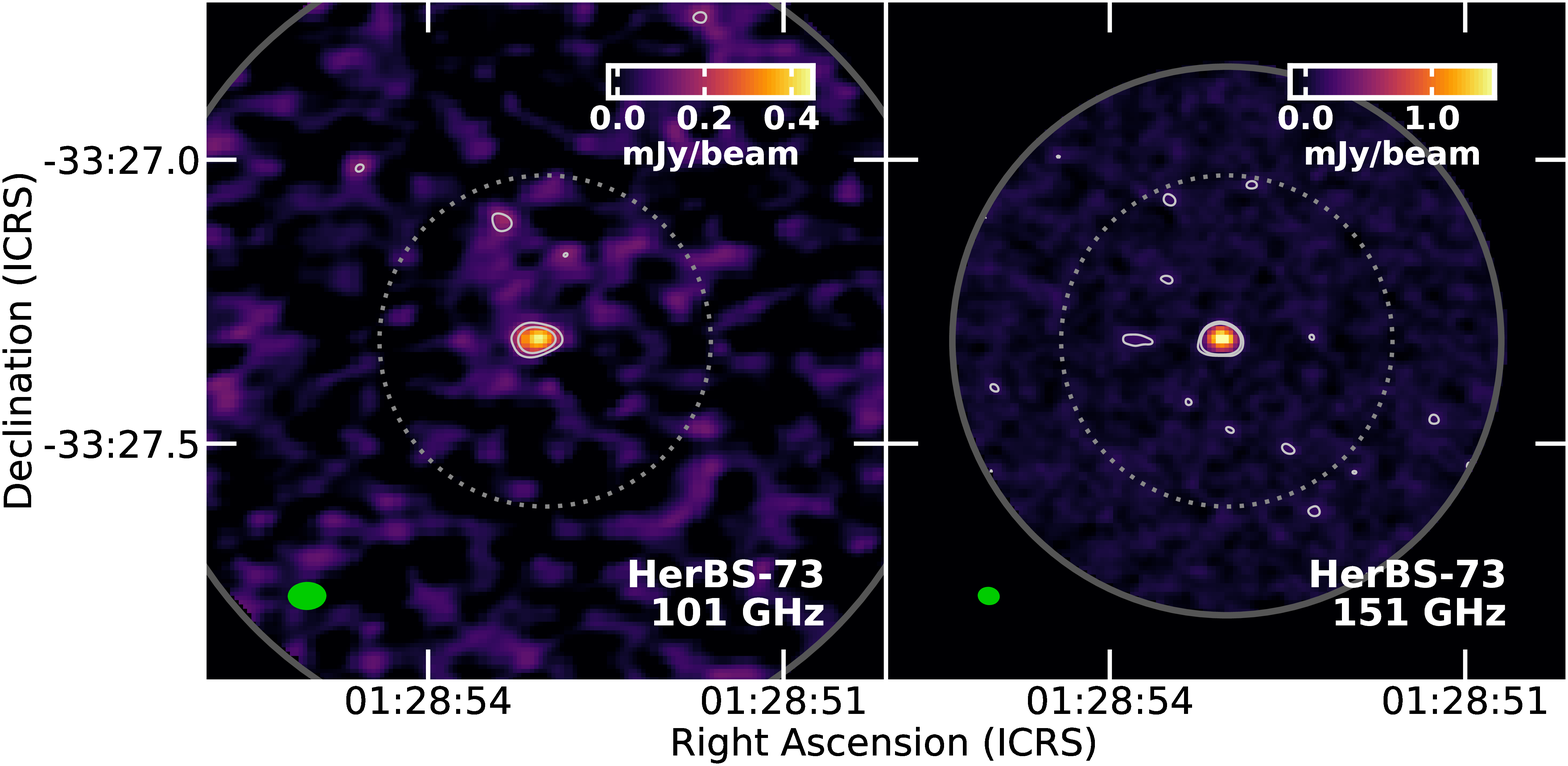}
			\vspace{0.5em}\\
			\includegraphics[width=7cm]{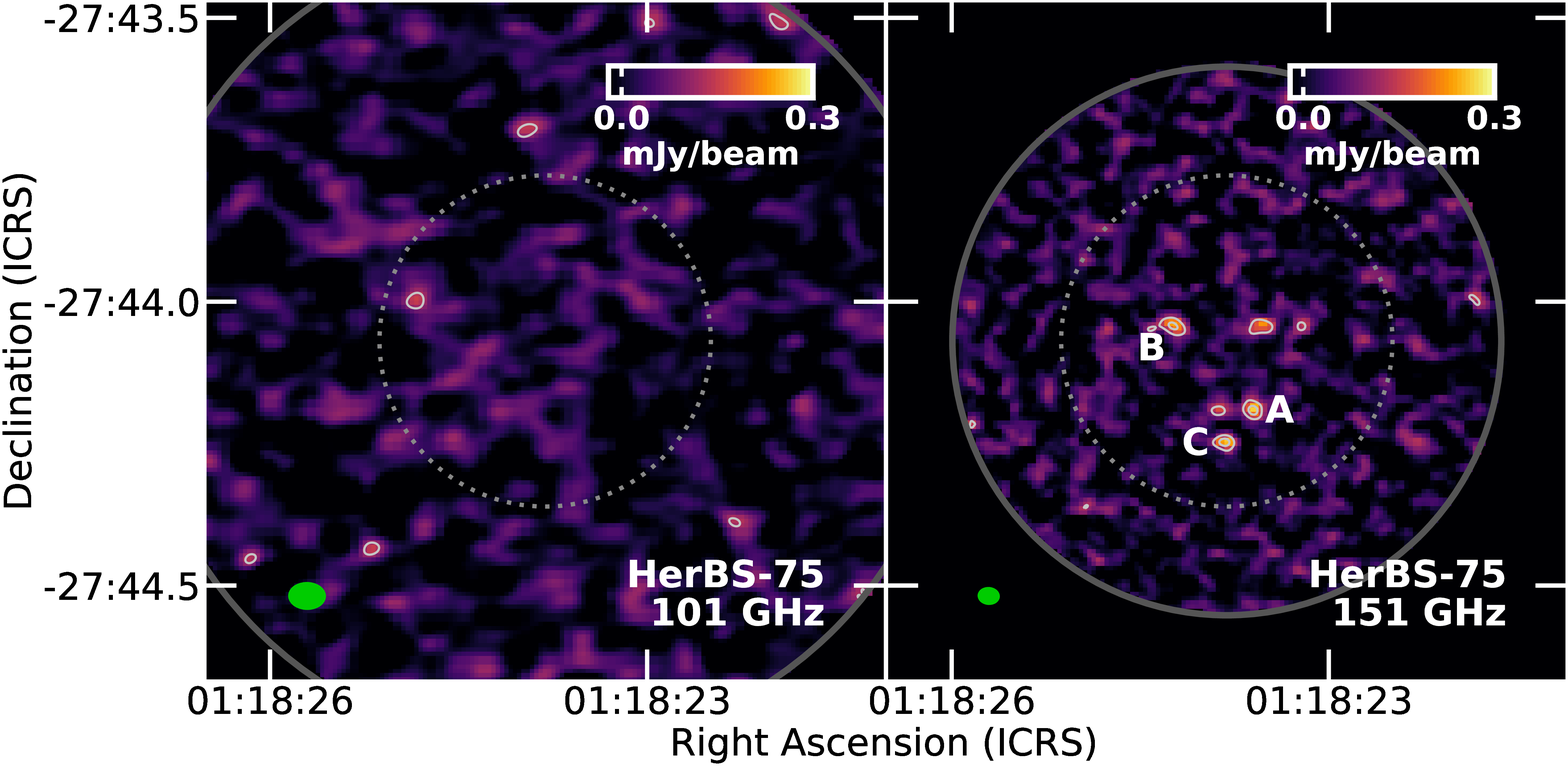}
			\ \ \ \ \ \
			\includegraphics[width=7cm]{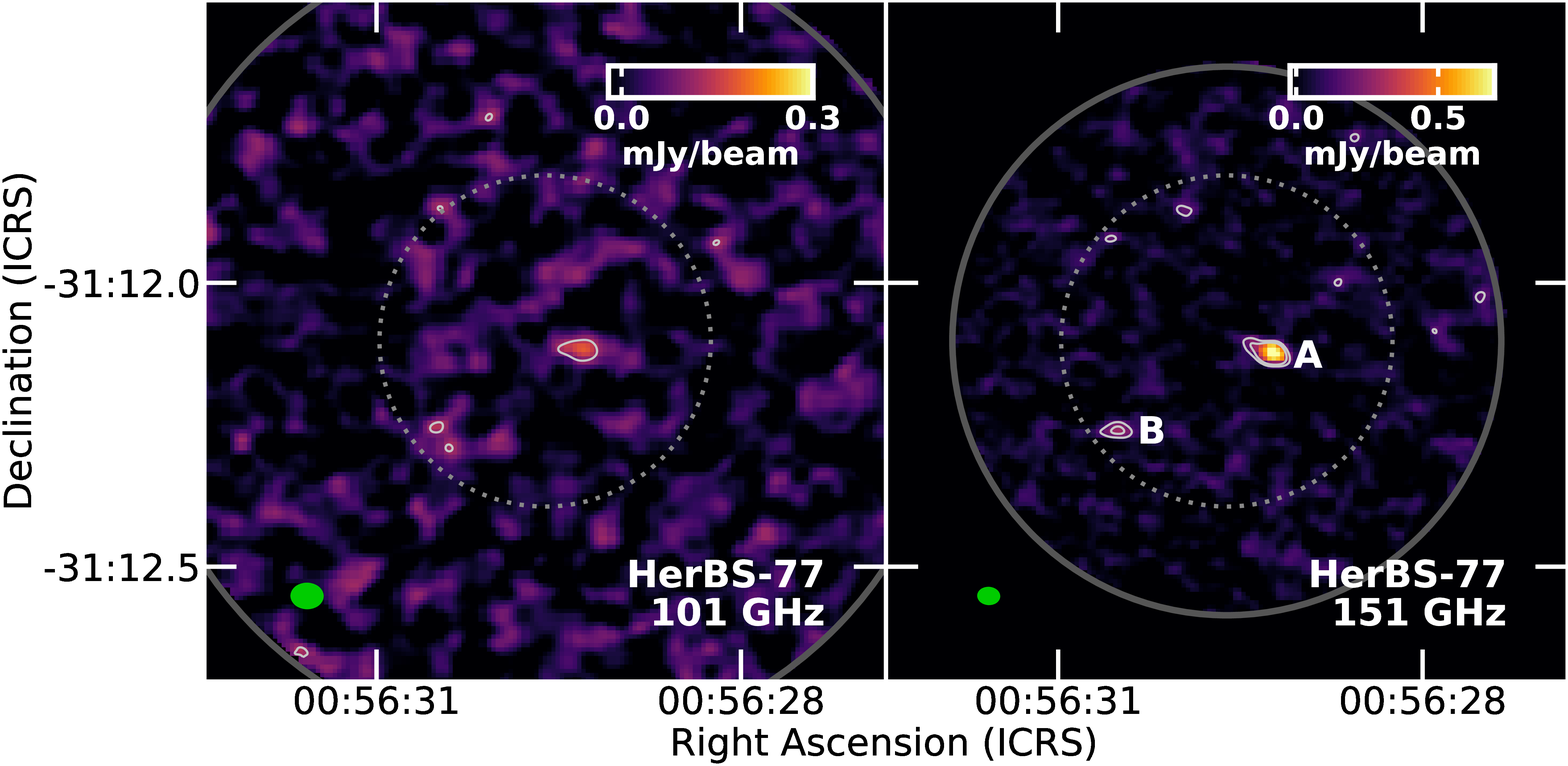}
			\vspace{0.5em}\\
			\includegraphics[width=7cm]{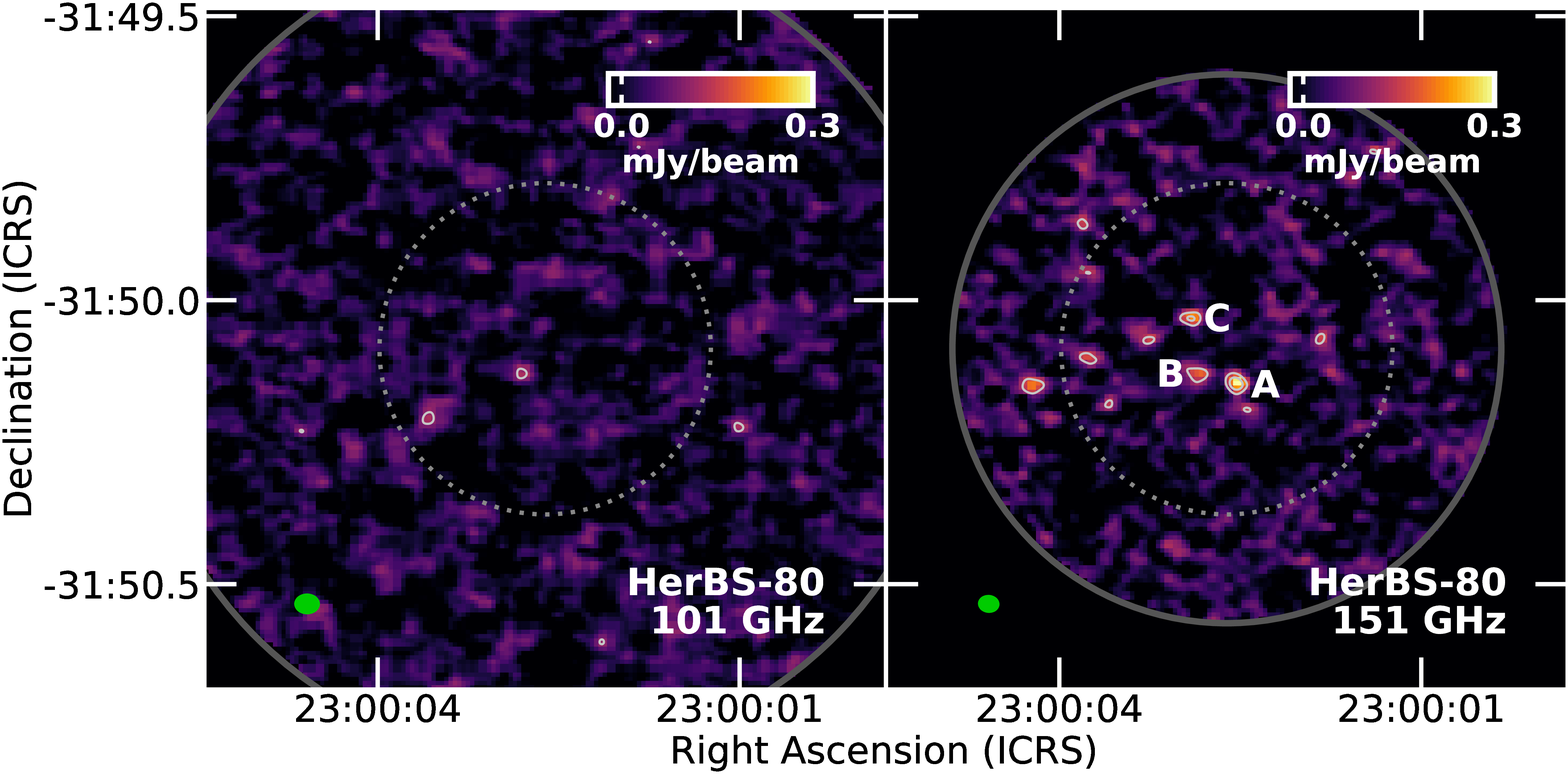}
			\ \ \ \ \ \
			\includegraphics[width=7cm]{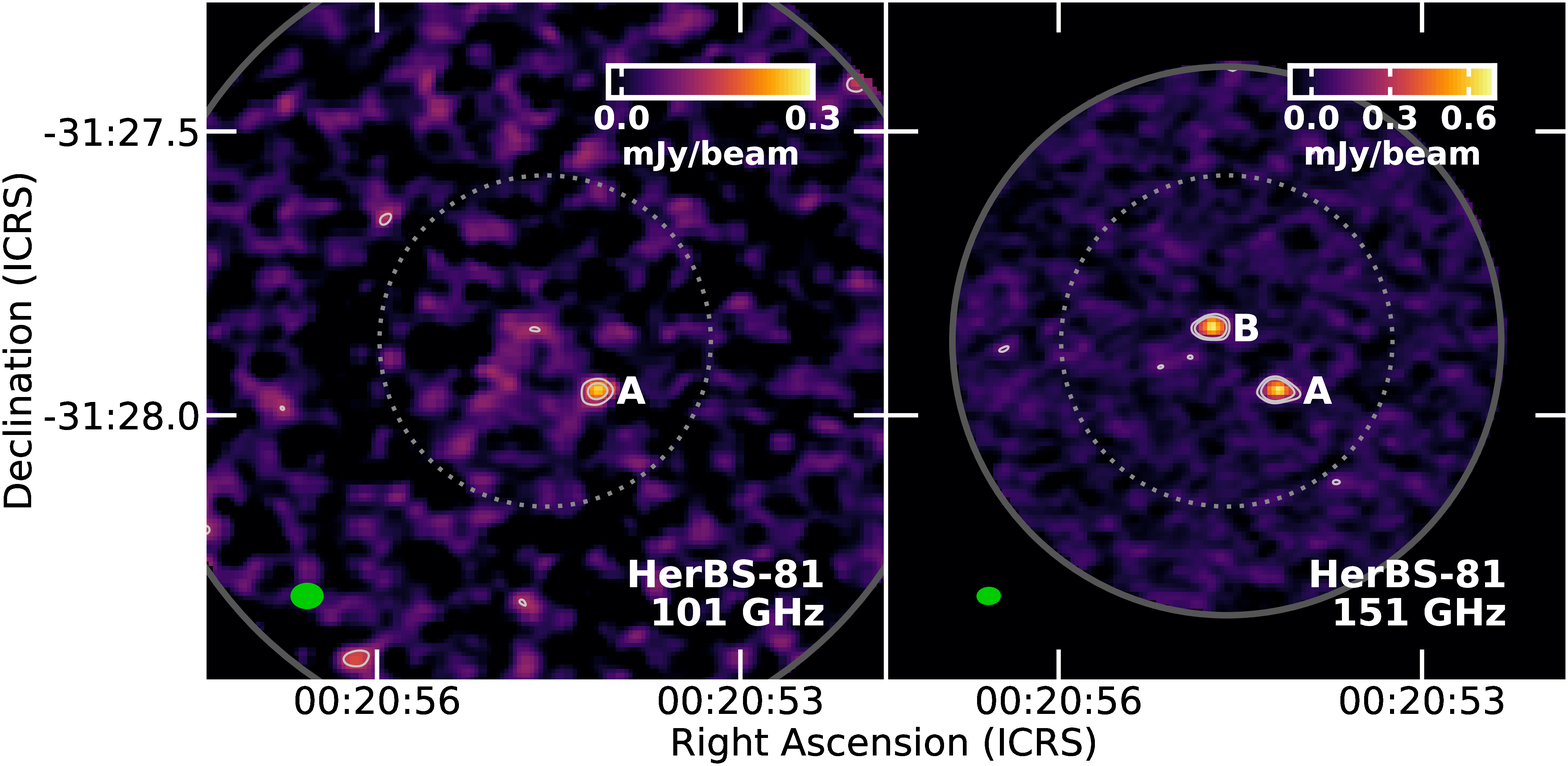}
			\vspace{0.5em}\\
			\includegraphics[width=7cm]{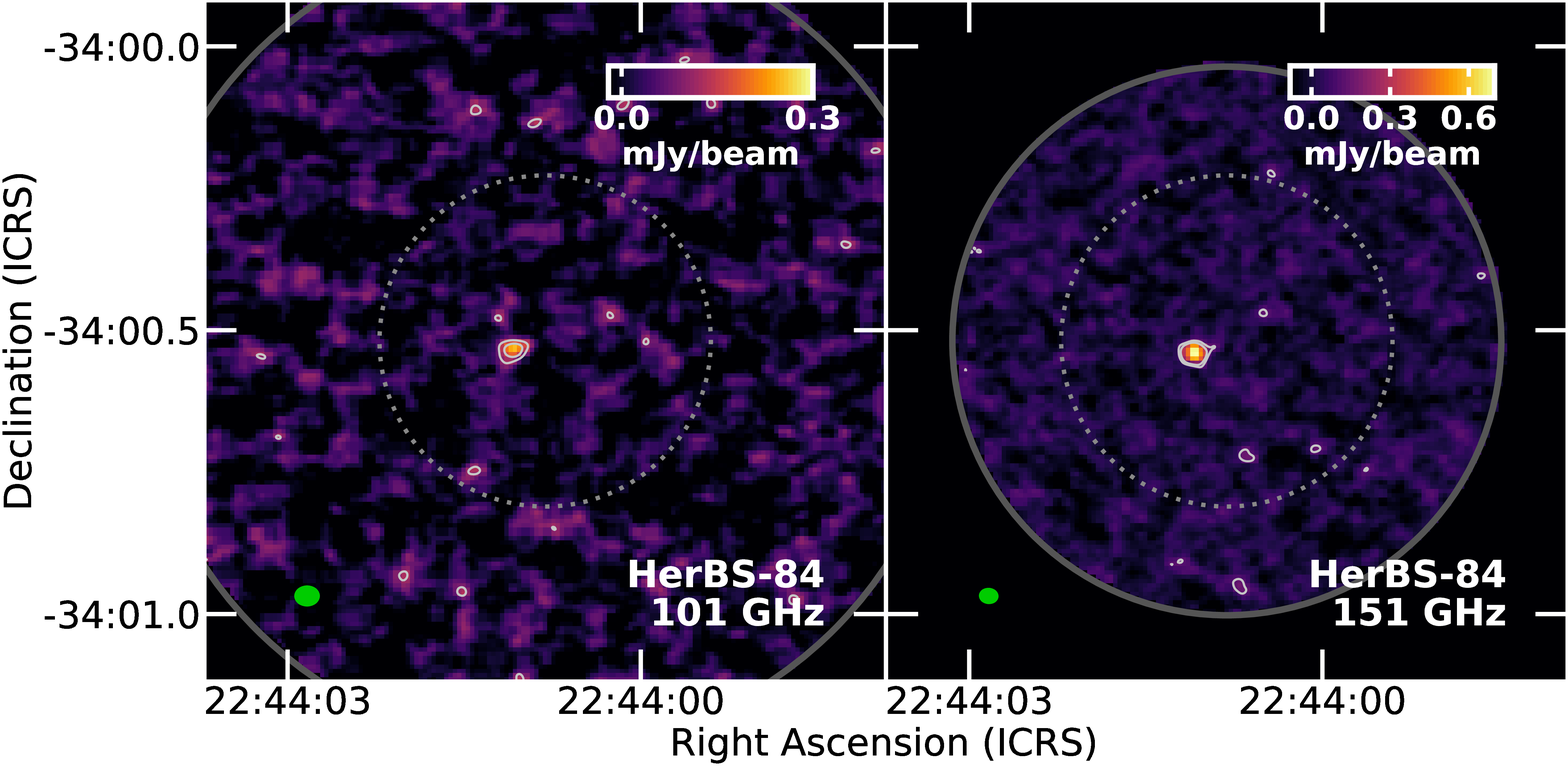}
			\ \ \ \ \ \
			\includegraphics[width=7cm]{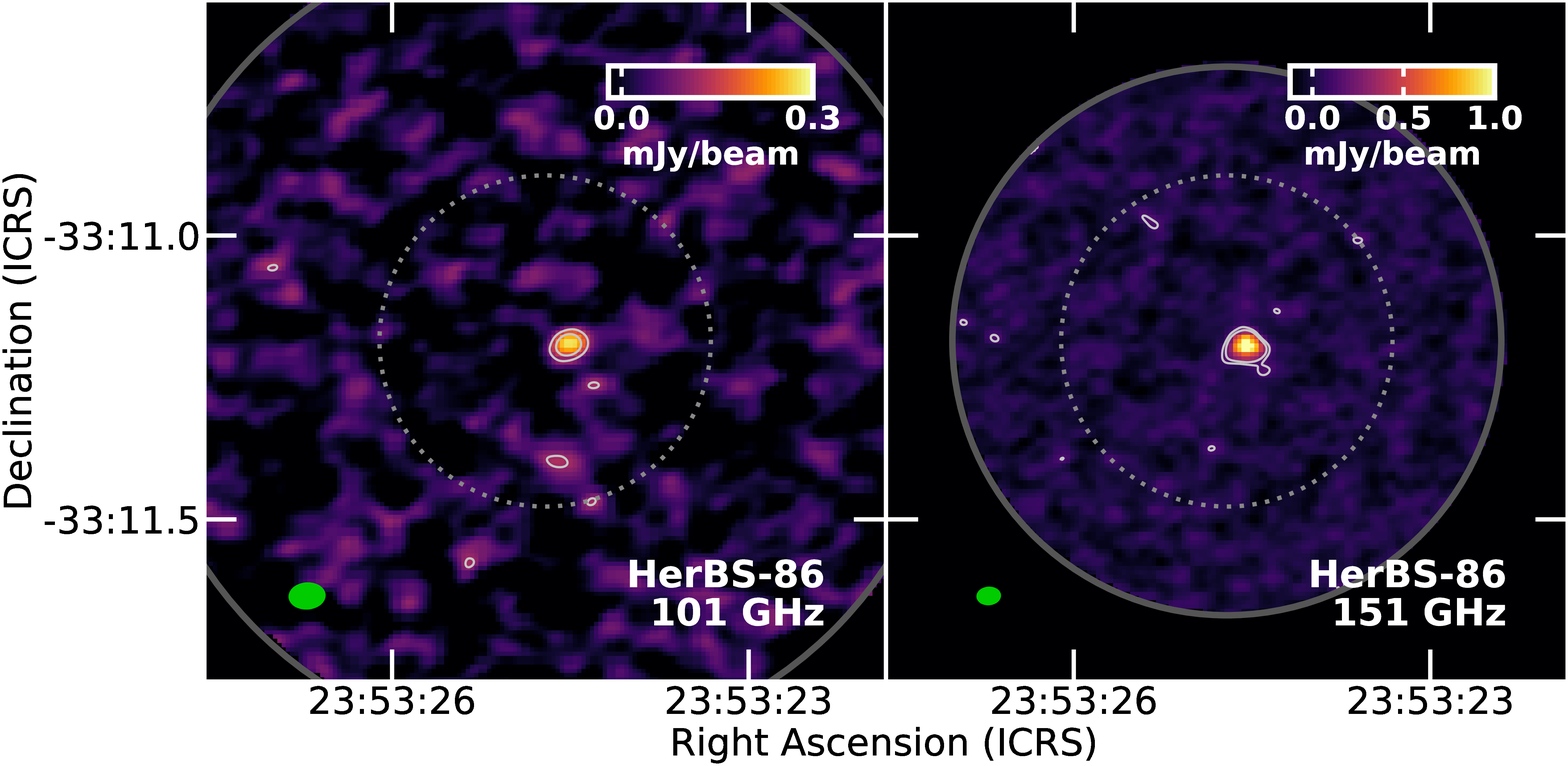}
			\vspace{0.5em}\\
			\includegraphics[width=7cm]{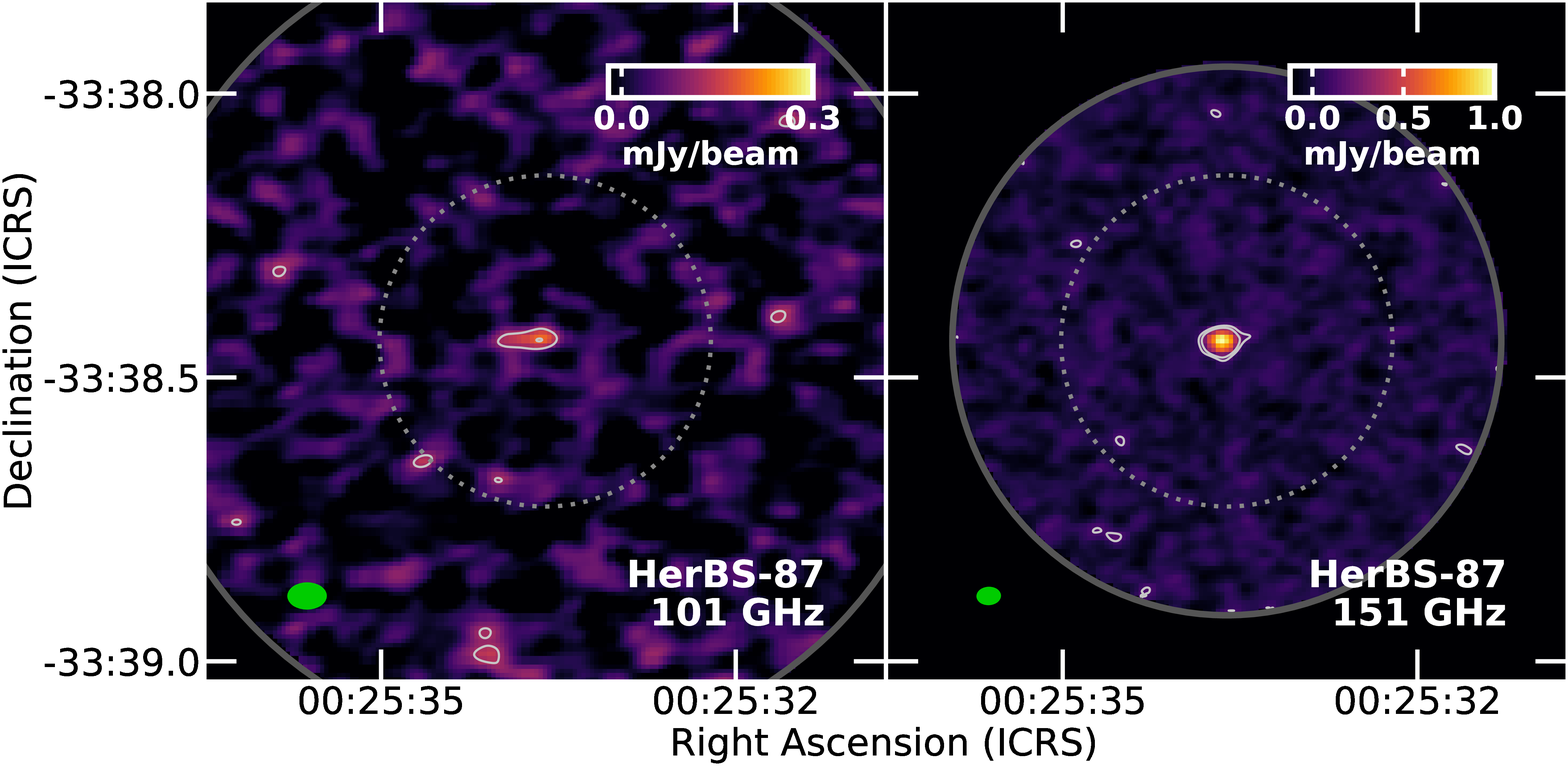}
			\ \ \ \ \ \
			\includegraphics[width=7cm]{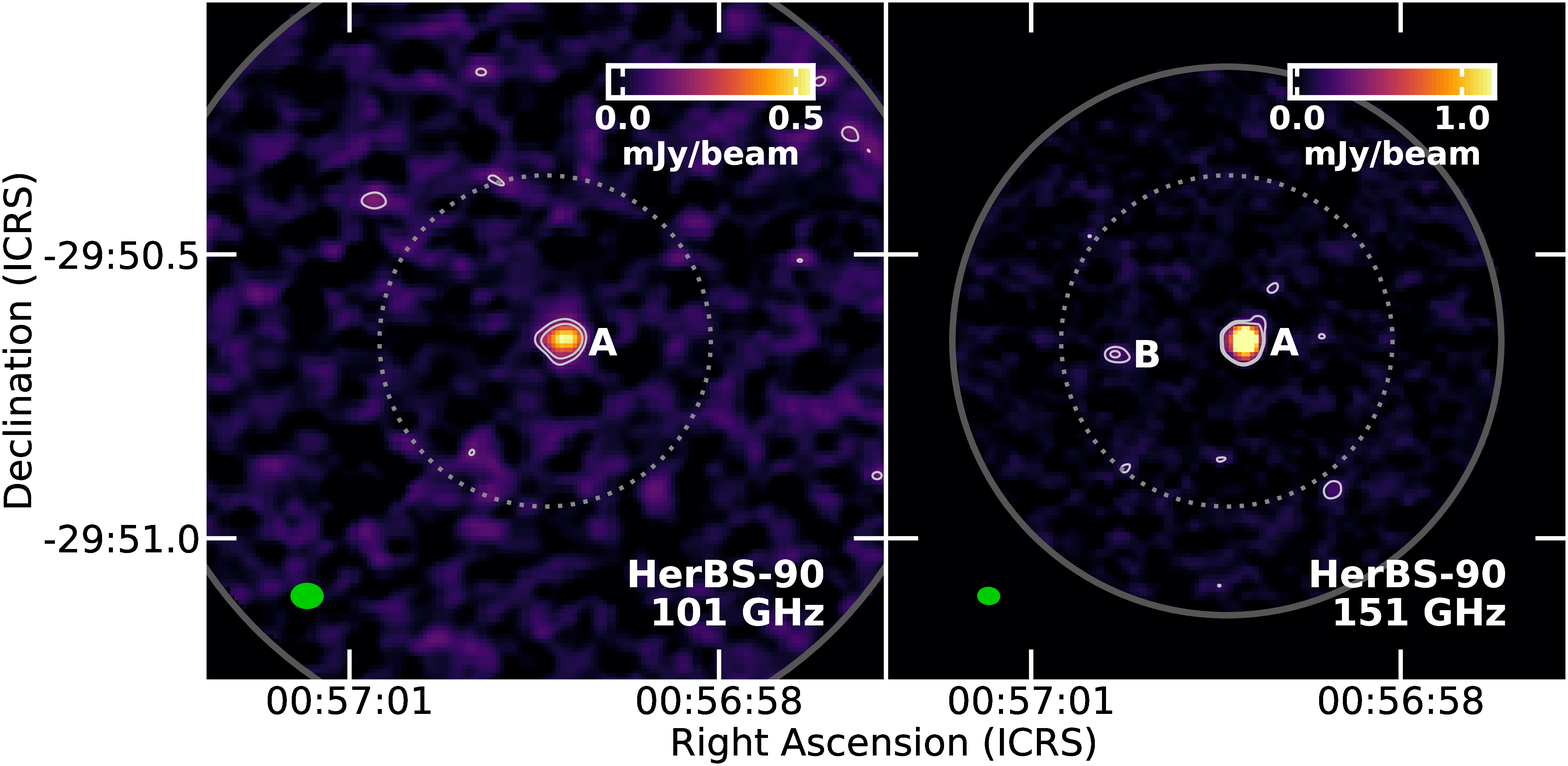}
		\end{center}
		\caption{Continued.}
	\end{figure*}
	
	\addtocounter{figure}{-1}
		
	\begin{figure*}
		\begin{center}
			\includegraphics[width=7cm]{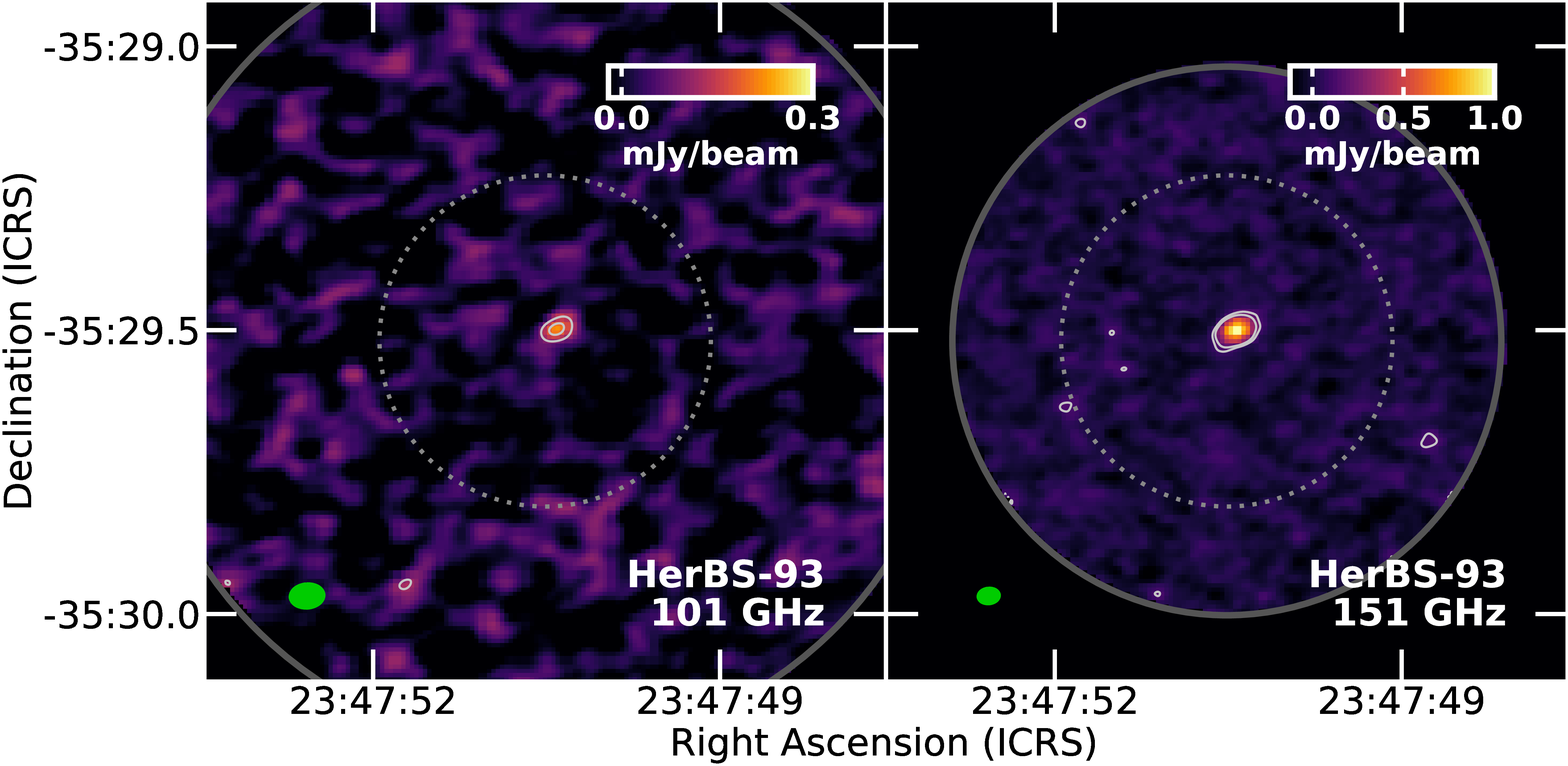}
			\ \ \ \ \ \
			\includegraphics[width=7cm]{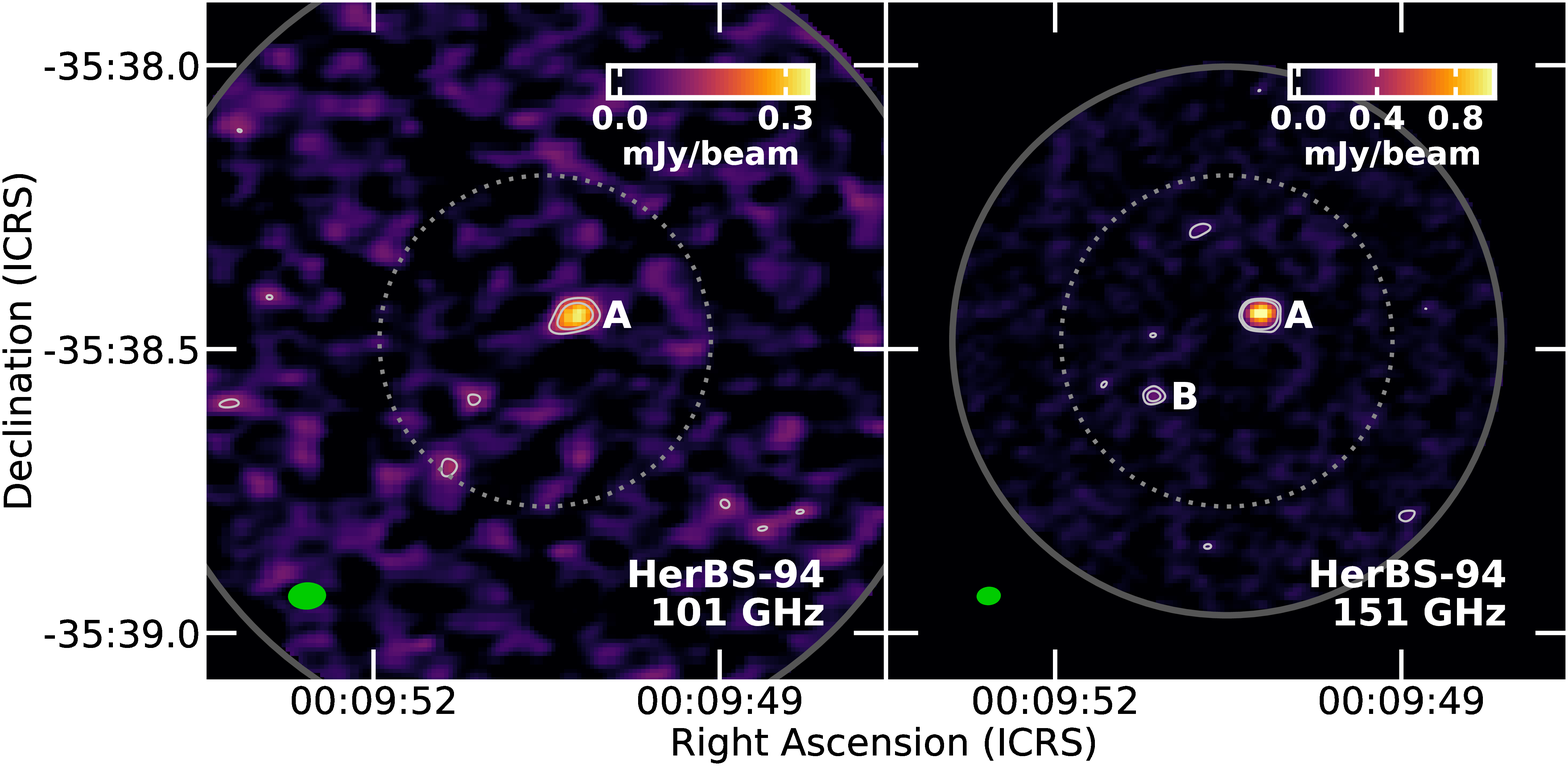}
			\vspace{0.5em}\\
			\includegraphics[width=7cm]{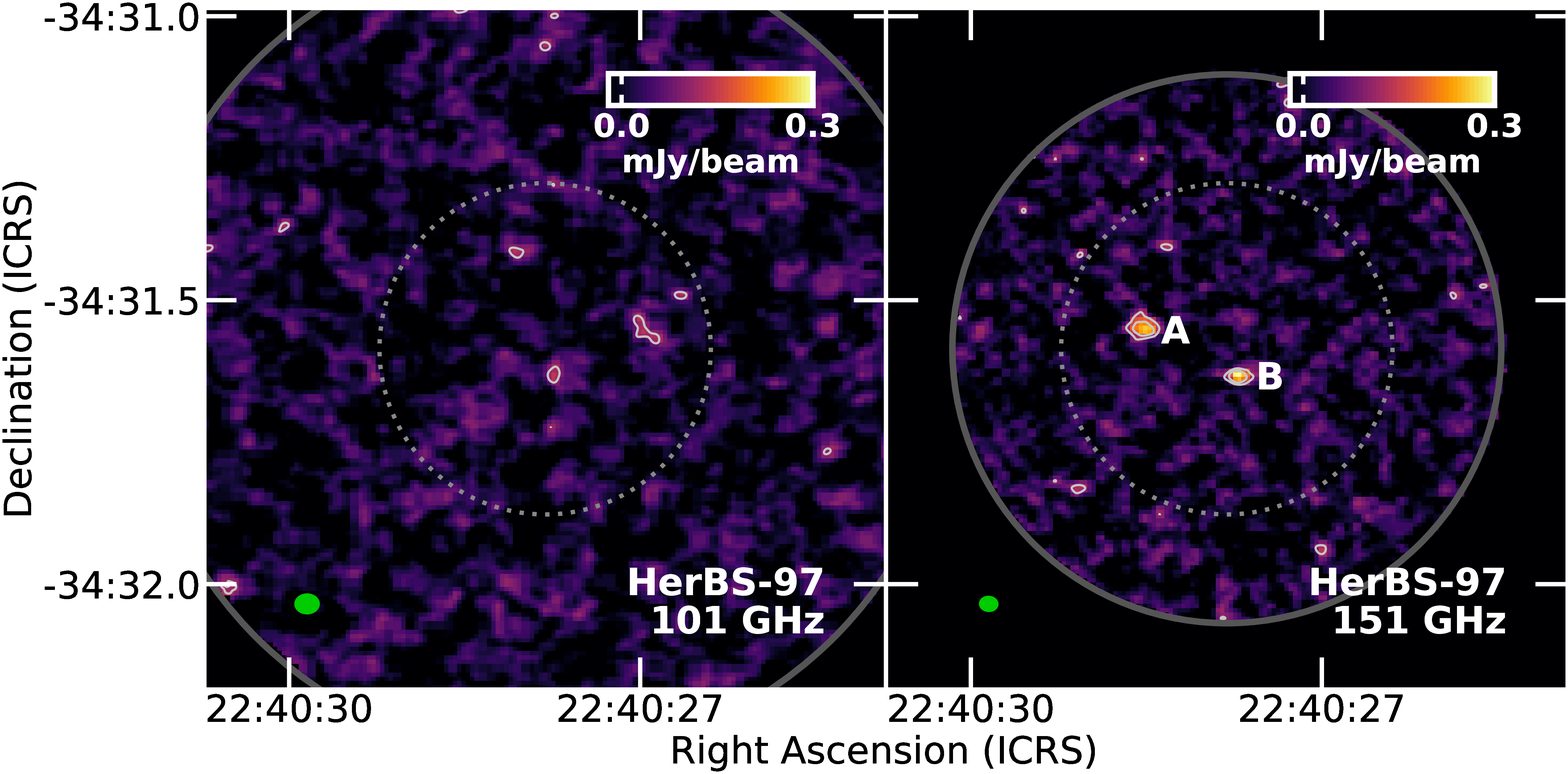}
			\ \ \ \ \ \
			\includegraphics[width=7cm]{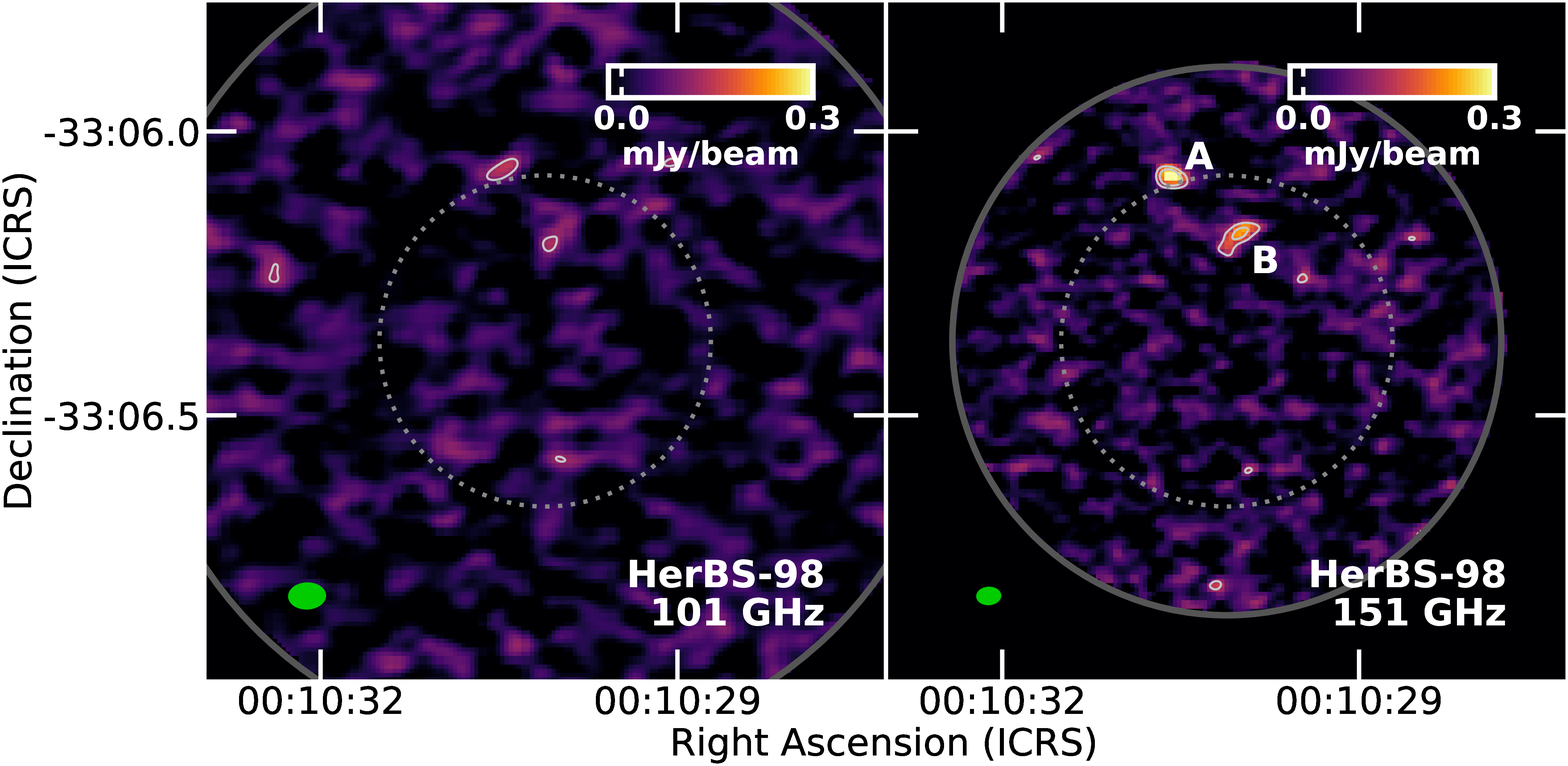}
			\vspace{0.5em}\\
			\includegraphics[width=7cm]{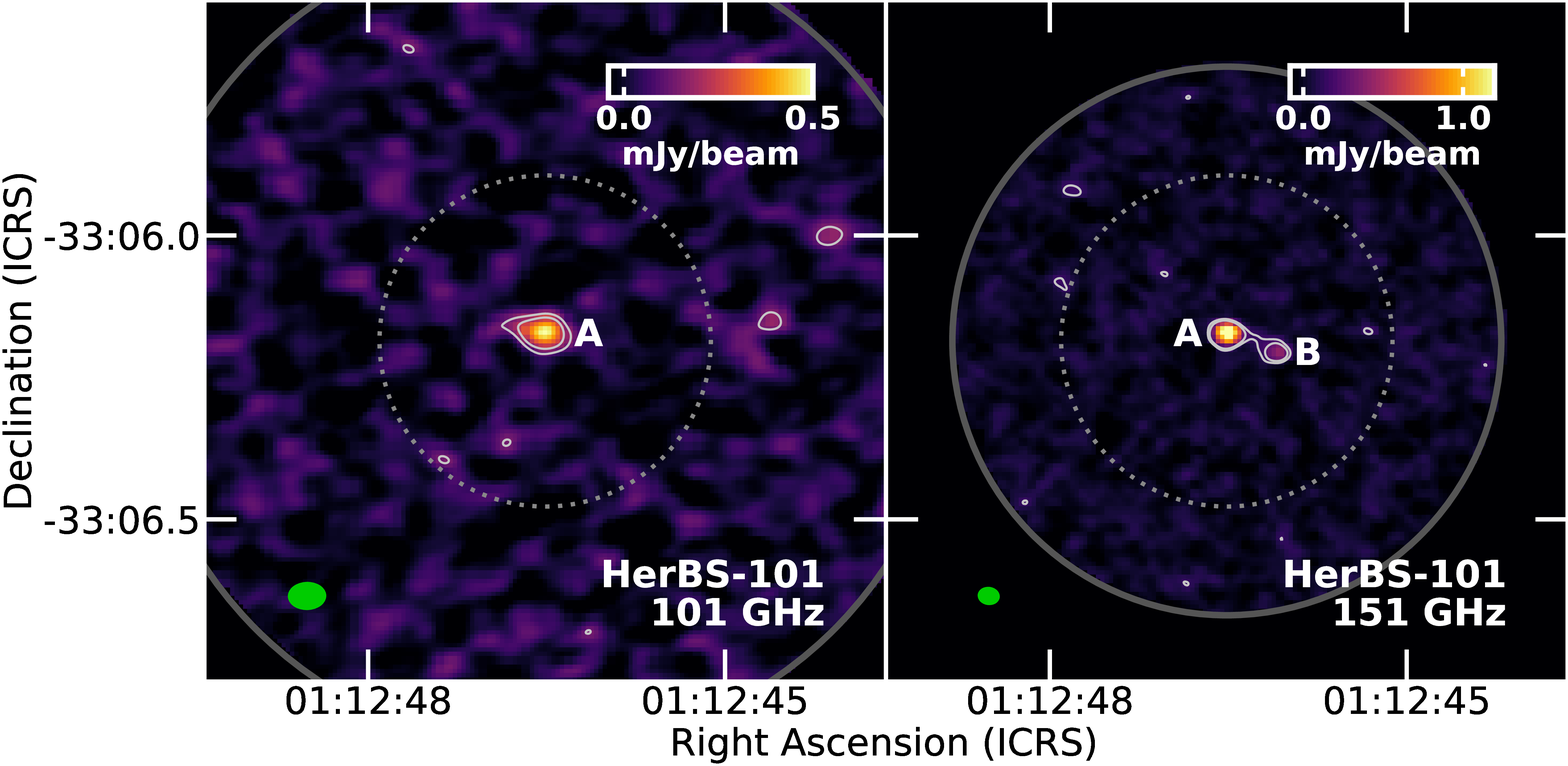}
			\ \ \ \ \ \
			\includegraphics[width=7cm]{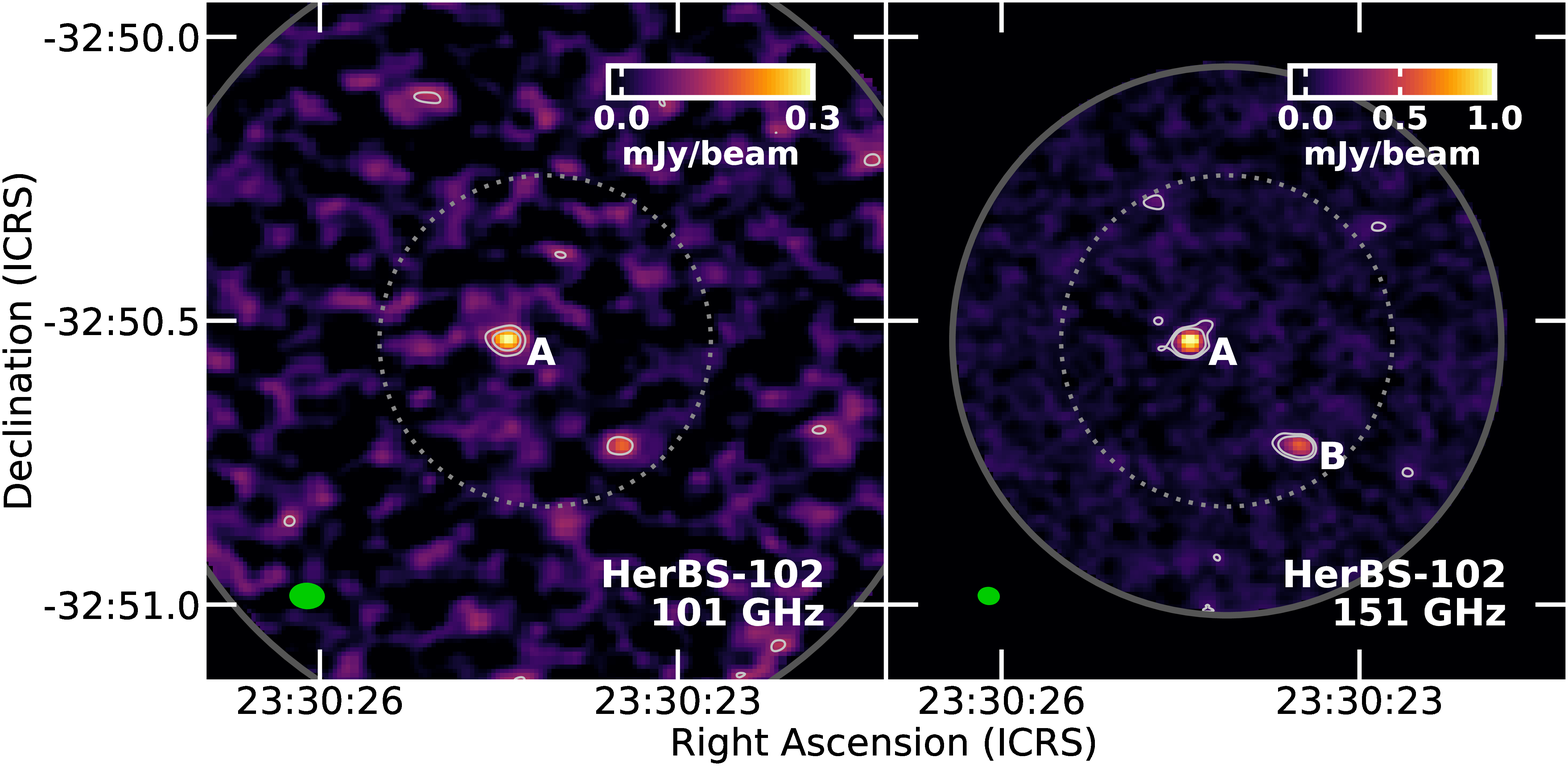}
			\vspace{0.5em}\\
			\includegraphics[width=7cm]{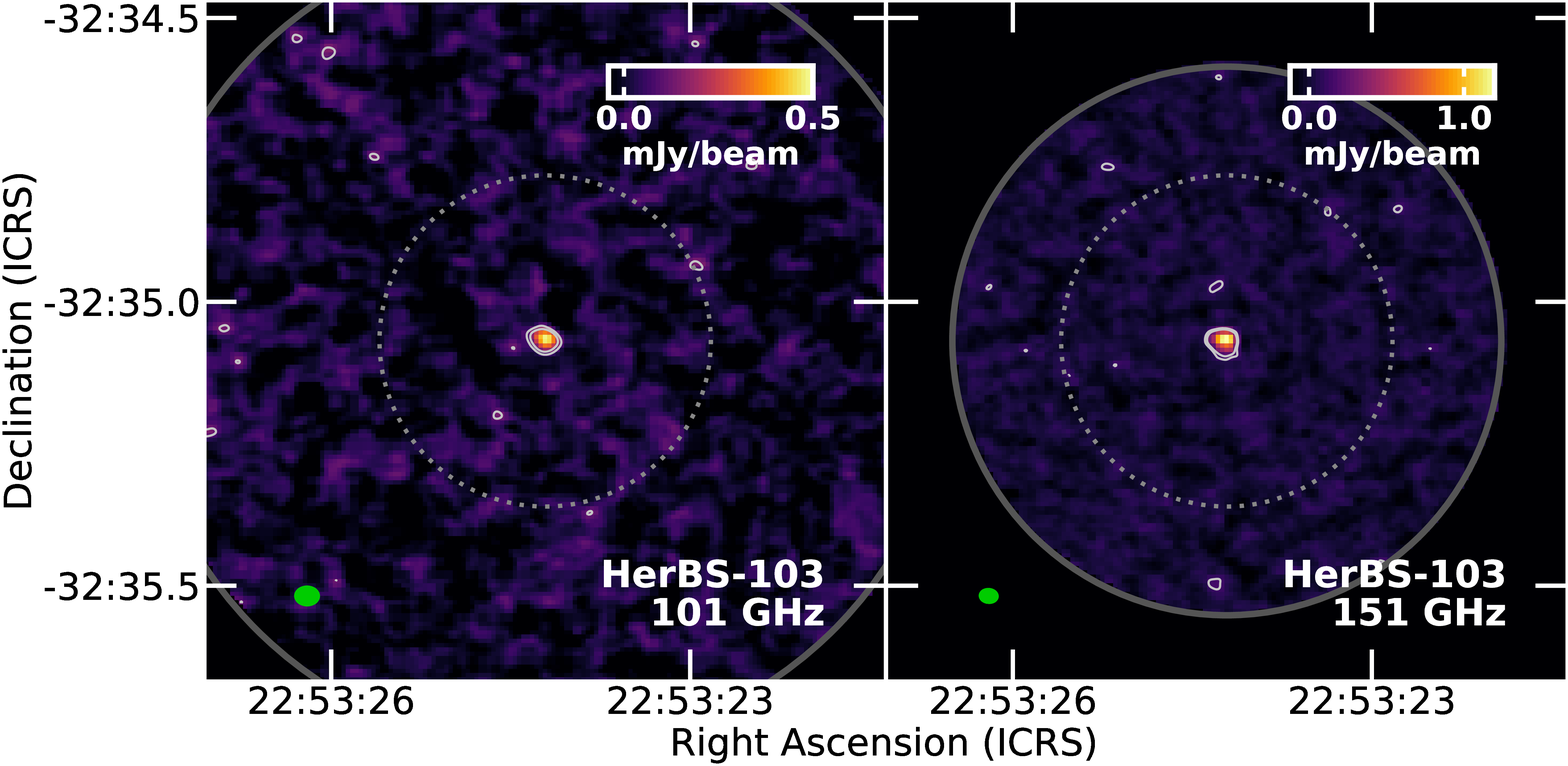}
			\ \ \ \ \ \
			\includegraphics[width=7cm]{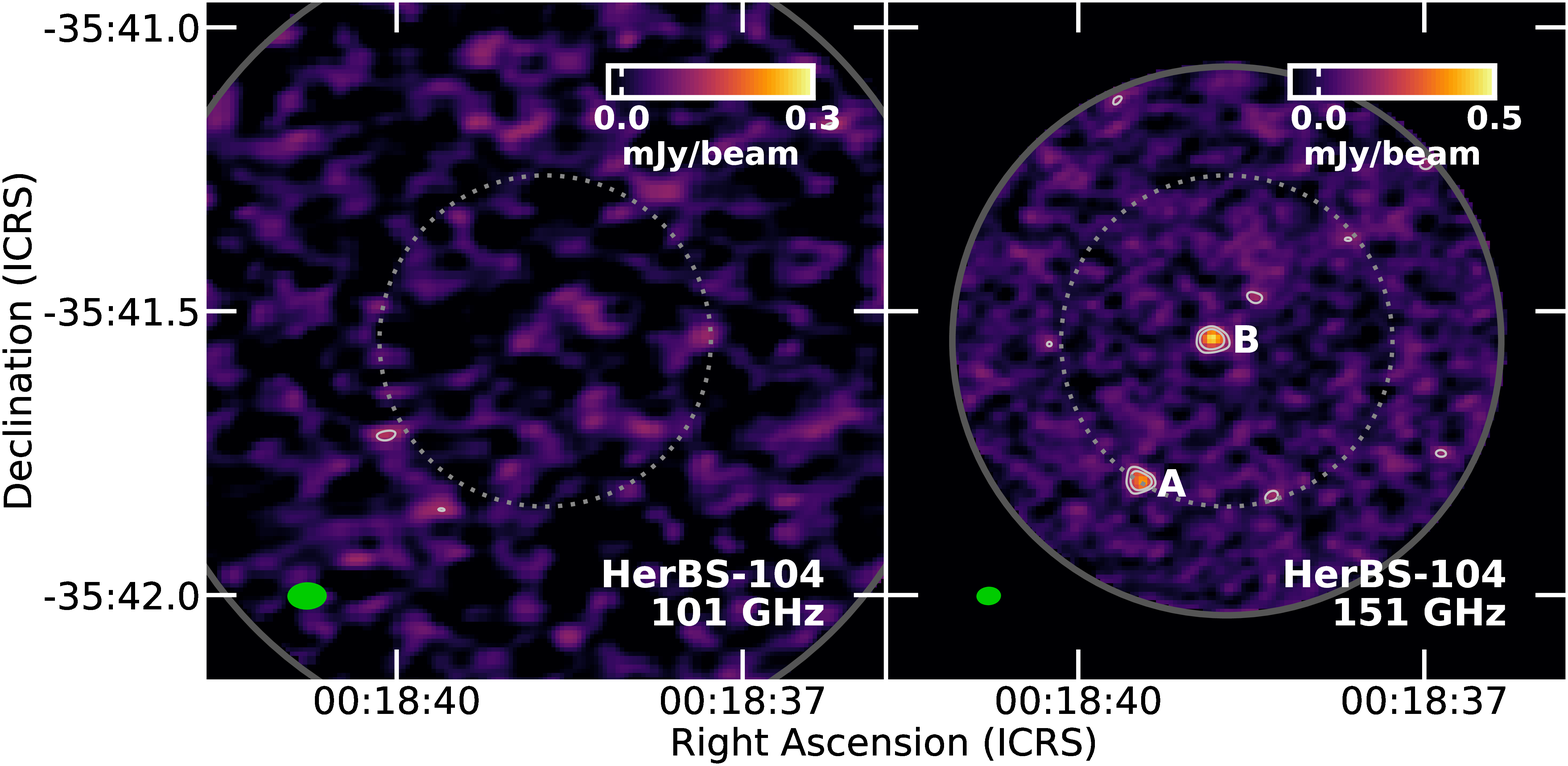}
			\vspace{0.5em}\\
			\includegraphics[width=7cm]{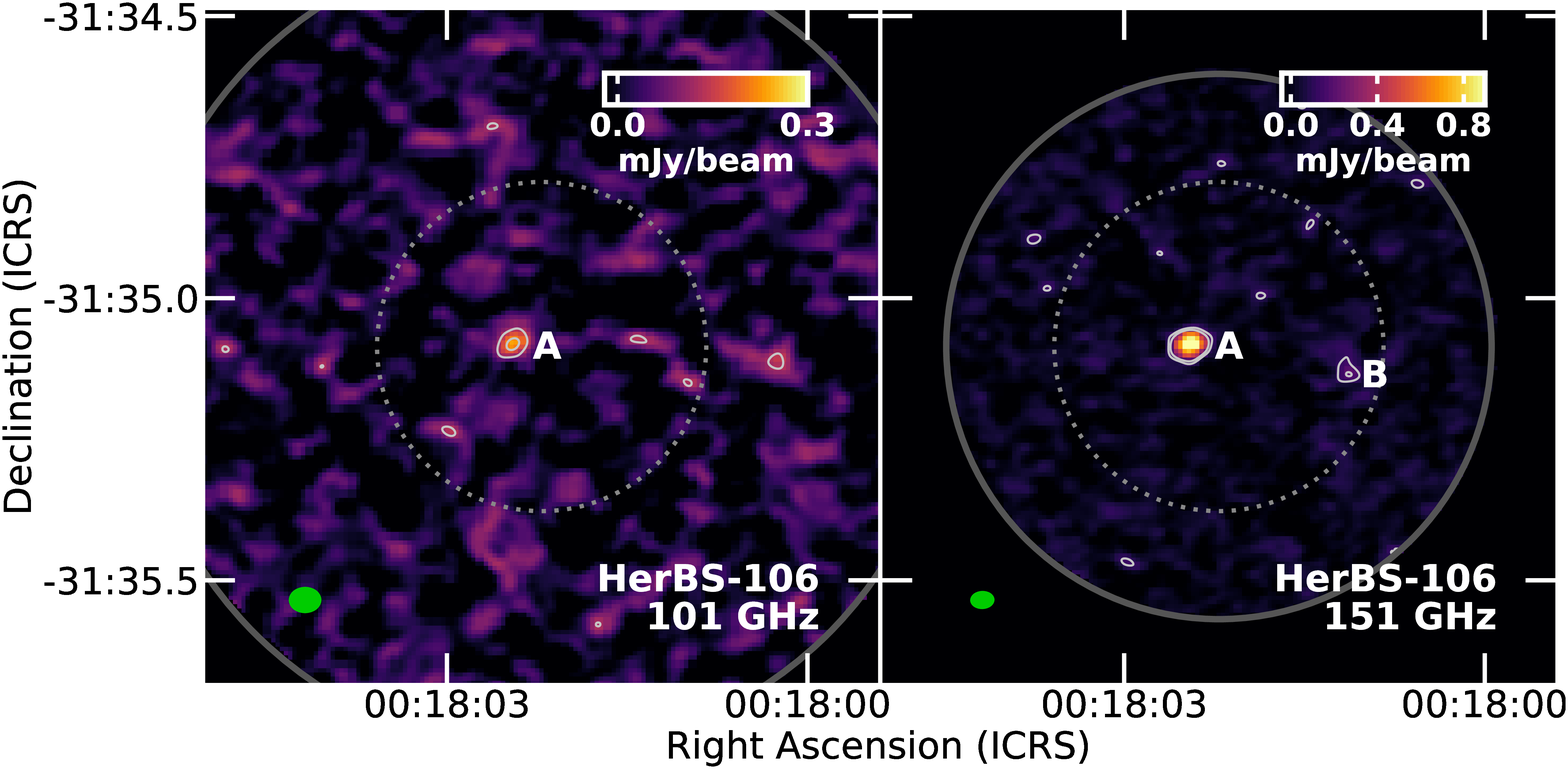}
			\ \ \ \ \ \
			\includegraphics[width=7cm]{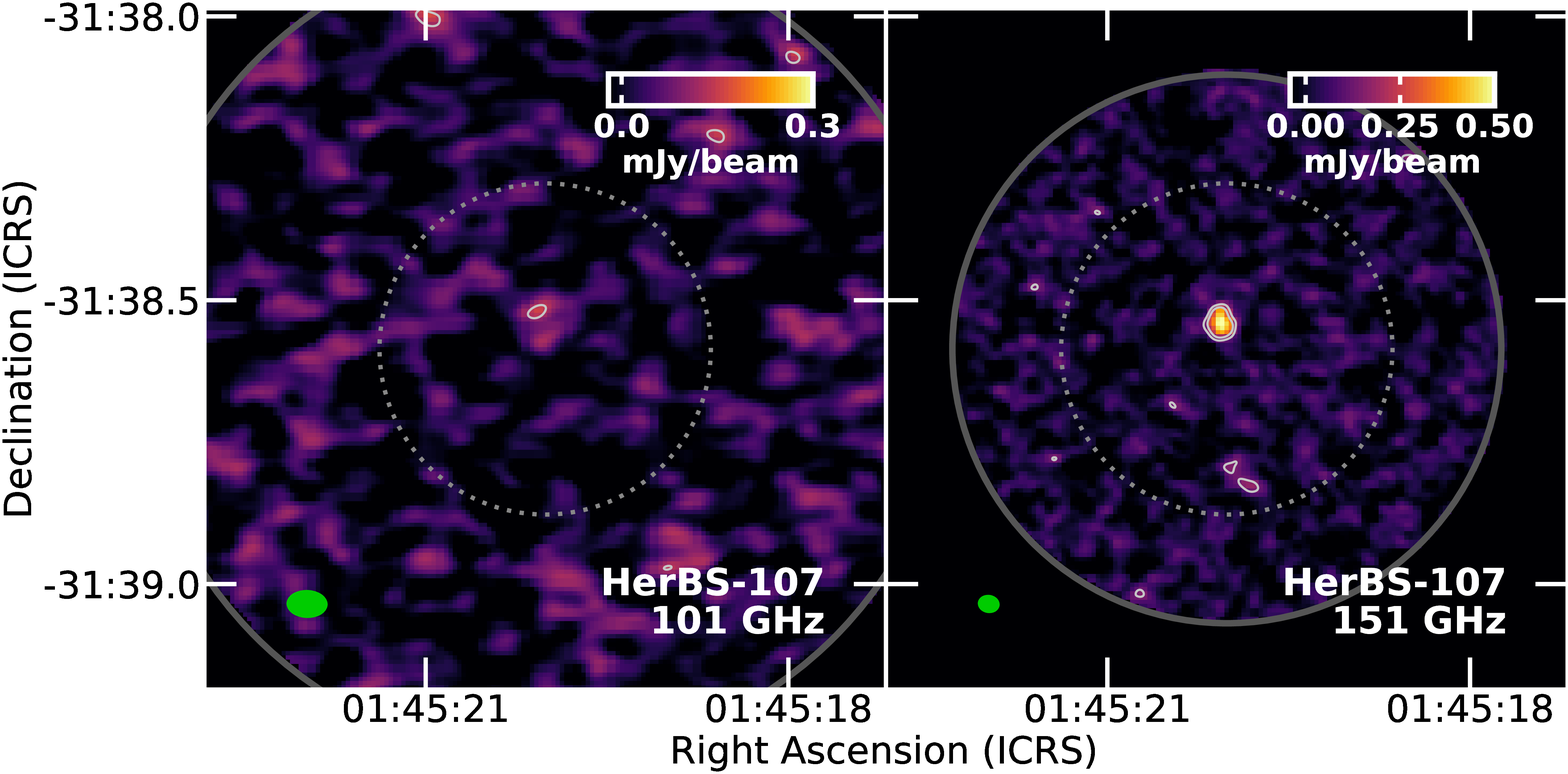}
			\vspace{0.5em}\\
			\includegraphics[width=7cm]{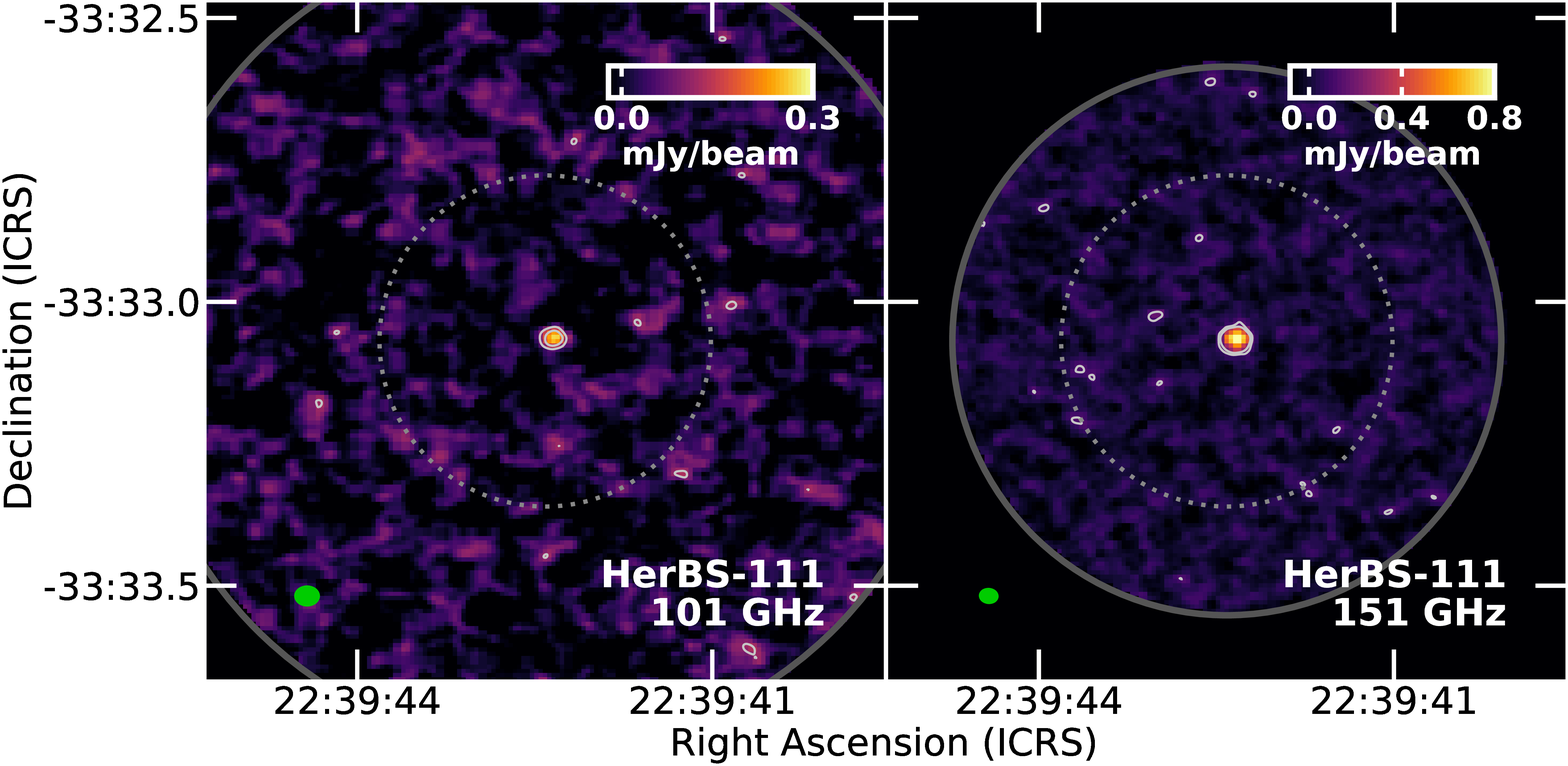}
			\ \ \ \ \ \
			\includegraphics[width=7cm]{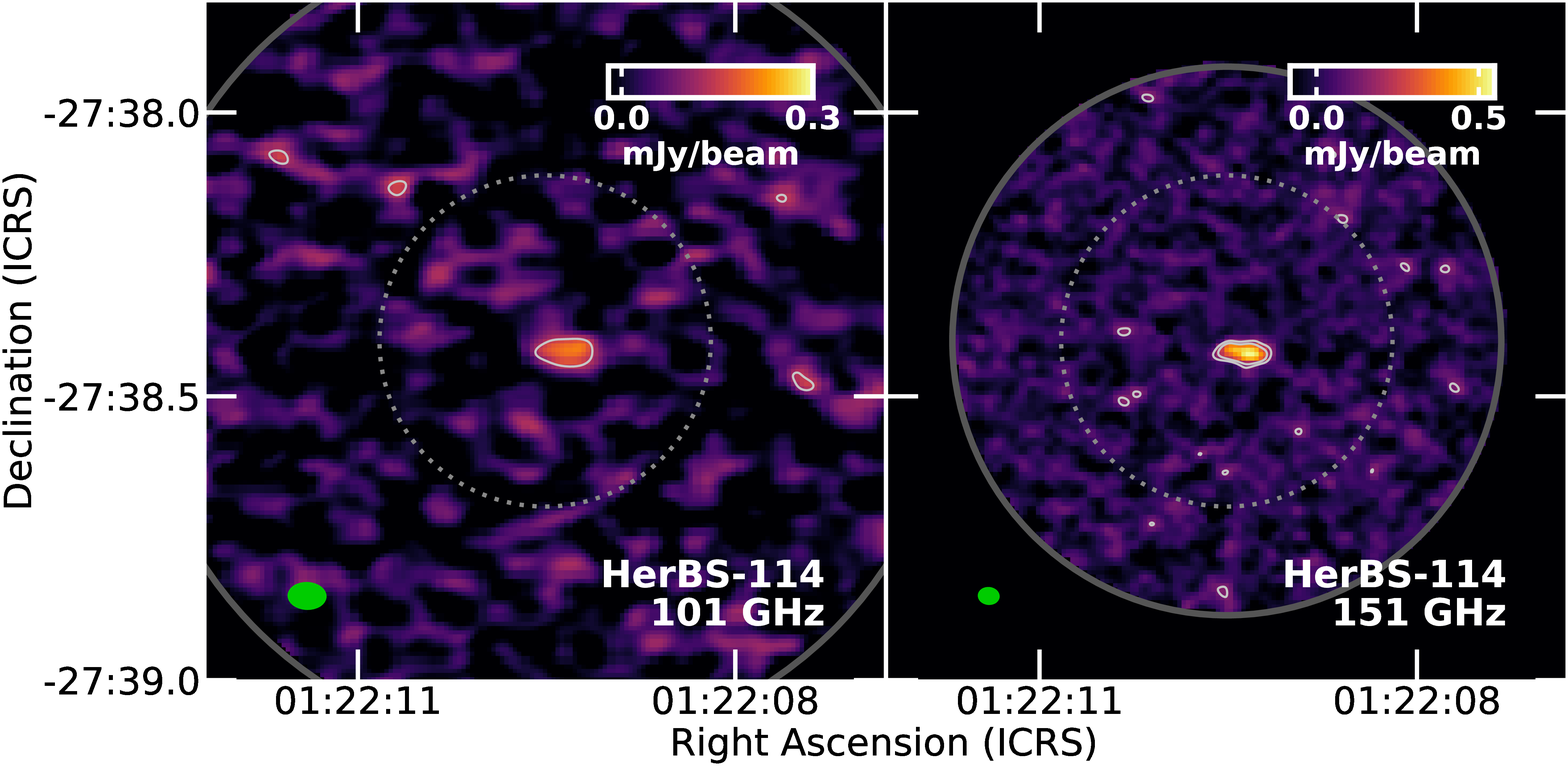}
		\end{center}
		\caption{Continued.}
	\end{figure*}
	
	\addtocounter{figure}{-1}
	
	\begin{figure*}
		\begin{center}
			\includegraphics[width=7cm]{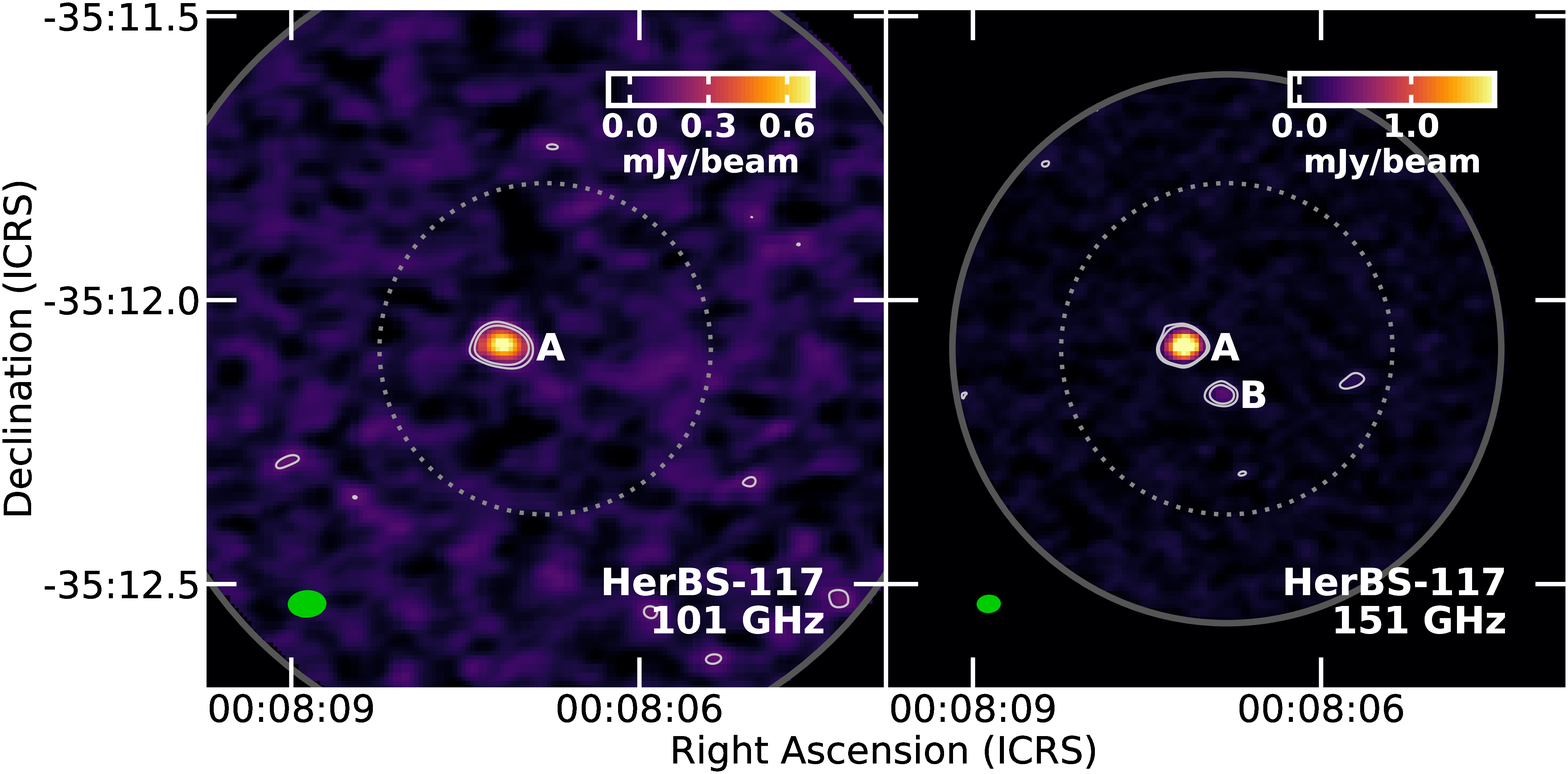}
			\ \ \ \ \ \
			\includegraphics[width=7cm]{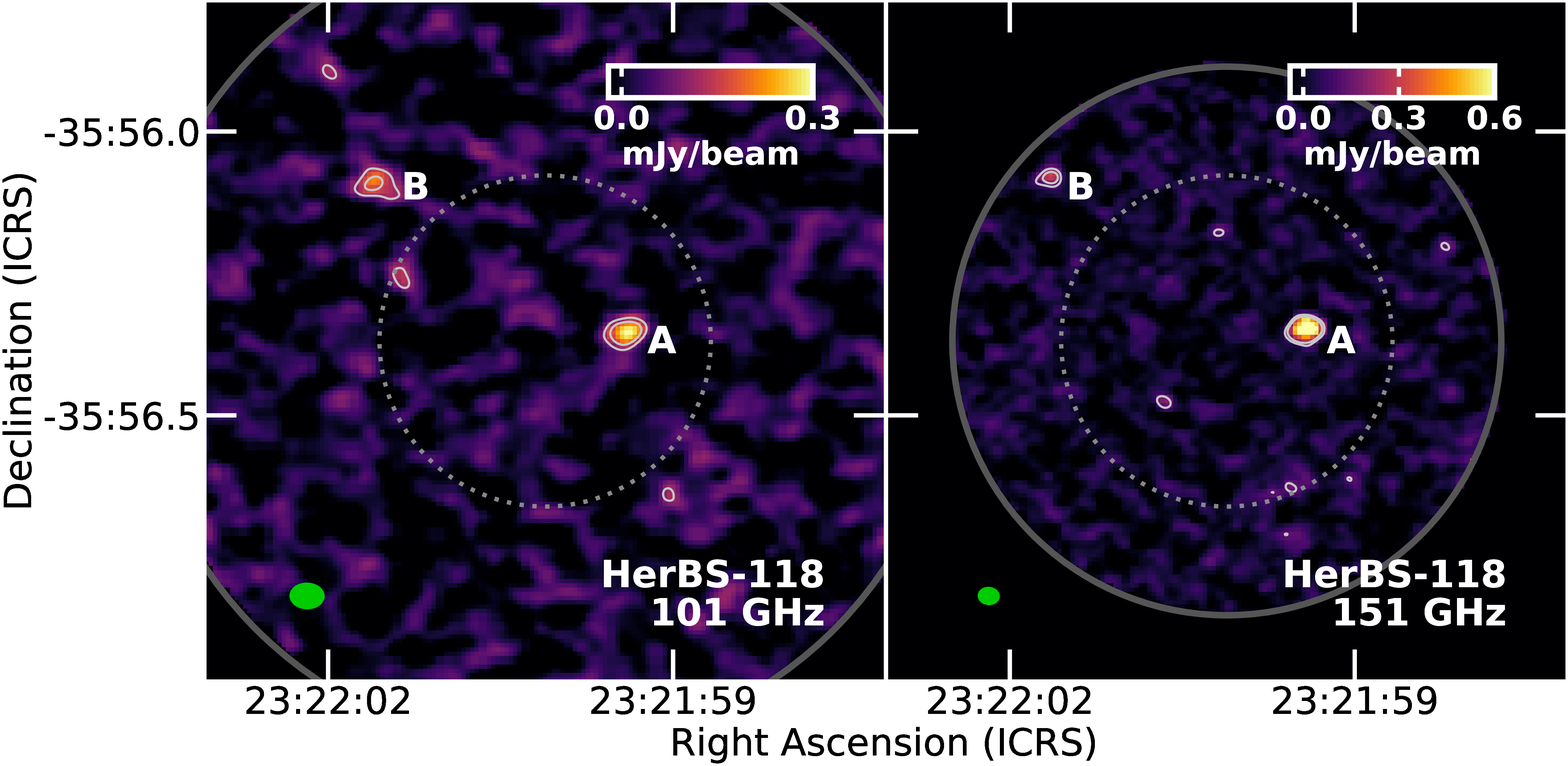}
			\vspace{0.5em}\\
			\includegraphics[width=7cm]{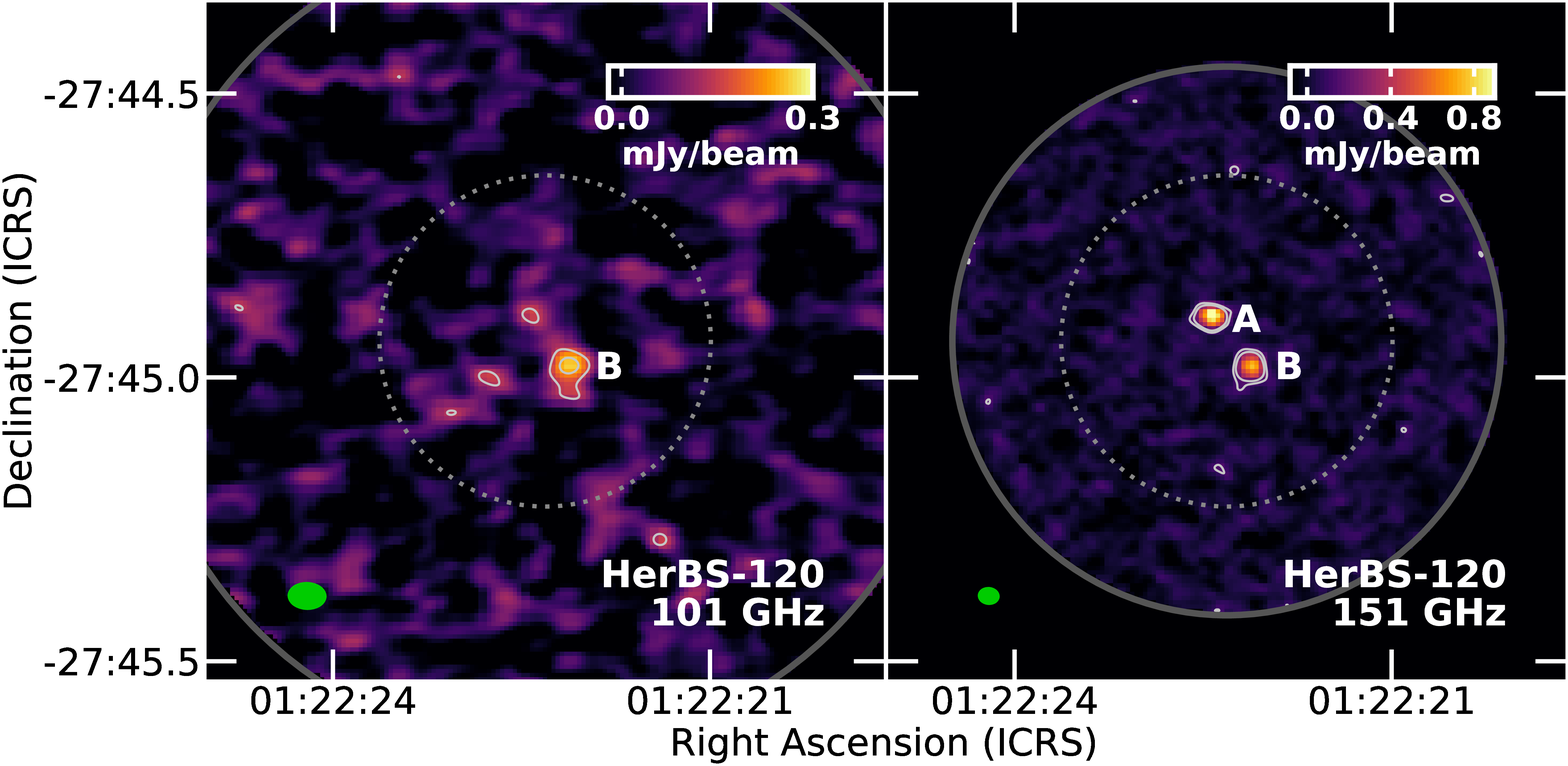}
			\ \ \ \ \ \
			\includegraphics[width=7cm]{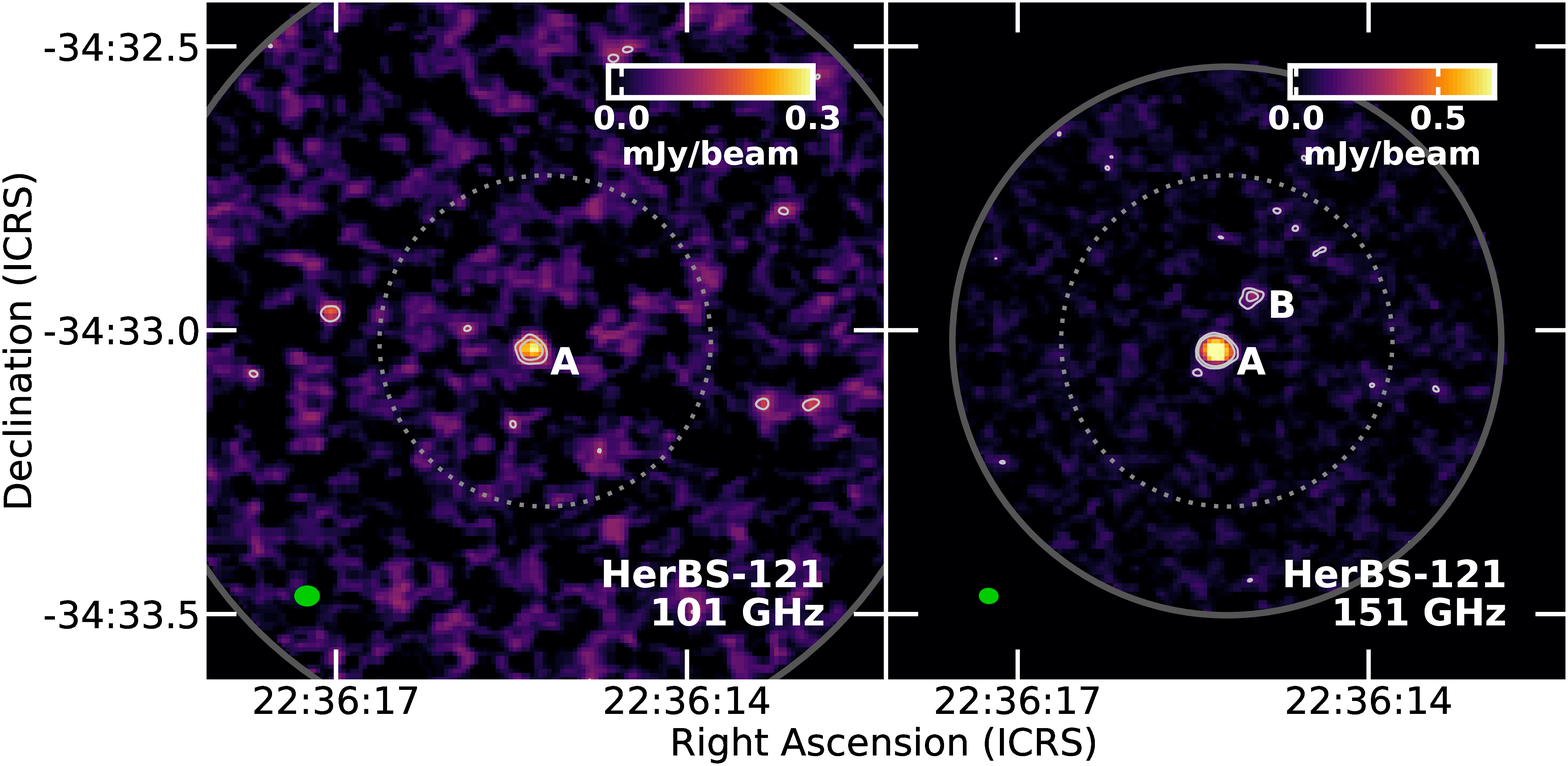}
			\vspace{0.5em}\\
			\includegraphics[width=7cm]{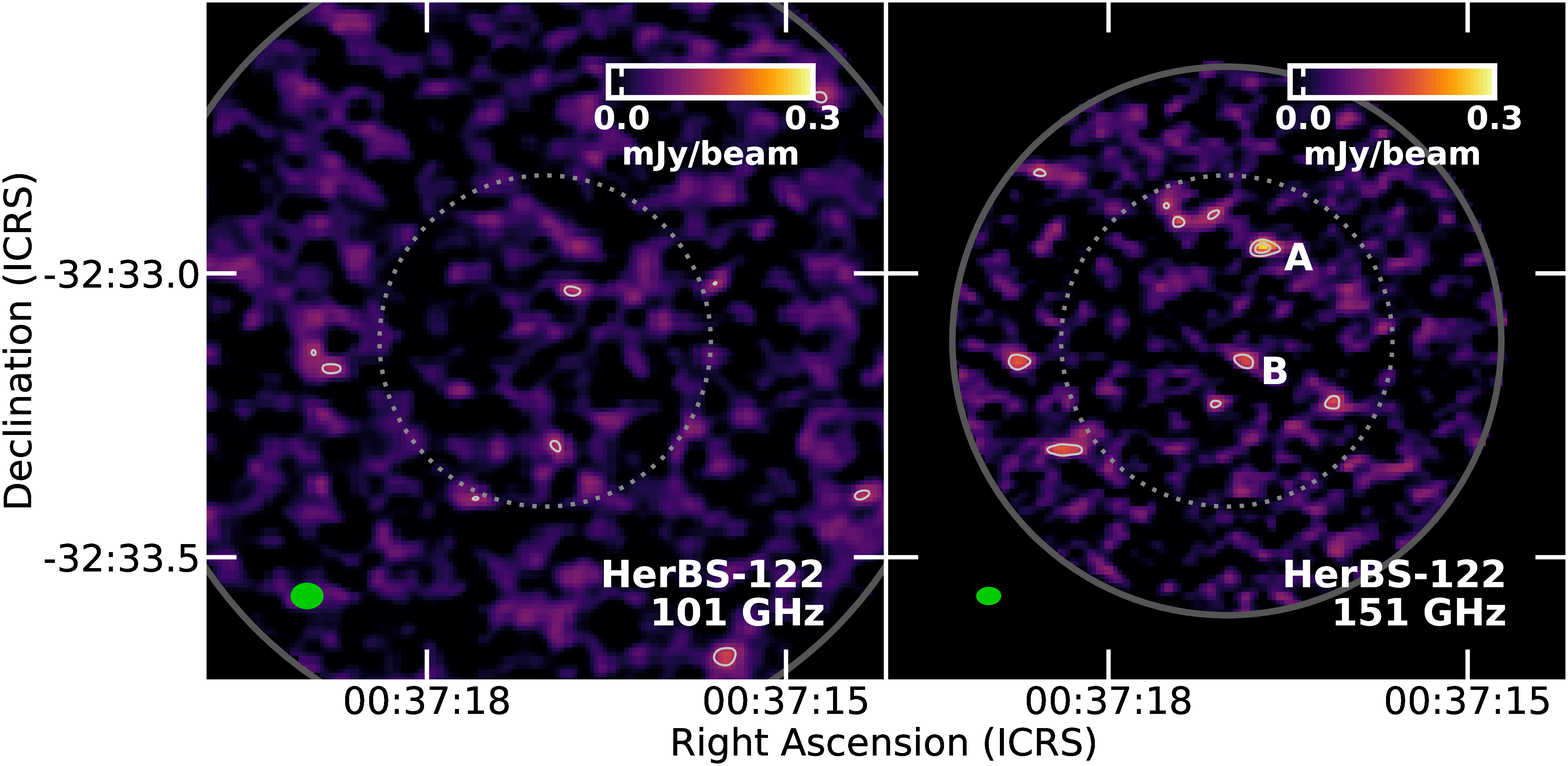}
			\ \ \ \ \ \
			\includegraphics[width=7cm]{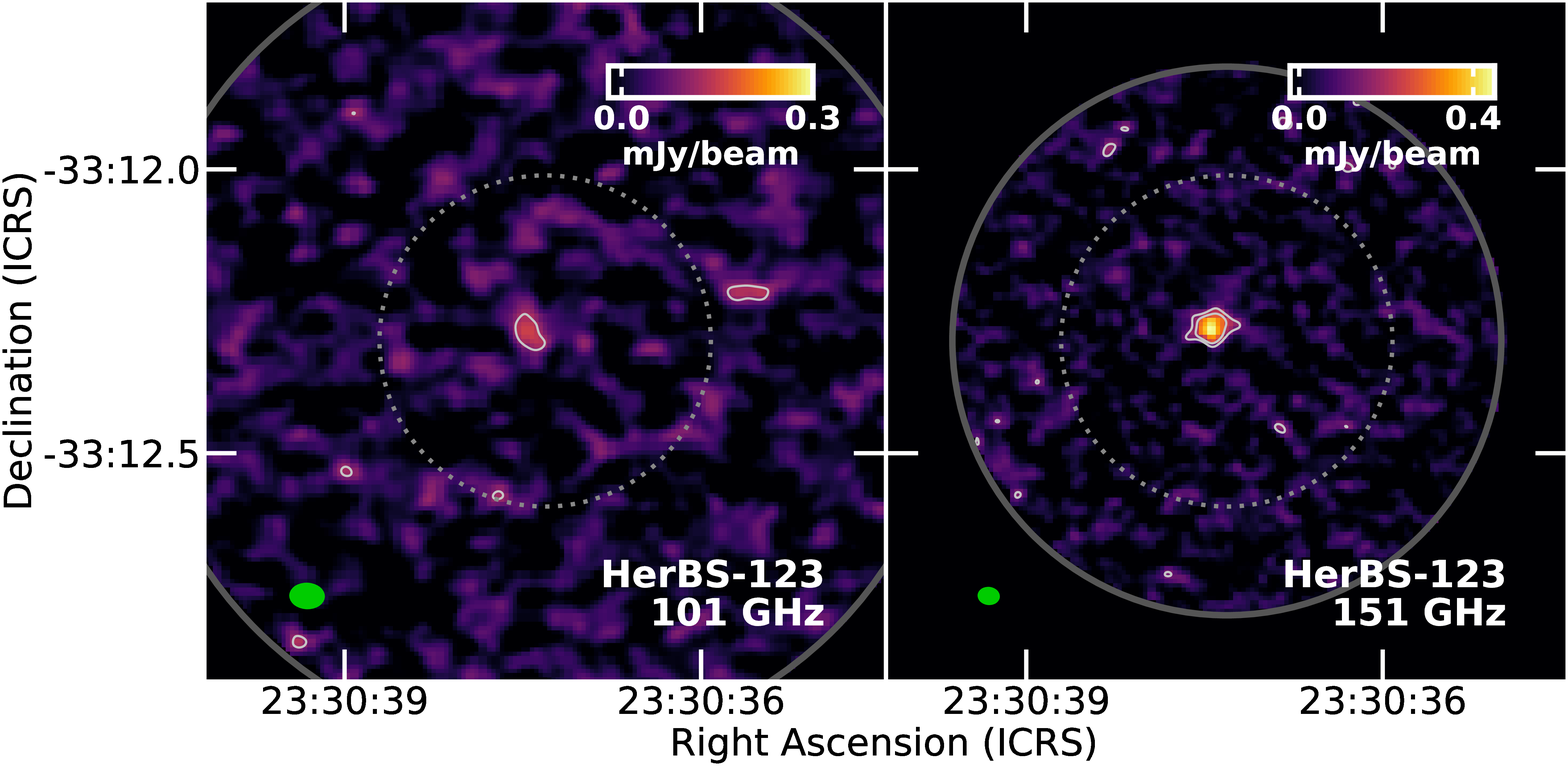}
			\vspace{0.5em}\\
			\includegraphics[width=7cm]{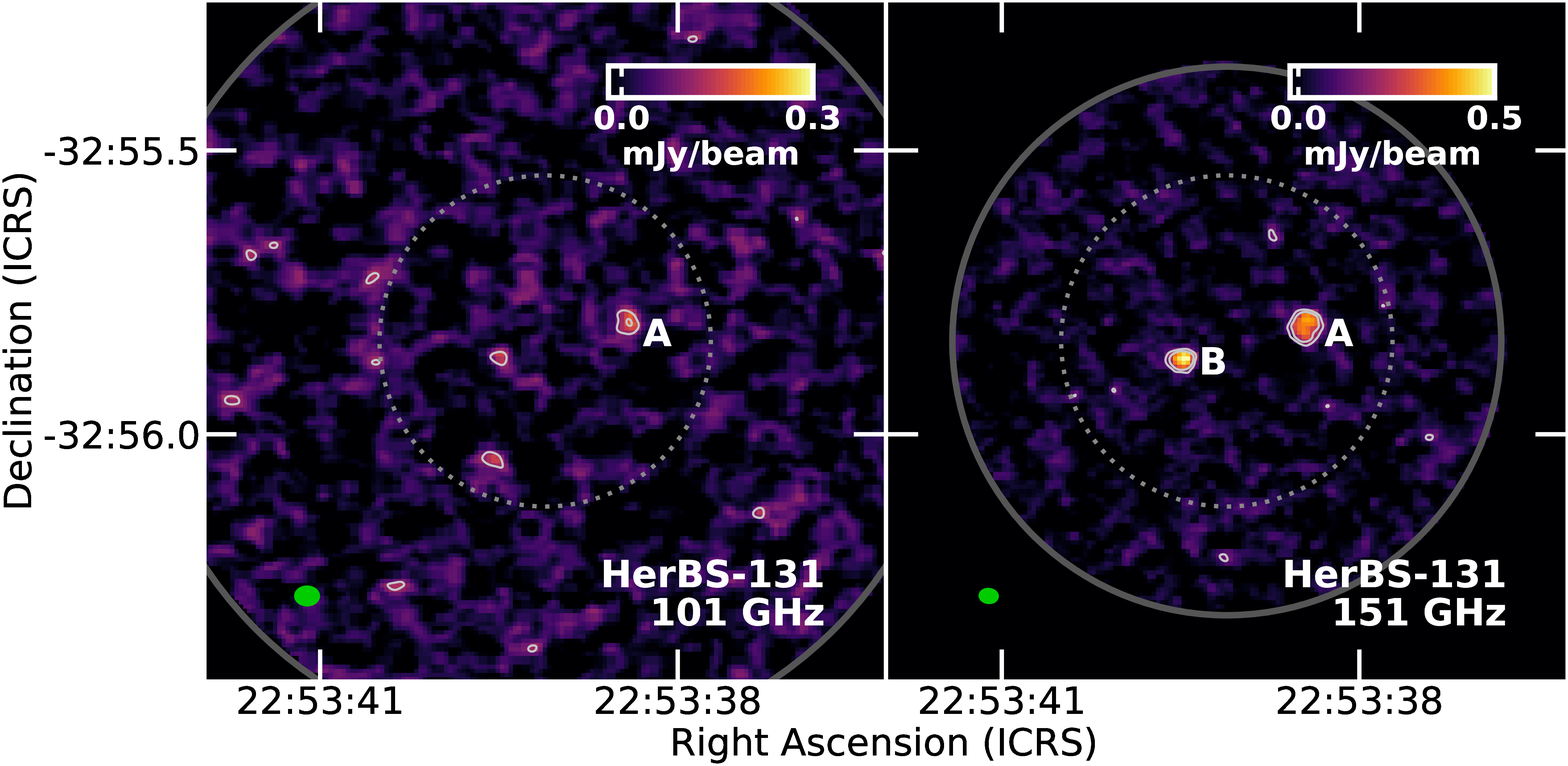}
			\ \ \ \ \ \
			\includegraphics[width=7cm]{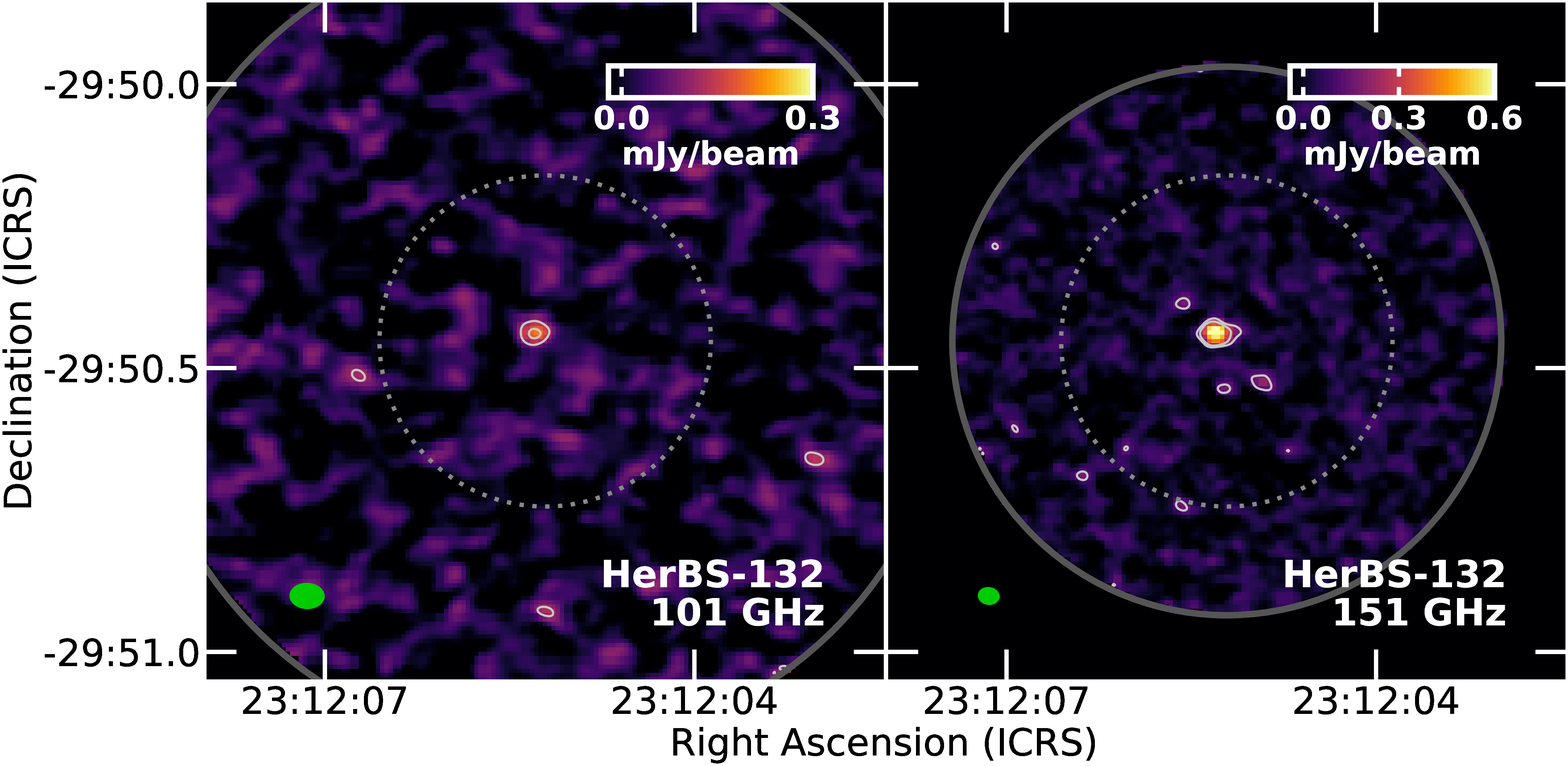}
			\vspace{0.5em}\\
			\includegraphics[width=7cm]{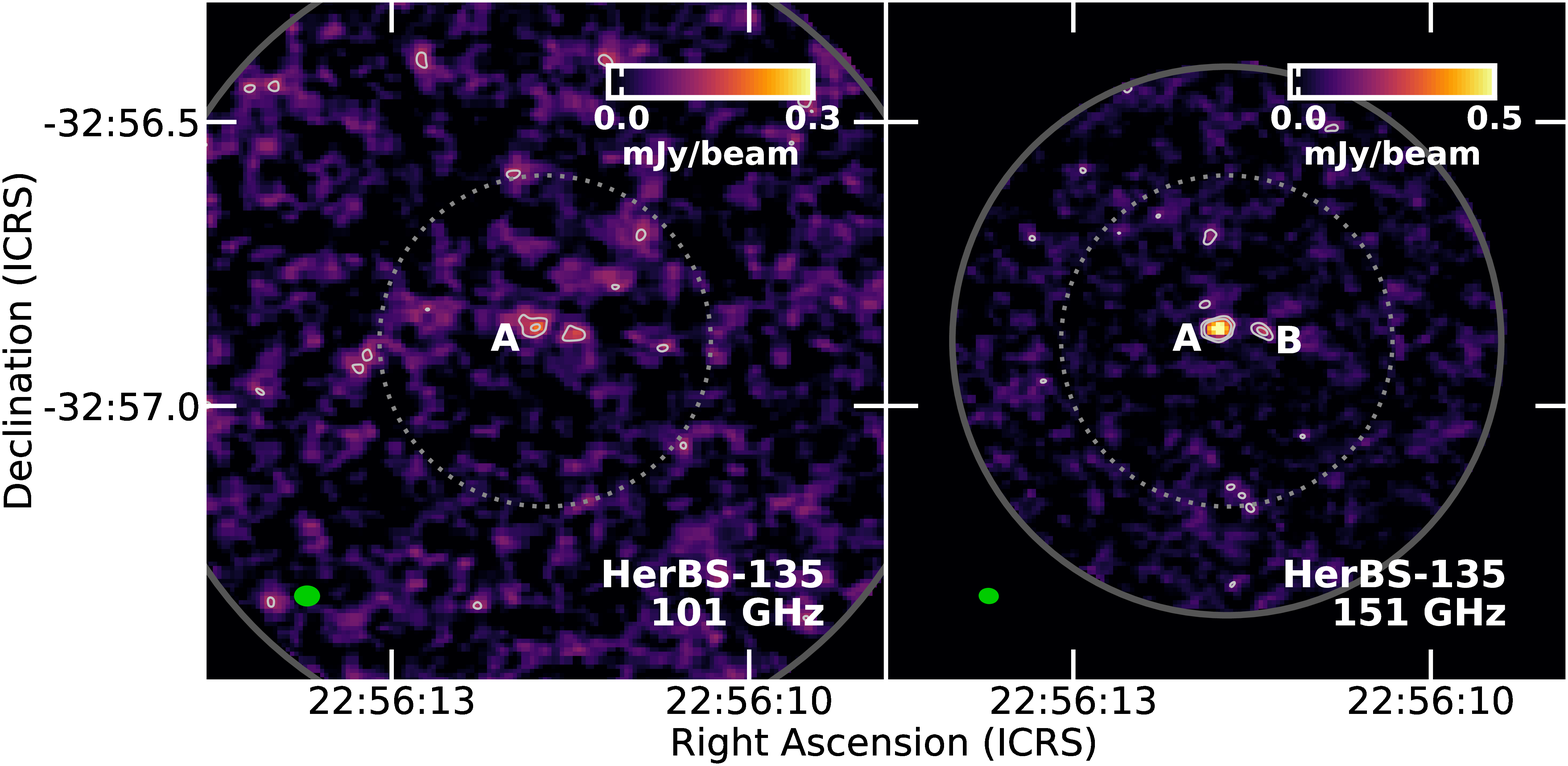}
			\ \ \ \ \ \
			\includegraphics[width=7cm]{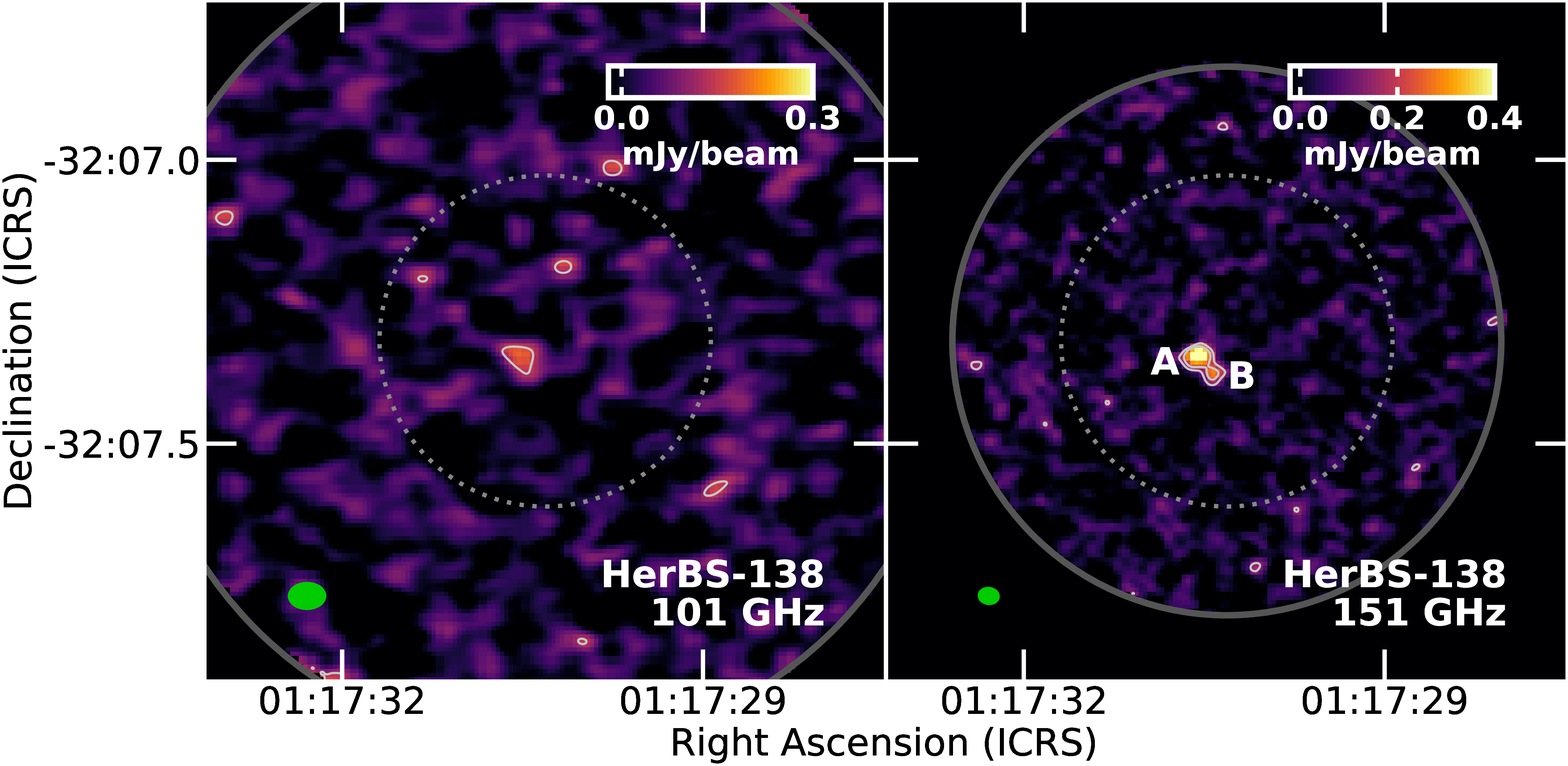}
			\vspace{0.5em}\\
			\includegraphics[width=7cm]{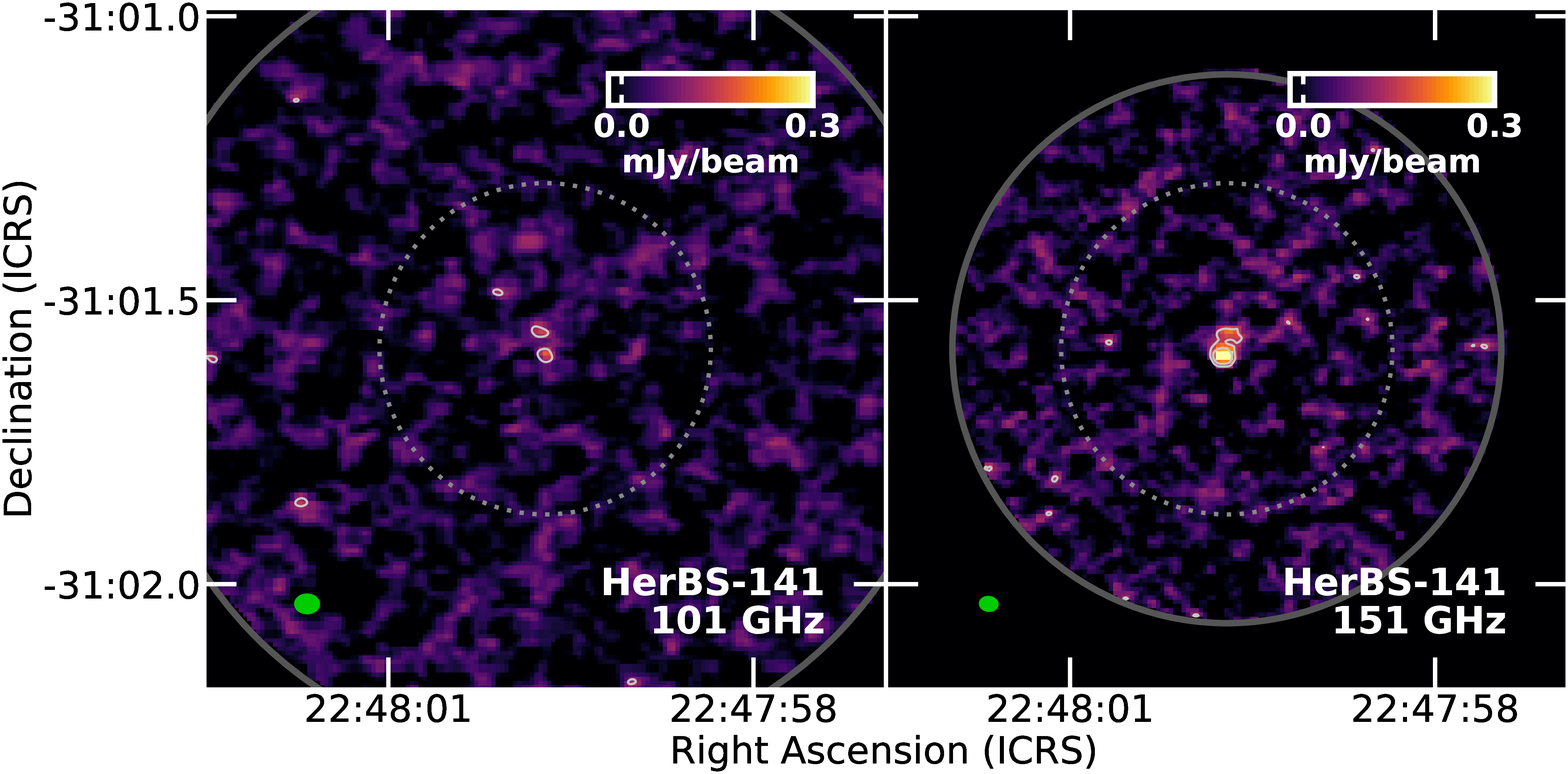}
			\ \ \ \ \ \
			\includegraphics[width=7cm]{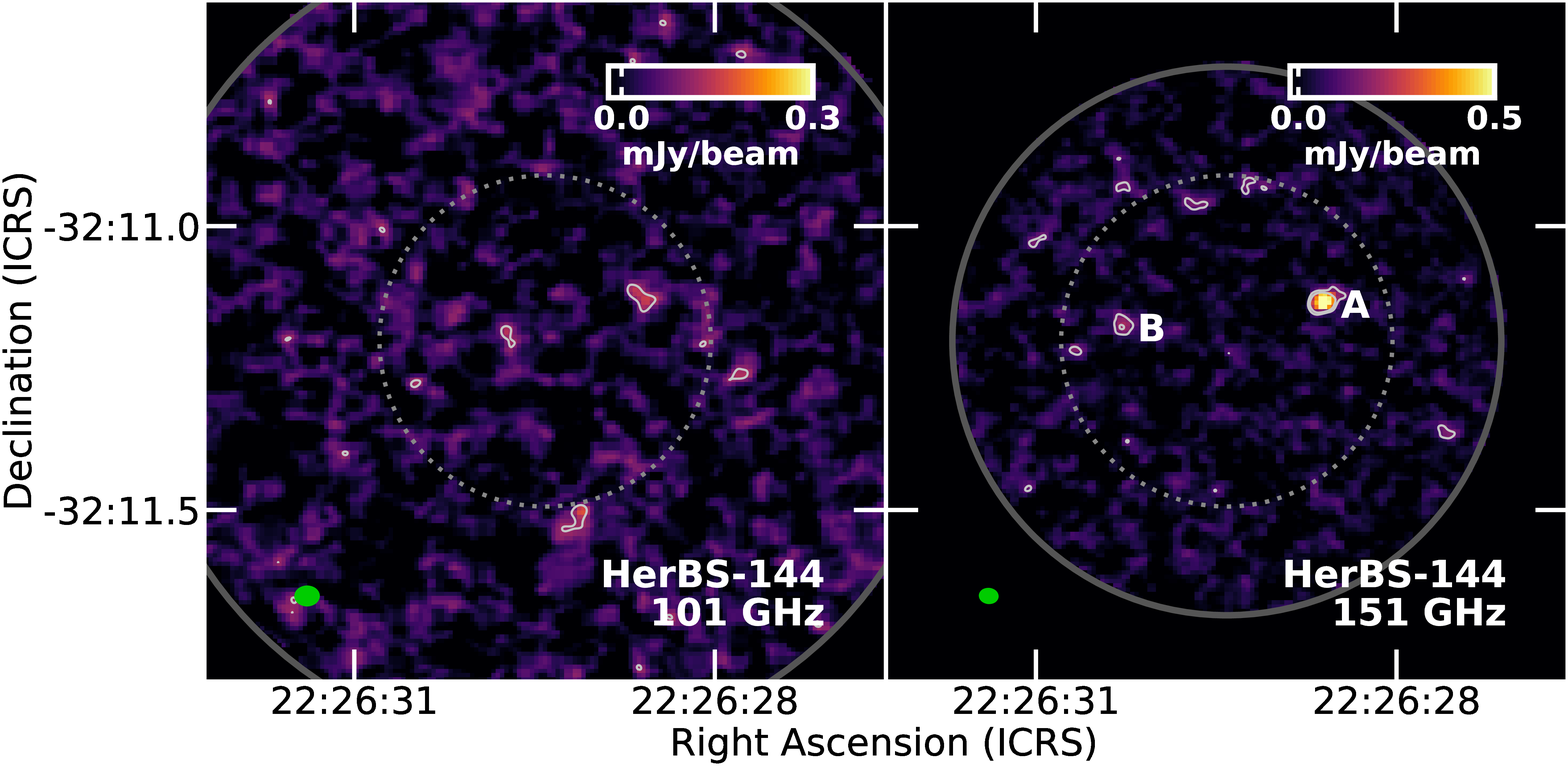}
		\end{center}
		\caption{Continued.}
	\end{figure*}
	
	\addtocounter{figure}{-1}
	
	\begin{figure*}
		\begin{center}
			\includegraphics[width=7cm]{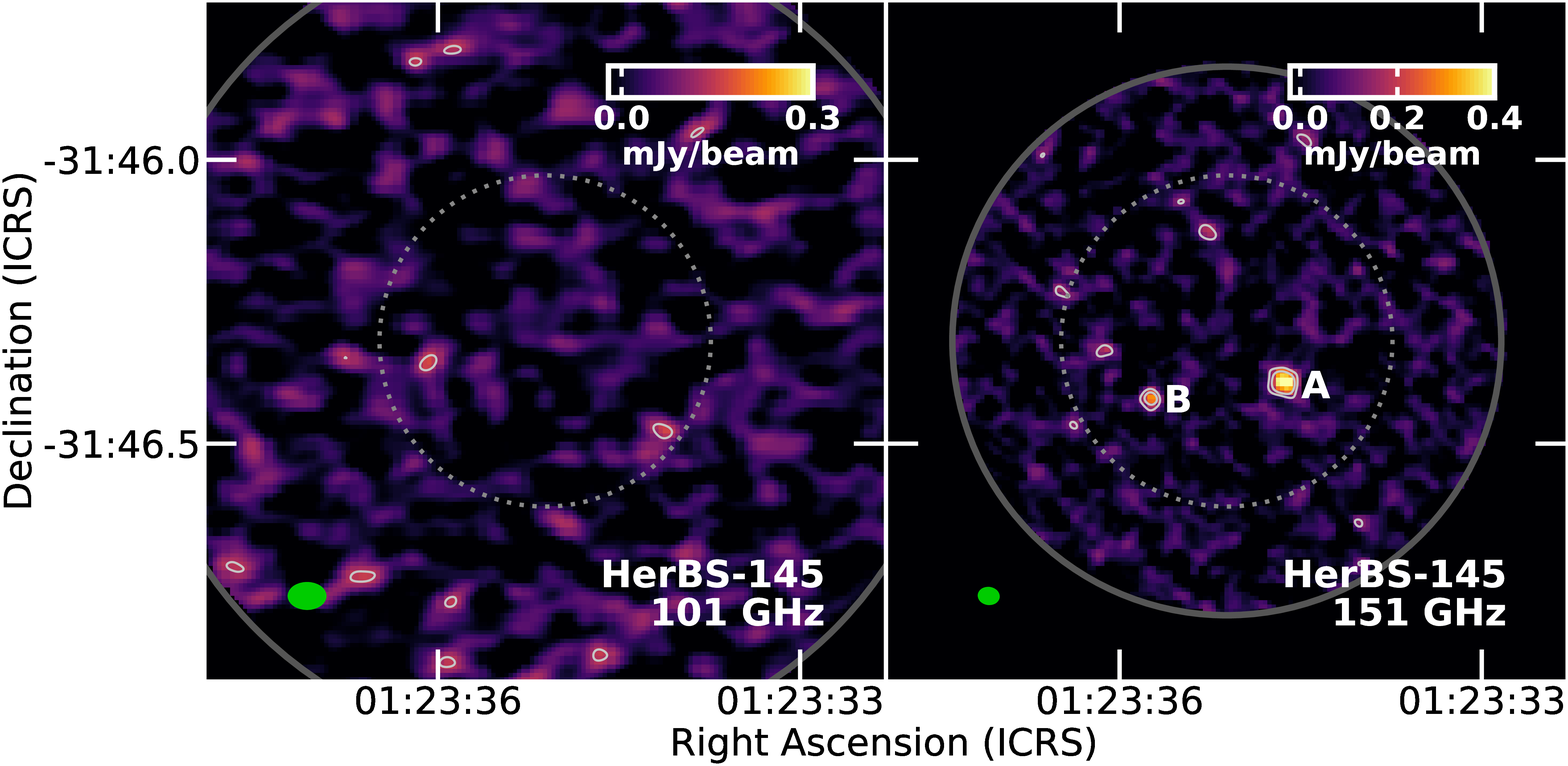}
			\ \ \ \ \ \
			\includegraphics[width=7cm]{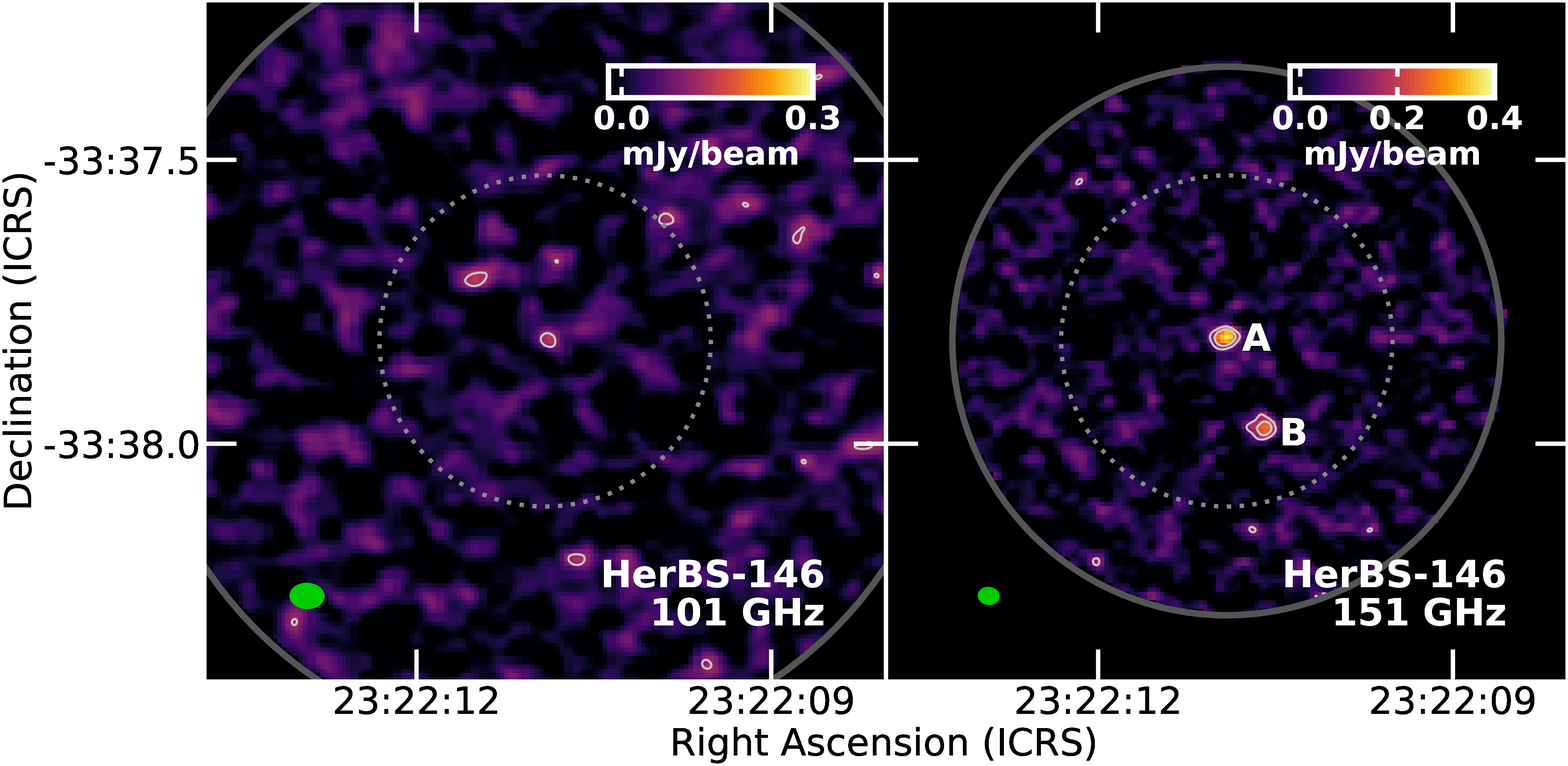}
			\vspace{0.5em}\\
			\includegraphics[width=7cm]{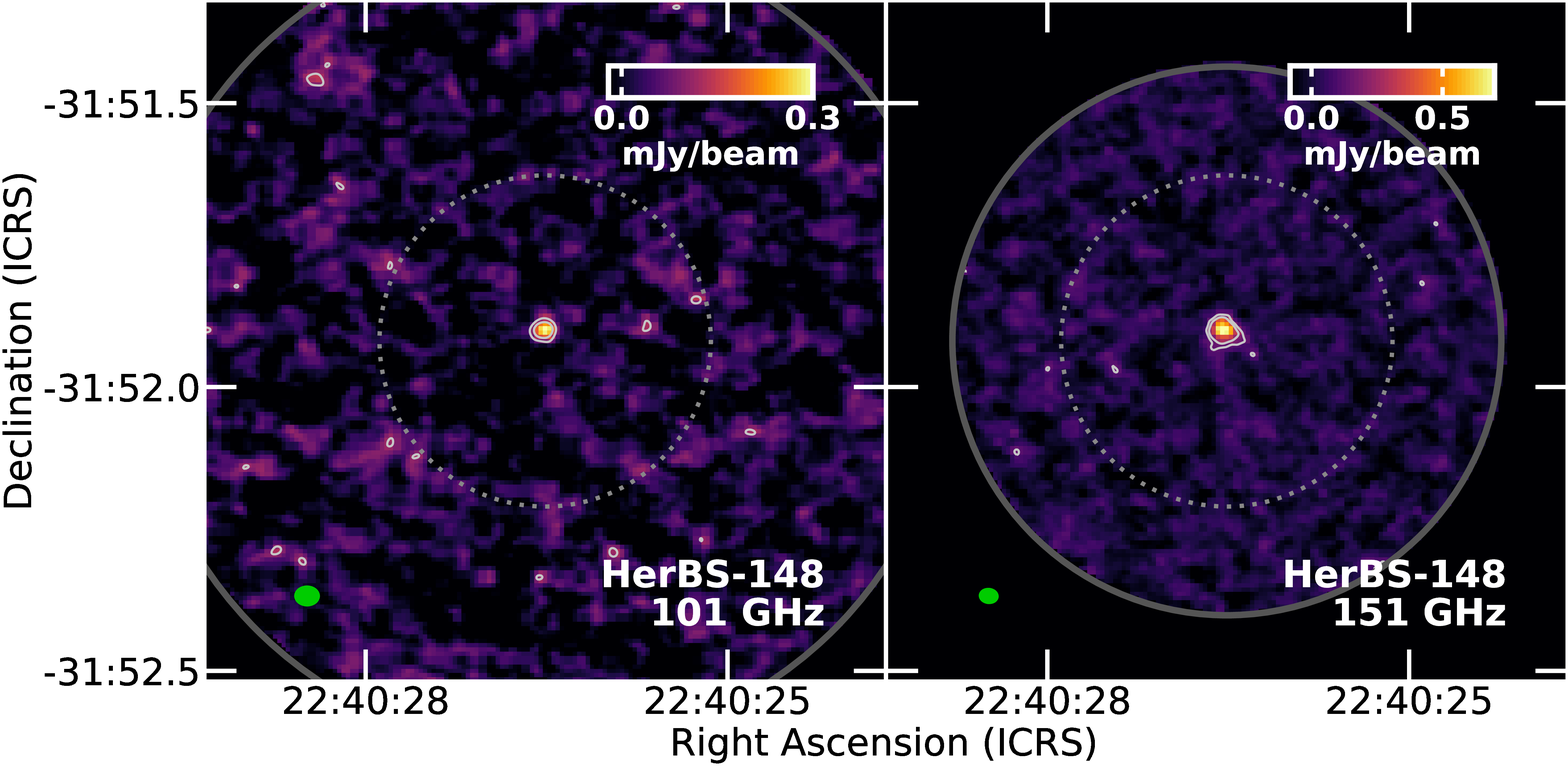}
			\ \ \ \ \ \
			\includegraphics[width=7cm]{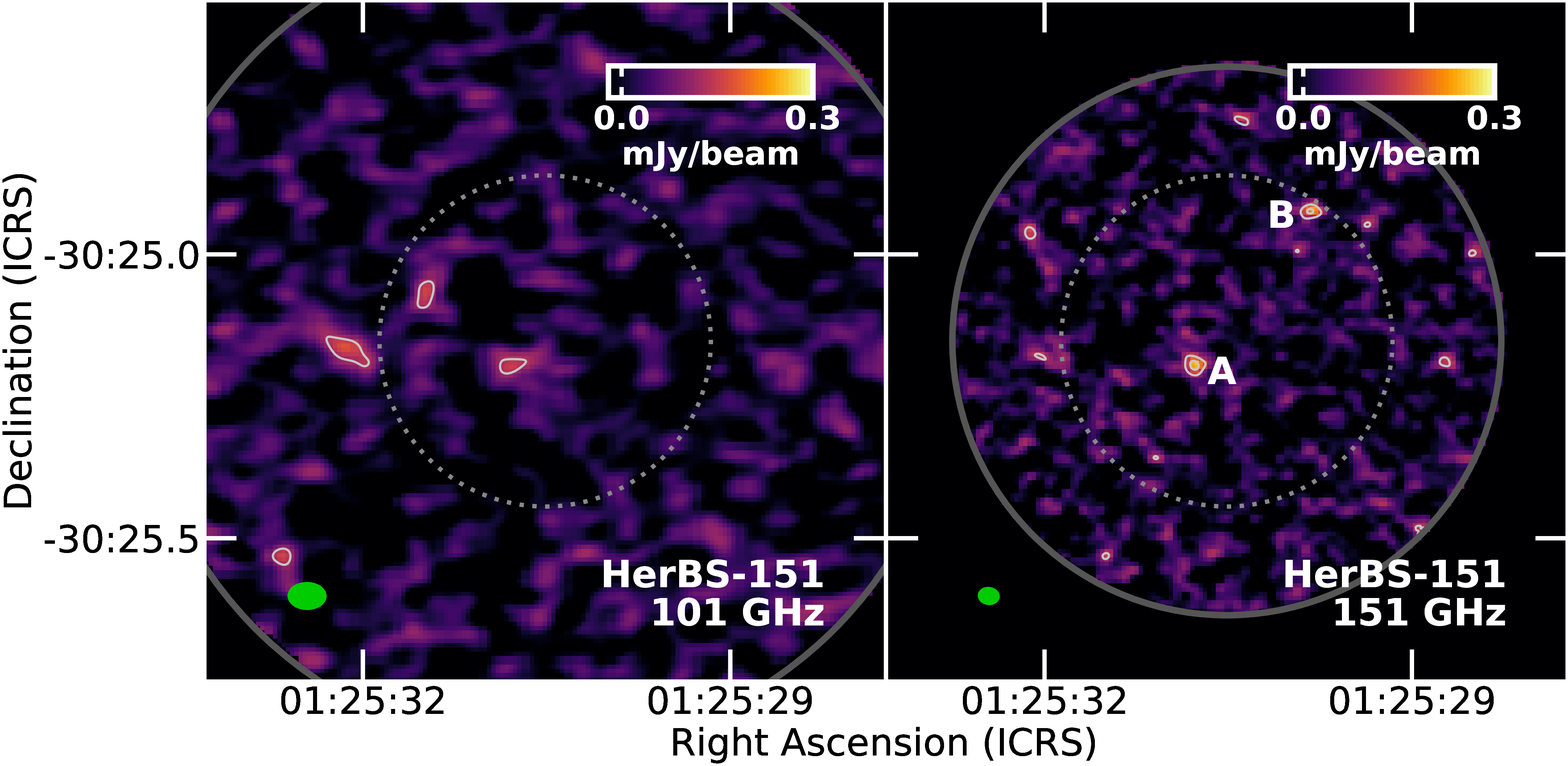}
			\vspace{0.5em}\\
			\includegraphics[width=7cm]{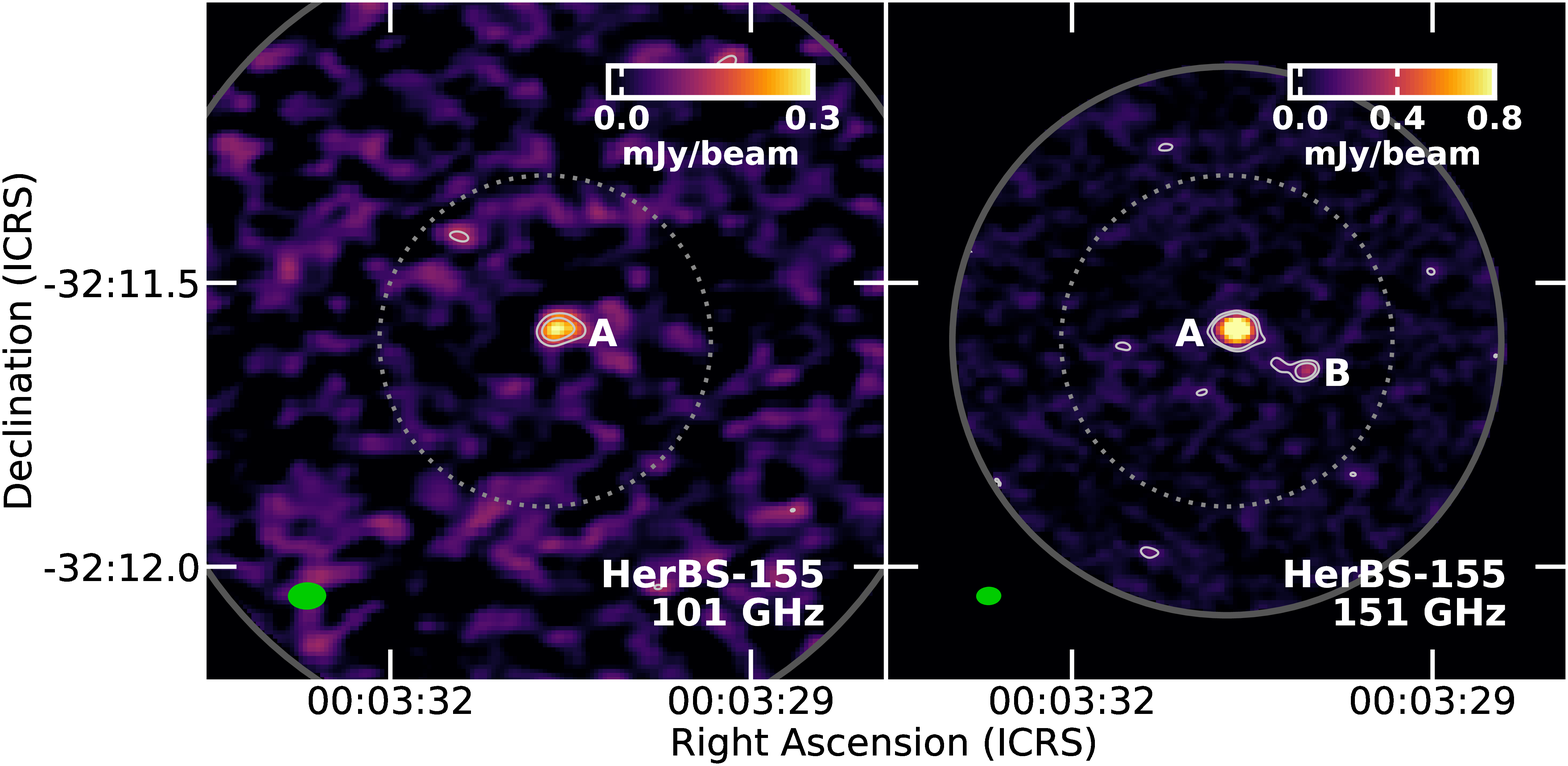}
			\ \ \ \ \ \
			\includegraphics[width=7cm]{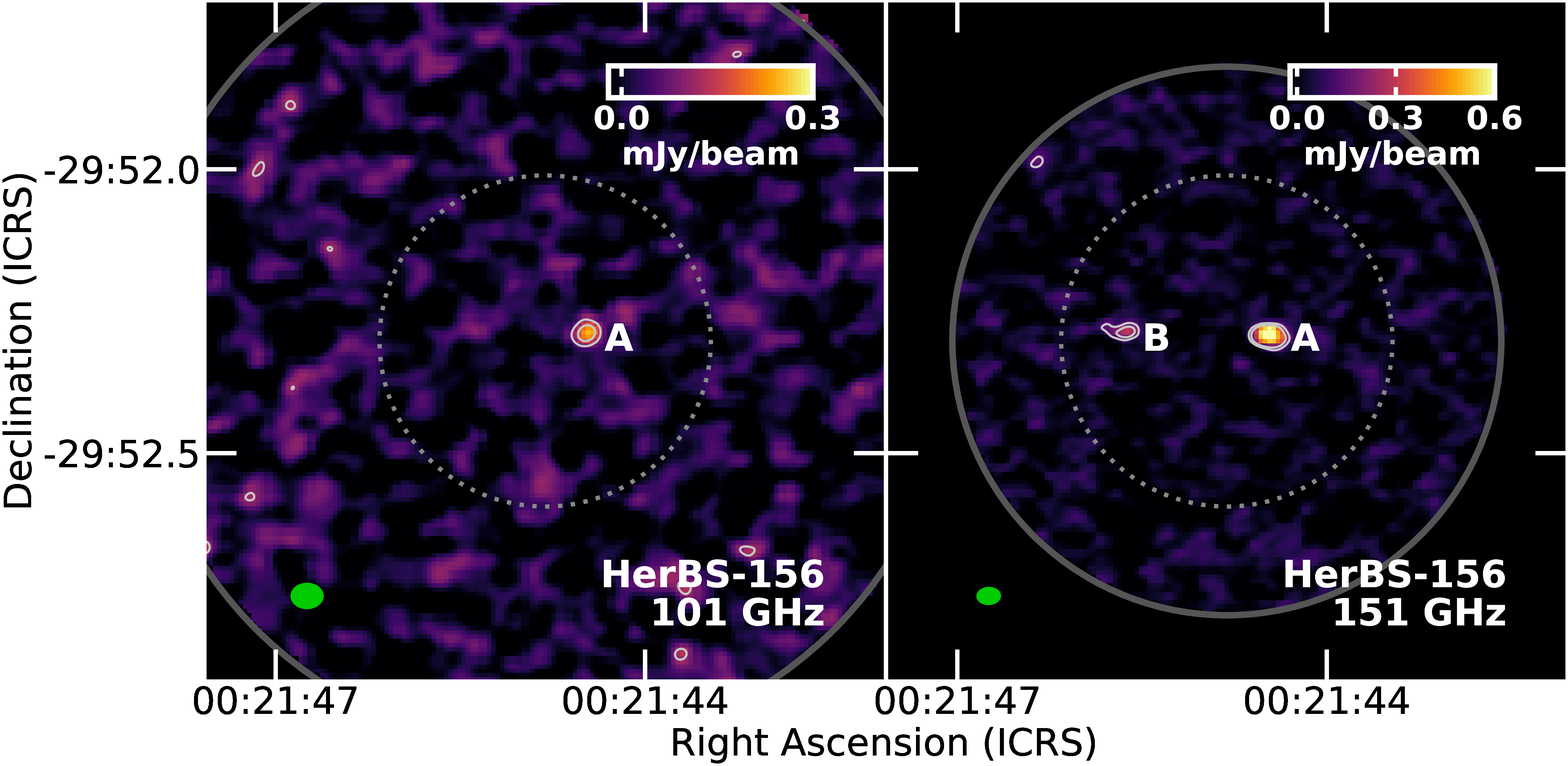}
			\vspace{0.5em}\\
			\includegraphics[width=7cm]{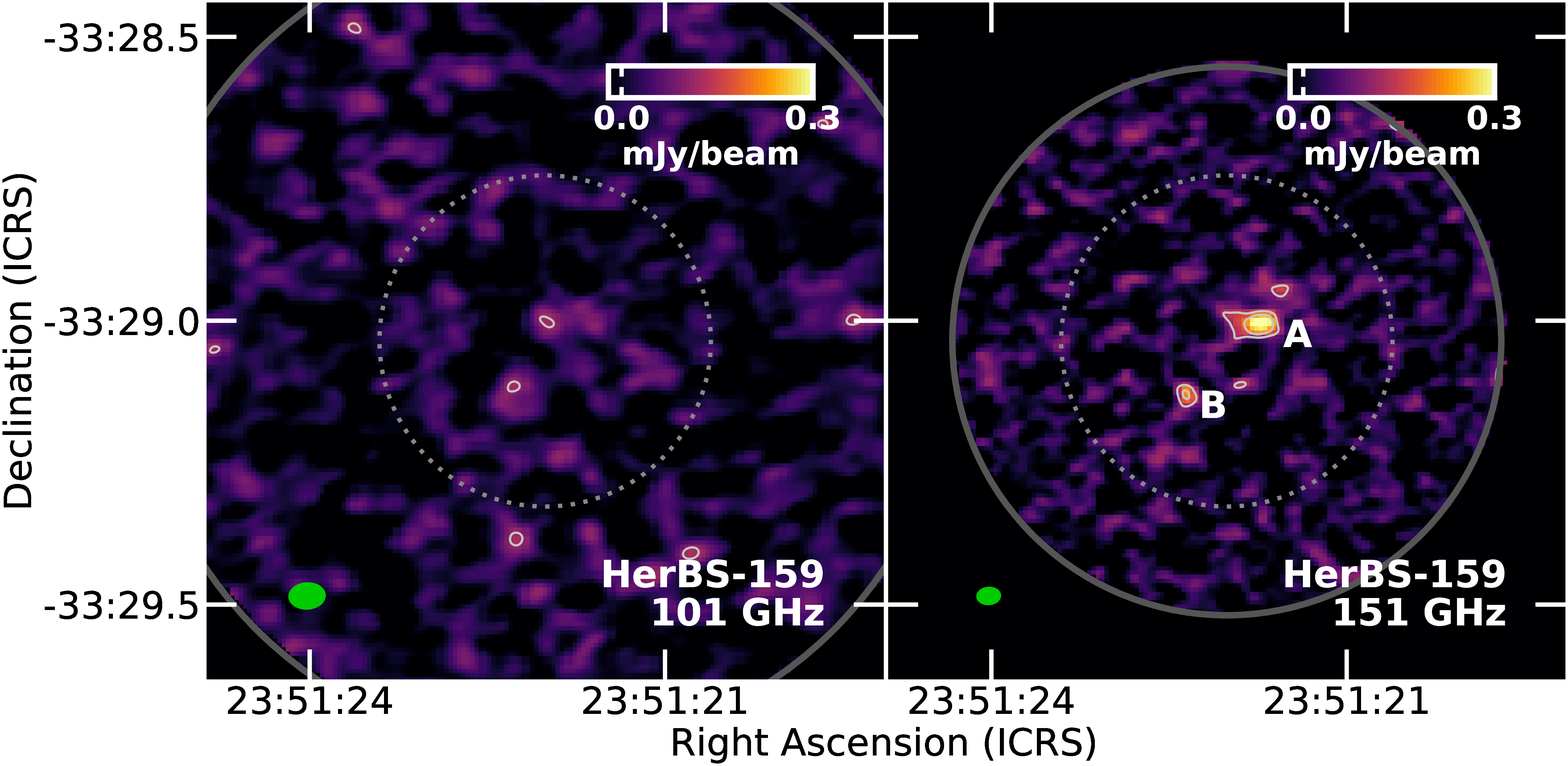}
			\ \ \ \ \ \
			\includegraphics[width=7cm]{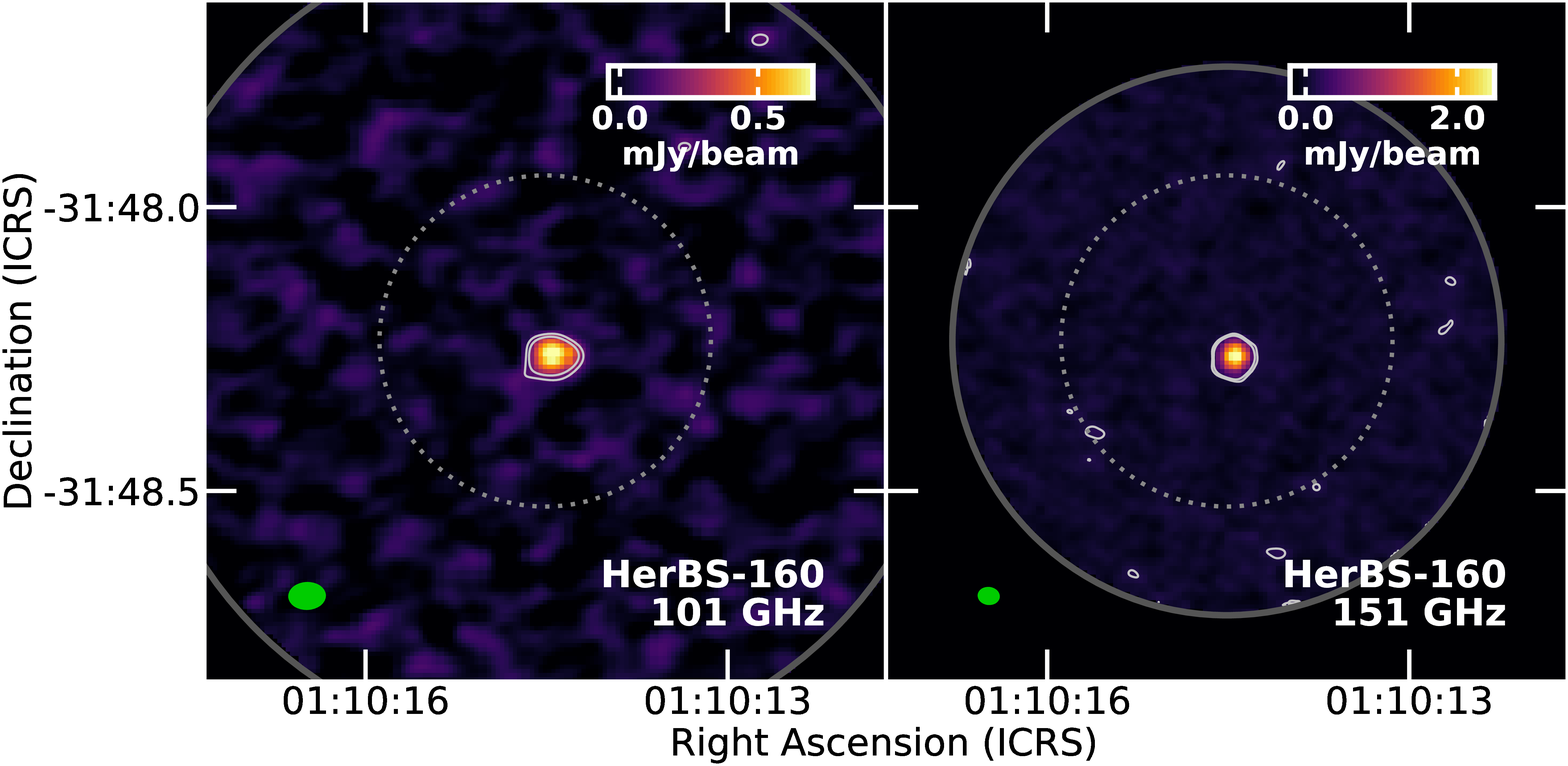}
			\vspace{0.5em}\\
			\includegraphics[width=7cm]{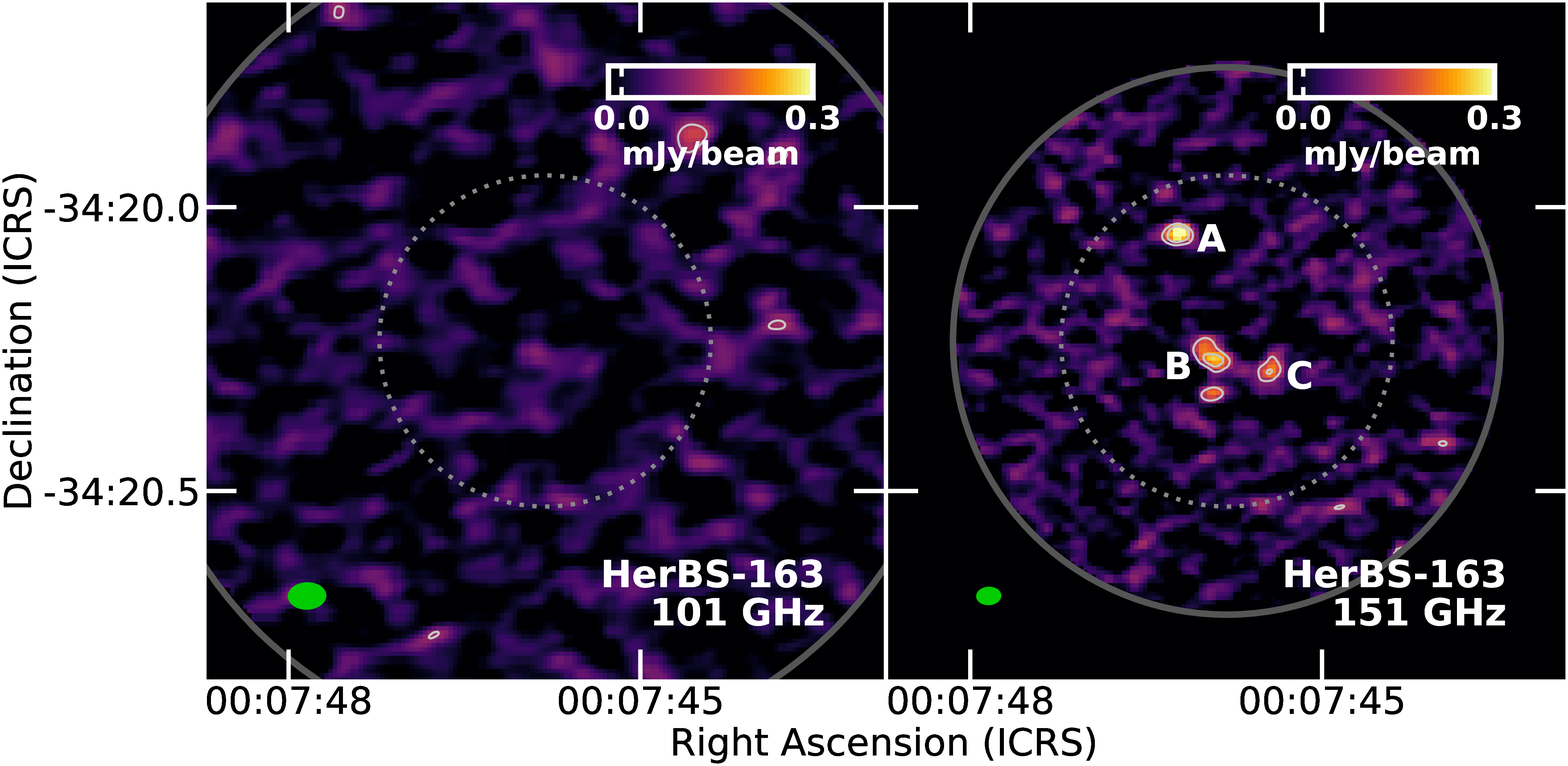}
			\ \ \ \ \ \
			\includegraphics[width=7cm]{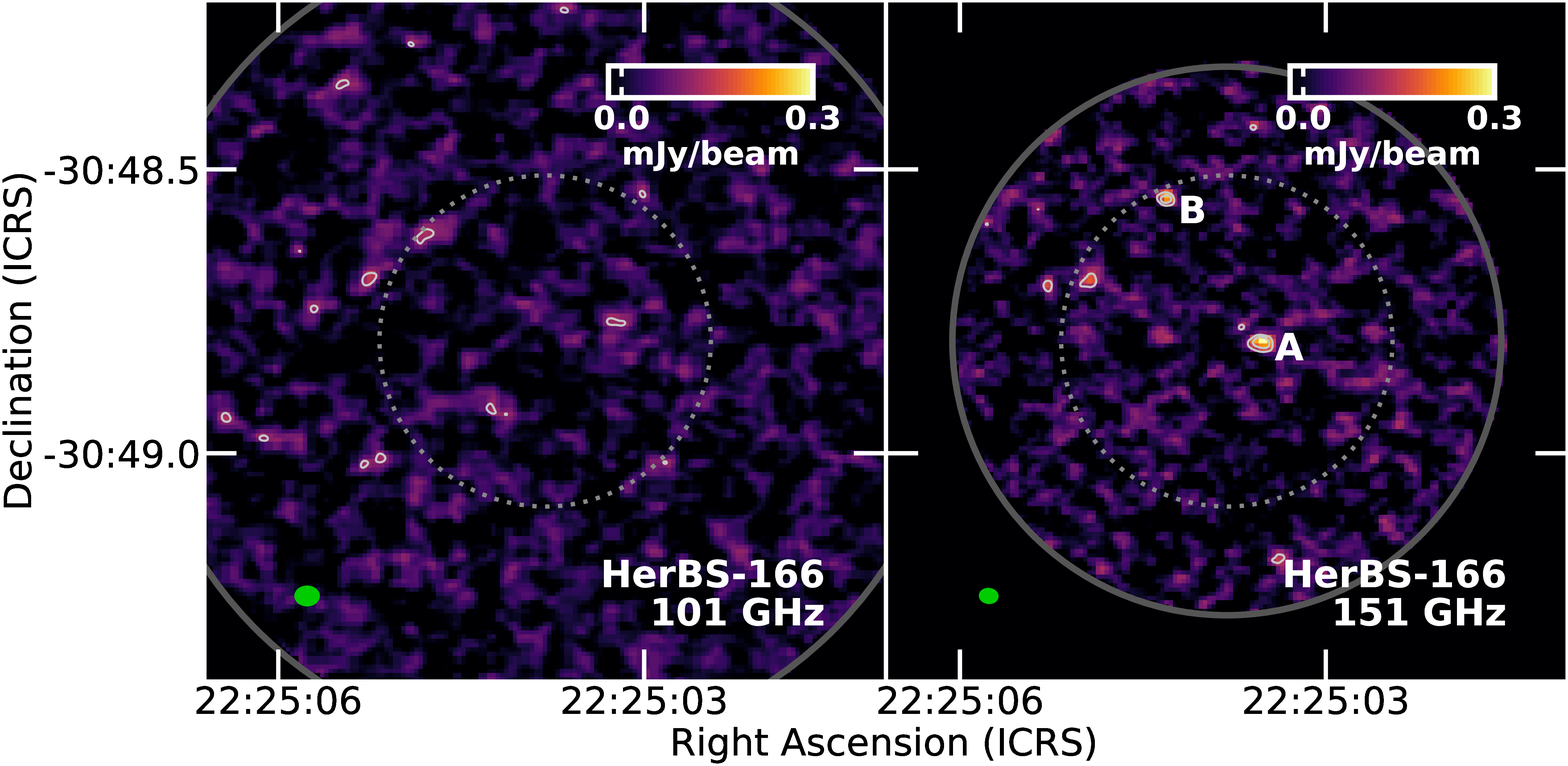}
			\vspace{0.5em}\\
			\includegraphics[width=7cm]{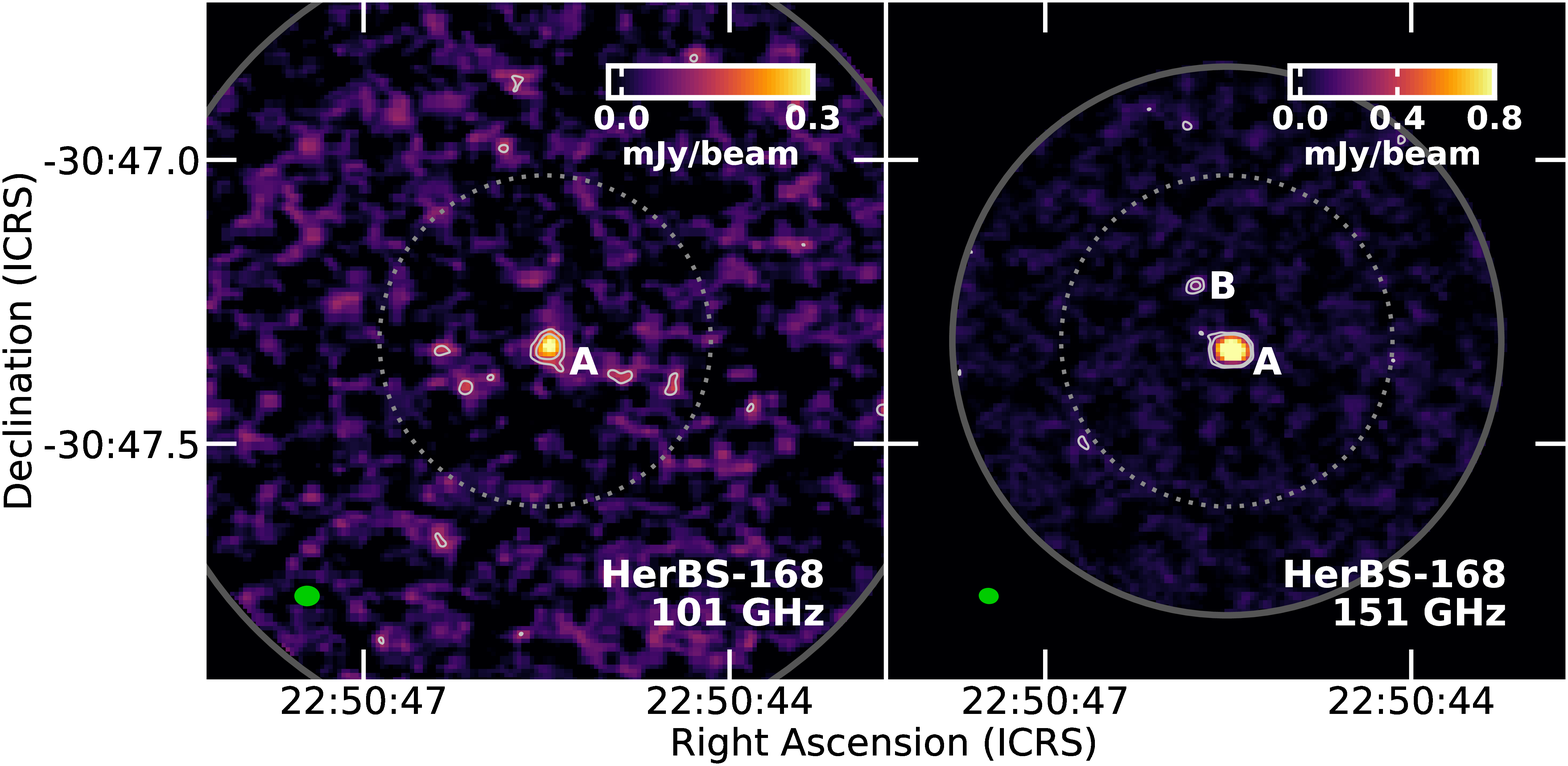}
			\ \ \ \ \ \
			\includegraphics[width=7cm]{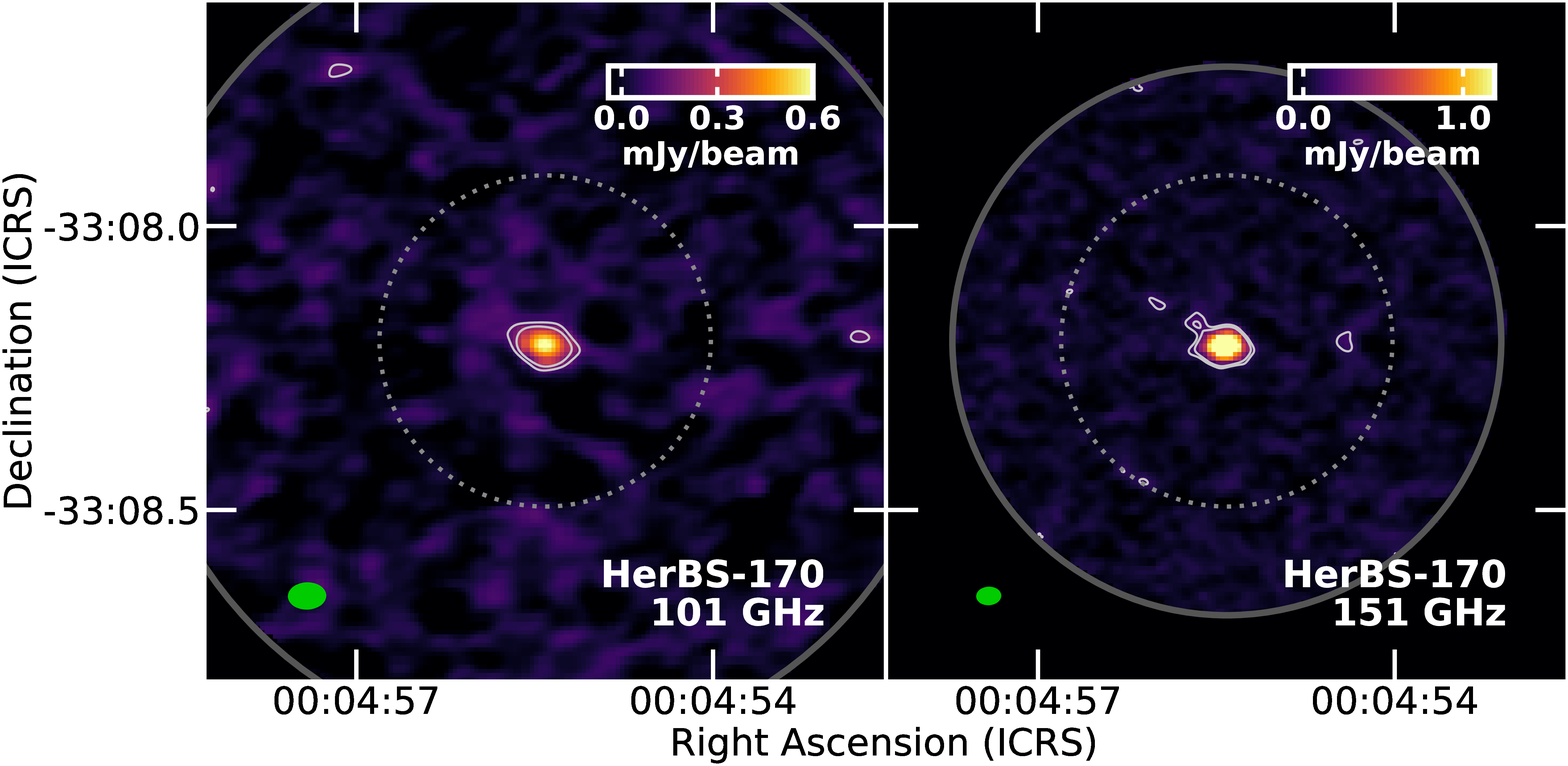}
		\end{center}
		\caption{Continued.}
	\end{figure*}
	
	\addtocounter{figure}{-1}
	
	\begin{figure*}
		\begin{center}
			\includegraphics[width=7cm]{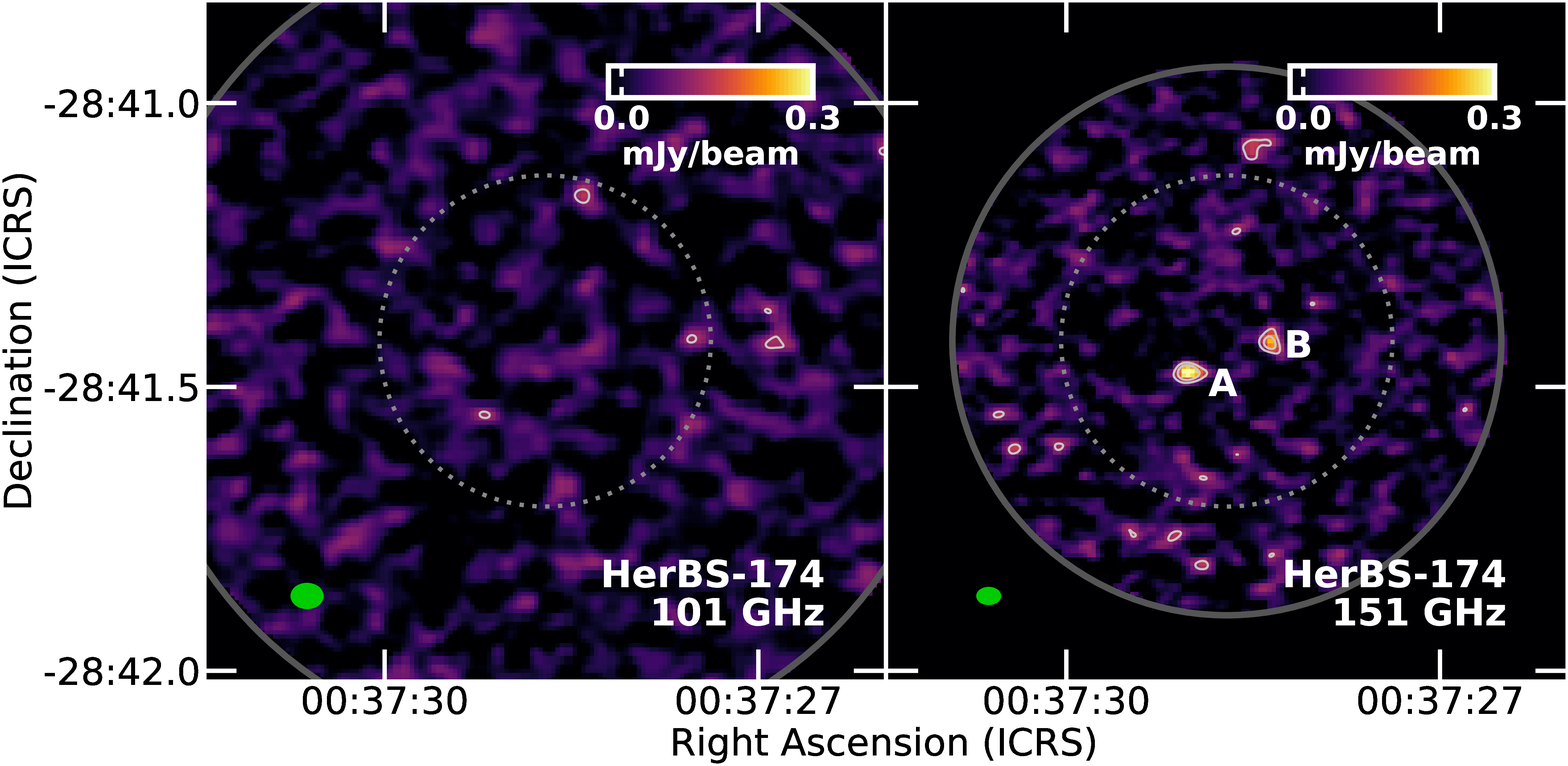}
			\ \ \ \ \ \
			\includegraphics[width=7cm]{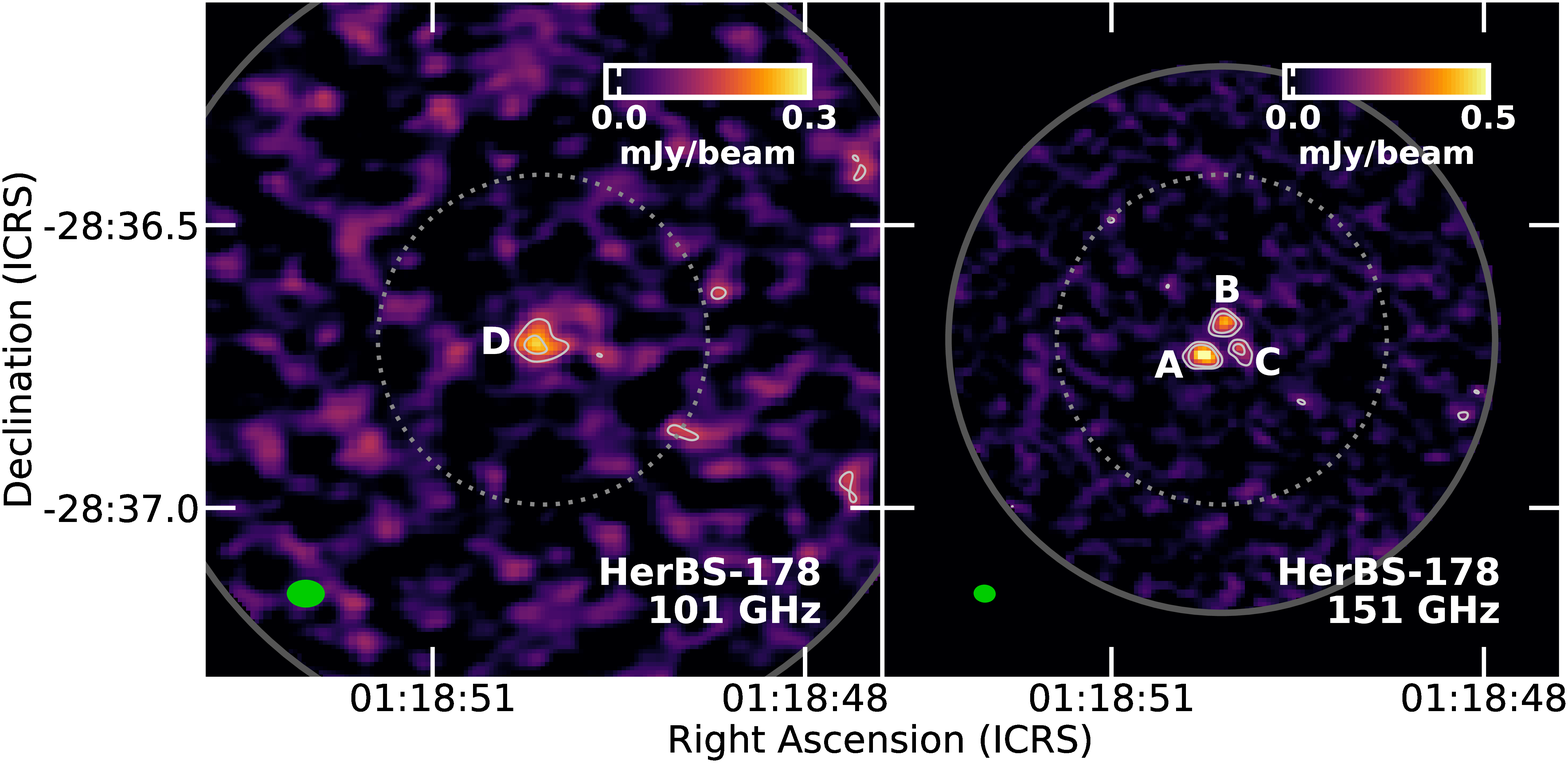}
			\vspace{0.5em}\\
			\includegraphics[width=7cm]{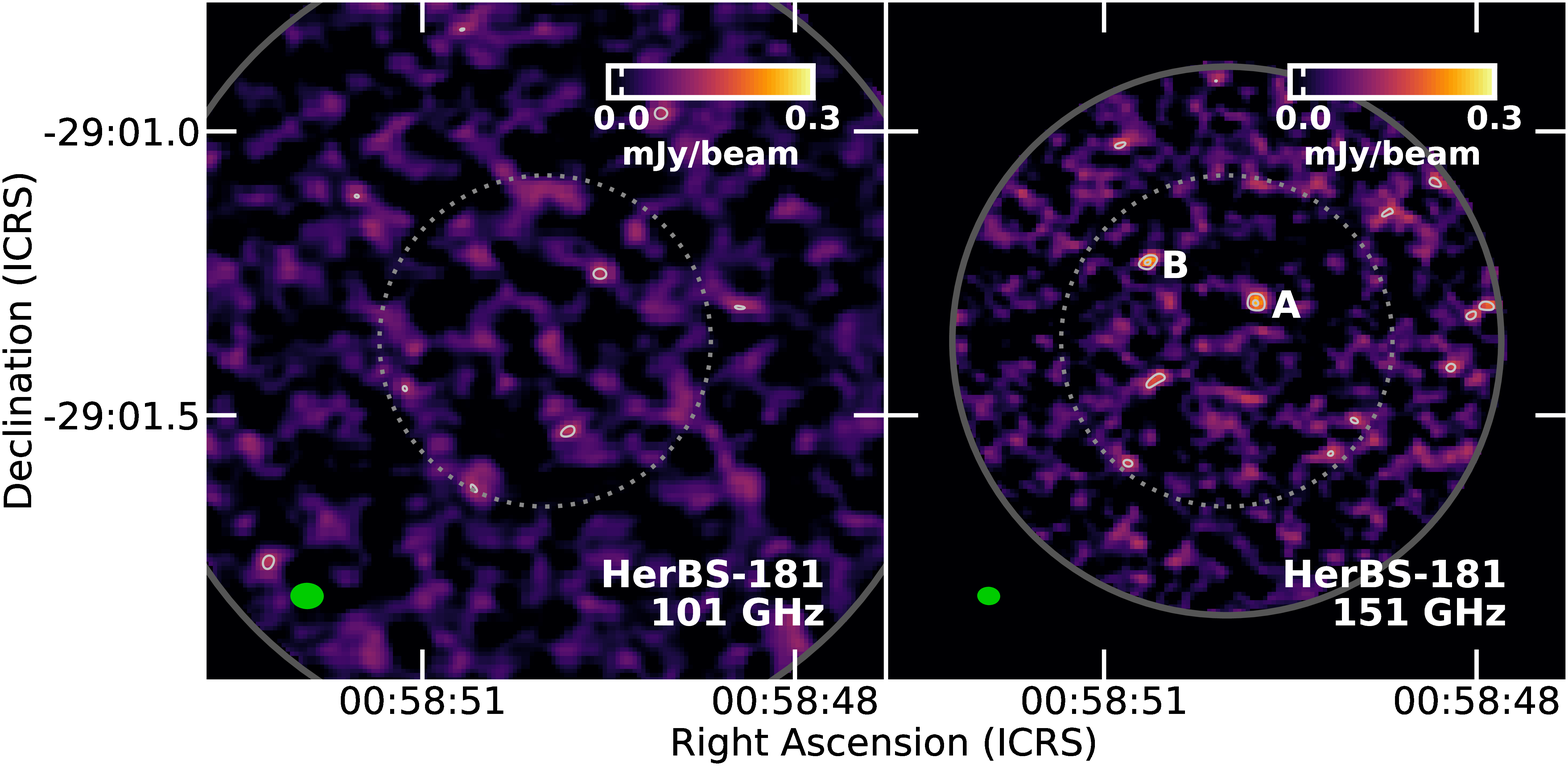}
			\ \ \ \ \ \
			\includegraphics[width=7cm]{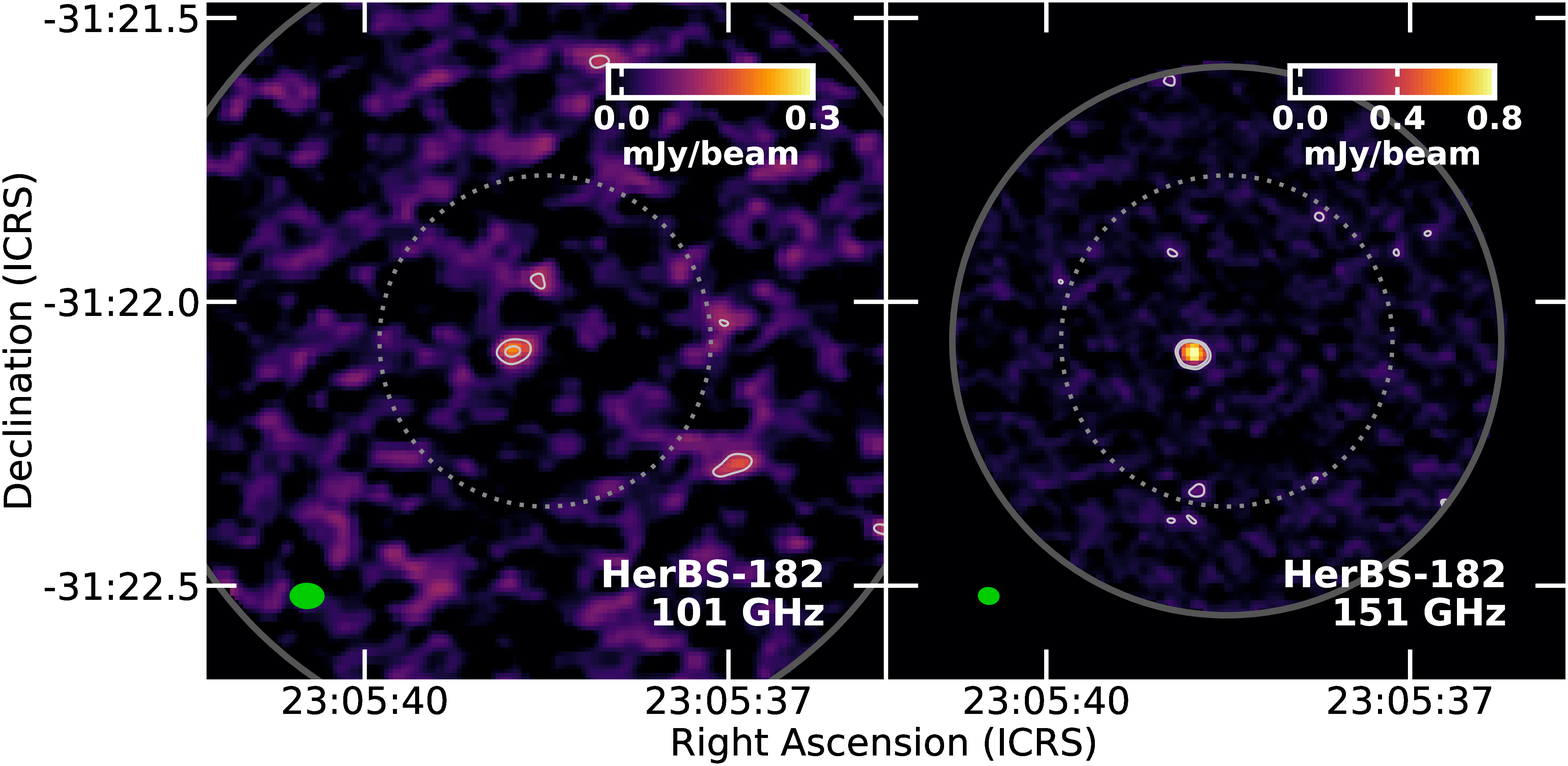}
			\vspace{0.5em}\\
			\includegraphics[width=7cm]{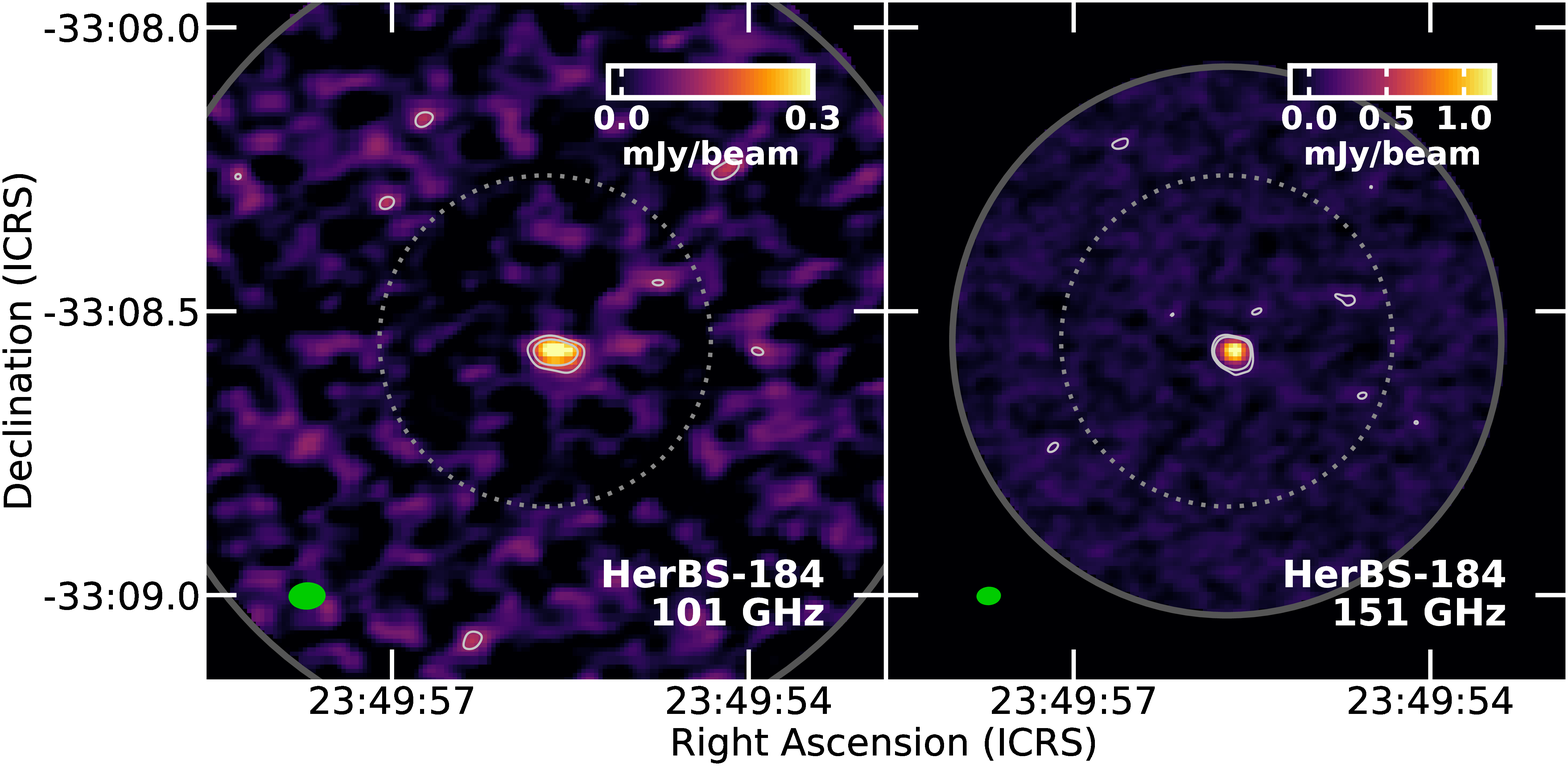}
			\ \ \ \ \ \
			\includegraphics[width=7cm]{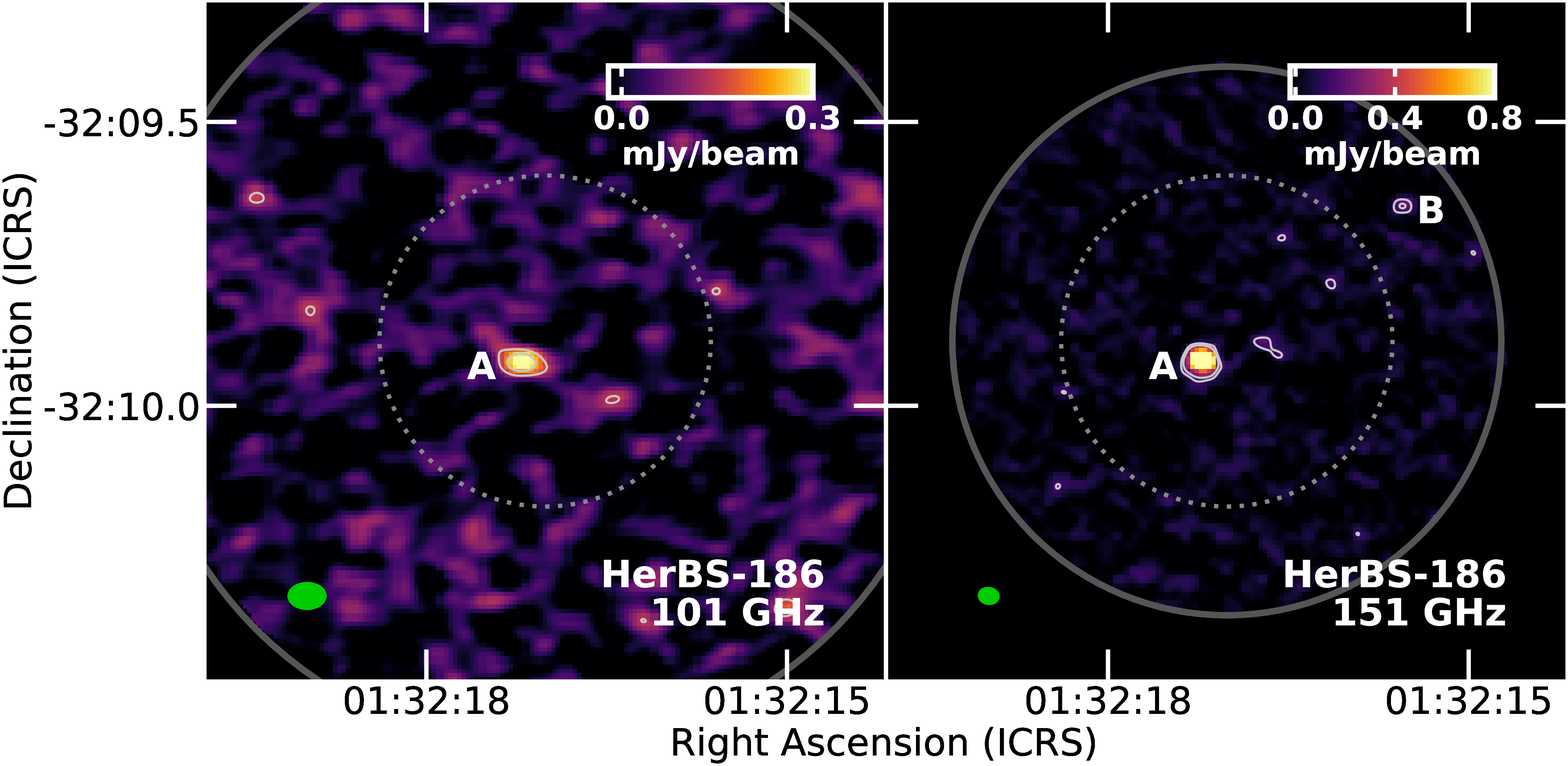}
			\vspace{0.5em}\\
			\includegraphics[width=7cm]{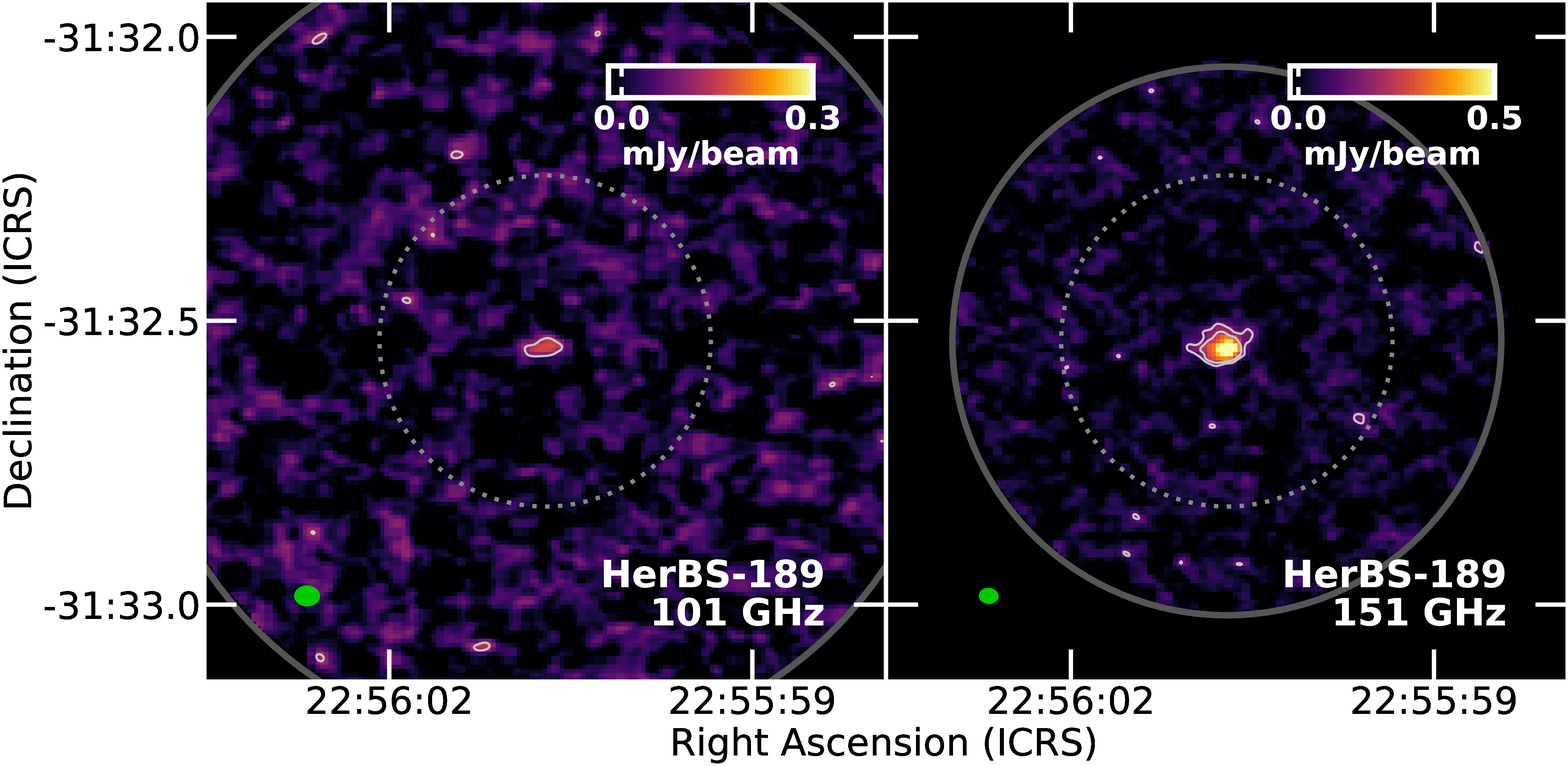}
			\ \ \ \ \ \
			\includegraphics[width=7cm]{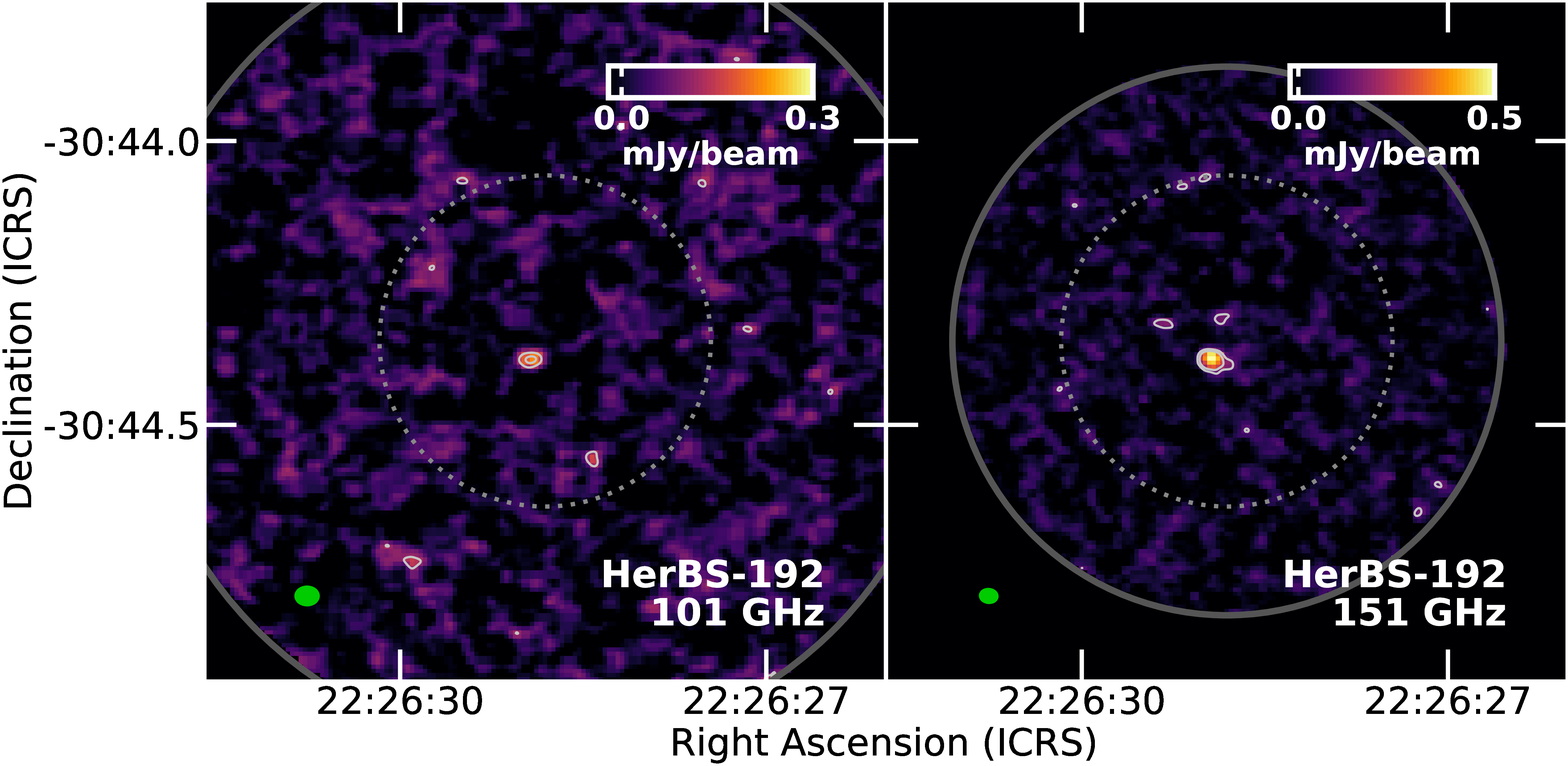}
			\vspace{0.5em}\\
			\includegraphics[width=7cm]{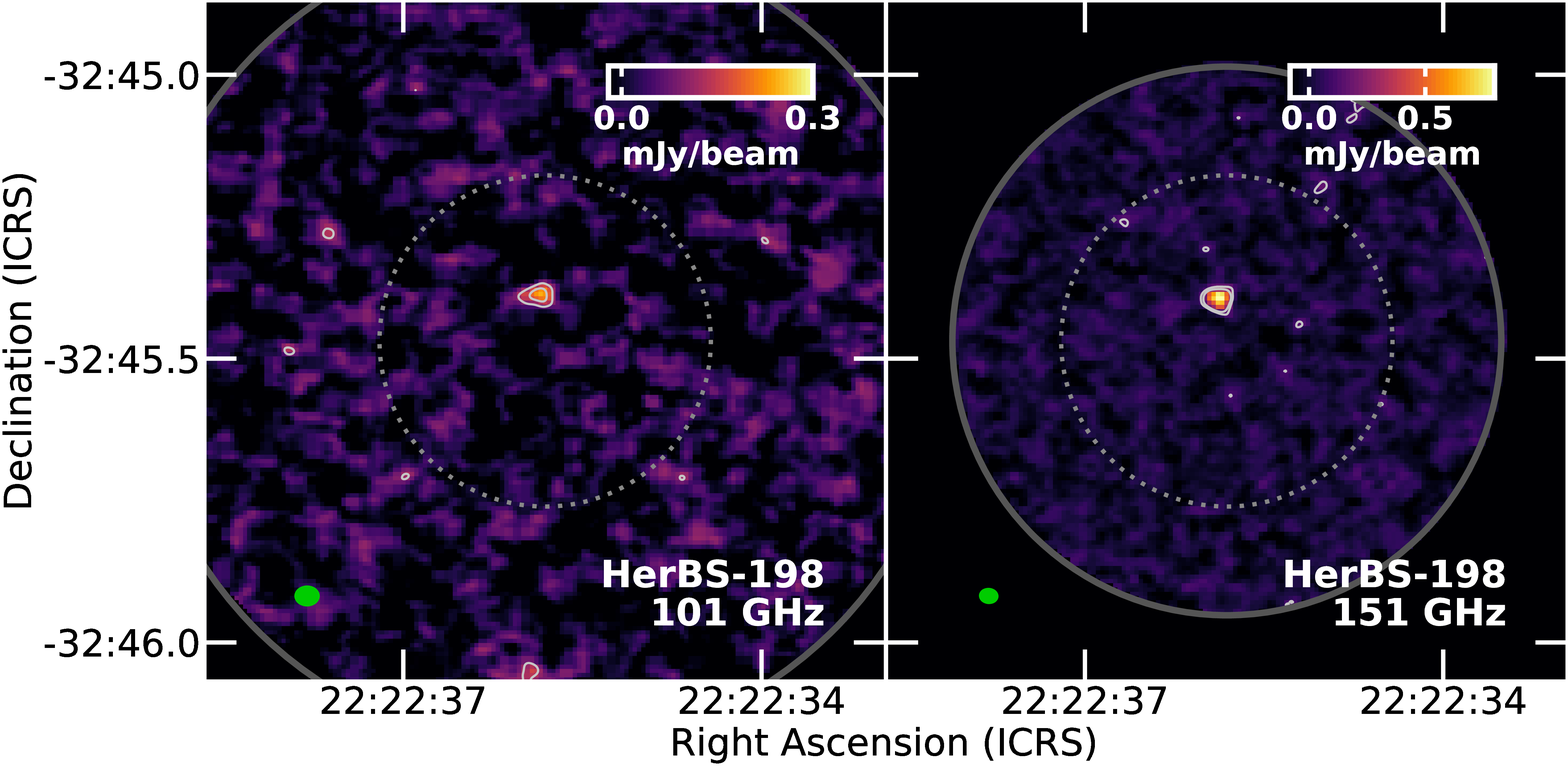}
			\ \ \ \ \ \
			\includegraphics[width=7cm]{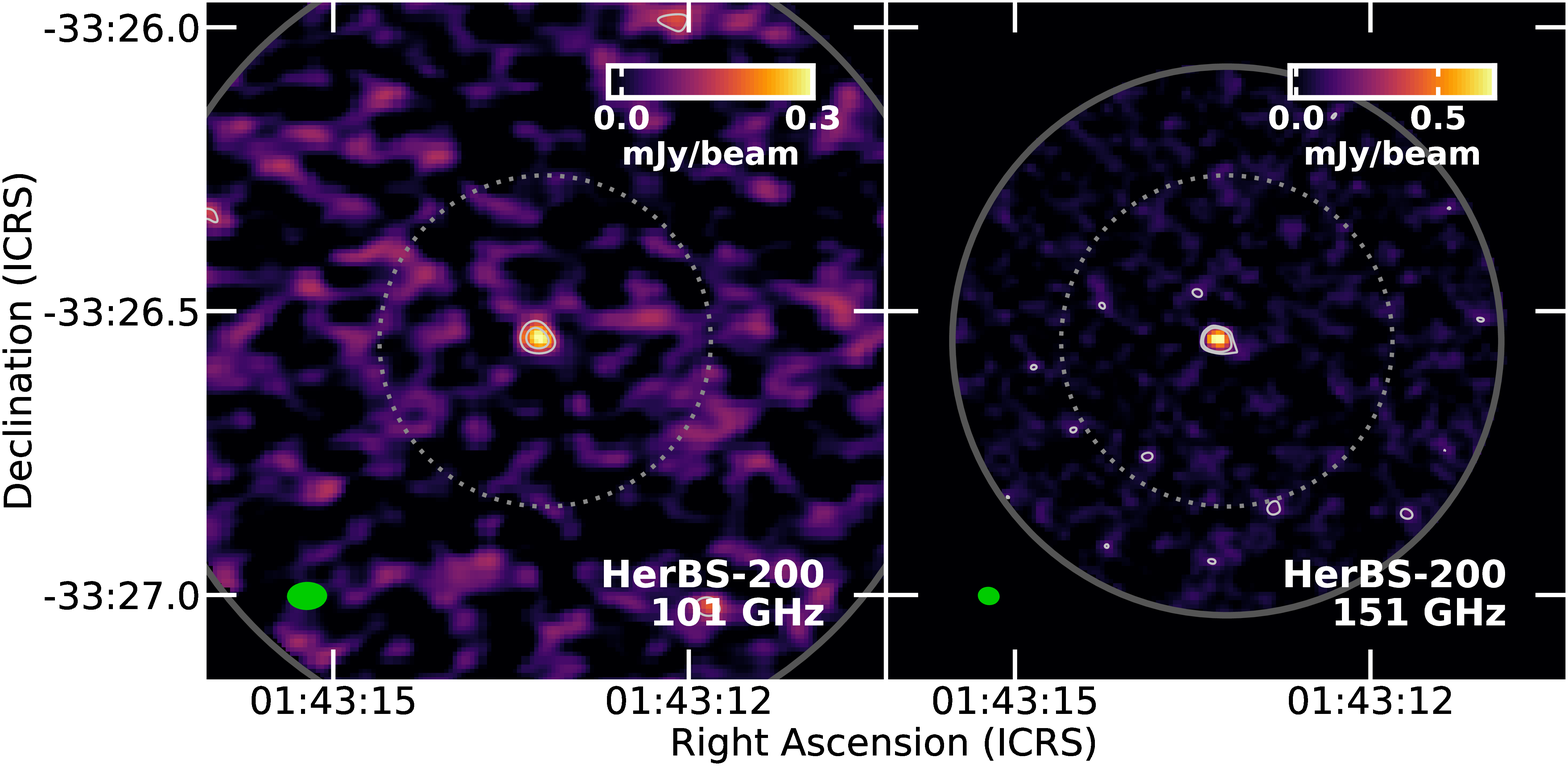}
			\vspace{0.5em}\\
			\includegraphics[width=7cm]{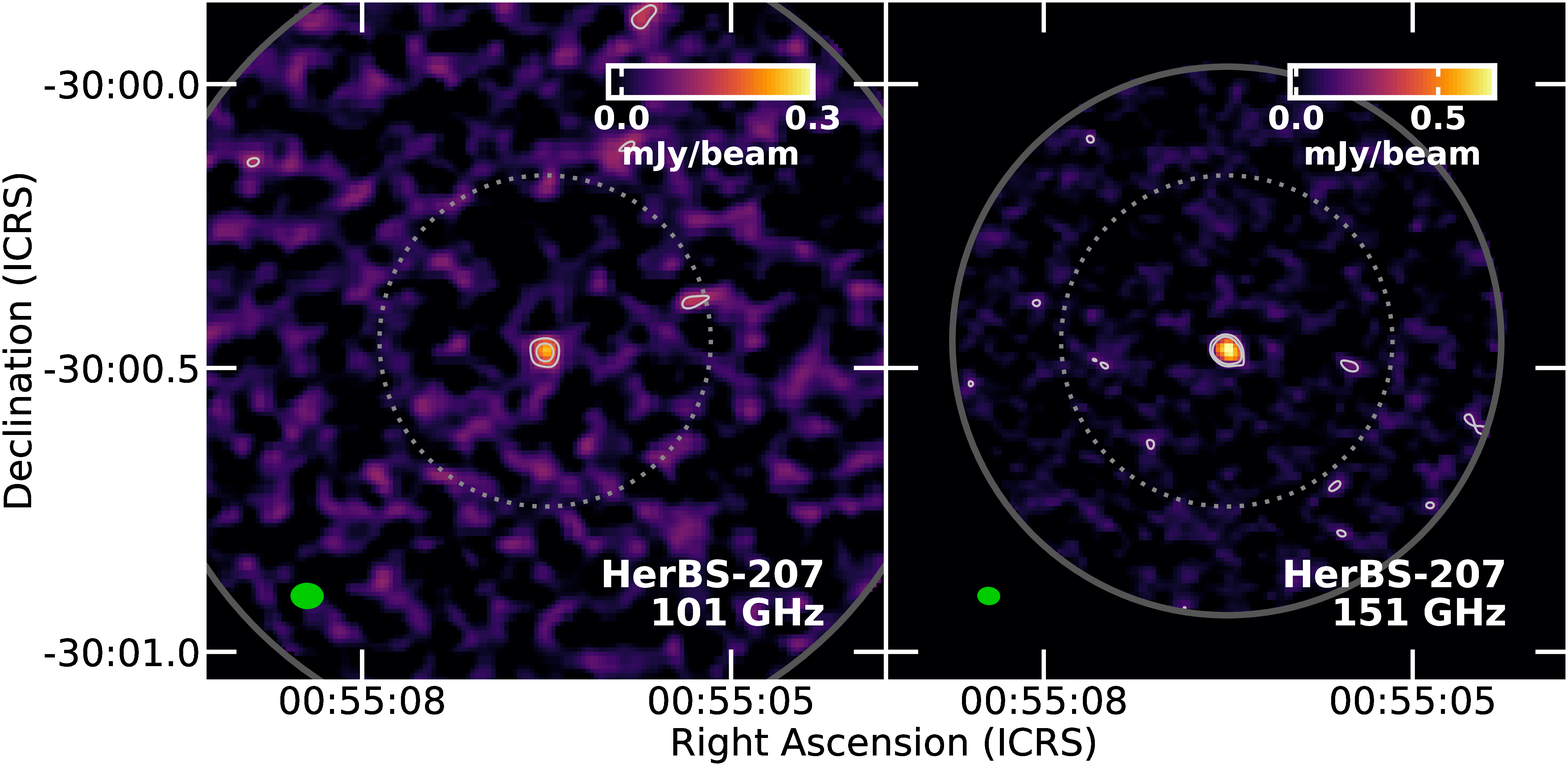}
			\ \ \ \ \ \
			\includegraphics[width=7cm]{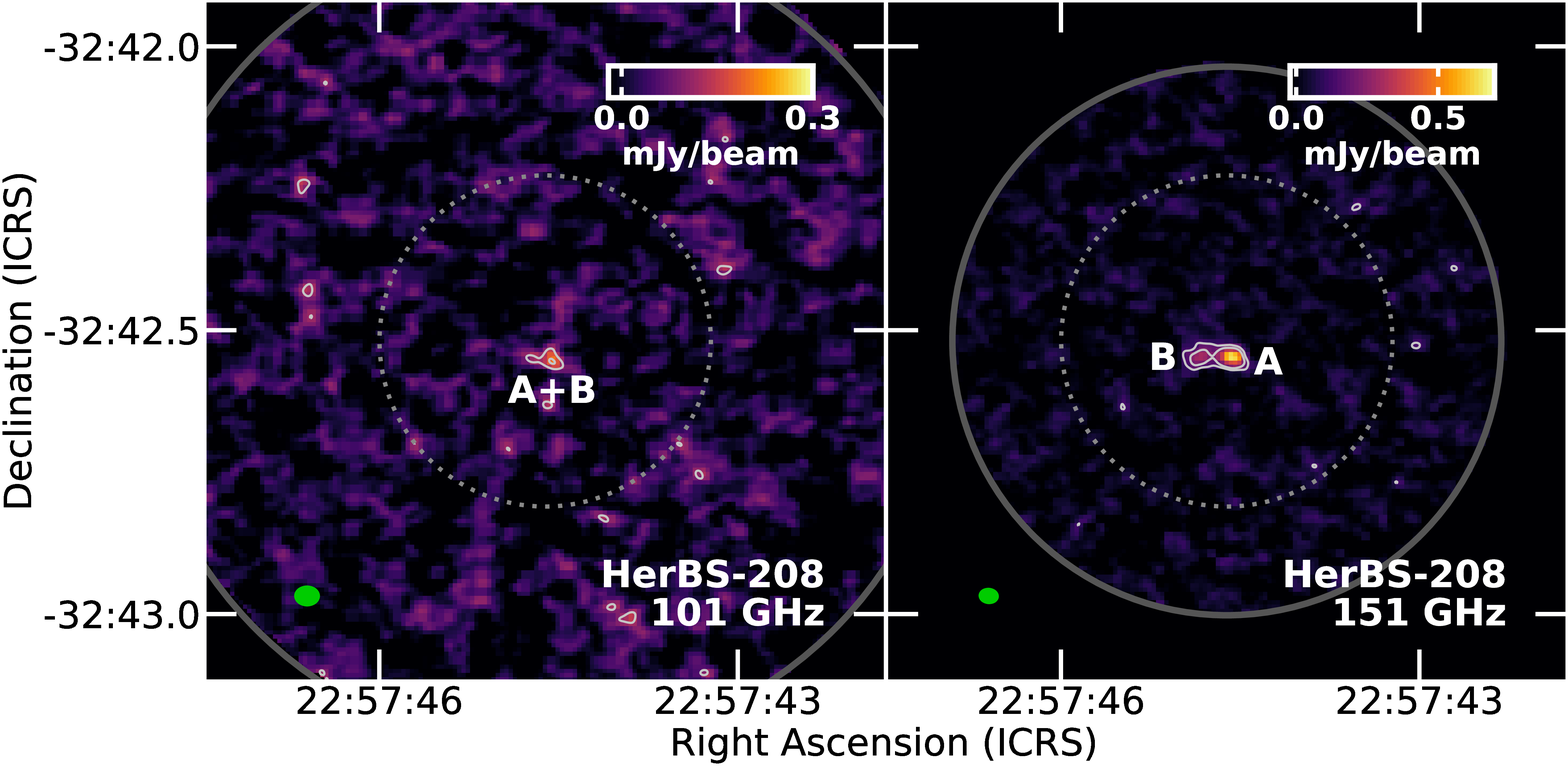}
		\end{center}
		\caption{Continued.}
	\end{figure*}
	
	\addtocounter{figure}{-1}
	
	\begin{figure*}
		\begin{center}
			\includegraphics[width=7cm]{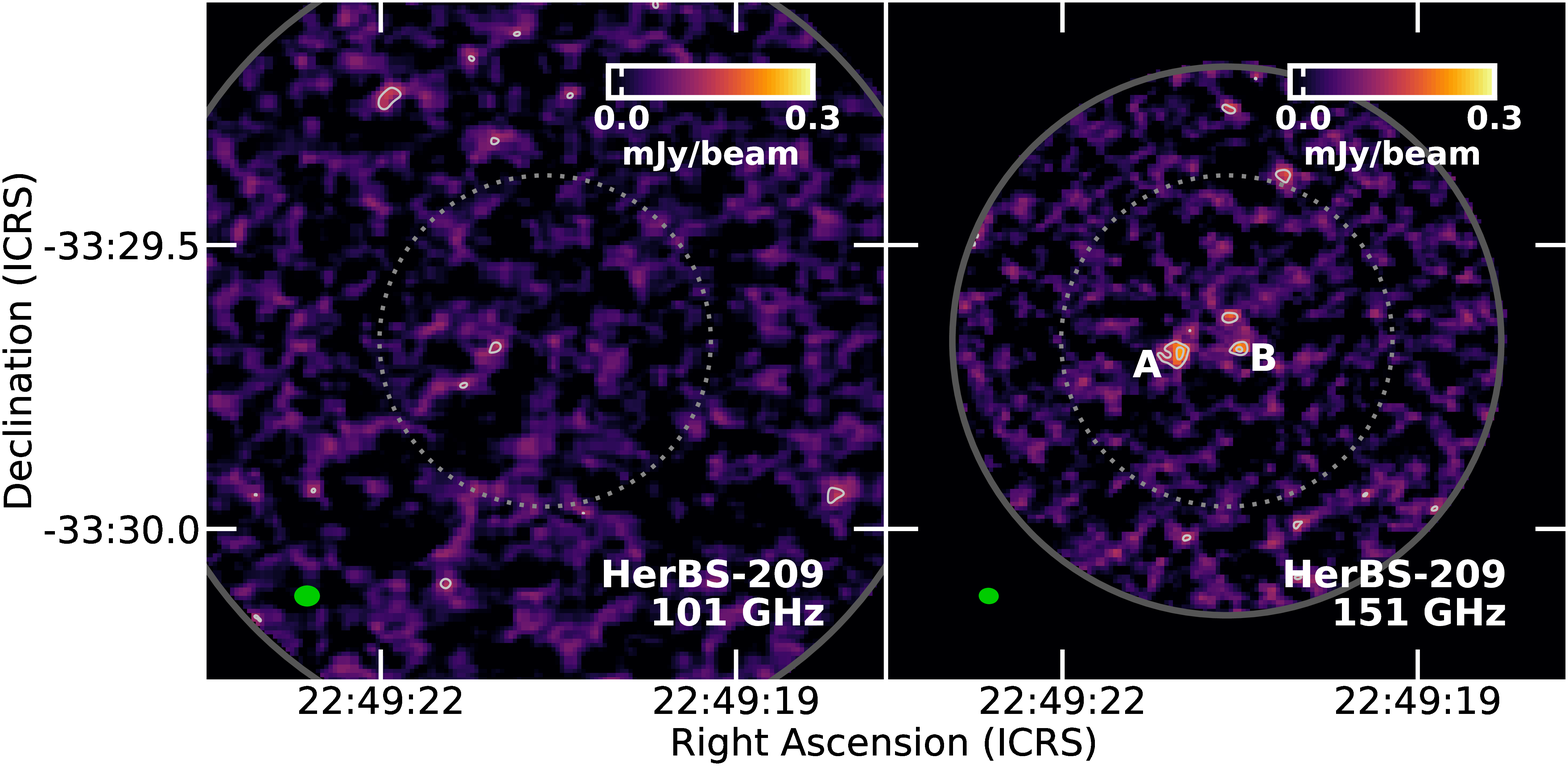}
		\end{center}
		\caption{Continued.}
	\end{figure*}

\bsp	
\label{lastpage}
\end{document}